\documentclass[traditabstract,longauth]{aa}

\usepackage[nonamebreak]{natbib}
\usepackage[stable]{footmisc}
\usepackage{txfonts}
\usepackage{placeins}
\usepackage{natbib}
\bibpunct{(}{)}{;}{a}{}{,} 
\usepackage{graphicx}
\usepackage{epstopdf}
\usepackage{ifthen}
\usepackage[breaklinks, colorlinks, citecolor=blue]{hyperref}
\usepackage{subfigure}
\usepackage[table,usenames,dvipsnames]{xcolor}

\def\WMAP{\textit{WMAP}}

\newcommand{\ncib}{\gamma^{\rm CIB}}
\newcommand{\camspec}{{\tt CamSpec}}
\newcommand{\plik}{{\tt Plik}}

\newcommand{\WP}{WP}
\newcommand{\highL}{highL}
\newcommand{\lensing}{lensing}
\newcommand{\planckonly}{\planck}
\newcommand{\plancklensing}{\planck+\lensing}
\newcommand{\HighL}{\highL}

\newcommand{\As}{A_{\rm s}}
\newcommand{\At}{A_{\rm t}}
\newcommand{\nt}{n_{\rm t}}
\newcommand{\ns}{n_{\rm s}}
\newcommand{\lcdm}{$\Lambda$CDM}
\newcommand{\rpivot}{r_{0.05}}
\newcommand{\rzerotwo}{r_{0.002}}
\newcommand{\Alens}{A_{\rm L}}
\newcommand{\Aphiphi}{A_{\rm L}^{\phi\phi}}

\newcommand{\nrun}{d \ns / d\ln k}
\newcommand{\zre}{z_{\text{re}}}
\newcommand{\yhe}{Y_{\text{P}}}
\newcommand{\zeq}{z_{\text{eq}}}
\newcommand{\nnu}{N_{\rm eff}}
\newcommand{\neff}{N_{\rm eff}}
\newcommand{\mnu}{\sum m_\nu}
\newcommand{\sumnu}{\sum m_\nu}
\newcommand{\mnusterile}{m_{\nu,\, \mathrm{sterile}}^{\mathrm{eff}}}
\newcommand{\meffsterile}{\mnusterile}

\newcommand{\Tsterile}{T_{\mathrm{s}}}

\newcommand{\zdrag}{z_{\rm drag}}
\newcommand{\rdrag}{r_{\rm drag}}
\newcommand{\zstar}{z_{\ast}}
\newcommand{\rstar}{r_{\ast}}
\newcommand{\rs}{r_{\rm s}}
\newcommand{\thetastar}{\theta_{\ast}}
\newcommand{\DAstar}{D_{\rm A}}

\newcommand{\Ailensbestfit}{A_i^{\mbox{\scriptsize{theory}}}}
\newcommand{\Alensbestfit}{A^{\mbox{\scriptsize{theory}}}}

\providecommand{\Planck}{\textit{Planck}}
\providecommand{\planck}{\Planck}

\providecommand{\gea}{\gtrsim}

\providecommand{\agt}{\gea}
\providecommand{\text}[1]{\rm{#1}}

\newcommand{\Mpc}{\text{Mpc}}

\providecommand{\muK}{\mu\rm{K}}

\newcommand{\muKsq}{(\mu\rm{K})^2}
\newcommand{\lmax}{l_{\text{max}}}
\newcommand{\lmin}{l_{\text{min}}}

\newcommand{\eV}{\,\text{eV}}

\providecommand{\Omk}{\Omega_K}
\providecommand{\Oml}{\Omega_{\Lambda}}
\providecommand{\Omtot}{\Omega_{\mathrm{tot}}}
\providecommand{\Omb}{\Omega_{\mathrm{b}}}
\providecommand{\Omc}{\Omega_{\mathrm{c}}}
\providecommand{\Omm}{\Omega_{\mathrm{m}}}
\providecommand{\omb}{\omega_{\mathrm{b}}}
\providecommand{\omc}{\omega_{\mathrm{c}}}
\providecommand{\omm}{\omega_{\mathrm{m}}}

\providecommand{\CAMB}{{\tt camb}}
\providecommand{\COSMOMC}{{\tt CosmoMC}}
\providecommand{\CMBFAST}{{\tt cmbfast}}
\providecommand{\COSMICS}{{\tt Cosmics}}
\providecommand{\CLASS}{{\tt class}}

\providecommand{\LCDM}{{$\rm{\Lambda CDM}$}}
\providecommand{\COSMOREC}{{\tt CosmoRec}}
\providecommand{\HYREC}{{\tt HyRec}}
\providecommand{\RECFAST}{{\tt recfast}}
\providecommand{\HALOFIT}{{\tt halofit}}

\newcommand{\begm}{\begin{pmatrix}}
\newcommand{\enm}{\end{pmatrix}}

\newcommand\ba{\begin{eqnarray}}
\newcommand\ea{\end{eqnarray}}
\newcommand\bea{\begin{eqnarray}}
\newcommand\eea{\end{eqnarray}}

\newcommand\be{\begin{equation}}
\newcommand\ee{\end{equation}}

\newcommand{\boldvec}[1]{{\mbox{\boldmath{$#1$}}}}

\newcommand{\vM}{\boldvec{M}}

\newcommand{\vx}{\boldvec{x}}

\newcommand{\clp}{\mathcal{P}}

\newcommand{\clr}{\mathcal{R}}

\def\pmb#1{\setbox0=\hbox{#1}%
    \kern-.025em\copy0\kern-\wd0
    \kern.05em\copy0\kern-\wd0
    \kern-.025em\raise.0433em\box0}
\def\ltsima{$\; \buildrel < \over \sim \;$}
\def\gtsima{$\; \buildrel > \over \sim \;$}
\def\simlt{\lower.5ex\hbox{\ltsima}}
\def\simgt{\lower.5ex\hbox{\gtsima}}

\def\p2Y{\;_2Y}
\def\m2Y{\;_{-2}Y}
\newcounter{parentequation}
\def\beglet{
  \addtocounter{equation}{1}%
  \setcounter{parentequation}{\value{equation}}%
  \setcounter{equation}{0}%
  \def\theequation{\arabic{parentequation}\alph{equation}}%
  \ignorespaces
}
\def\endlet{
  \setcounter{equation}{\value{parentequation}}%
  \def\theequation{\arabic{equation}}%
}

\def\setsymbol#1#2{\expandafter\def\csname #1\endcsname{#2}}
\def\getsymbol#1{\csname #1\endcsname}

\def\Planck{\textit{Planck}}

\def\all2013resultspapers{\nocite{planck2013-p01, planck2013-p02, planck2013-p02a, planck2013-p02d, planck2013-p02b, planck2013-p03, planck2013-p03c, planck2013-p03f, planck2013-p03d, planck2013-p03e, planck2013-p01a, planck2013-p06, planck2013-p03a, planck2013-pip88, planck2013-p08, planck2013-p11, planck2013-p12, planck2013-p13, planck2013-p14, planck2013-p15, planck2013-p05b, planck2013-p17, planck2013-p09, planck2013-p09a, planck2013-p20, planck2013-p19, planck2013-pipaberration, planck2013-p05, planck2013-p05a, planck2013-pip56, planck2013-p06b}}

\newbox\tablebox    \newdimen\tablewidth
\def\leaderfil{\leaders\hbox to 5pt{\hss.\hss}\hfil}
\def\endPlancktable{\tablewidth=\columnwidth 
    $$\hss\copy\tablebox\hss$$
    \vskip-\lastskip\vskip -2pt}
\def\endPlancktablewide{\tablewidth=\textwidth 
    $$\hss\copy\tablebox\hss$$
    \vskip-\lastskip\vskip -2pt}
\def\tablenote#1 #2\par{\begingroup \parindent=0.8em
    \abovedisplayshortskip=0pt\belowdisplayshortskip=0pt
    \noindent
    $$\hss\vbox{\hsize\tablewidth \hangindent=\parindent \hangafter=1 \noindent
    \hbox to \parindent{$^#1$\hss}\strut#2\strut\par}\hss$$
    \endgroup}
\def\doubleline{\vskip 3pt\hrule \vskip 1.5pt \hrule \vskip 5pt}

\def\L2{\ifmmode L_2\else $L_2$\fi}

\def\DeltaT{\ifmmode \Delta T\else $\Delta T$\fi}
\def\deltat{\ifmmode \Delta t\else $\Delta t$\fi}
\def\fknee{\ifmmode f_{\rm knee}\else $f_{\rm knee}$\fi}
\def\Fmax{\ifmmode F_{\rm max}\else $F_{\rm max}$\fi}
\def\solar{\ifmmode{\rm M}_{\mathord\odot}\else${\rm M}_{\mathord\odot}$\fi}
\def\Msolar{\ifmmode{\rm M}_{\mathord\odot}\else${\rm M}_{\mathord\odot}$\fi}
\def\Lsolar{\ifmmode{\rm L}_{\mathord\odot}\else${\rm L}_{\mathord\odot}$\fi}

\def\inv{\ifmmode^{-1}\else$^{-1}$\fi}
\def\mo{\ifmmode^{-1}\else$^{-1}$\fi}
\def\sup#1{\ifmmode ^{\rm #1}\else $^{\rm #1}$\fi}
\def\expo#1{\ifmmode \times 10^{#1}\else $\times 10^{#1}$\fi}
\def\,{\thinspace}
\def\lsim{\mathrel{\raise .4ex\hbox{\rlap{$<$}\lower 1.2ex\hbox{$\sim$}}}}
\def\gsim{\mathrel{\raise .4ex\hbox{\rlap{$>$}\lower 1.2ex\hbox{$\sim$}}}}

\let\gea=\gsim
\def\simprop{\mathrel{\raise .4ex\hbox{\rlap{$\propto$}\lower 1.2ex\hbox{$\sim$}}}}
\def\deg{\ifmmode^\circ\else$^\circ$\fi}
\def\pdeg{\ifmmode $\setbox0=\hbox{$^{\circ}$}\rlap{\hskip.11\wd0 .}$^{\circ}
          \else \setbox0=\hbox{$^{\circ}$}\rlap{\hskip.11\wd0 .}$^{\circ}$\fi}
\def\arcs{\ifmmode {^{\scriptstyle\prime\prime}}
          \else $^{\scriptstyle\prime\prime}$\fi}
\def\arcm{\ifmmode {^{\scriptstyle\prime}}
          \else $^{\scriptstyle\prime}$\fi}
\newdimen\sa  \newdimen\sb
\def\parcs{\sa=.07em \sb=.03em
     \ifmmode \hbox{\rlap{.}}^{\scriptstyle\prime\kern -\sb\prime}\hbox{\kern -\sa}
     \else \rlap{.}$^{\scriptstyle\prime\kern -\sb\prime}$\kern -\sa\fi}
\def\parcm{\sa=.08em \sb=.03em
     \ifmmode \hbox{\rlap{.}\kern\sa}^{\scriptstyle\prime}\hbox{\kern-\sb}
     \else \rlap{.}\kern\sa$^{\scriptstyle\prime}$\kern-\sb\fi}
\def\ra[#1 #2 #3.#4]{#1\sup{h}#2\sup{m}#3\sup{s}\llap.#4}
\def\dec[#1 #2 #3.#4]{#1\deg#2\arcm#3\arcs\llap.#4}
\def\deco[#1 #2 #3]{#1\deg#2\arcm#3\arcs}
\def\rra[#1 #2]{#1\sup{h}#2\sup{m}}

\def\dots{\relax\ifmmode \ldots\else $\ldots$\fi}
\def\WHzsr{\ifmmode $W\,Hz\mo\,sr\mo$\else W\,Hz\mo\,sr\mo\fi}
\def\mHz{\ifmmode $\,mHz$\else \,mHz\fi}
\def\GHz{\ifmmode $\,GHz$\else \,GHz\fi}
\def\mKs{\ifmmode $\,mK\,s$^{1/2}\else \,mK\,s$^{1/2}$\fi}
\def\muKs{\ifmmode \,\mu$K\,s$^{1/2}\else \,$\mu$K\,s$^{1/2}$\fi}
\def\muKRJs{\ifmmode \,\mu$K$_{\rm RJ}$\,s$^{1/2}\else \,$\mu$K$_{\rm RJ}$\,s$^{1/2}$\fi}
\def\muKHz{\ifmmode \,\mu$K\,Hz$^{-1/2}\else \,$\mu$K\,Hz$^{-1/2}$\fi}
\def\MJysr{\ifmmode \,$MJy\,sr\mo$\else \,MJy\,sr\mo\fi}
\def\MJysrmK{\ifmmode \,$MJy\,sr\mo$\,mK$_{\rm CMB}\mo\else \,MJy\,sr\mo\,mK$_{\rm CMB}\mo$\fi}
\def\microns{\ifmmode \,\mu$m$\else \,$\mu$m\fi}

\def\muK{\ifmmode \,\mu$K$\else \,$\mu$\hbox{K}\fi}
\def\microK{\ifmmode \,\mu$K$\else \,$\mu$\hbox{K}\fi}
\def\muW{\ifmmode \,\mu$W$\else \,$\mu$\hbox{W}\fi}
\def\kms{\ifmmode $\,km\,s$^{-1}\else \,km\,s$^{-1}$\fi}
\def\kmsMpc{\ifmmode $\,\kms\,Mpc\mo$\else \,\kms\,Mpc\mo\fi}
\setsymbol{LFI:center:frequency:70GHz:units}{70.3\,GHz}
\setsymbol{LFI:center:frequency:44GHz:units}{44.1\,GHz}
\setsymbol{LFI:center:frequency:30GHz:units}{28.5\,GHz}

\setsymbol{LFI:center:frequency:70GHz}{70.3}
\setsymbol{LFI:center:frequency:44GHz}{44.1}
\setsymbol{LFI:center:frequency:30GHz}{28.5}

\setsymbol{LFI:center:frequency:LFI18:Rad:M:units}{71.7\GHz}
\setsymbol{LFI:center:frequency:LFI19:Rad:M:units}{67.5\GHz}
\setsymbol{LFI:center:frequency:LFI20:Rad:M:units}{69.2\GHz}
\setsymbol{LFI:center:frequency:LFI21:Rad:M:units}{70.4\GHz}
\setsymbol{LFI:center:frequency:LFI22:Rad:M:units}{71.5\GHz}
\setsymbol{LFI:center:frequency:LFI23:Rad:M:units}{70.8\GHz}
\setsymbol{LFI:center:frequency:LFI24:Rad:M:units}{44.4\GHz}
\setsymbol{LFI:center:frequency:LFI25:Rad:M:units}{44.0\GHz}
\setsymbol{LFI:center:frequency:LFI26:Rad:M:units}{43.9\GHz}
\setsymbol{LFI:center:frequency:LFI27:Rad:M:units}{28.3\GHz}
\setsymbol{LFI:center:frequency:LFI28:Rad:M:units}{28.8\GHz}
\setsymbol{LFI:center:frequency:LFI18:Rad:S:units}{70.1\GHz}
\setsymbol{LFI:center:frequency:LFI19:Rad:S:units}{69.6\GHz}
\setsymbol{LFI:center:frequency:LFI20:Rad:S:units}{69.5\GHz}
\setsymbol{LFI:center:frequency:LFI21:Rad:S:units}{69.5\GHz}
\setsymbol{LFI:center:frequency:LFI22:Rad:S:units}{72.8\GHz}
\setsymbol{LFI:center:frequency:LFI23:Rad:S:units}{71.3\GHz}
\setsymbol{LFI:center:frequency:LFI24:Rad:S:units}{44.1\GHz}
\setsymbol{LFI:center:frequency:LFI25:Rad:S:units}{44.1\GHz}
\setsymbol{LFI:center:frequency:LFI26:Rad:S:units}{44.1\GHz}
\setsymbol{LFI:center:frequency:LFI27:Rad:S:units}{28.5\GHz}
\setsymbol{LFI:center:frequency:LFI28:Rad:S:units}{28.2\GHz}

\setsymbol{LFI:center:frequency:LFI18:Rad:M}{71.7}
\setsymbol{LFI:center:frequency:LFI19:Rad:M}{67.5}
\setsymbol{LFI:center:frequency:LFI20:Rad:M}{69.2}
\setsymbol{LFI:center:frequency:LFI21:Rad:M}{70.4}
\setsymbol{LFI:center:frequency:LFI22:Rad:M}{71.5}
\setsymbol{LFI:center:frequency:LFI23:Rad:M}{70.8}
\setsymbol{LFI:center:frequency:LFI24:Rad:M}{44.4}
\setsymbol{LFI:center:frequency:LFI25:Rad:M}{44.0}
\setsymbol{LFI:center:frequency:LFI26:Rad:M}{43.9}
\setsymbol{LFI:center:frequency:LFI27:Rad:M}{28.3}
\setsymbol{LFI:center:frequency:LFI28:Rad:M}{28.8}
\setsymbol{LFI:center:frequency:LFI18:Rad:S}{70.1}
\setsymbol{LFI:center:frequency:LFI19:Rad:S}{69.6}
\setsymbol{LFI:center:frequency:LFI20:Rad:S}{69.5}
\setsymbol{LFI:center:frequency:LFI21:Rad:S}{69.5}
\setsymbol{LFI:center:frequency:LFI22:Rad:S}{72.8}
\setsymbol{LFI:center:frequency:LFI23:Rad:S}{71.3}
\setsymbol{LFI:center:frequency:LFI24:Rad:S}{44.1}
\setsymbol{LFI:center:frequency:LFI25:Rad:S}{44.1}
\setsymbol{LFI:center:frequency:LFI26:Rad:S}{44.1}
\setsymbol{LFI:center:frequency:LFI27:Rad:S}{28.5}
\setsymbol{LFI:center:frequency:LFI28:Rad:S}{28.2}

\setsymbol{LFI:white:noise:sensitivity:70GHz:units}{134.7\muKs}
\setsymbol{LFI:white:noise:sensitivity:44GHz:units}{164.7\muKs}
\setsymbol{LFI:white:noise:sensitivity:30GHz:units}{143.4\muKs}

\setsymbol{LFI:white:noise:sensitivity:70GHz}{134.7}
\setsymbol{LFI:white:noise:sensitivity:44GHz}{164.7}
\setsymbol{LFI:white:noise:sensitivity:30GHz}{143.4}

\setsymbol{LFI:white:noise:sensitivity:LFI18:Rad:M:units}{512.0\muKs}
\setsymbol{LFI:white:noise:sensitivity:LFI19:Rad:M:units}{581.4\muKs}
\setsymbol{LFI:white:noise:sensitivity:LFI20:Rad:M:units}{590.8\muKs}
\setsymbol{LFI:white:noise:sensitivity:LFI21:Rad:M:units}{455.2\muKs}
\setsymbol{LFI:white:noise:sensitivity:LFI22:Rad:M:units}{492.0\muKs}
\setsymbol{LFI:white:noise:sensitivity:LFI23:Rad:M:units}{507.7\muKs}
\setsymbol{LFI:white:noise:sensitivity:LFI24:Rad:M:units}{462.2\muKs}
\setsymbol{LFI:white:noise:sensitivity:LFI25:Rad:M:units}{413.6\muKs}
\setsymbol{LFI:white:noise:sensitivity:LFI26:Rad:M:units}{478.6\muKs}
\setsymbol{LFI:white:noise:sensitivity:LFI27:Rad:M:units}{277.7\muKs}
\setsymbol{LFI:white:noise:sensitivity:LFI28:Rad:M:units}{312.3\muKs}
\setsymbol{LFI:white:noise:sensitivity:LFI18:Rad:S:units}{465.7\muKs}
\setsymbol{LFI:white:noise:sensitivity:LFI19:Rad:S:units}{555.6\muKs}
\setsymbol{LFI:white:noise:sensitivity:LFI20:Rad:S:units}{623.2\muKs}
\setsymbol{LFI:white:noise:sensitivity:LFI21:Rad:S:units}{564.1\muKs}
\setsymbol{LFI:white:noise:sensitivity:LFI22:Rad:S:units}{534.4\muKs}
\setsymbol{LFI:white:noise:sensitivity:LFI23:Rad:S:units}{542.4\muKs}
\setsymbol{LFI:white:noise:sensitivity:LFI24:Rad:S:units}{399.2\muKs}
\setsymbol{LFI:white:noise:sensitivity:LFI25:Rad:S:units}{392.6\muKs}
\setsymbol{LFI:white:noise:sensitivity:LFI26:Rad:S:units}{418.6\muKs}
\setsymbol{LFI:white:noise:sensitivity:LFI27:Rad:S:units}{302.9\muKs}
\setsymbol{LFI:white:noise:sensitivity:LFI28:Rad:S:units}{285.3\muKs}

\setsymbol{LFI:white:noise:sensitivity:LFI18:Rad:M}{512.0}
\setsymbol{LFI:white:noise:sensitivity:LFI19:Rad:M}{581.4}
\setsymbol{LFI:white:noise:sensitivity:LFI20:Rad:M}{590.8}
\setsymbol{LFI:white:noise:sensitivity:LFI21:Rad:M}{455.2}
\setsymbol{LFI:white:noise:sensitivity:LFI22:Rad:M}{492.0}
\setsymbol{LFI:white:noise:sensitivity:LFI23:Rad:M}{507.7}
\setsymbol{LFI:white:noise:sensitivity:LFI24:Rad:M}{462.2}
\setsymbol{LFI:white:noise:sensitivity:LFI25:Rad:M}{413.6}
\setsymbol{LFI:white:noise:sensitivity:LFI26:Rad:M}{478.6}
\setsymbol{LFI:white:noise:sensitivity:LFI27:Rad:M}{277.7}
\setsymbol{LFI:white:noise:sensitivity:LFI28:Rad:M}{312.3}
\setsymbol{LFI:white:noise:sensitivity:LFI18:Rad:S}{465.7}
\setsymbol{LFI:white:noise:sensitivity:LFI19:Rad:S}{555.6}
\setsymbol{LFI:white:noise:sensitivity:LFI20:Rad:S}{623.2}
\setsymbol{LFI:white:noise:sensitivity:LFI21:Rad:S}{564.1}
\setsymbol{LFI:white:noise:sensitivity:LFI22:Rad:S}{534.4}
\setsymbol{LFI:white:noise:sensitivity:LFI23:Rad:S}{542.4}
\setsymbol{LFI:white:noise:sensitivity:LFI24:Rad:S}{399.2}
\setsymbol{LFI:white:noise:sensitivity:LFI25:Rad:S}{392.6}
\setsymbol{LFI:white:noise:sensitivity:LFI26:Rad:S}{418.6}
\setsymbol{LFI:white:noise:sensitivity:LFI27:Rad:S}{302.9}
\setsymbol{LFI:white:noise:sensitivity:LFI28:Rad:S}{285.3}

\setsymbol{LFI:knee:frequency:70GHz:units}{29.5\mHz}
\setsymbol{LFI:knee:frequency:44GHz:units}{56.2\mHz}
\setsymbol{LFI:knee:frequency:30GHz:units}{113.7\mHz}

\setsymbol{LFI:knee:frequency:70GHz}{29.5}
\setsymbol{LFI:knee:frequency:44GHz}{56.2}
\setsymbol{LFI:knee:frequency:30GHz}{113.7}

\setsymbol{LFI:knee:frequency:LFI18:Rad:M:units}{16.3\mHz}
\setsymbol{LFI:knee:frequency:LFI19:Rad:M:units}{15.1\mHz}
\setsymbol{LFI:knee:frequency:LFI20:Rad:M:units}{18.7\mHz}
\setsymbol{LFI:knee:frequency:LFI21:Rad:M:units}{37.2\mHz}
\setsymbol{LFI:knee:frequency:LFI22:Rad:M:units}{12.7\mHz}
\setsymbol{LFI:knee:frequency:LFI23:Rad:M:units}{34.6\mHz}
\setsymbol{LFI:knee:frequency:LFI24:Rad:M:units}{46.2\mHz}
\setsymbol{LFI:knee:frequency:LFI25:Rad:M:units}{24.9\mHz}
\setsymbol{LFI:knee:frequency:LFI26:Rad:M:units}{67.6\mHz}
\setsymbol{LFI:knee:frequency:LFI27:Rad:M:units}{187.4\mHz}
\setsymbol{LFI:knee:frequency:LFI28:Rad:M:units}{122.2\mHz}
\setsymbol{LFI:knee:frequency:LFI18:Rad:S:units}{17.7\mHz}
\setsymbol{LFI:knee:frequency:LFI19:Rad:S:units}{22.0\mHz}
\setsymbol{LFI:knee:frequency:LFI20:Rad:S:units}{8.7\mHz}
\setsymbol{LFI:knee:frequency:LFI21:Rad:S:units}{25.9\mHz}
\setsymbol{LFI:knee:frequency:LFI22:Rad:S:units}{15.8\mHz}
\setsymbol{LFI:knee:frequency:LFI23:Rad:S:units}{129.8\mHz}
\setsymbol{LFI:knee:frequency:LFI24:Rad:S:units}{100.9\mHz}
\setsymbol{LFI:knee:frequency:LFI25:Rad:S:units}{38.9\mHz}
\setsymbol{LFI:knee:frequency:LFI26:Rad:S:units}{58.9\mHz}
\setsymbol{LFI:knee:frequency:LFI27:Rad:S:units}{104.4\mHz}
\setsymbol{LFI:knee:frequency:LFI28:Rad:S:units}{40.7\mHz}

\setsymbol{LFI:knee:frequency:LFI18:Rad:M}{16.3}
\setsymbol{LFI:knee:frequency:LFI19:Rad:M}{15.1}
\setsymbol{LFI:knee:frequency:LFI20:Rad:M}{18.7}
\setsymbol{LFI:knee:frequency:LFI21:Rad:M}{37.2}
\setsymbol{LFI:knee:frequency:LFI22:Rad:M}{12.7}
\setsymbol{LFI:knee:frequency:LFI23:Rad:M}{34.6}
\setsymbol{LFI:knee:frequency:LFI24:Rad:M}{46.2}
\setsymbol{LFI:knee:frequency:LFI25:Rad:M}{24.9}
\setsymbol{LFI:knee:frequency:LFI26:Rad:M}{67.6}
\setsymbol{LFI:knee:frequency:LFI27:Rad:M}{187.4}
\setsymbol{LFI:knee:frequency:LFI28:Rad:M}{122.2}
\setsymbol{LFI:knee:frequency:LFI18:Rad:S}{17.7}
\setsymbol{LFI:knee:frequency:LFI19:Rad:S}{22.0}
\setsymbol{LFI:knee:frequency:LFI20:Rad:S}{8.7}
\setsymbol{LFI:knee:frequency:LFI21:Rad:S}{25.9}
\setsymbol{LFI:knee:frequency:LFI22:Rad:S}{15.8}
\setsymbol{LFI:knee:frequency:LFI23:Rad:S}{129.8}
\setsymbol{LFI:knee:frequency:LFI24:Rad:S}{100.9}
\setsymbol{LFI:knee:frequency:LFI25:Rad:S}{38.9}
\setsymbol{LFI:knee:frequency:LFI26:Rad:S}{58.9}
\setsymbol{LFI:knee:frequency:LFI27:Rad:S}{104.4}
\setsymbol{LFI:knee:frequency:LFI28:Rad:S}{40.7}

\setsymbol{LFI:slope:70GHz:units}{$-1.03$\mHz}
\setsymbol{LFI:slope:44GHz:units}{$-0.89$\mHz}
\setsymbol{LFI:slope:30GHz:units}{$-0.87$\mHz}

\setsymbol{LFI:slope:70GHz}{$-1.03$}
\setsymbol{LFI:slope:44GHz}{$-0.89$}
\setsymbol{LFI:slope:30GHz}{$-0.87$}

\setsymbol{LFI:slope:LFI18:Rad:M:units}{$-1.04$\mHz}
\setsymbol{LFI:slope:LFI19:Rad:M:units}{$-1.09$\mHz}
\setsymbol{LFI:slope:LFI20:Rad:M:units}{$-0.69$\mHz}
\setsymbol{LFI:slope:LFI21:Rad:M:units}{$-1.56$\mHz}
\setsymbol{LFI:slope:LFI22:Rad:M:units}{$-1.01$\mHz}
\setsymbol{LFI:slope:LFI23:Rad:M:units}{$-0.96$\mHz}
\setsymbol{LFI:slope:LFI24:Rad:M:units}{$-0.83$\mHz}
\setsymbol{LFI:slope:LFI25:Rad:M:units}{$-0.91$\mHz}
\setsymbol{LFI:slope:LFI26:Rad:M:units}{$-0.95$\mHz}
\setsymbol{LFI:slope:LFI27:Rad:M:units}{$-0.87$\mHz}
\setsymbol{LFI:slope:LFI28:Rad:M:units}{$-0.88$\mHz}
\setsymbol{LFI:slope:LFI18:Rad:S:units}{$-1.15$\mHz}
\setsymbol{LFI:slope:LFI19:Rad:S:units}{$-1.00$\mHz}
\setsymbol{LFI:slope:LFI20:Rad:S:units}{$-0.95$\mHz}
\setsymbol{LFI:slope:LFI21:Rad:S:units}{$-0.92$\mHz}
\setsymbol{LFI:slope:LFI22:Rad:S:units}{$-1.01$\mHz}
\setsymbol{LFI:slope:LFI23:Rad:S:units}{$-0.95$\mHz}
\setsymbol{LFI:slope:LFI24:Rad:S:units}{$-0.73$\mHz}
\setsymbol{LFI:slope:LFI25:Rad:S:units}{$-1.16$\mHz}
\setsymbol{LFI:slope:LFI26:Rad:S:units}{$-0.79$\mHz}
\setsymbol{LFI:slope:LFI27:Rad:S:units}{$-0.82$\mHz}
\setsymbol{LFI:slope:LFI28:Rad:S:units}{$-0.91$\mHz}

\setsymbol{LFI:slope:LFI18:Rad:M}{$-1.04$}
\setsymbol{LFI:slope:LFI19:Rad:M}{$-1.09$}
\setsymbol{LFI:slope:LFI20:Rad:M}{$-0.69$}
\setsymbol{LFI:slope:LFI21:Rad:M}{$-1.56$}
\setsymbol{LFI:slope:LFI22:Rad:M}{$-1.01$}
\setsymbol{LFI:slope:LFI23:Rad:M}{$-0.96$}
\setsymbol{LFI:slope:LFI24:Rad:M}{$-0.83$}
\setsymbol{LFI:slope:LFI25:Rad:M}{$-0.91$}
\setsymbol{LFI:slope:LFI26:Rad:M}{$-0.95$}
\setsymbol{LFI:slope:LFI27:Rad:M}{$-0.87$}
\setsymbol{LFI:slope:LFI28:Rad:M}{$-0.88$}
\setsymbol{LFI:slope:LFI18:Rad:S}{$-1.15$}
\setsymbol{LFI:slope:LFI19:Rad:S}{$-1.00$}
\setsymbol{LFI:slope:LFI20:Rad:S}{$-0.95$}
\setsymbol{LFI:slope:LFI21:Rad:S}{$-0.92$}
\setsymbol{LFI:slope:LFI22:Rad:S}{$-1.01$}
\setsymbol{LFI:slope:LFI23:Rad:S}{$-0.95$}
\setsymbol{LFI:slope:LFI24:Rad:S}{$-0.73$}
\setsymbol{LFI:slope:LFI25:Rad:S}{$-1.16$}
\setsymbol{LFI:slope:LFI26:Rad:S}{$-0.79$}
\setsymbol{LFI:slope:LFI27:Rad:S}{$-0.82$}
\setsymbol{LFI:slope:LFI28:Rad:S}{$-0.91$}

\setsymbol{LFI:FWHM:70GHz:units}{13\parcm01}
\setsymbol{LFI:FWHM:44GHz:units}{27\parcm92}
\setsymbol{LFI:FWHM:30GHz:units}{32\parcm65}

\setsymbol{LFI:FWHM:70GHz}{13.01}
\setsymbol{LFI:FWHM:44GHz}{27.92}
\setsymbol{LFI:FWHM:30GHz}{32.65}

\setsymbol{LFI:FWHM:LFI18:units}{13\parcm39}
\setsymbol{LFI:FWHM:LFI19:units}{13\parcm01}
\setsymbol{LFI:FWHM:LFI20:units}{12\parcm75}
\setsymbol{LFI:FWHM:LFI21:units}{12\parcm74}
\setsymbol{LFI:FWHM:LFI22:units}{12\parcm87}
\setsymbol{LFI:FWHM:LFI23:units}{13\parcm27}
\setsymbol{LFI:FWHM:LFI24:units}{22\parcm98}
\setsymbol{LFI:FWHM:LFI25:units}{30\parcm46}
\setsymbol{LFI:FWHM:LFI26:units}{30\parcm31}
\setsymbol{LFI:FWHM:LFI27:units}{32\parcm65}
\setsymbol{LFI:FWHM:LFI28:units}{32\parcm66}

\setsymbol{LFI:FWHM:LFI18}{13.39}
\setsymbol{LFI:FWHM:LFI19}{13.01}
\setsymbol{LFI:FWHM:LFI20}{12.75}
\setsymbol{LFI:FWHM:LFI21}{12.74}
\setsymbol{LFI:FWHM:LFI22}{12.87}
\setsymbol{LFI:FWHM:LFI23}{13.27}
\setsymbol{LFI:FWHM:LFI24}{22.98}
\setsymbol{LFI:FWHM:LFI25}{30.46}
\setsymbol{LFI:FWHM:LFI26}{30.31}
\setsymbol{LFI:FWHM:LFI27}{32.65}
\setsymbol{LFI:FWHM:LFI28}{32.66}

\setsymbol{LFI:FWHM:uncertainty:LFI18:units}{0.170\arcm}
\setsymbol{LFI:FWHM:uncertainty:LFI19:units}{0.174\arcm}
\setsymbol{LFI:FWHM:uncertainty:LFI20:units}{0.170\arcm}
\setsymbol{LFI:FWHM:uncertainty:LFI21:units}{0.156\arcm}
\setsymbol{LFI:FWHM:uncertainty:LFI22:units}{0.164\arcm}
\setsymbol{LFI:FWHM:uncertainty:LFI23:units}{0.171\arcm}
\setsymbol{LFI:FWHM:uncertainty:LFI24:units}{0.652\arcm}
\setsymbol{LFI:FWHM:uncertainty:LFI25:units}{1.075\arcm}
\setsymbol{LFI:FWHM:uncertainty:LFI26:units}{1.131\arcm}
\setsymbol{LFI:FWHM:uncertainty:LFI27:units}{1.266\arcm}
\setsymbol{LFI:FWHM:uncertainty:LFI28:units}{1.287\arcm}

\setsymbol{LFI:FWHM:uncertainty:LFI18}{0.170}
\setsymbol{LFI:FWHM:uncertainty:LFI19}{0.174}
\setsymbol{LFI:FWHM:uncertainty:LFI20}{0.170}
\setsymbol{LFI:FWHM:uncertainty:LFI21}{0.156}
\setsymbol{LFI:FWHM:uncertainty:LFI22}{0.164}
\setsymbol{LFI:FWHM:uncertainty:LFI23}{0.171}
\setsymbol{LFI:FWHM:uncertainty:LFI24}{0.652}
\setsymbol{LFI:FWHM:uncertainty:LFI25}{1.075}
\setsymbol{LFI:FWHM:uncertainty:LFI26}{1.131}
\setsymbol{LFI:FWHM:uncertainty:LFI27}{1.266}
\setsymbol{LFI:FWHM:uncertainty:LFI28}{1.287}

\setsymbol{HFI:center:frequency:100GHz:units}{100\,GHz}
\setsymbol{HFI:center:frequency:143GHz:units}{143\,GHz}
\setsymbol{HFI:center:frequency:217GHz:units}{217\,GHz}
\setsymbol{HFI:center:frequency:353GHz:units}{353\,GHz}
\setsymbol{HFI:center:frequency:545GHz:units}{545\,GHz}
\setsymbol{HFI:center:frequency:857GHz:units}{857\,GHz}

\setsymbol{HFI:center:frequency:100GHz}{100}
\setsymbol{HFI:center:frequency:143GHz}{143}
\setsymbol{HFI:center:frequency:217GHz}{217}
\setsymbol{HFI:center:frequency:353GHz}{353}
\setsymbol{HFI:center:frequency:545GHz}{545}
\setsymbol{HFI:center:frequency:857GHz}{857}

\setsymbol{HFI:Ndetectors:100GHz}{8}
\setsymbol{HFI:Ndetectors:143GHz}{11}
\setsymbol{HFI:Ndetectors:217GHz}{12}
\setsymbol{HFI:Ndetectors:353GHz}{12}
\setsymbol{HFI:Ndetectors:545GHz}{3}
\setsymbol{HFI:Ndetectors:857GHz}{4}

\setsymbol{HFI:FWHM:Maps:100GHz:units}{9\parcm88}
\setsymbol{HFI:FWHM:Maps:143GHz:units}{7\parcm18}
\setsymbol{HFI:FWHM:Maps:217GHz:units}{4\parcm87}
\setsymbol{HFI:FWHM:Maps:353GHz:units}{4\parcm65}
\setsymbol{HFI:FWHM:Maps:545GHz:units}{4\parcm72}
\setsymbol{HFI:FWHM:Maps:857GHz:units}{4\parcm39}
\setsymbol{HFI:FWHM:Maps:100GHz}{9.88}
\setsymbol{HFI:FWHM:Maps:143GHz}{7.18}
\setsymbol{HFI:FWHM:Maps:217GHz}{4.87}
\setsymbol{HFI:FWHM:Maps:353GHz}{4.65}
\setsymbol{HFI:FWHM:Maps:545GHz}{4.72}
\setsymbol{HFI:FWHM:Maps:857GHz}{4.39}

\setsymbol{HFI:beam:ellipticity:Maps:100GHz}{1.15}
\setsymbol{HFI:beam:ellipticity:Maps:143GHz}{1.01}
\setsymbol{HFI:beam:ellipticity:Maps:217GHz}{1.06}
\setsymbol{HFI:beam:ellipticity:Maps:353GHz}{1.05}
\setsymbol{HFI:beam:ellipticity:Maps:545GHz}{1.14}
\setsymbol{HFI:beam:ellipticity:Maps:857GHz}{1.19}

\setsymbol{HFI:FWHM:Mars:100GHz:units}{9\parcm37}
\setsymbol{HFI:FWHM:Mars:143GHz:units}{7\parcm04}
\setsymbol{HFI:FWHM:Mars:217GHz:units}{4\parcm68}
\setsymbol{HFI:FWHM:Mars:353GHz:units}{4\parcm43}
\setsymbol{HFI:FWHM:Mars:545GHz:units}{3\parcm80}
\setsymbol{HFI:FWHM:Mars:857GHz:units}{3\parcm67}

\setsymbol{HFI:FWHM:Mars:100GHz}{9.37}
\setsymbol{HFI:FWHM:Mars:143GHz}{7.04}
\setsymbol{HFI:FWHM:Mars:217GHz}{4.68}
\setsymbol{HFI:FWHM:Mars:353GHz}{4.43}
\setsymbol{HFI:FWHM:Mars:545GHz}{3.80}
\setsymbol{HFI:FWHM:Mars:857GHz}{3.67}

\setsymbol{HFI:beam:ellipticity:Mars:100GHz}{1.18}
\setsymbol{HFI:beam:ellipticity:Mars:143GHz}{1.03}
\setsymbol{HFI:beam:ellipticity:Mars:217GHz}{1.14}
\setsymbol{HFI:beam:ellipticity:Mars:353GHz}{1.09}
\setsymbol{HFI:beam:ellipticity:Mars:545GHz}{1.25}
\setsymbol{HFI:beam:ellipticity:Mars:857GHz}{1.03}

\setsymbol{HFI:CMB:relative:calibration:100GHz}{$\lsim 1\%$}
\setsymbol{HFI:CMB:relative:calibration:143GHz}{$\lsim 1\%$}
\setsymbol{HFI:CMB:relative:calibration:217GHz}{$\lsim 1\%$}
\setsymbol{HFI:CMB:relative:calibration:353GHz}{$\lsim 1\%$}
\setsymbol{HFI:CMB:relative:calibration:545GHz}{}
\setsymbol{HFI:CMB:relative:calibration:857GHz}{}

\setsymbol{HFI:CMB:absolute:calibration:100GHz}{$\lsim 2\%$}
\setsymbol{HFI:CMB:absolute:calibration:143GHz}{$\lsim 2\%$}
\setsymbol{HFI:CMB:absolute:calibration:217GHz}{$\lsim 2\%$}
\setsymbol{HFI:CMB:absolute:calibration:353GHz}{$\lsim 2\%$}
\setsymbol{HFI:CMB:absolute:calibration:545GHz}{}
\setsymbol{HFI:CMB:absolute:calibration:857GHz}{}

\setsymbol{HFI:FIRAS:gain:calibration:accuracy:statistical:100GHz}{}
\setsymbol{HFI:FIRAS:gain:calibration:accuracy:statistical:143GHz}{}
\setsymbol{HFI:FIRAS:gain:calibration:accuracy:statistical:217GHz}{}
\setsymbol{HFI:FIRAS:gain:calibration:accuracy:statistical:353GHz}{2.5\%}
\setsymbol{HFI:FIRAS:gain:calibration:accuracy:statistical:545GHz}{1\%}
\setsymbol{HFI:FIRAS:gain:calibration:accuracy:statistical:857GHz}{0.5\%}

\setsymbol{HFI:FIRAS:gain:calibration:accuracy:systematic:100GHz}{}
\setsymbol{HFI:FIRAS:gain:calibration:accuracy:systematic:143GHz}{}
\setsymbol{HFI:FIRAS:gain:calibration:accuracy:systematic:217GHz}{}
\setsymbol{HFI:FIRAS:gain:calibration:accuracy:systematic:353GHz}{}
\setsymbol{HFI:FIRAS:gain:calibration:accuracy:systematic:545GHz}{7\%}
\setsymbol{HFI:FIRAS:gain:calibration:accuracy:systematic:857GHz}{7\%}

\setsymbol{HFI:FIRAS:zero:point:accuracy:100GHz:units}{0.8\MJysr}
\setsymbol{HFI:FIRAS:zero:point:accuracy:143GHz:units}{}
\setsymbol{HFI:FIRAS:zero:point:accuracy:217GHz:units}{}
\setsymbol{HFI:FIRAS:zero:point:accuracy:353GHz:units}{1.4\MJysr}
\setsymbol{HFI:FIRAS:zero:point:accuracy:545GHz:units}{2.2\MJysr}
\setsymbol{HFI:FIRAS:zero:point:accuracy:857GHz:units}{1.7\MJysr}

\setsymbol{HFI:FIRAS:zero:point:accuracy:100GHz}{0.8}
\setsymbol{HFI:FIRAS:zero:point:accuracy:143GHz}{}
\setsymbol{HFI:FIRAS:zero:point:accuracy:217GHz}{}
\setsymbol{HFI:FIRAS:zero:point:accuracy:353GHz}{1.4}
\setsymbol{HFI:FIRAS:zero:point:accuracy:545GHz}{2.2}
\setsymbol{HFI:FIRAS:zero:point:accuracy:857GHz}{1.7}

\setsymbol{HFI:unit:conversion:100GHz:units}{0.2415\MJysrmK}
\setsymbol{HFI:unit:conversion:143GHz:units}{0.3694\MJysrmK}
\setsymbol{HFI:unit:conversion:217GHz:units}{0.4811\MJysrmK}
\setsymbol{HFI:unit:conversion:353GHz:units}{0.2883\MJysrmK}
\setsymbol{HFI:unit:conversion:545GHz:units}{0.05826\MJysrmK}
\setsymbol{HFI:unit:conversion:857GHz:units}{0.002238\MJysrmK}

\setsymbol{HFI:unit:conversion:100GHz}{0.2415}
\setsymbol{HFI:unit:conversion:143GHz}{0.3694}
\setsymbol{HFI:unit:conversion:217GHz}{0.4811}
\setsymbol{HFI:unit:conversion:353GHz}{0.2883}
\setsymbol{HFI:unit:conversion:545GHz}{0.05826}
\setsymbol{HFI:unit:conversion:857GHz}{0.002238}

\setsymbol{HFI:colour:correction:alpha=-2:V1.01:100GHz}{0.9893}
\setsymbol{HFI:colour:correction:alpha=-2:V1.01:143GHz}{0.9759}
\setsymbol{HFI:colour:correction:alpha=-2:V1.01:217GHz}{1.0007}
\setsymbol{HFI:colour:correction:alpha=-2:V1.01:353GHz}{1.0028}
\setsymbol{HFI:colour:correction:alpha=-2:V1.01:545GHz}{1.0019}
\setsymbol{HFI:colour:correction:alpha=-2:V1.01:857GHz}{0.9889}

\setsymbol{HFI:colour:correction:alpha=0:V1.01:100GHz}{1.0008}
\setsymbol{HFI:colour:correction:alpha=0:V1.01:143GHz}{1.0148}
\setsymbol{HFI:colour:correction:alpha=0:V1.01:217GHz}{0.9909}
\setsymbol{HFI:colour:correction:alpha=0:V1.01:353GHz}{0.9888}
\setsymbol{HFI:colour:correction:alpha=0:V1.01:545GHz}{0.9878}
\setsymbol{HFI:colour:correction:alpha=0:V1.01:857GHz}{1.0014}

\providecommand{\sorthelp}[1]{}

\newcommand{\referee}[1]{{ #1}}

\begin{document}

\title{\Planck\ 2013 results. XVI. Cosmological parameters}
\titlerunning{Planck Cosmological Parameters}
\authorrunning{Planck Collaboration}
\author{The Planck team}
\institute{L2 \and Earth}
\date{Received 1 January 2013/Accepted 1 January 2013}

\author{\small
Planck Collaboration:
P.~A.~R.~Ade\inst{93}
\and
N.~Aghanim\inst{65}
\and
C.~Armitage-Caplan\inst{99}
\and
M.~Arnaud\inst{79}
\and
M.~Ashdown\inst{76, 6}
\and
F.~Atrio-Barandela\inst{19}
\and
J.~Aumont\inst{65}
\and
C.~Baccigalupi\inst{92}
\and
A.~J.~Banday\inst{102, 10}
\and
R.~B.~Barreiro\inst{72}
% ...
}
\institute{\small
APC, AstroParticule et Cosmologie, Universit\'{e} Paris Diderot, CNRS/IN2P3, CEA/lrfu, Observatoire de Paris, Sorbonne Paris Cit\'{e}, 10, rue Alice Domon et L\'{e}onie Duquet, 75205 Paris Cedex 13, France\\
\and
Aalto University Mets\"{a}hovi Radio Observatory, Mets\"{a}hovintie 114, FIN-02540 Kylm\"{a}l\"{a}, Finland\\
\and
African Institute for Mathematical Sciences, 6-8 Melrose Road, Muizenberg, Cape Town, South Africa\\
\and
Agenzia Spaziale Italiana Science Data Center, Via del Politecnico snc, 00133, Roma, Italy\\
\and
Agenzia Spaziale Italiana, Viale Liegi 26, Roma, Italy\\
\and
Astrophysics Group, Cavendish Laboratory, University of Cambridge, J J Thomson Avenue, Cambridge CB3 0HE, U.K.\\
\and
Astrophysics \& Cosmology Research Unit, School of Mathematics, Statistics \& Computer Science, University of KwaZulu-Natal, Westville Campus, Private Bag X54001, Durban 4000, South Africa\\
\and
Atacama Large Millimeter/submillimeter Array, ALMA Santiago Central Offices, Alonso de Cordova 3107, Vitacura, Casilla 763 0355, Santiago, Chile\\
\and
CITA, University of Toronto, 60 St. George St., Toronto, ON M5S 3H8, Canada\\
\and
CNRS, IRAP, 9 Av. colonel Roche, BP 44346, F-31028 Toulouse cedex 4, France\\
\and
California Institute of Technology, Pasadena, California, U.S.A.\\
\and
Centre for Theoretical Cosmology, DAMTP, University of Cambridge, Wilberforce Road, Cambridge CB3 0WA, U.K.\\
\and
Centro de Estudios de F\'{i}sica del Cosmos de Arag\'{o}n (CEFCA), Plaza San Juan, 1, planta 2, E-44001, Teruel, Spain\\
\and
Computational Cosmology Center, Lawrence Berkeley National Laboratory, Berkeley, California, U.S.A.\\
\and
Consejo Superior de Investigaciones Cient\'{\i}ficas (CSIC), Madrid, Spain\\
\and
DSM/Irfu/SPP, CEA-Saclay, F-91191 Gif-sur-Yvette Cedex, France\\
\and
DTU Space, National Space Institute, Technical University of Denmark, Elektrovej 327, DK-2800 Kgs. Lyngby, Denmark\\
\and
D\'{e}partement de Physique Th\'{e}orique, Universit\'{e} de Gen\`{e}ve, 24, Quai E. Ansermet,1211 Gen\`{e}ve 4, Switzerland\\
\and
Departamento de F\'{\i}sica Fundamental, Facultad de Ciencias, Universidad de Salamanca, 37008 Salamanca, Spain\\
\and
Departamento de F\'{\i}sica, Universidad de Oviedo, Avda. Calvo Sotelo s/n, Oviedo, Spain\\
\and
Department of Astronomy and Astrophysics, University of Toronto, 50 Saint George Street, Toronto, Ontario, Canada\\
\and
Department of Astrophysics/IMAPP, Radboud University Nijmegen, P.O. Box 9010, 6500 GL Nijmegen, The Netherlands\\
\and
Department of Electrical Engineering and Computer Sciences, University of California, Berkeley, California, U.S.A.\\
\and
Department of Physics \& Astronomy, University of British Columbia, 6224 Agricultural Road, Vancouver, British Columbia, Canada\\
\and
Department of Physics and Astronomy, Dana and David Dornsife College of Letter, Arts and Sciences, University of Southern California, Los Angeles, CA 90089, U.S.A.\\
\and
Department of Physics and Astronomy, University College London, London WC1E 6BT, U.K.\\
\and
Department of Physics and Astronomy, University of Sussex, Brighton BN1 9QH, U.K.\\
\and
Department of Physics, Florida State University, Keen Physics Building, 77 Chieftan Way, Tallahassee, Florida, U.S.A.\\
\and
Department of Physics, Gustaf H\"{a}llstr\"{o}min katu 2a, University of Helsinki, Helsinki, Finland\\
\and
Department of Physics, Princeton University, Princeton, New Jersey, U.S.A.\\
\and
Department of Physics, University of California, Berkeley, California, U.S.A.\\
\and
Department of Physics, University of California, One Shields Avenue, Davis, California, U.S.A.\\
\and
Department of Physics, University of California, Santa Barbara, California, U.S.A.\\
\and
Department of Physics, University of Illinois at Urbana-Champaign, 1110 West Green Street, Urbana, Illinois, U.S.A.\\
\and
Dipartimento di Fisica e Astronomia G. Galilei, Universit\`{a} degli Studi di Padova, via Marzolo 8, 35131 Padova, Italy\\
\and
Dipartimento di Fisica e Scienze della Terra, Universit\`{a} di Ferrara, Via Saragat 1, 44122 Ferrara, Italy\\
\and
Dipartimento di Fisica, Universit\`{a} La Sapienza, P. le A. Moro 2, Roma, Italy\\
\and
Dipartimento di Fisica, Universit\`{a} degli Studi di Milano, Via Celoria, 16, Milano, Italy\\
\and
Dipartimento di Fisica, Universit\`{a} degli Studi di Trieste, via A. Valerio 2, Trieste, Italy\\
\and
Dipartimento di Fisica, Universit\`{a} di Roma Tor Vergata, Via della Ricerca Scientifica, 1, Roma, Italy\\
\and
Discovery Center, Niels Bohr Institute, Blegdamsvej 17, Copenhagen, Denmark\\
\and
Dpto. Astrof\'{i}sica, Universidad de La Laguna (ULL), E-38206 La Laguna, Tenerife, Spain\\
\and
European Southern Observatory, ESO Vitacura, Alonso de Cordova 3107, Vitacura, Casilla 19001, Santiago, Chile\\
\and
European Space Agency, ESAC, Planck Science Office, Camino bajo del Castillo, s/n, Urbanizaci\'{o}n Villafranca del Castillo, Villanueva de la Ca\~{n}ada, Madrid, Spain\\
\and
European Space Agency, ESTEC, Keplerlaan 1, 2201 AZ Noordwijk, The Netherlands\\
\and
Finnish Centre for Astronomy with ESO (FINCA), University of Turku, V\"{a}is\"{a}l\"{a}ntie 20, FIN-21500, Piikki\"{o}, Finland\\
\and
Haverford College Astronomy Department, 370 Lancaster Avenue, Haverford, Pennsylvania, U.S.A.\\
\and
Helsinki Institute of Physics, Gustaf H\"{a}llstr\"{o}min katu 2, University of Helsinki, Helsinki, Finland\\
\and
INAF - Osservatorio Astrofisico di Catania, Via S. Sofia 78, Catania, Italy\\
\and
INAF - Osservatorio Astronomico di Padova, Vicolo dell'Osservatorio 5, Padova, Italy\\
\and
INAF - Osservatorio Astronomico di Roma, via di Frascati 33, Monte Porzio Catone, Italy\\
\and
INAF - Osservatorio Astronomico di Trieste, Via G.B. Tiepolo 11, Trieste, Italy\\
\and
INAF Istituto di Radioastronomia, Via P. Gobetti 101, 40129 Bologna, Italy\\
\and
INAF/IASF Bologna, Via Gobetti 101, Bologna, Italy\\
\and
INAF/IASF Milano, Via E. Bassini 15, Milano, Italy\\
\and
INFN, Sezione di Bologna, Via Irnerio 46, I-40126, Bologna, Italy\\
\and
INFN, Sezione di Roma 1, Universit\`{a} di Roma Sapienza, Piazzale Aldo Moro 2, 00185, Roma, Italy\\
\and
IPAG: Institut de Plan\'{e}tologie et d'Astrophysique de Grenoble, Universit\'{e} Joseph Fourier, Grenoble 1 / CNRS-INSU, UMR 5274, Grenoble, F-38041, France\\
\and
ISDC Data Centre for Astrophysics, University of Geneva, ch. d'Ecogia 16, Versoix, Switzerland\\
\and
IUCAA, Post Bag 4, Ganeshkhind, Pune University Campus, Pune 411 007, India\\
\and
Imperial College London, Astrophysics group, Blackett Laboratory, Prince Consort Road, London, SW7 2AZ, U.K.\\
\and
Infrared Processing and Analysis Center, California Institute of Technology, Pasadena, CA 91125, U.S.A.\\
\and
Institut N\'{e}el, CNRS, Universit\'{e} Joseph Fourier Grenoble I, 25 rue des Martyrs, Grenoble, France\\
\and
Institut Universitaire de France, 103, bd Saint-Michel, 75005, Paris, France\\
\and
Institut d'Astrophysique Spatiale, CNRS (UMR8617) Universit\'{e} Paris-Sud 11, B\^{a}timent 121, Orsay, France\\
\and
Institut d'Astrophysique de Paris, CNRS (UMR7095), 98 bis Boulevard Arago, F-75014, Paris, France\\
\and
Institute for Space Sciences, Bucharest-Magurale, Romania\\
\and
Institute of Astronomy and Astrophysics, Academia Sinica, Taipei, Taiwan\\
\and
Institute of Astronomy, University of Cambridge, Madingley Road, Cambridge CB3 0HA, U.K.\\
\and
Institute of Theoretical Astrophysics, University of Oslo, Blindern, Oslo, Norway\\
\and
Instituto de Astrof\'{\i}sica de Canarias, C/V\'{\i}a L\'{a}ctea s/n, La Laguna, Tenerife, Spain\\
\and
Instituto de F\'{\i}sica de Cantabria (CSIC-Universidad de Cantabria), Avda. de los Castros s/n, Santander, Spain\\
\and
Istituto di Fisica del Plasma, CNR-ENEA-EURATOM Association, Via R. Cozzi 53, Milano, Italy\\
\and
Jet Propulsion Laboratory, California Institute of Technology, 4800 Oak Grove Drive, Pasadena, California, U.S.A.\\
\and
Jodrell Bank Centre for Astrophysics, Alan Turing Building, School of Physics and Astronomy, The University of Manchester, Oxford Road, Manchester, M13 9PL, U.K.\\
\and
Kavli Institute for Cosmology Cambridge, Madingley Road, Cambridge, CB3 0HA, U.K.\\
\and
LAL, Universit\'{e} Paris-Sud, CNRS/IN2P3, Orsay, France\\
\and
LERMA, CNRS, Observatoire de Paris, 61 Avenue de l'Observatoire, Paris, France\\
\and
Laboratoire AIM, IRFU/Service d'Astrophysique - CEA/DSM - CNRS - Universit\'{e} Paris Diderot, B\^{a}t. 709, CEA-Saclay, F-91191 Gif-sur-Yvette Cedex, France\\
\and
Laboratoire Traitement et Communication de l'Information, CNRS (UMR 5141) and T\'{e}l\'{e}com ParisTech, 46 rue Barrault F-75634 Paris Cedex 13, France\\
\and
Laboratoire de Physique Subatomique et de Cosmologie, Universit\'{e} Joseph Fourier Grenoble I, CNRS/IN2P3, Institut National Polytechnique de Grenoble, 53 rue des Martyrs, 38026 Grenoble cedex, France\\
\and
Laboratoire de Physique Th\'{e}orique, Universit\'{e} Paris-Sud 11 \& CNRS, B\^{a}timent 210, 91405 Orsay, France\\
\and
Lawrence Berkeley National Laboratory, Berkeley, California, U.S.A.\\
\and
Max-Planck-Institut f\"{u}r Astrophysik, Karl-Schwarzschild-Str. 1, 85741 Garching, Germany\\
\and
McGill Physics, Ernest Rutherford Physics Building, McGill University, 3600 rue University, Montr\'{e}al, QC, H3A 2T8, Canada\\
\and
MilliLab, VTT Technical Research Centre of Finland, Tietotie 3, Espoo, Finland\\
\and
National University of Ireland, Department of Experimental Physics, Maynooth, Co. Kildare, Ireland\\
\and
Niels Bohr Institute, Blegdamsvej 17, Copenhagen, Denmark\\
\and
Observational Cosmology, Mail Stop 367-17, California Institute of Technology, Pasadena, CA, 91125, U.S.A.\\
\and
Optical Science Laboratory, University College London, Gower Street, London, U.K.\\
\and
SB-ITP-LPPC, EPFL, CH-1015, Lausanne, Switzerland\\
\and
SISSA, Astrophysics Sector, via Bonomea 265, 34136, Trieste, Italy\\
\and
School of Physics and Astronomy, Cardiff University, Queens Buildings, The Parade, Cardiff, CF24 3AA, U.K.\\
\and
School of Physics and Astronomy, University of Nottingham, Nottingham NG7 2RD, U.K.\\
\and
Space Research Institute (IKI), Russian Academy of Sciences, Profsoyuznaya Str, 84/32, Moscow, 117997, Russia\\
\and
Space Sciences Laboratory, University of California, Berkeley, California, U.S.A.\\
\and
Special Astrophysical Observatory, Russian Academy of Sciences, Nizhnij Arkhyz, Zelenchukskiy region, Karachai-Cherkessian Republic, 369167, Russia\\
\and
Stanford University, Dept of Physics, Varian Physics Bldg, 382 Via Pueblo Mall, Stanford, California, U.S.A.\\
\and
Sub-Department of Astrophysics, University of Oxford, Keble Road, Oxford OX1 3RH, U.K.\\
\and
Theory Division, PH-TH, CERN, CH-1211, Geneva 23, Switzerland\\
\and
UPMC Univ Paris 06, UMR7095, 98 bis Boulevard Arago, F-75014, Paris, France\\
\and
Universit\'{e} de Toulouse, UPS-OMP, IRAP, F-31028 Toulouse cedex 4, France\\
% ...
}

\abstract{ {\bf Abstract:} This paper presents the first cosmological
results based on \planck\ measurements of the cosmic microwave
background (CMB) temperature and lensing-potential power spectra. We
find that the \planck\ spectra at high multipoles ($\ell \gea 40$) are
extremely well described by the standard spatially-flat
six-parameter \lcdm\ cosmology with a power-law spectrum of adiabatic
scalar perturbations.  Within the context of this cosmology,
the \planck\ data determine the cosmological parameters to high
precision: the angular size of the sound horizon at recombination, the
physical densities of baryons and cold dark matter, and the scalar
spectral index are estimated to be $\theta_\ast = (1.04147 \pm
0.00062)\times10^{-2}$, $\Omega_{\rm b}h^2 = 0.02205 \pm 0.00028$,
$\Omega_{\rm c}h^2 = 0.1199 \pm 0.0027$, and $n_{\rm s} = 0.9603 \pm
0.0073$, respectively
\referee{(Note that in this abstract we quote 68\% errors on measured parameters
and 95\% upper limits on other parameters.)} For this cosmology, we find a low
value of the Hubble constant,
$H_0 = (67.3 \pm 1.2) \, {\rm km}\,{\rm s}^{-1}\,{\rm Mpc}^{-1}$,
and a high value of the matter density
parameter, $\Omega_{\rm m} = 0.315 \pm 0.017$. These values are in
tension with recent direct measurements of $H_0$ and the
magnitude-redshift relation for Type Ia supernovae, but are in
excellent agreement with geometrical constraints from baryon acoustic
oscillation (BAO) surveys.  Including curvature, we find that the
Universe is consistent with spatial flatness to percent level
precision using \planck\ CMB data alone.  We use high-resolution CMB
data together with \planck\ to provide greater control on
extragalactic foreground components in an investigation of extensions
to the six-parameter \lcdm\ model. We present selected results from a
large grid of cosmological models, using a range of additional
astrophysical data sets in addition to \planck\ and high-resolution
CMB data. None of these models are  favoured over the standard
six-parameter \lcdm\ cosmology. The deviation of the scalar spectral
index from unity is insensitive to the addition of tensor
modes and to  changes in the matter content of the Universe.
We find an upper limit of $r_{0.002} < 0.11$ on the
tensor-to-scalar ratio.  There is no evidence for additional
neutrino-like relativistic particles beyond the three families of
neutrinos in the standard model. Using BAO and CMB data, we find
$N_{\rm eff}=3.30 \pm 0.27$ for the effective number of relativistic
degrees of freedom, and an upper limit of $0.23 \, {\rm eV}$ for the
sum of neutrino masses. Our results are in excellent agreement with
big bang nucleosynthesis and the standard value of
$N_{\rm eff}=3.046$.  We find no evidence for dynamical dark energy; using BAO
and CMB data, the dark energy equation of state parameter is
constrained to be $w=-1.13 ^{+0.13}_{-0.10}$.  We also use
the \planck\ data to set limits on a possible variation of the
fine-structure constant, dark matter annihilation and primordial
magnetic fields. Despite the success of the six-parameter \lcdm\ model
in describing the \planck\ data at high multipoles, we note that this
cosmology does not provide a good fit to the temperature power
spectrum at low multipoles.  The unusual shape of the spectrum 
in the multipole range $20 \la \ell \la 40$
was seen previously in the \WMAP\ data and is a real
feature of the primordial CMB anisotropies. The poor fit to the 
spectrum at low multipoles is not of decisive significance, but is an
``anomaly'' in an otherwise self-consistent analysis of the \planck\
temperature data.}

\keywords{Cosmology: observations -- Cosmology: theory
 -- cosmic microwave background -- cosmological parameters}

\date{\vspace{-0.2in} \today}
\titlerunning{Cosmological parameters}
\maketitle


\section{Introduction} \label{sec:introduction}

The discovery of the cosmic microwave background (CMB) by
\citet{Penzias:1965} established the modern paradigm of the
hot big bang cosmology. Almost immediately after this seminal discovery,
searches began for  anisotropies in the CMB -- the primordial
signatures of the fluctuations that grew to form the structure that we see
today\footnote{For a good review of the early history of CMB studies see 
\citet{Peebles:09}.}.  \referee{After a number of earlier detections}, convincing evidence for a dipole anisotropy was
reported by \citet{Smoot:77}, but despite many attempts, 
the detection of higher-order anisotropies proved elusive until
the first results from the \textit{Cosmic Background Explorer}
({\it COBE\/};~\citealt{Smoot:92}).
The {\it COBE\/} results established the
existence of a nearly scale-invariant spectrum of primordial fluctuations on
angular scales larger than $7^\circ$, consistent with the predictions of
inflationary cosmology, and stimulated a new generation of precision
measurements of the CMB of which this set of papers forms a part.  

CMB anisotropies are widely recognized as one of the most powerful
probes of cosmology and early-Universe physics.  Given a set of
initial conditions and assumptions concerning the background
cosmology, the angular power spectrum of the CMB anisotropies can be
computed numerically to high precision using linear perturbation
theory (see Sect.~\ref{sec:model}). The combination of precise
experimental measurements and accurate theoretical predictions can be
used to set tight constraints on cosmological parameters. The
influential  results from the \textit{Wilkinson Microwave Anisotropy Probe}
(\WMAP) satellite~\citep{Bennett:03,Spergel:03}, following on from
earlier ground-based and sub-orbital experiments\footnote{It is worth
highlighting here the pre-\WMAP\ constraints on the geometry of the Universe
by the BOOMERang (Balloon Observations of Millimetric Extragalactic Radiation
and Geomagnetics;~\citealt{deBernardis:00}) and MAXIMA (Millimeter-wave
Anisotropy Experiment Imaging Array;~\citealt{Balbi:00}) experiments, for
example.},
demonstrated the power of this approach, which has been followed by all
subsequent CMB experiments.

\begin{figure*}
\centering
\includegraphics[width=180mm, angle=0]{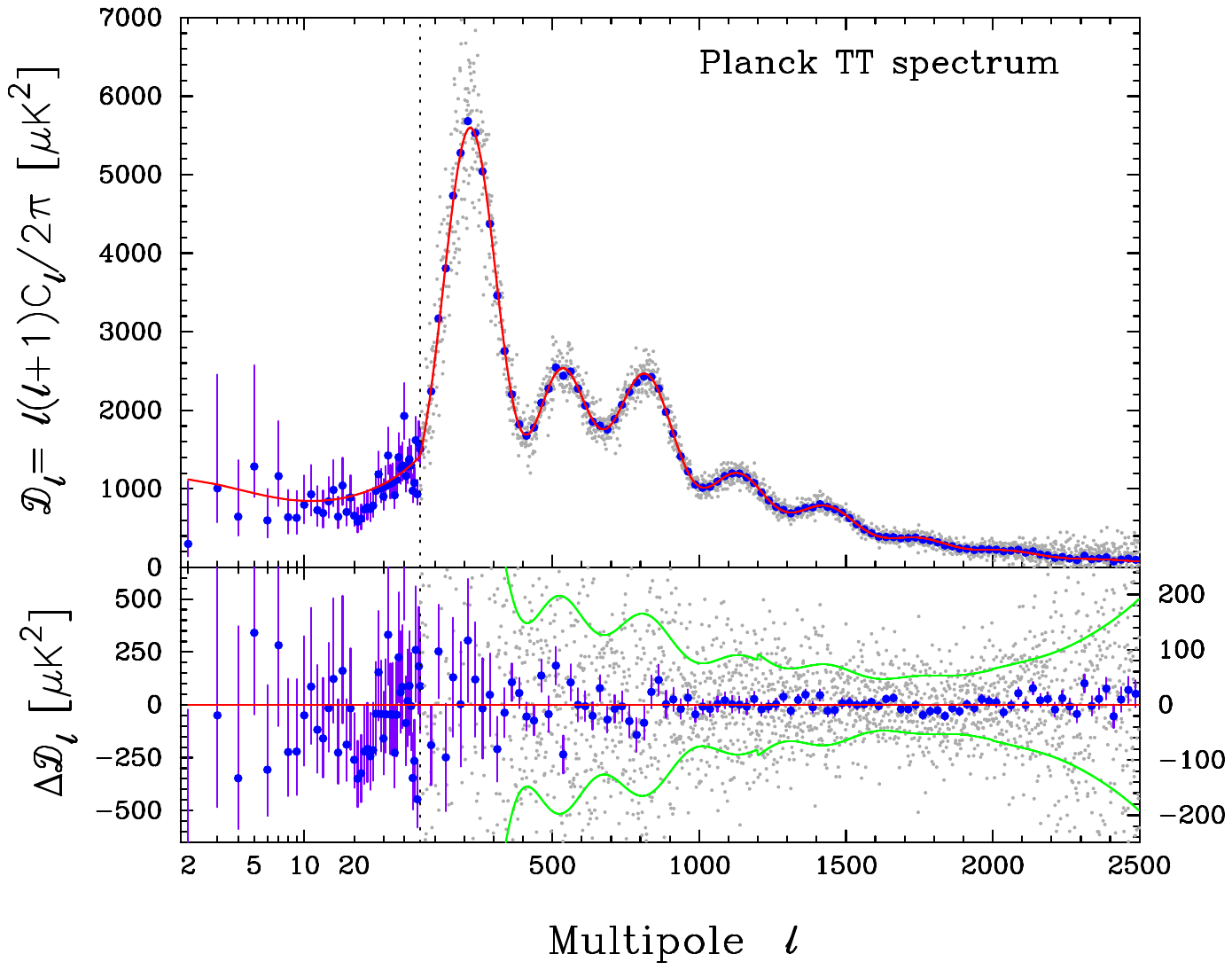}
\caption{\planck\ foreground-subtracted temperature power spectrum
(with foreground and other ``nuisance'' parameters
fixed to their best-fit values for the base \lcdm\ model). The power
spectrum at low multipoles ($\ell=2$--$49$, plotted on a
logarithmic multipole scale) is determined by the {\tt Commander}
algorithm applied to the  \planck\ maps in the frequency range $30$--$353$\,GHz
over 91\% of the sky.  This is used to construct a low-multipole temperature
likelihood using a Blackwell-Rao estimator, as described
in~\citet{planck2013-p08}. The asymmetric error bars show 68\% confidence
limits and include the contribution from uncertainties in foreground
subtraction. At multipoles $50 \le \ell \le 2500$
(plotted on a linear multipole scale) we show the best-fit
The CMB spectrum computed from the {\tt CamSpec} likelihood
(see~\citealt{planck2013-p08}) after removal of unresolved foreground
components. \referee{This spectrum is averaged over the frequency range
$100$--$217\,{\rm GHz}$ using frequency-dependent diffuse sky cuts (retaining
58\% of the sky at 100 GHz and 37\% of the sky at $143$ and $217\,{\rm GHz}$)
and is sample-variance limited to $\ell \sim 1600$.} The light grey points
show the power spectrum multipole-by-multipole. The blue points show averages
in bands of width $\Delta \ell \approx 31$ together with $1\,\sigma$ errors
computed from the diagonal components of the band-averaged covariance matrix
(which includes contributions from beam and foreground uncertainties). The red
line shows the temperature spectrum for the best-fit base \lcdm\ cosmology.
The lower panel shows the power spectrum residuals with respect to this
theoretical model. The green lines show the $\pm 1\,\sigma$ errors on the
individual power spectrum estimates at high multipoles computed from the
{\tt CamSpec} covariance matrix. Note the change in vertical scale in the
lower panel at $\ell=50$.}
\label{TTspec}
\end{figure*}

\planck\footnote{\Planck\ (\url{http://www.esa.int/Planck}) is a project of
the European Space Agency (ESA) with instruments provided by two scientific
consortia funded by ESA member states (in particular the lead countries France
and Italy), with contributions from NASA (USA) and telescope reflectors
provided by a collaboration between ESA and a scientific consortium led and
funded by Denmark.} is the third-generation space mission, following
\textit{COBE} and
\WMAP, dedicated to measurements of the CMB anistropies. The primary
aim of \planck~\citep{planck2005-bluebook} is to measure the temperature
and polarization anisotropies with micro-Kelvin sensitivity per
resolution element over the entire sky. The wide frequency coverage of
\planck\ (30--857\,GHz) was chosen to provide accurate discrimination of
Galactic emission from the primordial anisotropies and to enable a
broad range of ancilliary science, such as detections of galaxy
clusters, extragalactic point sources and 
the properties of Galactic dust emission.
This paper, one of a set
associated with the 2013 release of data from the
\Planck\ mission~\citep{planck2013-p01}, describes the first cosmological
parameter results from the \Planck\ temperature power spectrum.

The results from \WMAP\ (see~\citealt{bennett2012}
and~\citealt{hinshaw2012} for the final nine-year \WMAP\ results)
together with those from high-resolution ground-based CMB experiments
(e.g., \citealt{Reichardt:12,Story:12,Sievers:13}) are remarkably
consistent with the predictions of a ``standard'' cosmological
model. This model is based upon a spatially-flat, expanding Universe
whose dynamics are governed by General Relativity and whose
constituents are dominated by cold dark matter (CDM) and a
cosmological constant ($\Lambda$) at late times.  The primordial seeds
of structure formation are Gaussian-distributed adiabatic fluctuations
with an almost scale-invariant spectrum.  This
model (which will be referred to as the base \lcdm\ model in this
paper) is described by only six key parameters. Despite its
simplicity, the base \lcdm\ model has proved to be successful in
describing a wide range of cosmological data in addition to the CMB,
including the Type Ia supernovae magnitude-distance relation, baryon
acoustic oscillation measurements, the large-scale clustering of
galaxies and cosmic shear (as reviewed in Sect.~\ref{sec:datasets}).

Nevertheless, there have been some suggestions of new physics beyond 
that assumed in the base \lcdm\ model. Examples include various large-angle
``anomalies'' 
in the CMB (as reviewed by the \WMAP\ team in~\citealt{Bennett:11}) and hints
of new physics, such as additional relativistic particles, that might steepen
the high multipole ``damping tail'' of the CMB temperature power
spectrum~\citep{dunkley11,hou12}. Furthermore, developments in
early-Universe cosmology over the last 20 years or so have led to a
rich phenomenology (see e.g.,~\citealt{Baumann:09} for a review). It is
easy to construct models that  preserve the main features of simple
single-field inflationary models, but lead to distinctive observational
signatures such as non-Gaussianity, isocurvature modes or topological
defects.

A major goal of the \planck\ experiment is to test the \lcdm\ model to
high precision and identify areas of tension.  From previous CMB experiments
and other cosmological probes, we know that any departures from the standard
six-parameter \lcdm\ cosmology  are likely to be small and challenging to
detect. \planck, with
its combination of high sensitivity, wide frequency range and all-sky
coverage, is uniquely well-suited to this challenge.

The focus of this paper is to investigate cosmological constraints
from the temperature power spectrum measured by \planck.
Figure~\ref{TTspec} summarizes some important aspects of the
\planck\ temperature power spectrum; we plot this as ${\cal D}_\ell
\equiv \ell(\ell+1)C_\ell/2\pi$ (a notation we will use throughout
this paper) versus multipole $\ell$.  The temperature likelihood used
in this paper is a hybrid: over the multipole range $\ell=2$--$49$,
the likelihood is based on a component-separation algorithm applied to
91\% of the sky~\citep{planck2013-p06,planck2013-p08}.  The likelihood
at higher multipoles is constructed from cross-spectra over the
frequency range 100--217\,GHz, as discussed
in~\citet{planck2013-p08}. It is important to recognize that
unresolved foregrounds (and other factors such as beams and calibration
uncertainties) need to be modelled to high precision to achieve the
science goals of this paper. There is therefore no unique
``\planck\ primordial temperature spectrum''.  Figure 1 is based on a
full likelihood solution for foreground and other ``nuisance''
parameters \emph{assuming a cosmological model}. A change in the
cosmology will lead to small changes in the \planck\ primordial CMB
power spectrum because of differences in the foreground
solution. Neverthess, Fig.~\ref{TTspec} provides a good illustration
of the precision achieved by \planck. The precision is so high that
conventional power spectrum plots (shown in the upper panel of
Fig.~\ref{TTspec}) are usually uninformative. We therefore place high
weight in this paper on plots of \emph{residuals} with respect to the
best-fit model (shown in the lower panel). Figure~\ref{TTspec} also
serves to illustrate the highly interconnected nature of this series
of papers. The temperature likelihood used in this paper utilizes data
from both the \planck\ Low Frequency Instrument (LFI) and High
Frequency Instrument (HFI). The data-processing chains for these two
instruments and beam calibrations are described
in~\citet{planck2013-p02}, \citet{planck2013-p03}, and associated papers
\citep{planck2013-p02a,planck2013-p02d,planck2013-p02b,planck2013-p03c,planck2013-p03f,planck2013-p03d,planck2013-p03e}.
 Component
separation is described in~\citet{planck2013-p06} and the temperature
power spectrum and likelihood, as used in this paper, are described
in~\citet{planck2013-p08}.  \citet{planck2013-p08} also presents a
detailed analysis of the robustness of the likelihood to various
choices, such as frequency ranges and sky masks (and also compares the
likelihood to results from an independent likelihood code based on
different assumptions, see also Appendix \ref{app:test}). Consistency of the \planck\ maps across
frequencies is demonstrated in~\citet{planck2013-p01a}, and the level
of consistency with \WMAP\ is assessed.

This paper is closely linked to other papers reporting cosmological results in this series.
We make heavy use of the gravitational lensing power spectrum and likelihood
estimated from an analysis of the 4-point function of the \planck\ maps~\citep{planck2013-p12}. The present
paper concentrates on simple parameterizations of the spectrum of primordial
fluctuations. Tests of specific models of inflation, isocurvature modes, broken
scale-invariance etc.\ are discussed in~\citet{planck2013-p17}. Here, we
assume throughout that the initial fluctuations are Gaussian and statistically
isotropic. Precision tests of non-Gaussianity, from \planck\ estimates of the
3- and 4-point functions of the temperature anisotropies, are presented in~\citet{planck2013-p09a}. Tests of isotropy and additional tests of non-Gaussianity 
using \planck\ data are discussed in~\citet{planck2013-p09} and
\citet{planck2013-p19}.

The outline of the paper is as follows.
In Sect.~\ref{sec:model} we define our notation and cosmological
parameter choices. This section also summarizes aspects
of the  Markov chain Monte Carlo (MCMC) sampler used in this paper and of the
CMB Boltzmann code used to predict theoretical temperature power spectra.
Section~\ref{sec:planck_lcdm} presents results on cosmological parameters
using \planck\ data alone.  For this data release we do not use \planck\
polarization data in the likelihood,
and we therefore rely on \WMAP\ polarization data at low multipoles to constrain
the optical depth, $\tau$,  from reionization. An interesting aspect
of Sect.~\ref{sec:planck_lcdm}  is to assess whether CMB gravitational lensing
measurements from \planck\ 
can be used to constrain the optical depth without the
use of  \WMAP\ polarization measurements.

Section~\ref{sec:highell} introduces additional CMB temperature data
from high-resolution experiments. This section presents a detailed
description of how we have modified the \planck\ model for unresolved
foreground and ``nuisance'' parameters introduced
in~\citet{planck2013-p08} to enable the \planck\ spectra to be used
together with those from other CMB experiments. Combining high-resolution CMB
experiments with \planck\ mitigates the effects of unresolved
foregrounds which, as we will show, can affect cosmological parameters
(particularly for extensions to the base \lcdm\ model) if the
foreground parameters are allowed too much
freedom. Section~\ref{sec:highell} ends with a detailed analysis of
whether the base \lcdm\ model provides an acceptable fit to the CMB
temperature power spectra from \planck\ and other experiments.

It is well known that certain cosmological parameter combinations are
highly degenerate using CMB power spectrum measurements alone~\citep{Zaldarriaga_cosmo:97,Efstathiou:99,Howlett:2012mh}. These degeneracies can be broken by combining
with other cosmological data (though the \planck\ lensing analysis
does help to break the principal ``geometrical'' degeneracy,  as discussed
in Sect.~\ref{sec:lensing}). Section~\ref{sec:datasets} discusses additional
``astrophysical'' data that are
used in combination with \planck. Since the \planck\ temperature data
are so precise, we have been selective in the additional data sets that
we have chosen to use. Section~\ref{sec:datasets} discusses our rationale for making these choices.

Having made a thorough investigation of the base \lcdm\ model,
Sect.~\ref{sec:grid} describes extended models, including models
with non-power-law spectral indices, tensor modes, curvature,
additional relativistic species, neutrino masses and dynamical dark energy. This
section also discusses constraints on models with annihilating dark matter, primordial magnetic
fields and a time-variable fine-structure constant.

Finally, we present our conclusions in Sect.~\ref{sec:conclusions}.
Appendix~\ref{app:wmap} compares the \planck\ and \WMAP\ base \lcdm\ cosmologies. Appendix~\ref{app:spt} contrasts the \planck\ best-fit \lcdm\ cosmology with that determined recently by combining data from the South Pole Telescope with \WMAP~\citep{Story:12}. Appendix~\ref{app:test} discusses the dependence of our results for extended models on foreground modelling and likelihood choices, building on the discussion in~\citet{planck2013-p08} for the base \lcdm\ model.

\referee{Since the appearance of the first draft of this paper, there have
  been a number of developments that affect both the \Planck\ data and
  some of the constraints from supplementary astrophysical data used
  in this paper.

 The primary developments are as follows. [1] After the submission of
 this paper, we discovered a minor error in the ordering of the beam
 transfer functions applied to each of the \camspec\ $217\times 217\,{\rm GHz}$
 cross-spectra before their coaddition to form a single spectrum. Correcting for this error changes the mean
 $217\times217\,{\rm GHz}$ spectrum by a smooth function with an amplitude of a
 few $\muKsq$. An extensive analysis of a revised likelihood showed
 that this error has negligible impact on cosmological parameters and
 that it is absorbed by small shifts in the foreground
 parameters. Since the effect is so minor, we have decided not to
 change any of the numbers in this paper and not to revise the public
 version of the \camspec\ likelihood.  [2] The foreground-corrected
 $217\times217\,{\rm GHz}$ spectrum shows a small negative residual (or ``dip'')
 with respect to the best-fit base \LCDM\ theoretical model at multipoles $\ell
 \approx 1800$. This can be seen most clearly in
 Fig. \ref{PlanckandHighL} in this paper. After submission of this
 paper we found  evidence that this feature is  
a  residual systematic in the data
 associated with incomplete 4\,K line removal (see \citealt{planck2013-p03} for a discussion of
the 4\,K line removal algorithm). The 4\,K lines, at specific
 frequencies in the detector timelines, are caused by an
 electromagnetic-interference/electromagnetic-compatibility (EMI-EMC)
 problem between the $^4{\rm He}$ Joule-Thomson (4\,K) cooler drive electronics and the read-out
 electronics. This interference is time-variable.  Tests in which we
 have applied more stringent flagging of 4\,K lines show that the
 $\ell=1800$ feature is reduced to negligible levels in all sky
 surveys, including Survey 1 in which the effect is strongest. 
 The 2014 \Planck\ data release will include improvements in
 the 4\,K line removal.  It is important to emphasise that this
 systematic is a small effect.  Analysis of cosmological parameters,
 removing the multipole range around $\ell = 1800$ (and also analysis
 of the full mission data, where the effect is diluted by the
 additional sky surveys) shows that the impact of this feature on
 cosmological parameters is small (i.e., less than half a standard
 deviation) even for extensions to the base \LCDM\ cosmology. Some quantitiative tests of the impact of this systematic on cosmology are summarized in 
Appendix \ref{app:test}. [3]
 An error  was found in the dark energy model used for theoretical predictions
with equation of state $w\ne -1$, leading to few-percent $C_\ell$ errors at very low multipoles in
extreme models with $w\agt -0.5$. We have checked, using the corrected October
2013 \CAMB\ version, that this propagates to only a very small error on
marginalized parameters and that the results presented in this
paper are consistent to within the stated numerical accuracy.
   [4] After
  this paper was submitted, \cite{Humphreys:2013} presented the final
  results of a long-term campaign to establish a new geometric maser
  distance to NGC4258. Their revised distance of $(7.60 \pm 0.23) \, {\rm
    Mpc}$ leads to a lowering of the Hubble constant, based on the
  Cepheid distance scale, to $H_0 = (72.0 \pm 3.0) \, {\rm km}\, {\rm
    s}^{-1}\, {\rm Mpc}^{-1}$,  partially alleviating the tension between
  the \citet{Riess:2011yx} results and the \Planck\ results on $H_0$
  discussed  in Sect.~\ref{sec:hubble} and subsequent sections. [5]
  In a recent paper, \citet{Betoule:2013} present results from an
  extensive programme that improves the photometric calibrations of
  the SDSS and SNLS supernovae surveys. An analysis of the SDSS-II and SNLS
supernovae samples,  including revisions  to the photometric calibrations, 
  favours a higher value
  of $\Omega_{\rm m}=0.295\pm 0.034$  for  the  base \LCDM\ model,
  consistent with the \Planck\ results
  discussed in Sect.~\ref{subsec:datasetsSNe} \citep{Betoule:2014}.

  A detailed discussion of the impact of the changes discussed here on
  cosmology will be deferred until the \Planck\ 2014 data release,
  which will include improvements to the low-level data processing
  and, by which time, improved complementary astrophysical data sets
  (such as a revised SNLS compilation) should be available to us.  In
  revising this paper, we have taken the view that this, and other
  \Planck\ papers in this 2013 release, should be regarded as a
  snapshot of the \Planck\ analysis as it was in early 2013.  We have
  therefore kept revisions to a minimum. Nevertheless, readers of this
  paper, and users of products from the \Planck\ Legacy
  Archive\footnote{\url{http://www.sciops.esa.int/index.php?project=planck&page=Planck_Legacy_Archive}}
  (such as parameter tables and MCMC chains), should be aware of
  developments since the first submission of this paper.

}

\section{Model, parameters, and methodology} \label{sec:model}

\begin{table*}[tmb] 
\begingroup 
\newdimen\tblskip \tblskip=5pt
\caption{Cosmological parameters used in our analysis. For each, we give the symbol, prior range, value taken in the base \lcdm\ cosmology (where appropriate), and summary definition (see text for details).
The top block contains parameters with uniform priors that are varied in the
MCMC chains. The ranges of these priors are listed in square brackets.
The lower blocks define various derived parameters.}
\label{tab:params}
\vskip -5mm
\footnotesize 
\setbox\tablebox=\vbox{ %
\newdimen\digitwidth 
\setbox0=\hbox{\rm 0}
\digitwidth=\wd0
\catcode`*=\active
\def*{\kern\digitwidth}
\newdimen\signwidth
\setbox0=\hbox{+}
\signwidth=\wd0
\catcode`!=\active
\def!{\kern\signwidth}
\halign{\hbox to 2.7cm{#\leaderfil}\tabskip=0.4cm& \hfil#\hfil\tabskip=0.6cm&
 \hfil#\hfil\tabskip=0.6cm&  #\hfil\tabskip=0pt\cr
\noalign{\doubleline}
\omit\hfil Parameter\hfil&\omit\hfil Prior range\hfil&\omit\hfil Baseline\hfil&\omit\hfil Definition\hfil\cr
\noalign{\vskip 3pt\hrule\vskip 3pt}
$\omb \equiv \Omb h^2$& $[0.005, 0.1]$ & \dots& Baryon density today\cr
$\omc \equiv \Omc h^2$& $[0.001, 0.99]$& \dots& Cold dark matter density today\cr
$100\theta_{\mathrm{MC}}$ & $[0.5, 10.0]$ & \dots& $100\,{\times}$ approximation to $\rstar/D_{\rm A}$ (CosmoMC)\cr
$\tau                $&   $[0.01, 0.8]$ & \dots& Thomson scattering optical depth due to reionization\cr
$\Omk            $&  $[-0.3, 0.3]$ & 0& Curvature parameter today with $\Omtot= 1 - \Omk$\cr
$\mnu        $& $[0, 5]$ & $0.06$ & The sum of neutrino masses in eV\cr
$\mnusterile$&  $[0, 3]$ &0&Effective mass of sterile neutrino in eV\cr
$w_0                $& $[-3.0, -0.3]$ & $-1$& Dark energy equation of state$^{a}$, $w(a) = w_0 + (1-a) w_a$\cr
$w_a                 $& $[-2, 2]$ & 0&  As above (perturbations modelled using PPF)\cr
$\neff       $& $[0.05, 10.0]$ & 3.046& Effective number of neutrino-like relativistic degrees of freedom (see text)\cr
$\yhe                $& $[0.1, 0.5]$ & BBN& Fraction of baryonic mass in helium\cr
$\Alens        $& $[0,  10]$& 1& Amplitude of the lensing power relative to the physical value\cr
$\ns           $& $[0.9, 1.1]$ & \dots& Scalar spectrum power-law index ($k_0 = 0.05\Mpc^{-1}$)\cr
$\nt           $& $\nt = -r_{0.05}/8$ & Inflation& Tensor spectrum power-law index ($k_0 = 0.05\Mpc^{-1}$)\cr
$\nrun$&   $[-1, 1]$ & 0& Running of the spectral index\cr
$\ln(10^{10}\As) $& $[2.7, 4.0]$ & \dots& Log power of the primordial curvature perturbations ($k_0 = 0.05\,\Mpc^{-1}$)\cr
$\rpivot          $& $[0, 2]$ & 0& Ratio of tensor primordial power to curvature power at $k_0 = 0.05\,\Mpc^{-1}$\cr
\noalign{\vskip 3pt\hrule\vskip 3pt}
$\Oml      $&     & \dots& Dark energy density divided by the critical density today\cr
$t_0                 $&  & \dots& Age of the Universe today (in Gyr)\cr
$\Omm     $&  & \dots& Matter density (inc.\ massive neutrinos) today divided by the critical density\cr
$\sigma_8            $&   & \dots& RMS matter fluctuations today in linear theory\cr
$\zre                $&   & \dots& Redshift at which Universe is half reionized\cr
$H_0                 $&[20,100] & \dots& Current expansion rate in $\rm{km}\, \rm{s}^{-1}\Mpc^{-1}$\cr
$r_{0.002}           $&   & 0& Ratio of tensor primordial power to curvature power at $k_0 = 0.002\,\Mpc^{-1}$\cr
$10^9 \As      $&    & \dots& $10^9\,\times$
dimensionless curvature power spectrum at $k_0 = 0.05\,\Mpc^{-1}$\cr
$\omm\equiv\Omm h^2$&  & \dots& Total matter density today (inc.\ massive neutrinos)\cr
\noalign{\vskip 3pt\hrule\vskip 3pt}
$\zstar            $&   & \dots& Redshift for which the optical depth equals unity (see text)\cr
$\rstar=\rs(\zstar) $&    & \dots& Comoving size of the sound horizon at $z = z_\ast$\cr
$100\theta_\ast         $&    & \dots& $100\,\times$ angular size of
sound horizon at $z=\zstar$ ($\rstar/D_{\rm A})$ \cr
$\zdrag       $&     & \dots& Redshift at which baryon-drag optical depth equals unity (see text)\cr
$\rdrag=\rs(\zdrag)$&   & \dots& Comoving size of the sound horizon at $z = \zdrag$\cr
$k_{\rm D}           $&   & \dots& Characteristic damping comoving wavenumber ($\Mpc^{-1}$)\cr
$100\theta_{\rm D}      $&   & \dots& $100\,\times$ angular extent of photon diffusion at last scattering (see text)\cr
$\zeq          $&   & \dots& Redshift of matter-radiation equality (massless neutrinos)\cr
$100\theta_{\rm eq}     $&  & \dots& $100\,\times$ angular size of the comoving horizon at matter-radiation equality\cr
$\rdrag/D_{\mathrm{V}}(0.57)$ &   &\dots& BAO distance
ratio at $z=0.57$ (see Sect.~\ref{sec:BAO})\cr
\noalign{\vskip 3pt\hrule\vskip 3pt}}}
\endPlancktablewide 
\tablenote \textit{a} For dynamical dark energy models with constant equation of state, we denote the equation of state by $w$ and adopt the same prior as for $w_0$.\par
\endgroup
\end{table*}

\medskip

\subsection{Theoretical model}

We shall treat anisotropies in the CMB as small fluctuations about a
Friedmann-Robertson-Walker metric whose evolution is described by
General Relativity.  We shall not consider modified gravity scenarios
or ``active'' sources of fluctuations such as cosmic defects.  The
latter are discussed in ~\citet{planck2013-p20}.  Under our assumptions, the
evolution of the perturbations can be computed accurately using a CMB
Boltzmann code once the initial conditions, ionization history and
constituents of the Universe are specified.  We discuss each of these
in this section, establishing our notation.  Our conventions are
consistent with those most commonly adopted in the field and in
particular with those used in the
\CAMB\footnote{\url{http://camb.info}} Boltzmann code
\citep{Lewis:1999bs}, which is the default code used in this paper.

\subsubsection{Matter and radiation content}

We adopt the usual convention of writing the Hubble constant at the
present day as $H_0=100\,h\,{\rm km}\,{\rm s}^{-1}\,{\rm Mpc}^{-1}$.
For our baseline model, we assume that the cold dark matter is
pressureless, stable and non-interacting, with a physical density
$\omega_{\rm c} \equiv\Omc h^2$.  The baryons, with density
$\omega_{\rm b} \equiv \Omb h^2$, are assumed to consist almost
entirely of hydrogen and helium; we parameterize the mass fraction in
helium by $\yhe$.  The process of standard big bang nucleosynthesis
(BBN) can be accurately modelled, and gives a predicted relation
between $\yhe$, the photon-baryon ratio, and the expansion rate (which
depends on the number of relativistic degrees of freedom).  By default
we use interpolated results from the {\tt PArthENoPE} BBN code
\citep{Pisanti:2007hk} to set $\yhe$, following \cite{Hamann:2011ge},
which for the \Planck\ best-fitting base model
(assuming no additional relativistic
components and negligible neutrino degeneracy)
gives $\yhe = 0.2477$.
We shall compare our results with the predictions of
BBN in Sect.~\ref{subsec:BBN}.

The photon temperature today is well measured to be
$T_0 = 2.7255\pm 0.0006\,\mathrm{K}$ \citep{Fixsen:2009ug};
we adopt $T_0 = 2.7255\,\mathrm{K}$ as our fiducial value.
We assume full thermal equilibrium prior to neutrino decoupling.
The decoupling of the neutrinos is nearly, but not entirely, complete
by the time of electron-positron annihilation.
This leads to a slight heating of the neutrinos in addition to that
expected for the photons and hence to a small departure from the
thermal equilibrium prediction $T_\gamma = (11/4)^{1/3} T_\nu$ between the
photon temperature $T_\gamma$ and the neutrino temperature $T_\nu$.
We account for the additional energy density in neutrinos by
assuming that they have a thermal distribution with an effective
energy density
\be
  \rho_\nu = \nnu\,
    \frac{7}{8}\left(\frac{4}{11}\right)^{4/3}\rho_\gamma,
    \label{def:Neff}
\ee
with $\nnu =3.046$ in the baseline model~\citep{Mangano:2001iu,Mangano:2005cc}.
This density is divided equally between three neutrino species
while they remain relativistic.

In our baseline model we assume a minimal-mass normal hierarchy for the
neutrino masses, accurately approximated for current cosmological data
as a single massive eigenstate with $m_\nu=0.06\, \eV$ ($\Omega_\nu h^2
\approx \sum m_\nu/93.04\,\eV \approx 0.0006$; corrections and
uncertainties at the ${\rm meV}$ level are well below the accuracy
required here).  This is consistent with global fits to recent
oscillation and other data \citep{Tortola:2012te}, but is not the only
possibility.  We discuss more general neutrino mass constraints
in Sect.~\ref{sec:neutrino}.

We shall also consider the possibility of extra radiation, beyond that
included in the Standard Model. We model this as
additional massless neutrinos contributing to the total $\nnu$
determining the radiation density as in Eq.~(\ref{def:Neff}).  We keep
the mass model and heating consistent with the baseline model at
$\nnu=3.046$, so there is one massive neutrino with $\nnu^{(\rm
  massive)}=3.046/3 \approx 1.015$, and massless neutrinos with
$\nnu^{(\rm massless)}=\nnu-1.015$.  In the case where $\nnu < 1.015$
we use one massive eigenstate with reduced temperature.

\subsubsection{Ionization history}

To make accurate predictions for the CMB power spectra, the background
ionization history has to be calculated to high accuracy.  Although
the main processes that lead to recombination at $z\approx 1090$ are well
understood, cosmological parameters from \planck\ can be sensitive to
sub-percent differences in the ionization fraction
$x_{\rm e}$ \citep{Hu:1995fqa,Lewis:2006ym,RubinoMartin:2009ry,Shaw:2011ez}.
The process of recombination takes the Universe from a state of fully
ionized hydrogen and helium in the early Universe, through to the
completion of recombination with residual fraction $x_{\rm e} \sim 10^{-4}$.
Sensitivity of the CMB power spectrum to $x_{\rm e}$ enters through changes
to the sound horizon at recombination, from changes in the timing of
recombination, and to the detailed shape of the recombination transition,
which affects the thickness of the last-scattering surface and hence the
amount of small-scale diffusion (Silk) damping, polarization, and line-of-sight
averaging of the perturbations.

Since the pioneering work of~\cite{Peebles:68} and \cite{Zeldovich:69},
which identified the main physical processes involved in recombination,
there has been significant progress in numerically modelling the many
relevant atomic transitions and processes that can affect the details
of the recombination process \citep{Hu:1995fqa,Seager:1999km,Wong:2007ym,Hirata:2007sp,
Switzer:2007sn,RubinoMartin:2009ry,Grin:2009ik,Chluba:2010ca,AliHaimoud:2010ym,AliHaimoud:2010dx}.
In recent years a consensus has emerged between the results of two
multi-level atom codes
\HYREC\footnote{\url{http://www.sns.ias.edu/~yacine/hyrec/hyrec.html}}~\citep{Switzer:2007sn,Hirata:2008ny,AliHaimoud:2010dx},
and
\COSMOREC\footnote{\url{http://www.chluba.de/CosmoRec/}}~\citep{Chluba:2010fy,Chluba:2010ca},
demonstrating agreement at a level better than that required for
\planck\ (differences less that $4\times 10^{-4}$ in the predicted
temperature power spectra on small scales).

These recombination codes are remarkably fast,  given the complexity of the calculation.
However, the recombination history can be computed even more rapidly by
using the simple effective three-level atom model developed by
\cite{Seager:1999km} and implemented in the \RECFAST\
code\footnote{\url{http://www.astro.ubc.ca/people/scott/recfast.html}},
with appropriately chosen small correction functions calibrated to the
full numerical results \citep{Wong:2007ym,RubinoMartin:2009ry,Shaw:2011ez}.
We use \RECFAST\ in our baseline parameter analysis, with correction functions
adjusted so that the predicted power spectra $C_\ell$ agree with those from
the latest versions of \HYREC\ (January 2012) and \COSMOREC\ (v2) to
better than $ 0.05\%$\footnote{The updated \RECFAST\ used here in the baseline model is
publicly available as version 1.5.2 and is the default in \CAMB\ as of
October 2012.}.
We have confirmed, using importance sampling, that cosmological parameter
constraints using \RECFAST\ are consistent with those using \COSMOREC\ at
the $0.05\,\sigma$ level.
Since the results of the \planck\ parameter analysis are crucially dependent
on the accuracy of the recombination history, we have also checked, following~\cite{Lewis:2006ym},
that there is no strong evidence for simple deviations from the assumed history. However, we note that any deviation from
the assumed history could significantly shift parameters compared to the results presented here and we have not
performed a detailed sensitivity analysis.

The background recombination model should accurately capture the
ionization history until the Universe is reionized at late times via
ultra-violet photons from stars and/or active galactic nuclei.  We
approximate reionization as being relatively sharp, with the mid-point
parameterized by a redshift $\zre$ (where $x_{\rm e}=f/2$) and width
parameter $\Delta z_{\rm re}=0.5$.  Hydrogen reionization and the
first reionization of helium are assumed to occur simultaneously, so
that when reionization is complete $x_{\rm e }= f \equiv
1+f_{\rm{He}}\approx 1.08$ \citep{Lewis:2008wr}, where $f_{\rm He}$ is the
helium-to-hydrogen ratio by number.  In this
parameterization, the optical depth is almost independent of $\Delta
z_{\rm re}$ and the only impact of the specific functional form on
cosmological parameters comes from very small changes to the shape of
the polarization power spectrum on large angular scales.  The second reionization of
helium (i.e., ${\rm He}^{+}\to{\rm He}^{++}$) produces
very small changes to the power spectra ($\Delta\tau \sim 0.001$,
where $\tau$ is the optical depth to Thomson scattering) and does not
need to be modelled in detail.  We include the second reionization of
helium at a fixed redshift of $z=3.5$ (consistent with observations of
Lyman-$\alpha$ forest lines in quasar spectra, e.g.,~\citealt{Becker:11}),
which is sufficiently accurate for the parameter analyses described in
this paper.

\subsubsection{Initial conditions}

In our baseline model we assume  purely adiabatic scalar perturbations
at very early times, with a (dimensionless) curvature power spectrum parameterized by
\begin{equation}
  \clp_\clr(k) = \As
    \left(\frac{k}{k_0}\right)^{\ns-1+(1/2)(\nrun) \ln(k/k_0)}, \label{PS1}
\end{equation}
with $n_{\rm s}$ and $\nrun$ taken to be constant.  For most of this paper we
shall assume no ``running'', i.e., a power-law spectrum with $d \ns / d\ln k = 0$.
The pivot scale, $k_0$, is chosen to be $k_0=0.05\,\Mpc^{-1}$, roughly in
the middle of the logarithmic range of scales probed by \planck.
With this choice, $\ns$ is not strongly degenerate with the amplitude
parameter $\As$.

The amplitude of the small-scale linear CMB power spectrum is proportional to
$e^{-2\tau}A_{\rm s}$.  Because \planck\ measures this amplitude very
accurately there is  a tight linear constraint between $\tau$ and
$\ln A_{\rm s}$ (see Sect.~\ref{subsec:tau}).
For this reason we usually use $\ln A_{\rm s}$ as a base parameter with
a flat prior, which has a significantly more Gaussian posterior than
$A_{\rm s}$. A linear parameter redefinition then also allows the degeneracy
between $\tau$ and $A_{\rm s}$ to be explored efficiently.
(The degeneracy between $\tau$ and $A_{\rm s}$ is broken by the relative
amplitudes of large-scale temperature and polarization CMB anisotropies and by
the non-linear effect of CMB lensing.)

We shall also consider  extended models with a significant amplitude of primordial
gravitational waves (tensor modes). Throughout this paper, the (dimensionless) tensor mode spectrum
is parameterized as a power-law with\footnote{For a transverse-traceless spatial tensor $H_{ij}$, the tensor
  part of the metric is $ds^2 = a^2[d\eta^2 - (\delta_{ij}+2H_{ij})
  dx^i dx^j]$, and $\clp_{\rm t}$ is defined so that $\clp_{\rm t}(k)
  = \partial_{\ln k} \langle 2 H_{ij} 2H^{ij}\rangle$.}
\be
 \clp_{\rm t}(k)=\At \left(\frac{k}{k_0}\right)^{\nt} .
\ee
We define $\rpivot \equiv A_{\rm t}/A_{\rm s}$, the primordial tensor-to-scalar ratio at
$k=k_0$.
Our constraints are only weakly sensitive to the tensor spectral index,
$n_{\rm t}$ (which is assumed to be close to zero), and we adopt the theoretically motivated
single-field inflation consistency relation $n_{\rm t}=-\rpivot /8$,
rather than varying $n_{\rm t}$ independently.
We put a flat prior on $\rpivot$, but also report the constraint at
$k=0.002\,\Mpc^{-1}$ (denoted $r_{0.002}$),
which is closer to the scale at which there is some sensitivity to tensor modes in the large-angle temperature power spectrum. Most previous CMB experiments
have reported constraints on $r_{0.002}$.
For further discussion of the tensor-to-scalar ratio and its
implications for inflationary models see~\citet{planck2013-p17}.

\subsubsection{Dark energy}

In our baseline model we assume that the dark energy is a cosmological
constant with current density parameter $\Oml$. When considering a dynamical
dark energy component,  we parameterize the equation of state either as a constant $w$ or as a function of the cosmological scale factor, $a$, with
\be
  w(a) \equiv \frac{p}{\rho} = w_0 + (1-a)w_a,  \label{DE0}
\ee
and assume that the dark energy does not interact with other constituents
other than through gravity. Since this model allows the equation of
state to cross below $-1$, a single-fluid model cannot be used
self-consistently.  We therefore use the parameterized post-Friedmann (PPF)
model of~\citet{Fang:2008sn}. For models with
$w>-1$, the PPF model agrees with fluid
models to significantly better accuracy than required for the
results reported in this paper.

\subsubsection{Power spectra}

Over the last decades there has been significant progress in improving
the accuracy, speed and generality of the numerical calculation of the
CMB power spectra given an ionization history and set of cosmological
parameters \citep[see e.g.,][]{Bond:87,Sugiyama:95,Ma:1995ey,
Hu:1995fqa,Seljak:1996is,Hu:1997hp,Zaldarriaga:1997va,Lewis:1999bs,
Lesgourgues:2011rh}.
Our baseline
numerical Boltzmann code is \CAMB\footnote{\url{http://camb.info}}
\citep[March 2013;][]{Lewis:1999bs}, a parallelized line-of-sight code
developed from \CMBFAST\ \citep{Seljak:1996is} and \COSMICS\
\citep{Bertschinger:1995er,Ma:1995ey}, which calculates the lensed CMB
temperature and polarization power spectra.  The code has been
publicly available for over a decade and has been very well tested
(and improved) by the community.  Numerical stability and accuracy of
the calculation at the sensitivity of \planck\ has been explored in
detail \citep{Hamann:2009yy,Lesgourgues:2011rg,Howlett:2012mh},
demonstrating that the raw numerical precision is sufficient for
numerical errors on parameter constraints from \planck\ to be less
than $10\%$ of the statistical error around the assumed cosmological
model.  (For the high multipole CMB data at $\ell>2000$ introduced in
Sect. \ref{sec:highell}, the default \CAMB\ settings are adequate
because the power spectra of these experiments are dominated by
unresolved foregrounds and have large errors at high multipoles.)  To
test the potential impact of \CAMB\ errors, we importance-sample a
subset of samples from the posterior parameter space using higher
accuracy settings.  This confirms that differences purely due to
numerical error in the theory prediction are less than $10\%$ of the
statistical error
for all parameters, both with and without inclusion of CMB data at
high multipoles.
We also performed additional tests of the robustness and accuracy of our
 results by reproducing a fraction of them with the independent
 Boltzmann code \CLASS\ \citep{Lesgourgues:2011re,Blas:2011rf}.

In the parameter analysis, information from CMB lensing enters in two
ways. Firstly, all the CMB power spectra are modelled using the lensed
spectra, which includes the approximately 5\% smoothing effect on
the acoustic peaks due to lensing. Secondly, for some results we
include the \planck\ lensing likelihood, which encapsulates the
lensing information in the (mostly squeezed-shape) CMB trispectrum via
a lensing potential power spectrum \citep{planck2013-p12}.
The theoretical predictions for the lensing potential power spectrum are
calculated by \CAMB, optionally with corrections for the non-linear matter power spectrum, along with the (non-linear) lensed
CMB power spectra. For the \planck\ temperature power spectrum, corrections
to the lensing effect due to non-linear structure growth can be neglected,
however the impact on the lensing potential reconstruction is important.
We use the \HALOFIT\ model \citep{Smith:2002dz} as updated by
\cite{Takahashi:2012em} to model the impact of non-linear growth on the
theoretical prediction for the lensing potential power.

\subsection{Parameter choices}

\subsubsection{Base parameters}

The first section of Table~\ref{tab:params}
lists our base parameters that have flat priors
when they are varied, along with their default values in the baseline model.
When parameters are varied, unless otherwise stated, prior ranges are
chosen to be much larger than the posterior, and hence do not affect
the results of parameter estimation. In addition to these priors,
we impose a ``hard'' prior on the Hubble constant of $[20, 100]
\ {\rm km}\, {\rm s}^{-1}\, {\rm Mpc}^{-1}$.

\subsubsection{Derived parameters}

Matter-radiation equality $\zeq$ is defined as the redshift at
which $\rho_\gamma+\rho_\nu = \rho_{\rm c}+\rho_{\rm b}$
(where $\rho_\nu$ approximates massive neutrinos as massless).

The redshift of last-scattering, $z_\ast$, is defined so that the optical
depth to Thomson scattering from $z=0$ (conformal time $\eta = \eta_0$)
to $z=z_\ast$ is unity, assuming no reionization.  The optical depth is
given by
\begin{equation}
  \tau(\eta) \equiv \int_{\eta_0}^\eta \dot\tau\ d\eta^\prime,
\end{equation}
where $\dot\tau = - an_{\rm e}\sigma_{\rm T}$ (and $n_{\rm e}$ is the density of free electrons and $\sigma_{\rm T}$ is the Thomson
cross section).  We define the angular scale of the sound horizon at last-scattering,
$\theta_\ast =r_{\rm s}(z_\ast)/D_{\rm A}(z_\ast)$, where $r_{\rm s}$ is
the sound horizon
\begin{equation}
  r_{\rm s}(z) = \int_0^{\eta(z)}  \frac{d\eta^\prime}{\sqrt{3(1+R)}},
\end{equation}
with $R \equiv 3 \rho_{\rm b}/(4\rho_\gamma)$. The parameter $\theta_{\mathrm{MC}}$
in Table 1 is an approximation to $\theta_\ast$ that is used in \COSMOMC\ and
is based on fitting formulae given in \citet{Hu:1995en}.

Baryon velocities decouple from the photon dipole when Compton
drag balances the gravitational force,  which happens at
$\tau_{\rm d} \sim 1$, where \citep{Hu:1995en}
\begin{equation}
  \tau_{\rm d}(\eta) \equiv \int^\eta_{\eta_0} \dot\tau\ d\eta^\prime/R.
\end{equation}
Here, again, $\tau$ is from recombination only, without reionization
contributions.
We define a drag redshift $z_{\rm drag}$, so that
$\tau_{\rm d}(\eta(z_{\rm drag})) = 1$.
The sound horizon at the drag epoch is an important scale that is often
used in studies of baryon acoustic oscillations; we denote this as
$r_{\rm drag}=r_{\rm s}(z_{\rm drag})$.  We compute $z_{\rm drag}$ and
$r_{\rm drag}$ numerically from \CAMB\ (see Sect.~\ref{sec:BAO} for details of application to BAO data).

The characteristic wavenumber for damping, $k_{\rm D}$, is given by
\begin{equation}
k_{\rm D}^{-2}(\eta) = -\frac{1}{6} \int_0^\eta d\eta^\prime
  \ \frac{1}{\dot\tau}\ \frac{R^2 + 16 (1+R)/15}{\left(1+R\right)^2}.
\end{equation}
We define the angular damping scale, $\theta_{\rm D} = \pi/(k_{\rm D} D_{\rm A})$, where $D_{\rm A}$
is the comoving angular diameter distance to $z_\ast$.

For our purposes, the normalization of the power spectrum is most conveniently
given by $A_{\rm s}$.  However, the alternative measure
$\sigma_8$ is often used in the literature, particularly in studies of large-scale
structure.  By definition, $\sigma_8$ is the rms fluctuation in total matter
(baryons + CDM + massive neutrinos) in $8\,h^{-1}$~Mpc spheres at
$z=0$, computed in linear theory.  It is related to the dimensionless matter power spectrum,
$\clp_{\rm m}$, by
\begin{equation}
  \sigma_R^2 = \int \frac{dk}{k}\ \clp_{\rm m}(k)
  \left[ \frac{3j_1(kR)}{kR} \right]^2,
\end{equation}
where $R=8\,h^{-1}$Mpc and $j_1$ is the spherical Bessel function of order 1.

In addition, we compute
$\Omega_{\rm m}h^3$ (a well-determined combination orthogonal to the acoustic
scale degeneracy in flat models; see e.g., \citealt{Percival:2002gq} and~\citealt{Howlett:2012mh}),
$10^9 A_{\rm s} e^{-2\tau}$ (which determines the small-scale linear
CMB anisotropy power),
$r_{0.002}$ (the ratio of the tensor to primordial curvature
power at $k = 0.002 \ {\rm Mpc}^{-1}$),
$\Omega_\nu h^2$ (the physical density in massive neutrinos), and the value of $\yhe$ from the BBN consistency condition.

\subsection{Likelihood}
\label{subsec:likelihood}

\cite{planck2013-p08} describes the \Planck\ temperature likelihood in
detail.  Briefly, at high multipoles ($\ell \ge 50$) we use the $100$,
$143$ and $217$ GHz temperature maps
\citep[constructed using {\tt HEALPix}][]{gorski2005} to form a high multipole
likelihood following the {\tt CamSpec} methodology described in
\cite{planck2013-p08}. Apodized Galactic masks, including an apodized
point source mask, are applied to individual detector/detector-set
maps at each frequency. The masks are carefully chosen to limit
contamination from diffuse Galactic emission to low levels (less than
$20\, \mu{\rm K}^2$ at all multipoles \referee{used in the likelihood) before
correction for Galactic dust emission}.\footnote{\referee{As described in
\citet{planck2013-p08}, we use spectra calculated
on different masks to isolate the contribution of Galactic dust at each
frequency, which we subtract from the $143\times143$,
$143\times 217$ and $217\times 217$ power spectra (i.e., the correction is
applied to the power spectra, not in the map domain). 
The Galactic dust templates are shown in Fig.~\ref{PlanckandHighL} and are
less than $5\,\muKsq$ at high multipoles
for the $217\times217$ spectrum and negligible at lower frequencies.
The residual contribution from Galactic dust after correction in the
$217\times217$ spectrum is smaller than $0.5\,\muKsq$ and smaller than the
errors from other sources such as beam uncertainties.}}
Thus we retain $57.8\%$ of
the sky at $100$ GHz and $37.3\%$ of the sky at $143$ and $217$
GHz. Mask-deconvolved and beam-corrected cross-spectra
\citep[following][]{Hietal02} are computed
for all detector/detector-set combinations and compressed to form
averaged $100\times100$, $143\times143$, $143\times217$ and
$217\times217$ pseudo-spectra (note that we do not retain the
$100\times143$ and $100\times 217$ cross-spectra in the likelihood).
Semi-analytic covariance matrices for these pseudo-spectra
\citep{Efstathiou:04} are used to form a high-multipole likelihood in
a fiducial Gaussian likelihood approximation
\citep{DABJK00,2008PhRvD..77j3013H}.

At low multipoles ($2 \le \ell \le 49$) the temperature likelihood is
based on a Blackwell-Rao estimator applied to Gibbs samples computed
by the {\tt Commander} algorithm \citep{Eriksen:08} from \planck\ maps
in the frequency range $30$--$353$ GHz over 91\% of the sky. The likelihood
at low multipoles therefore accounts for errors in foreground cleaning.

Detailed consistency tests of both the high- and low-multipole components
of the temperature likelihood are presented in \cite{planck2013-p08}.
The high-multipole \planck\ likelihood requires a number of additional
parameters to describe unresolved foreground components and other
``nuisance'' parameters (such as beam eigenmodes). The model adopted
for \planck\ is described in \cite{planck2013-p08}. A self-contained
account is given in Sect. \ref{sec:highell} which generalizes the model
to allow matching of the \planck\ likelihood to the likelihoods from
high-resolution CMB experiments. A complete list of the foreground
and  nuisance parameters is given in Table~\ref{tab:fgparams}.

\subsection{Sampling and confidence intervals}

We sample from the space of possible cosmological parameters with
Markov Chain Monte Carlo (MCMC) exploration using
\COSMOMC~\citep{Lewis:2002ah}.
This uses a Metropolis-Hastings algorithm to generate chains of samples
for a set of cosmological parameters, and also allows for importance
sampling of results to explore the impact of small changes in the analysis.
The set of parameters is internally orthogonalized to allow efficient
exploration of parameter degeneracies, and the baseline cosmological
parameters are chosen following \citet{Kosowsky:2002zt}, so that the
linear orthogonalisation allows efficient exploration of the main
geometric degeneracy \citep{Bond:97}.
The code \referee{has been thoroughly
tested by the community} and has recently been extended to sample efficiently
large numbers of ``fast'' parameters by use of a speed-ordered Cholesky
parameter rotation and a fast-parameter ``dragging'' scheme described
by~\cite{Neal04} and \cite{Lewis:2013hha}.

For our main cosmological parameter runs we execute eight chains until
they are converged, and the tails of the distribution are well enough
explored for the confidence intervals for each parameter to be evaluated
consistently in the last half of each chain.
We check that the spread in the means between chains is small
compared to the standard deviation, using the standard Gelman and Rubin
\citep{Gelman92} criterion $R-1 < 0.01$ in the least-converged
orthogonalized parameter. This is sufficient for reliable importance sampling
in most cases. We perform separate runs when the posterior volumes differ
enough that importance sampling is unreliable. Importance-sampled and extended
data-combination chains used for this paper satisfy $R-1 < 0.1$, and in almost
all cases are closer to 0.01.
We discard the first $30\%$ of each chain as burn in, where the
chains may be still converging and the sampling may be significantly
non-Markovian.  This is
due to the way \COSMOMC\ learns an accurate orthogonalisation
and proposal distribution for the parameters from the sample covariance
of previous samples.

From the samples, we generate estimates of the posterior mean of each
parameter of interest, along with a confidence interval.
We generally quote $68\%$ limits in the case of two-tail limits, so that
$32\%$ of samples are outside the limit range, and
there are $16\%$ of samples in each tail. For parameters where
the tails are significantly different shapes, we instead quote the interval
between extremal points with approximately equal marginalized probability density.
For parameters with prior bounds we either quote one-tail limits or
no constraint, depending on whether the posterior is significantly non-zero at the prior boundary. Our one-tail limits are always $95\%$ limits.
For parameters with nearly symmetric distribution we sometimes quote the mean and standard deviation ($\pm 1\,\sigma$).
The samples can also be used to estimate one,
two and three-dimensional marginalized parameter posteriors.
We use variable-width Gaussian kernel density estimates in all cases.

We have also performed an alternative analysis to the one described above, using an independent statistical method based on frequentist
profile likelihoods \citep{ProfileLik}.
This gives fits and error bars for the baseline cosmological parameters in excellent
agreement for both \Planck\ and \Planck\ combined with high-resolution CMB experiments,
consistent with the Gaussian form of the posteriors found from full parameter space sampling.

In addition to posterior means, we also quote maximum-likelihood
parameter values. These are generated using the {\tt BOBYQA} bounded
minimization
routine\footnote{\url{http://www.damtp.cam.ac.uk/user/na/NA_papers/NA2009_06.pdf}}.
Precision is limited by stability of the convergence, and values
quoted are typically reliable to within $\Delta \chi^2 \sim 0.6$,
which is the same order as differences arising from numerical errors
in the theory calculation.  For poorly constrained parameters the
actual value of the best-fit parameters is not very numerically stable
and should not be over-interpreted; in particular, highly degenerate
parameters in extended models and the foreground model can give many
apparently different solutions within this level of
accuracy.  The best-fit values should be interpreted as giving typical
theory and foreground power spectra that fit the data well, but are
generally non-unique at the numerical precision used; they are however
generally significantly better fits than any of the samples in the
parameter chains.  Best-fit values are useful for assessing residuals,
and differences between the best-fit and posterior means also help to
give an indication of the effect of asymmetries, parameter-volume and
prior-range effects on the posterior samples. We have cross-checked a
small subset of the best-fits with the widely used {\tt MINUIT}
software \citep{Minuit}, which can give somewhat more stable results.

\begin{figure*}
\begin{center}
\includegraphics[width=17cm]{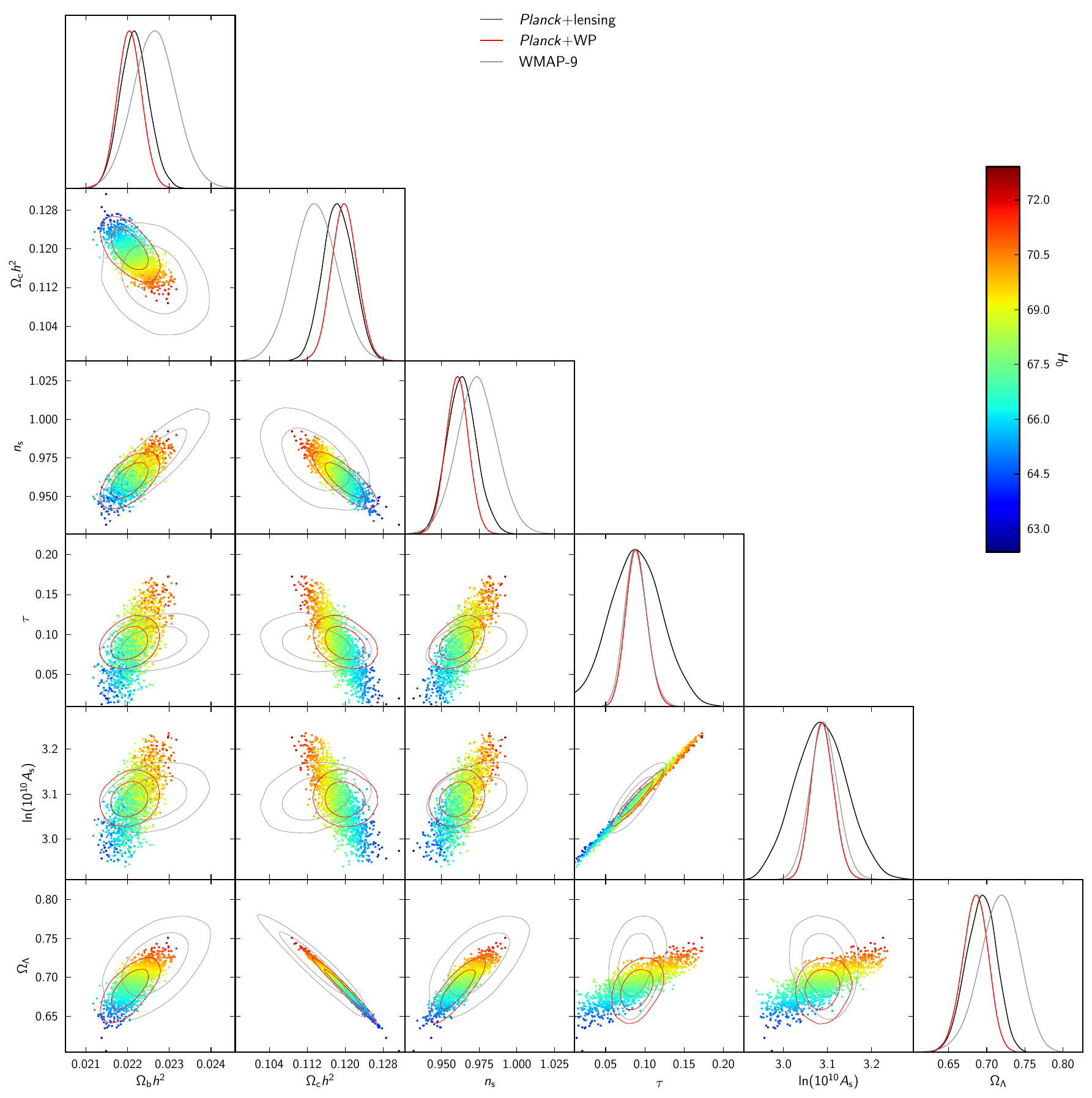}
\end{center}
\caption {Comparison of the base \LCDM\ model parameters for 
\plancklensing\ only
(colour-coded samples), and the 68\% and 95\% constraint contours adding \WMAP\
low-$\ell$ polarization (\WP; red contours), compared to \WMAP-9
(\citealt{bennett2012}; grey contours).
}
\label{fig:planckOnlyTriangle}
\end{figure*}

\section{ Constraints on the parameters of the base \lcdm\ model
 from \planck\ }
\label{sec:planck_lcdm}

In this section we discuss parameter constraints from \planck\ alone in the
\lcdm\ model. \planck\ provides a precision measurement of seven acoustic
peaks in the CMB temperature power spectrum.  The range of scales
probed by \planck\ is
sufficiently large that many parameters can be determined accurately
\emph{without} using low-$\ell$ polarization information to constrain
the optical depth, or indeed without using any other astrophysical data.

However, because the data are reaching the limit of astrophysical
confusion, interpretation of the peaks at higher multipoles requires a
reliable model for unresolved foregrounds. We model these here
parametrically, as described in
\cite{planck2013-p08},
and marginalize over the parameters with wide priors. We give a
detailed discussion of consistency of the foreground model in
Sect.~\ref{sec:highell}, making use of other high-$\ell$ CMB
observations, although as we shall see the parameters of the base $\Lambda$CDM
model have a weak sensitivity to foregrounds.

As foreground modelling is not especially critical for the base
$\Lambda$CDM model, we have decided to present the \Planck\ constraints
early in this paper, ahead of the detailed descriptions of the foreground
model, supplementary high-resolution CMB data sets, and additional
astrophysical data sets.  The reader can therefore gain a feel for some of the
key \planck\ results before being exposed to the lengthier discussions
of Sects.~\ref{sec:highell} and \ref{sec:datasets}, which are essential for the
analysis of extensions to the base $\Lambda$CDM cosmology presented in
Sect.~\ref{sec:grid}.

In addition to the temperature power spectrum measurement, the
\Planck\ lensing reconstruction (discussed in more detail in
Sect.~\ref{sec:lensing} and \citealt{planck2013-p12}) provides a
different probe of the perturbation amplitudes and geometry at late times. CMB
lensing can break degeneracies inherent in the temperature data alone,
especially the geometric degeneracy in non-flat models, providing a
strong constraint on spatial curvature using only CMB data.  The
lensing reconstruction constrains the matter fluctuation amplitude,
and hence the accurate measurement of the temperature anisotropy power
can be used together with the lensing reconstruction to infer the
relative suppression of the temperature anisotropies due to the finite optical depth to reionization.
 The large-scale polarization from
nine years of \WMAP\ observations~\citep{bennett2012} gives a
constraint on the optical depth consistent with the \planck\
temperature and lensing spectra.  Nevertheless, the \WMAP\
polarization constraint is somewhat tighter, so by including it we can
further improve constraints on some parameters.

We therefore also consider the combination of the \planck\
temperature power spectrum
with a \WMAP\
polarization low-multipole likelihood \citep{bennett2012} at $\ell \le 23$ (denoted \WP), as discussed in
\citet{planck2013-p08}\footnote{The \WP\ likelihood is based on the
\WMAP\ likelihood module as distributed
at \url{http://lambda.gsfc.nasa.gov} .}.
We refer to this CMB data combination as \Planck+\WP.

Table~\ref{LCDMparams} summarizes our constraints on cosmological
parameters from the \planck\ temperature power spectrum alone (labelled
``\planck''), from \planck\ in combination with
\planck\ lensing (\plancklensing) and with \WMAP\ low-$\ell$ polarization
(\Planck+\WP). Figure~\ref{fig:planckOnlyTriangle} shows a selection of
corresponding constraints on pairs of parameters and fully
marginalized one-parameter constraints compared to the final results
from \WMAP ~\citep{bennett2012}.

\begin{table*}
\begin{center}
\caption{Cosmological parameter values for the six-parameter
base  \LCDM\ model.
Columns 2 and 3 give results for the \Planck\ temperature power spectrum
data alone. Columns 4 and 5 combine the \Planck\ temperature
data  with \planck\ lensing, and
columns 6 and 7 include \WMAP\ polarization at low multipoles.
We give best
fit parameters \referee{(i.e. the parameters that maximise the overall likelihood
for each data combination)}  as well as $68\%$ confidence limits for constrained parameters.
The first six parameters  have flat priors. The remainder
are derived parameters as discussed in Sect. \ref{sec:model}.
 Beam, calibration parameters, and foreground parameters 
(see Sect. \ref{sec:highell}) are not listed for brevity.
Constraints on foreground parameters for \planck+\WP\ are given later in Table~\ref{LCDMForegroundparams}.
\label{LCDMparams}
}

\begingroup
\newdimen\tblskip \tblskip=5pt
\nointerlineskip
\vskip -3mm
\footnotesize
\setbox\tablebox=\vbox{
    \newdimen\digitwidth
    \setbox0=\hbox{\rm 0}
    \digitwidth=\wd0
    \catcode`"=\active
    \def"{\kern\digitwidth}
    \newdimen\signwidth
    \setbox0=\hbox{+}
    \signwidth=\wd0
    \catcode`!=\active
    \def!{\kern\signwidth}
\halign{
\hbox to 1.3in{$#$\leaderfil}\tabskip=1.5em&\hfil$#$\hfil&\hfil$#$\hfil&\hfil$#$\hfil&\hfil$#$\hfil&\hfil$#$\hfil&\hfil$#$\hfil\tabskip=0pt\cr
\noalign{\doubleline}
\multispan1\hfil \hfil&\multispan2\hfil \planckonly\hfil&\multispan2\hfil \plancklensing\hfil&\multispan2\hfil \Planck+\WP\hfil\cr
\noalign{\vskip -3pt}
\omit&\multispan2\hrulefill&\multispan2\hrulefill&\multispan2\hrulefill\cr
\omit\hfil Parameter\hfil&\omit\hfil Best fit\hfil&\omit\hfil 68\% limits\hfil&\omit\hfil Best fit\hfil&\omit\hfil 68\% limits\hfil&\omit\hfil Best fit\hfil&\omit\hfil 68\% limits\hfil\cr
\noalign{\vskip 3pt\hrule\vskip 5pt}
\Omega_{\mathrm{b}} h^2&0.022068&0.02207\pm 0.00033&0.022242&0.02217\pm 0.00033&0.022032&0.02205\pm 0.00028\cr
\noalign{\vskip 3pt}
\Omega_{\mathrm{c}} h^2&0.12029&0.1196\pm 0.0031&0.11805&0.1186\pm 0.0031&0.12038&0.1199\pm 0.0027\cr
\noalign{\vskip 3pt}
100\theta_{\mathrm{MC}}&1.04122&1.04132\pm 0.00068&1.04150&1.04141\pm 0.00067&1.04119&1.04131\pm 0.00063\cr
\noalign{\vskip 3pt}
\tau&0.0925&0.097\pm 0.038&0.0949&0.089\pm 0.032&0.0925&0.089^{+0.012}_{-0.014}\cr
\noalign{\vskip 3pt}
n_\mathrm{s}&0.9624&0.9616\pm 0.0094&0.9675&0.9635\pm 0.0094&0.9619&0.9603\pm 0.0073\cr
\noalign{\vskip 3pt}
\ln(10^{10} A_\mathrm{s})&3.098&3.103\pm 0.072&3.098&3.085\pm 0.057&3.0980&3.089^{+0.024}_{-0.027}\cr
\noalign{\vskip 5pt\hrule\vskip 3pt}
\Omega_\Lambda&0.6825&0.686\pm 0.020&0.6964&0.693\pm 0.019&0.6817&0.685^{+0.018}_{-0.016}\cr
\noalign{\vskip 3pt}
\Omega_{\mathrm{m}}&0.3175&0.314\pm 0.020&0.3036&0.307\pm 0.019&0.3183&0.315^{+0.016}_{-0.018}\cr
\noalign{\vskip 3pt}
\sigma_8&0.8344&0.834\pm 0.027&0.8285&0.823\pm 0.018&0.8347&0.829\pm 0.012\cr
z_{\mathrm{re}}&11.35&11.4^{+4.0}_{-2.8}&11.45&10.8^{+3.1}_{-2.5}&11.37&11.1\pm 1.1\cr
\noalign{\vskip 3pt}
H_0&67.11&67.4\pm 1.4&68.14&67.9\pm 1.5&67.04&67.3\pm 1.2\cr
\noalign{\vskip 3pt}
10^9 A_{\mathrm{s}}&2.215&2.23\pm
 0.16&2.215&2.19^{+0.12}_{-0.14}&2.215&2.196^{+0.051}_{-0.060}\cr
\noalign{\vskip 3pt}
\Omega_{\mathrm{m}} h^2&0.14300&0.1423\pm 0.0029&0.14094&0.1414\pm 0.0029&0.14305&0.1426\pm 0.0025\cr
\noalign{\vskip 3pt}
\Omega_{\mathrm{m}} h^3&0.09597&0.09590\pm 0.00059&0.09603&0.09593\pm 0.00058&0.09591&0.09589\pm 0.00057\cr
\noalign{\vskip 3pt}
Y_{\mathrm{P}}&0.247710&0.24771\pm 0.00014&0.247785&0.24775\pm 0.00014&0.247695&0.24770\pm 0.00012\cr
\noalign{\vskip 3pt}
\mathrm{Age}/\mathrm{Gyr}&13.819&13.813\pm 0.058&13.784&13.796\pm 0.058&13.8242&13.817\pm 0.048\cr
\noalign{\vskip 3pt}
z_\ast&1090.43&1090.37\pm 0.65&1090.01&1090.16\pm 0.65&1090.48&1090.43\pm 0.54\cr
\noalign{\vskip 3pt}
r_\ast&144.58&144.75\pm 0.66&145.02&144.96\pm 0.66&144.58&144.71\pm 0.60\cr
\noalign{\vskip 3pt}
100\theta_\ast&1.04139&1.04148\pm 0.00066&1.04164&1.04156\pm 0.00066&1.04136&1.04147\pm 0.00062\cr
\noalign{\vskip 3pt}
z_{\mathrm{drag}}&1059.32&1059.29\pm 0.65&1059.59&1059.43\pm 0.64&1059.25&1059.25\pm 0.58\cr
\noalign{\vskip 3pt}
r_{\mathrm{drag}}&147.34&147.53\pm 0.64&147.74&147.70\pm 0.63&147.36&147.49\pm 0.59\cr
\noalign{\vskip 3pt}
k_{\mathrm{D}}&0.14026&0.14007\pm 0.00064&0.13998&0.13996\pm 0.00062&0.14022&0.14009\pm 0.00063\cr
\noalign{\vskip 3pt}
100\theta_{\mathrm{D}}&0.161332&0.16137\pm 0.00037&0.161196&0.16129\pm 0.00036&0.161375&0.16140\pm 0.00034\cr
\noalign{\vskip 3pt}
z_{\mathrm{eq}}&3402&3386\pm 69&3352&3362\pm 69&3403&3391\pm 60\cr
\noalign{\vskip 3pt}
100\theta_{\mathrm{eq}}&0.8128&0.816\pm 0.013&0.8224&0.821\pm 0.013&0.8125&0.815\pm 0.011\cr
\noalign{\vskip 3pt}
r_{\mathrm{drag}}/D_{\mathrm{V}}(0.57)&0.07130&0.0716\pm 0.0011&0.07207&0.0719\pm 0.0011&0.07126&0.07147\pm 0.00091\cr
\noalign{\vskip 5pt\hrule\vskip 3pt}
} 
} 
\endPlancktable
\endgroup

\end{center}
\end{table*}

\subsection{Acoustic scale}

The characteristic angular size of the fluctuations in the CMB is called
the acoustic scale. It is determined by the comoving size of the sound
horizon at the time of last-scattering, $\rs(z_\ast)$, and the angular diameter
distance at which we are observing the fluctuations, $\DAstar(z_\ast)$. With
accurate measurement of seven acoustic peaks, \planck\ determines the observed
angular size $\thetastar = \rs/\DAstar$ to better than $0.1\%$ precision
at $1\,\sigma$:
\be
\thetastar = (1.04148\pm 0.00066)\times 10^{-2} = 0.596724^\circ \pm 0.00038^\circ.
\ee
Since this parameter is constrained by the \emph{positions} of the peaks
but not their amplitudes, it is quite robust; the measurement is very stable
to changes in data combinations and the assumed cosmology. Foregrounds, beam
uncertainties, or any systematic effects which only contribute a smooth
component to the observed spectrum will not substantially affect the frequency
of the oscillations, and hence this determination is likely to be \planck's
most robust precision measurement. The situation is analogous to 
baryon acoustic oscillations measurements in large-scale structure
surveys (see Sect. \ref{sec:BAO}),  but the CMB acoustic
measurement has the advantage that it is based on observations of the Universe
when the fluctuations were very accurately linear, so second and higher-order effects are expected to be  negligible\footnote{Note, however, that \planck's measurement of $\theta_\ast$ is now so accurate that $O(10^{-3})$ effects from aberration due to the relative motion between our frame and the CMB rest-frame are becoming non-negligible; see~\cite{planck2013-pipaberration}. The statistical anisotropy induced would lead to dipolar variations at the $10^{-3}$ level in $\theta_\ast$ determined locally on small regions of the sky. For \planck, we average over many such regions and we expect that the residual effect (due to asymmetry in the Galactic mask) on the marginalised values of other parameters is negligible.}.

\begin{figure}
\begin{center}
\includegraphics[width=8.8cm]{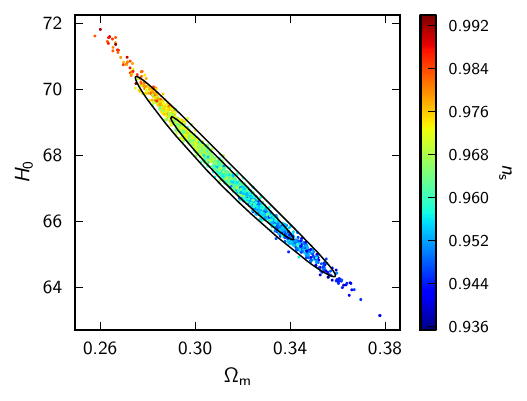}
\end{center}
\caption {Constraints in the $\Omega_{\rm m}$--$H_0$ plane.
Points show samples from the \Planck-only posterior, coloured by the
corresponding value of the spectral index $\ns$. The contours (68\% and 95\%)
show the
improved constraint from \plancklensing+\WP.
The degeneracy direction is significantly shortened by including \WP,
but the well-constrained direction of constant $\Omm h^3$
(set by the acoustic scale), is determined almost equally accurately from
\planck\ alone.
\label{fig:omm-h-degeneracy}
}
\end{figure}

The tight constraint on $\thetastar$ also implies tight constraints on 
some combinations of the cosmological parameters
that determine $\DAstar$ and $\rs$.
The sound horizon $\rs$ depends on the physical matter density parameters, and $\DAstar$ depends on the late-time evolution and geometry.
Parameter combinations that fit the \planck\ data must be constrained to be close to a surface of constant $\thetastar$.
This surface depends on the model that is assumed. For the  base
\LCDM\ model,  the main parameter dependence is approximately described by a $0.3\%$ constraint in the three-dimensional $\Omm$--$h$--$\Omb h^2$ subspace:
\be
\Omm h^{3.2}(\Omb h^2)^{-0.54}  = 0.695 \pm  0.002 \quad \mbox{(68\%; \planck)}.
\ee
Reducing further to a two-dimensional subspace gives a $0.6\%$ constraint on the combination
\be
\Omm h^3 = 0.0959\pm 0.0006 \quad \mbox{(68\%; \planck)}.
\ee
The principle component analysis direction is actually $\Omm h^{2.93}$
but this is conveniently close to $\Omm h^3$ and gives a similar constraint. The simple form is a coincidence of the \LCDM\ cosmology, error model, and
particular parameter values of the model~\citep{Percival:2002gq,Howlett:2012mh}.
The degeneracy between $H_0$ and $\Omm$ is illustrated in
Fig.~\ref{fig:omm-h-degeneracy}: parameters are constrained to lie in a
narrow strip where $\Omm h^3$ is nearly constant, but the orthogonal
direction is much more poorly constrained. The degeneracy direction involves
consistent changes in the $H_0$, $\Omm$, and $\Omb h^2$
parameters, so that the ratio of the sound horizon and angular 
diameter distance remains nearly constant. Changes in the density
parameters, however, also have other effects on the power spectrum and the spectral
index $\ns$ also changes to compensate. The
degeneracy is not exact; its extent is much more sensitive to other details
of the power spectrum shape. Additional data can  help further to
restrict the degeneracy. Figure~\ref{fig:omm-h-degeneracy}
 shows that adding \WMAP\ polarization
has almost no effect on the $\Omm h^3$ measurement, but shrinks the
orthogonal direction slightly from $\Omm h^{-3} = 1.03 \pm 0.13 $ to
$\Omm h^{-3} = 1.04 \pm 0.11$.

\subsection{Hubble parameter and dark energy density}
\label{subsec:LCDM_hubble}

The Hubble constant, $H_0$, and matter density parameter, $\Omm$, are 
only tightly  constrained in the combination $\Omm h^3$ discussed above, but
the extent of the degeneracy is limited by the effect of $\Omm h^2$ on the relative heights of the acoustic peaks. The  projection of the constraint ellipse
shown in Fig.~\ref{fig:omm-h-degeneracy}  onto the axes therefore yields useful
marginalized constraints on $H_0$ and $\Omm$ (or equivalently $\Omega_\Lambda$)
separately. We find the $2\%$ constraint on $H_0$:
\be
H_0 = (67.4\pm 1.4)\,  {\rm km}\,{\rm s}^{-1} \,{\rm Mpc}^{-1}
 \quad \mbox{(68\%; \planckonly)}.
\ee
The corresponding constraint on the dark energy density parameter is
\be
\Oml = 0.686\pm 0.020 \quad \mbox{(68\%; \planckonly)},
\ee
and for  the physical matter density we find
\be
\Omm h^2 = 0.1423\pm 0.0029 \quad \mbox{(68\%; \planckonly)}.
\ee

Note that these indirect constraints are highly model dependent. The 
data only measure accurately the acoustic scale, and the relation to underlying
expansion parameters (e.g., via the angular-diameter distance) depends on the
assumed cosmology, including the shape of the
primordial fluctuation spectrum. Even small
changes in model assumptions can change $H_0$ noticeably; for example, if we
neglect the $0.06\eV$ neutrino mass expected in the minimal hierarchy,
and instead take $\mnu=0$, the Hubble parameter constraint shifts to
\be
H_0 = (68.0\pm 1.4)\, {\rm km}\, {\rm s}^{-1} \, {\rm Mpc}^{-1}
 \,\,\, \mbox{(68\%; \planck,  $\mnu=0$)}.
\ee

\subsection{Matter densities}

\planck\ can measure the matter densities in baryons and dark matter
from the relative heights of the acoustic peaks.
However, as discussed above, there is a
partial degeneracy with the spectral index and other parameters that
limits the precision of the determination. 
With \planck\ there are now enough well measured peaks
that the extent of the degeneracy is limited, giving $\Omb h^2$ to an accuracy
of  $1.5\%$ without any additional data:
\be
\Omb h^2 = 0.02207\pm 0.00033 \quad \mbox{(68\%; \planckonly)}.
\ee
Adding \WMAP\ polarization information shrinks the errors by only $10\%$.

The dark matter density is slightly less accurately measured at around
3\%:
\be
\Omc h^2 = 0.1196\pm 0.0031 \quad \mbox{(68\%; \planckonly)}.
\ee

\subsection{Optical depth}
\label{subsec:tau}

\begin{figure*}
\begin{center}
\includegraphics[width=18cm]{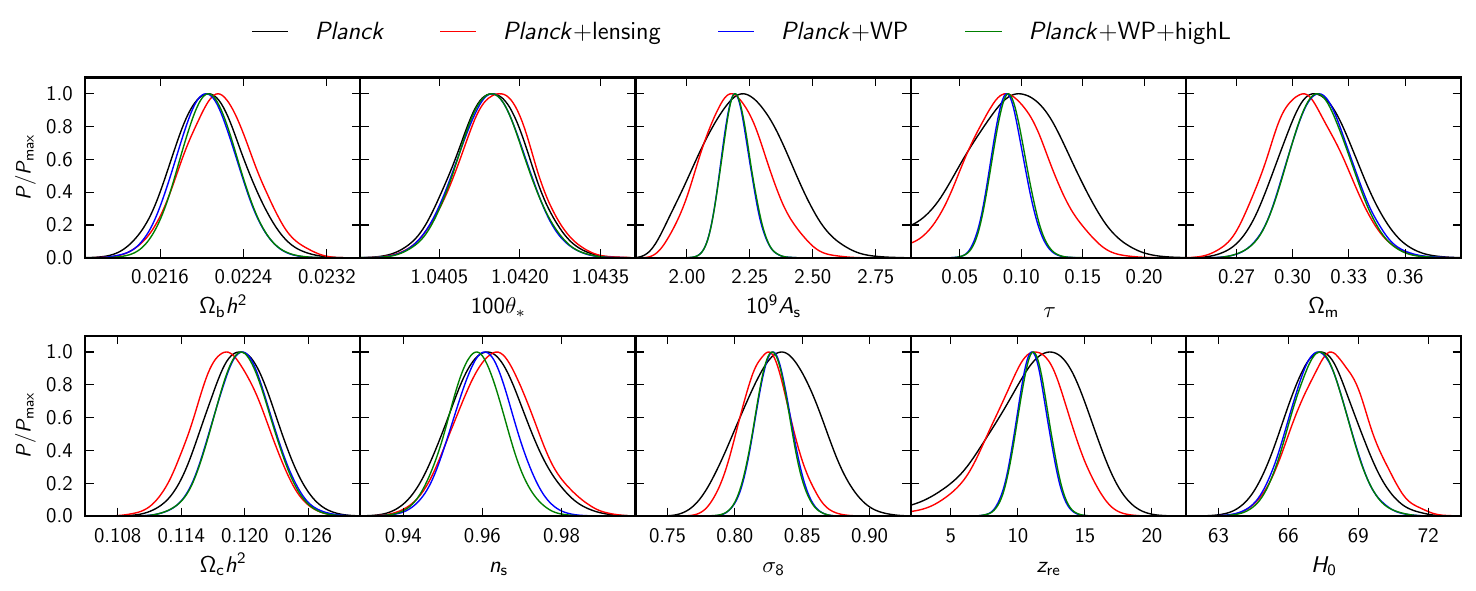}
\end{center}
\caption {Marginalized constraints on parameters of the base \LCDM\ model
for various data combinations.}
\label{fig:margeLCDM}
\end{figure*}

Small-scale fluctuations in the CMB are damped by Thomson scattering from free
electrons produced  at reionization. This scattering suppresses the
amplitude of the acoustic peaks  by $e^{-2\tau}$
on scales that correspond to perturbation modes with wavelength
smaller than the Hubble radius at reionization.
\planck\ measures the small-scale power spectrum with
high precision, and hence  accurately constrains the damped
amplitude $e^{-2\tau}\As$. With only unlensed temperature power
spectrum data, there is a large degeneracy between $\tau$ and $\As$,
which is weakly broken only by the power in large-scale modes that were
still super-Hubble scale at reionization.  However, lensing depends on the
actual amplitude of the matter fluctuations along the line of
sight. \planck\ accurately measures many acoustic peaks in the
\emph{lensed} temperature power spectrum, where the amount of lensing
smoothing depends on the fluctuation amplitude. Furthermore
\planck's lensing potential reconstruction  provides a more direct
measurement of the amplitude, independently of the optical depth. The
combination of the temperature data and \planck's lensing
reconstruction can therefore determine the optical depth $\tau$
relatively well. The combination gives 
\be 
\tau = 0.089 \pm 0.032 \quad \mbox{(68\%; \plancklensing)}.  
\ee 
As shown in Fig.~\ref{fig:margeLCDM} this provides marginal confirmation (just
under $2\,\sigma$) that the total optical depth is significantly
higher than would be obtained from sudden reionization at $z\sim 6$,
and is  consistent with the {\it WMAP}-9 constraint,  $\tau =0.089\pm
0.014$,  from large-scale polarization~\citep{bennett2012}. The
large-scale $E$-mode polarization measurement is very challenging
because it is a small signal relative to
polarized  Galactic emission on large scales, so this \planck\
polarization-free result is a valuable cross-check. The posterior for
the $\planck$ temperature power spectrum measurement alone also
consistently peaks at $\tau \sim 0.1$, where the constraint on the
optical depth is coming from the amplitude of the lensing smoothing
effect and (to a lesser extent) the relative power between small and large
scales.

Since lensing constrains the underlying fluctuation amplitude, the
matter density perturbation power is also well determined:
\be
\sigma_8 = 0.823 \pm 0.018 \quad \mbox{(68\%; \plancklensing)}.
\ee
Much of the residual uncertainty is caused by the
degeneracy with the optical depth.  Since the
small-scale temperature power spectrum more directly fixes
$\sigma_8e^{-\tau}$,  this combination is  tightly constrained:
\be
\sigma_8e^{-\tau}=0.753\pm 0.011 \quad \mbox{(68\%; \plancklensing)}.
\ee
The estimate of  $\sigma_8$ is significantly improved to
$\sigma_8 = 0.829\pm0.012$ by using the \WMAP\ polarization data to constrain
the optical depth, and is not strongly degenerate with $\Omm$. (We
shall see in Sect. \ref{subsec:additional} that the \planck\ results
are discrepant with recent estimates of combinations of $\sigma_8$
and $\Omm$ from cosmic shear measurements and counts of rich clusters 
of galaxies.)

\subsection{Spectral index}

The scalar spectral index defined in Eq.~(\ref{PS1})
is measured by \planck\ data alone to 1\% accuracy:
\be
\ns = 0.9616\pm 0.0094 \quad \mbox{(68\%; \planckonly)}. \label{AL1}
\ee
Since the optical depth $\tau$ affects the relative power between large scales (that are unaffected by scattering at reionization) and intermediate and small scales (that have their power suppressed by $e^{-2\tau}$), there is a partial degeneracy with $\ns$.
Breaking the degeneracy between $\tau$ and $n_s$ using \WMAP\ 
polarization  leads to a small improvement in the constraint:
\be
\ns = 0.9603\pm 0.0073 \quad \mbox{(68\%; \planck+\WP)}. \label{AL2}
\ee
Comparing Eqs.~(\ref{AL1}) and (\ref{AL2}), it is evident that the \planck\
temperature spectrum spans a wide enough range of multipoles
to give a highly significant detection of a 
deviation of the scalar spectral index from exact scale invariance
(at least in the base \lcdm\ cosmology) independent of \WMAP\
polarization information.

One might worry that the spectral index parameter is degenerate with
foreground parameters, since these act to increase smoothly the
amplitudes of the temperature power spectra at high multipoles.  The spectral index is
therefore liable to potential systematic errors if the foreground
model is poorly constrained.  Figure~\ref{fig:margeLCDM} shows the
marginalized constraints on the \LCDM\ parameters for various
combinations of data, including adding high-resolution CMB
measurements.  As will be discussed in Sect.~\ref{sec:highell},  the use of high-resolution CMB provides tighter constraints on the foreground
parameters (particularly ``minor'' foreground components) than from \planck\
data alone.  However, the small shifts in the means and widths of the
distributions shown in Fig.~\ref{fig:margeLCDM} indicate that, for the
base \lcdm\ cosmology, the errors on the cosmological parameters are
not limited by foreground uncertainties when considering \planck\
alone. The effects of foreground modelling assumptions and likelihood choices on
constraints on $\ns$ are discussed in Appendix~\ref{app:test}.

\newcommand{\tSZCIB}{{\mbox{\scriptsize{tSZ$\times$CIB}}}}

\section{ \planck\ combined with high-resolution CMB experiments: the base \lcdm\ model}
\label{sec:highell}

\begin{table*}[tmb]                 
\begingroup
\newdimen\tblskip \tblskip=5pt
\caption{Summary of the CMB temperature data sets used in this analysis.}                          
\label{Datasets}         
\nointerlineskip
\vskip -3mm
\footnotesize
\setbox\tablebox=\vbox{
   \newdimen\digitwidth
   \setbox0=\hbox{\rm 0}
   \digitwidth=\wd0
   \catcode`*=\active
   \def*{\kern\digitwidth}
   \newdimen\signwidth
   \setbox0=\hbox{+}
   \signwidth=\wd0
   \catcode`!=\active
   \def!{\kern\signwidth}
\halign{\hbox to 1.5in{#\leaderfil}\tabskip 2.2em&
        \hfil#\hfil&
        \hfil#\hfil&
        \hfil#\hfil&
        \hfil#\hfil&
        \hfil#\hfil&
        \hfil#\hfil&
        \hfil#\hfil&
        \hfil#\hfil&
        \hfil#\hfil\tabskip=0pt\cr                
\noalign{\doubleline}
\omit&Frequency&Area&$\ell_{\rm min}$&$\ell_{\rm max}$&$S_{\rm cut}^{\mbox{\scriptsize{\it a}}}$&$\nu_{\rm CMB}$ &$\nu_{\rm tSZ}$&$\nu_{\rm Radio}$&$\nu_{\rm IR}$\cr
\omit\hfil Experiment\hfil&[GHz]&[deg$^2$]&&&[mJy]&[GHz]&[GHz]&[GHz]&[GHz]\cr
\noalign{\vskip 4pt\hrule\vskip 6pt}
\planck&100&23846&**50&*1200& \dots&100.0&103.1&\dots&\dots\cr
\planck&143&15378&**50&*2000&\dots&143.0&145.1&\dots&146.3\cr
\planck&217&15378&*500&*2500&\dots&217.0&\dots&\dots&225.7\cr
\noalign{\vskip 4pt\hrule\vskip 6pt}
ACT (D13)&148&**600&*540&9440&15.0&148.4&146.9&147.6&149.7\cr
ACT (D13)&218&**600&1540&9440&15.0&218.3&220.2&217.6&219.6\cr
\noalign{\vskip 4pt\hrule\vskip 6pt}
SPT-high  (R12)&*95&**800&2000&10000&*6.4&*95.0&*97.6&*95.3&*97.9\cr
SPT-high  (R12)&150&**800&2000&10000&*6.4&150.0&152.9&150.2&153.8\cr
SPT-high  (R12)& 220&**800&2000&10000&*6.4&220.0&218.1&214.1&219.6\cr
\noalign{\vskip 3pt\hrule\vskip 4pt}}}
\endPlancktablewide                 
\tablenote {\textit a} Flux-density cut applied to the map by the point-source mask.
For \planck\ the point-source mask is based on a composite of sources
identified in the 100--353\,GHz maps, so there is no simple flux cut.\par
\endgroup
\end{table*}                        

The previous section adopted a foreground model with relatively loose
priors on its parameters. As discussed there and in~\citet{planck2013-p08},
for the base
\lcdm\ model, the cosmological parameters are relatively weakly correlated with
the parameters of the foreground model and so we expect that the cosmological
results reported in Sect.~\ref{sec:planck_lcdm} are robust. Fortunately, we can
get an additional handle on unresolved foregrounds, particularly ``minor''
components such as the kinetic SZ effect, by combining the \planck\ data
with data from high-resolution CMB experiments. The consistency of results
obtained with \planck\ data alone and \planck\ data combined with
high-resolution
CMB data gives added confidence to our cosmological results, particularly when
we come to investigate extensions to the base \lcdm\ cosmology
(Sect.~\ref{sec:grid}). In this section, we review the high-resolution CMB
data (hereafter, usually denoted \highL) that we combine with \planck\ and
then discuss how the foreground model
is adapted (with additional ``nuisance'' parameters) to handle multiple CMB
data sets. We then discuss the results of an MCMC analysis
of the base \lcdm\ model combining \planck\ data with the high-$\ell$ data.

\subsection{Overview of the high-$\ell$ CMB data sets}

The Atacama Cosmology Telescope (ACT) mapped the sky from 2007 to 2010
in two distinct regions, the equatorial stripe (ACTe) along
the celestial equator, and the southern stripe (ACTs) along declination
$-55\deg$, observing in total about $600\,\mathrm{deg}^2$.  The ACT
data sets at $148$ and $218\,\mathrm{GHz}$ are presented
in~\citet[hereafter D13]{2013arXiv1301.1037D} and cover the angular
scales $540<\ell<9440$ at $148\,\mathrm{GHz}$ and $1540<\ell<9440$ at
$218\,\mathrm{GHz}$.  Beam  errors are included in the
released covariance matrix. We include the ACT $148\times148$ spectra for $\ell \ge 1000$, and the
ACT $148\times218$ and $218\times 218$ spectra for $\ell \ge
1500$. The inclusion
of ACT spectra to $\ell =1000$ improves the accuracy of the
inter-calibration parameters between the high-$\ell$ experiments
and \planck. 

The South Pole Telescope observed a region of sky over the period
2007--10. Spectra are reported in~\citet[hereafter K11]{keisler11}
and~\citet[hereafter S12]{Story:12} for angular scales $650<\ell<3000$
at $150\,\mathrm{GHz}$, and in~\citet[hereafter R12]{Reichardt:12} for
angular scales $2000<\ell<10000$ at 95, 150 and $220\,\mathrm{GHz}$.
Beam errors are included in the released covariance
matrices used to form the SPT likelihood. The parameters of the base
\lcdm\ cosmology derived from the \textit{WMAP}-7+S12 data and 
(to a lesser extent) from K11 are in tension with \planck. Since the S12 spectra have
provided the strongest CMB constraints on cosmological parameters
prior to \planck, this discrepancy merits a more detailed analysis, which
is presented in appendix~\ref{app:spt}. The S12 and K11 data are not
used in combination with \planck\ in this paper.  Since the primary purpose of
including high-$\ell$ CMB data is to provide stronger constraints on
foregrounds, we use only the R12 SPT data at $\ell > 2000$  in
combination with \planck. We ignore any correlations between ACT/SPT and
\planck\ spectra over the overlapping multipole ranges.

Table \ref{Datasets} summarizes some key features of the CMB data sets
used in this paper. 

\begin{figure}
\centering
\includegraphics[width=75mm,angle=0]{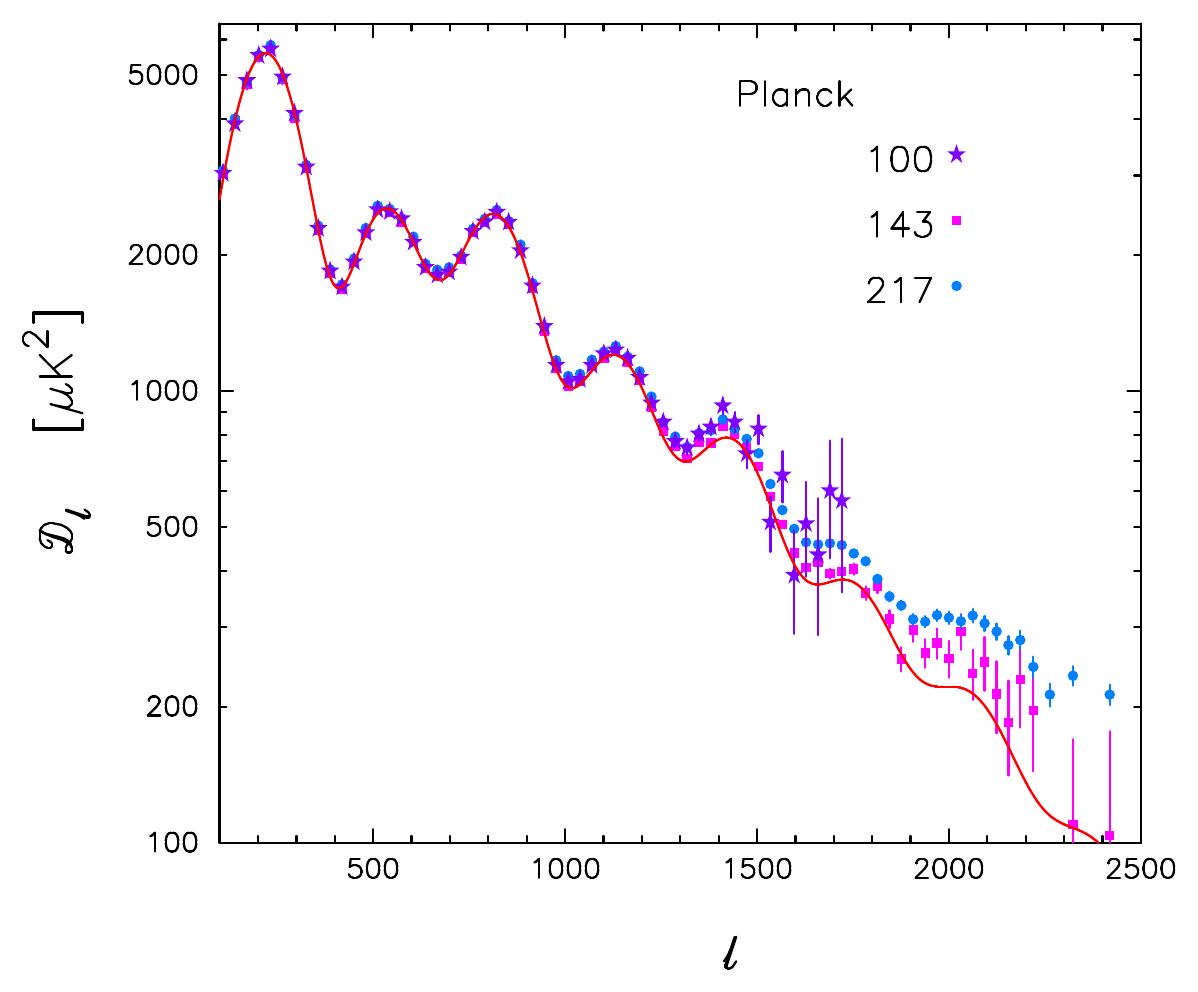}
\\
\vspace{\baselineskip}
\includegraphics[width=75mm,angle=0]{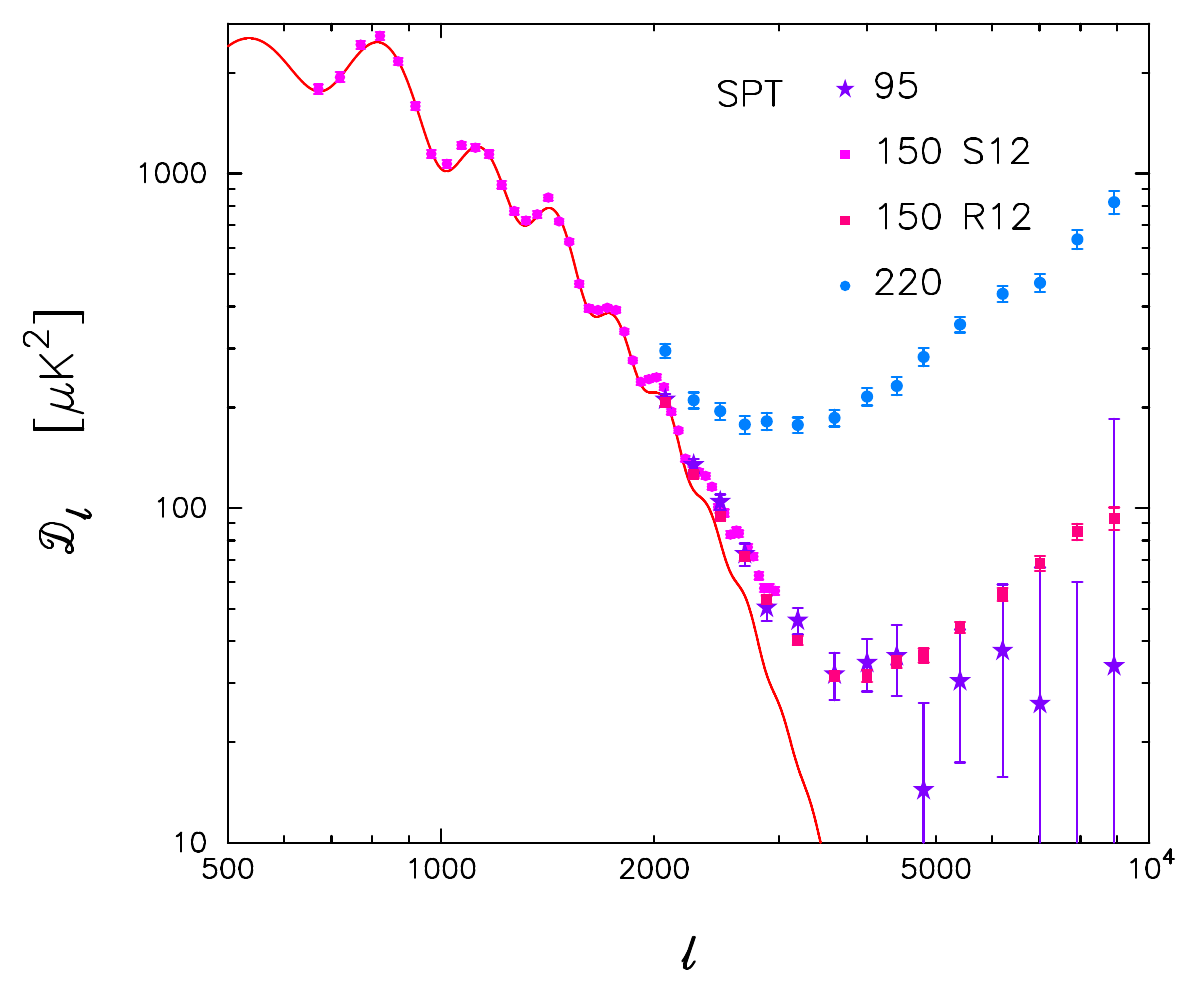}
\\
\vspace{\baselineskip}
\includegraphics[width=75mm,angle=0]{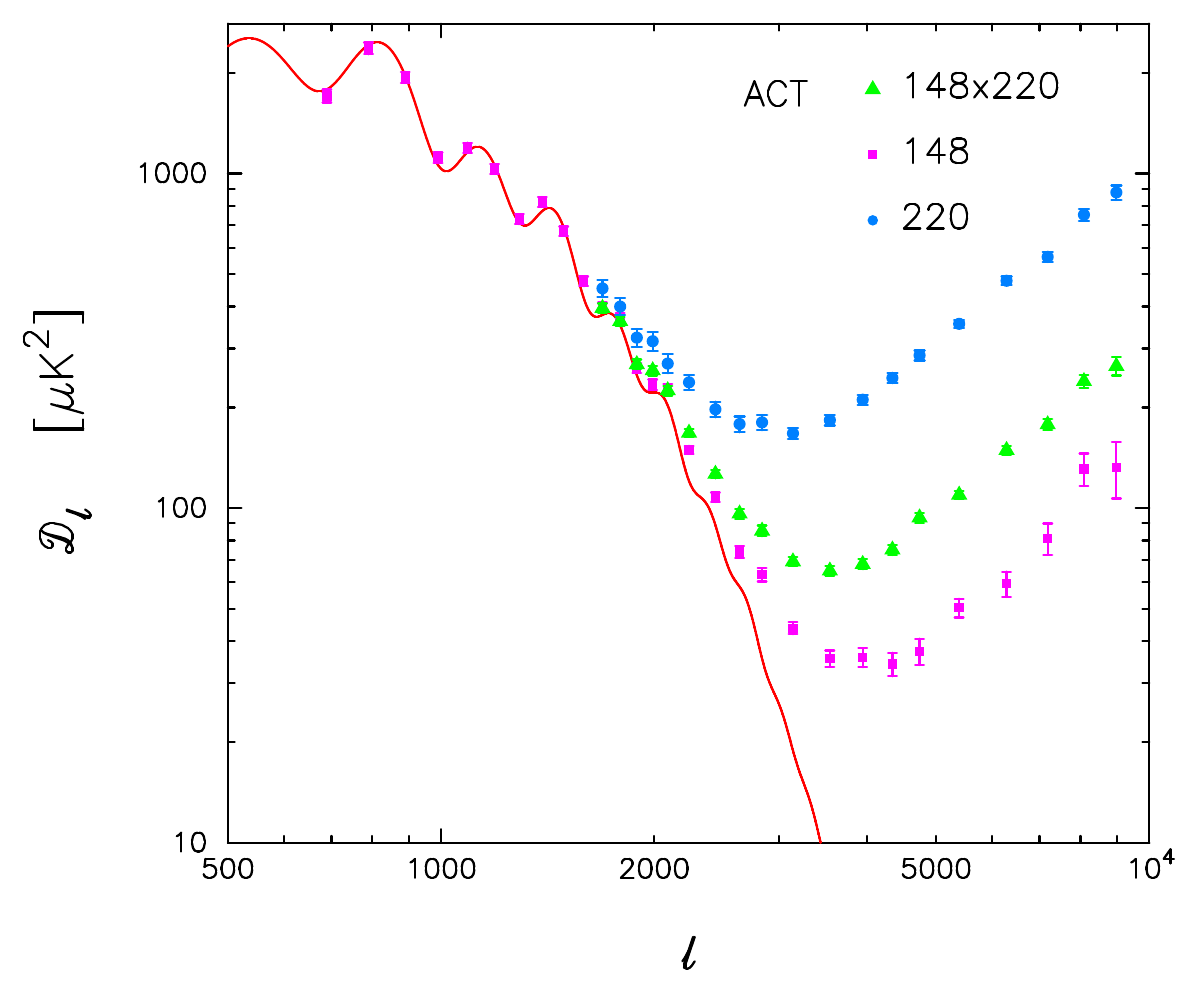}
\caption {\textit{Top}: \planck\ spectra at $100$, $143$ and
$217\,\mathrm{GHz}$ without subtraction of foregrounds. \textit{Middle}:
SPT spectra from R12 at 95, 150 and $220\,\mathrm{GHz}$, recalibrated to
\planck\ using the best-fit calibration, as discussed in the text.
The S12 SPT spectrum at $150\,\mathrm{GHz}$ is also shown, but without any
calibration correction. This spectrum is discussed in detail in
Appendix~\ref{app:spt}, but is not used elsewhere in this paper. 
\textit{Bottom}: ACT spectra (weighted averages of the equatorial and
southern fields) from D13 at $148$ and $220\,\mathrm{GHz}$,
and the $148\times 220\,\mathrm{GHz}$ 
cross-spectrum, with no extragalactic foreground corrections, recalibrated
to the \planck\ spectra as discussed in the text.  The solid line in each panel
shows the best-fit 
base \lcdm\ model from the combined \planck+WP+highL\ fits listed in Table~\ref{LCDMForegroundparams}.}
\label{Planck+ACT+SPT}
\end{figure}

\subsection{Model of unresolved foregrounds and ``nuisance'' parameters}

The model for unresolved foregrounds used in the \planck\ likelihood
is described in detail in~\citet{planck2013-p08}. Briefly, the model
includes power spectrum templates for clustered extragalactic point
sources (the cosmic infra-red background, hereafter CIB), thermal
(tSZ) and kinetic (kSZ) Sunyaev-Zeldovich contributions, and the
cross-correlation (tSZ$\times$CIB) between infra-red galaxies and the
thermal Sunyaev-Zeldovich effect. The model also includes amplitudes
for the Poisson contributions from radio and infra-red galaxies. The
templates are described in~\citet{planck2013-p08} and are kept fixed
here. (Appendix \ref{app:test} discusses briefly a few tests showing
the impact of varying some aspects of the foreground model.) The model
for unresolved foregrounds is similar to the models developed by the
ACT and the SPT teams \citep[e.g., R12;][]{Dunkley:13}.  The main
difference is in the treatment of the Poisson contribution from radio
and infra-red galaxies. In the ACT and SPT analyses, spectral models
are assumed for radio and infra-red galaxies. The Poisson point source
contributions can then be described by an amplitude for each
population, assuming either fixed spectral parameters or solving for
them. In addition, one can add additional parameters to describe the
decorrelation of the point source amplitudes with frequency (see e.g.,
\citealt{Millea:12}).  The \planck\ model assumes free
amplitudes for the point sources at each frequency, together with
appropriate correlation coefficients between frequencies. The model is
adapted to handle the ACT and SPT data as discussed later in this section.

Figure~\ref{Planck+ACT+SPT} illustrates the importance of unresolved
foregrounds in interpreting the power spectra of the three CMB data
sets.  The upper panel of Fig.~\ref{Planck+ACT+SPT} shows the \planck\
temperature spectra at $100$, $143$, and $217\,\mathrm{GHz}$, without
corrections for unresolved foregrounds (to avoid overcrowding, we have
not plotted the $143\times 217$ spectrum). The solid (red) lines show
the best-fit base \lcdm\ CMB spectrum corresponding to the combined
Planck+ACT+SPT+\WMAP\ polarization likelihood analysis, with parameters
listed in Table~\ref{LCDMForegroundparams}. The
middle panel shows the SPT spectra at $95$, $150$ and
$220\,\mathrm{GHz}$ from S12 and R12. In this figure, we have
recalibrated the R12 power spectra to match \planck\ using calibration
parameters derived from a full likelihood analysis of the base \lcdm\
model.  The S12 spectrum plotted is  exactly as
tabulated in S12, i.e., we have not recalibrated this spectrum to
\planck. (The consistency of the  S12 spectrum
with the theoretical model is discussed in further detail in 
 Appendix~\ref{app:spt}.) The lower panel of Fig.~\ref{Planck+ACT+SPT} shows the ACT spectra
from D13, recalibrated to \planck\ with calibration coefficients
determined from a joint likelihood analysis. The power spectra plotted
are an average of the ACTe and ACTs spectra, and include the
  small Galactic dust corrections described in D13.

  The small-scale SPT (R12) and ACT (D13) data are dominated by the
  extragalactic foregrounds and hence are highly effective in
  constraining the multi-parameter foreground model. In contrast,
  \planck\ has limited angular resolution and therefore limited
  ability to constrain unresolved foregrounds. \planck\ is sensitive
  to the Poisson point source contribution at each frequency and to
  the CIB contribution at $217\,\mathrm{GHz}$. \planck\ has some
  limited sensitivity to the tSZ amplitude from the
  $100\,\mathrm{GHz}$ channel (and almost no sensitivity at
  $143\,\mathrm{GHz}$).  The remaining foreground contributions are
  poorly constrained by \planck\ and highly degenerate with each other
  in a \planck-alone analysis.  The main gain in combining \planck\
  with the high-resolution ACT and SPT data is in breaking some of the
  degeneracies between foreground parameters which are poorly
  determined from \planck\ data alone.

  An important extension of the foreground parameterization described
  here over that developed in~\citet{planck2013-p08} concerns the use
  of effective frequencies. Different experiments (and different
  detectors within a frequency band) have non-identical bandpasses
  \citep{planck2013-p03d} and this needs to be taken into account in
  the foreground modelling.  Consider, for example, the amplitude of
  the CIB template at $217\,\mathrm{GHz}$, $A^{\rm CIB}_{217}$,
  introduced in~\citet{planck2013-p08}. The effective frequency for a
  dust-like component for the averaged $217\,\mathrm{GHz}$ spectrum
  used in the \planck\ likelihood is $225.7\,\mathrm{GHz}$.  To avoid
  cumbersome notation, we solve for the CIB amplitude $A^{\rm
    CIB}_{217}$ \emph{at the CMB effective frequency of }
  $217\,\mathrm{GHz}$.  The actual amplitude measured in the \planck\
  $217\,\mathrm{GHz}$ band is $1.33 A^{\rm CIB}_{217}$, reflecting the
  different effective frequencies of a dust-like component compared to
  the blackbody primordial CMB (see Eq.~\ref{CIB1} below).  With
  appropriate effective frequencies, the single amplitude $A^{\rm
    CIB}_{217}$ can be used to parameterize the CIB contributions to
  the ACT and SPT power spectra in their respective $218$ and
  $220\,\mathrm{GHz}$ bands. A similar methodology is applied to match
  the tSZ amplitudes for each experiment.

The relevant effective frequencies for the foreground parameterization
discussed below are listed in Table~\ref{Datasets}. For the high resolution experiments,
these are as quoted in R12 and \citet{Dunkley:13}. For \planck\ these effective frequencies were
computed from the individual HFI bandpass measurements~\citep{planck2013-p03d},
and vary by a few percent from detector to detector. The numbers quoted
in Table~\ref{Datasets} are based on an 
approximate average of the individual detector bandpasses 
using the weighting scheme for individual detectors/detector-sets
applied in  the {\tt CamSpec} likelihood. (The resulting bandpass
correction factors for the tSZ and CIB amplitudes should be
accurate to better than $5\%$.) \referee{Note that all temperatures in this section are in
thermodynamic units.}

The ingredients of the foreground model and associated ``nuisance'' parameters are summarized in the following paragraphs.

\paragraph{Calibration factors:} To combine the \planck,
ACT and SPT likelihoods it is important to incorporate relative
calibration factors,
since the absolute calibrations of ACT and SPT have large errors
(e.g., around $3.5\%$ \referee{in power} for the SPT $150\,\mathrm{GHz}$ channel). We introduce three
{\it map} calibration parameters $y^{\rm SPT}_{95}$, $y^{\rm SPT}_{150}$ and $y^{\rm SPT}_{220}$ 
to rescale the R12 SPT spectra. These factors
rescale the cross-spectra at frequencies $\nu_i$ and $\nu_j$ as
\be
C_\ell^{\nu_i \times  \nu_j} \rightarrow y_{\nu_i}^{{\rm SPT}} y_{\nu_j}^{{\rm SPT}} C_\ell^{\nu_i \times \nu_j}.
\ee
In the analysis of
ACT, we solve for different map calibration factors for the ACTe and ACTs
spectra, $y^{\mathrm{ACTe}}_{148}$, $y^{\mathrm{ACTs}}_{148}$, $y^{\mathrm{ACTe}}_{218}$, and
$y^{\mathrm{ACTs}}_{218}$. In addition,
we solve for the $100\times 100$ and
$217\times 217$ \planck\ {\it power-spectrum} calibration factors $c_{100}$ and $c_{217}$,
with priors as described in~\citet{planck2013-p08}; see also Table~\ref{tab:fgparams}.
(The use of map calibration factors for ACT and SPT follows the conventions
 adopted by the ACT and SPT teams, while for the \planck\ power spectrum analysis
we have consistently used power-spectrum calibration factors.)

In a joint parameter analysis of \Planck+ACT+SPT, the inclusion of these
calibration parameters leads to
recalibrations that match the ACT, SPT and \planck\ $100\,\mathrm{GHz}$ and
$217\,\mathrm{GHz}$ channels to the calibration of the \planck\ $143\times 143$
spectrum (which, in turn, is linked to the calibration of the HFI
143-5 detector, as described in~\citealt{planck2013-p08}). It is worth mentioning
here that the \planck\ $143\times143\,\mathrm{GHz}$ spectrum is 2.5\%
{\it lower} than the \WMAP-9 combined V+W power spectrum
\citep{hinshaw2012}. This calibration offset between \planck\ HFI channels
and \WMAP\ is discussed in more detail in ~\citet{planck2013-p01a} and in 
Appendix~\ref{app:wmap}. 

\paragraph{Poisson point source amplitudes:} To avoid any possible
biases in modelling a mixed population of sources (synchrotron+dusty
galaxies) with differing spectra, we solve for each of the
Poisson point source amplitudes as free parameters. Thus, for \planck\
we solve for $A^{\rm PS}_{100}$, $A^{\rm PS}_{143}$, and $A^{\rm
  PS}_{217}$, giving the amplitude of the Poisson point source
contributions to ${\cal D}_{3000}$ for the $100\times100$,
$143\times143$, and $217\times 217$ spectra. The units of $A^{\rm PS}_\nu$
are therefore $\mu{\rm K}^2$. The Poisson point source contribution
to the $143\times 217$ spectrum is expressed as a correlation
coefficient, $r^{\rm PS}_{143\times217}$:
\be {\cal
  D}^{143\times217}_{3000} = r^{\rm PS}_{143\times217}\sqrt{A^{\rm
    PS}_{143} A^{\rm PS}_{217}}.
\ee
Note that we do not use the \Planck\ $100\times143$ and $100\times217$ spectra
in the likelihood, and so we do not include correlation
coefficients $r^{\rm PS}_{100\times143}$ or $r^{\rm PS}_{100\times217}$.
(These spectra carry little additional information on the primordial
CMB, but would require additional foreground parameters had we included
them in the likelihood.)

In an analogous way, the point source amplitudes for ACT and SPT are
characterized by the amplitudes $A^{\rm PS, ACT}_{148}$, $A^{\rm PS,
  ACT}_{217}$, $A^{\rm PS, SPT}_{95}$, $A^{\rm PS, SPT}_{150}$, and
$A^{\rm PS, SPT}_{220}$ (all in units of $\mu{\rm K}^2$) and three
correlation coefficients $r^{\rm PS}_{95\times150}$, $r^{\rm
  PS}_{95\times220}$, and $r^{\rm PS}_{150\times220}$. The last of
these correlation coefficients is common to ACT and SPT.

\paragraph{Kinetic SZ:} The kSZ template used here is from
\citet{Trac:11}. We solve for the amplitude $A^{\rm kSZ}$
(in units of $\mu{\rm K}^2$):
\begin{equation}
    {\cal D}^{\rm kSZ}_\ell = A^{\rm kSZ} {{\cal D}^{\rm kSZ\, template}_{\ell} \over
{\cal D}^{\rm kSZ\, template}_{3000}}. \label{SZ5}
\end{equation}

\paragraph{Thermal SZ:} We use the $\epsilon=0.5$ tSZ template from
\citet{Efstathiou:12} normalized to a frequency of $143\,$GHz.

For cross-spectra between frequencies
$\nu_i$ and $\nu_j$, the tSZ template is normalized as
\begin{equation}
{\cal D}_\ell^{\mathrm{tSZ}_{\nu_i\times \nu_j}}  =
A^{\rm tSZ}_{143}\frac{f(\nu_i)f(\nu_j)}{f^2(\nu_0)}
{{\cal D}_{\ell}^{\rm{tSZ\, template}} \over {\cal D}_{3000}^{\rm{tSZ\, template}}}, \label{SPTF1}
\end{equation}
where $\nu_0$ is the reference frequency of $143\,\mathrm{GHz}$,
 ${\cal D}_{\ell}^{\rm{tSZ \ template}}$ is the template spectrum at $143\,\mathrm{GHz}$, and
\begin{equation}
 f(\nu) =  \left ( x{ e^x + 1 \over e^x -1 } - 4\right ),
 \quad {\rm with}\ x = {h\nu \over k_{\rm B} T_{\rm CMB}}. \label{SPTF2}
\end{equation}
The tSZ contribution is therefore characterized by the
amplitude $A^{\rm tSZ}_{143}$ in units of $\mu{\rm K}^2$.

We neglect the tSZ contribution for any spectra involving the \planck\
$217\,$GHz, ACT $218\,$GHz, and SPT $220\,$GHz channels,
since the tSZ effect has
a null point at $\nu = 217 \,{\rm GHz}$. (For \planck\ the bandpasses of
the $217\,$GHz
detectors see less than $0.1\%$ of the $143\,$GHz tSZ power.)

\paragraph{Cosmic infrared background:} The CIB contributions are neglected in the  \planck\
$100\,$GHz and SPT $95\,$GHz bands and in any cross-spectra involving these
frequencies. The CIB power spectra at  higher frequencies
are characterized by three amplitude parameters and a spectral index,
\beglet
\begin{eqnarray}
{\cal  D}^{{\rm CIB}_{143\times143}}_\ell & = & A^{\rm CIB}_{143} \left ( {\ell \over 3000} \right )^{\gamma^{\rm CIB}}, \\ \label{CIB0a}
{\cal D}^{{\rm CIB}_{217\times217}}_\ell & = & A^{\rm CIB}_{217} \left ( {\ell \over 3000} \right )^{\gamma^{\rm CIB}},  \\ \label{CIB0b}
{\cal  D}^{{\rm CIB}_{143\times217}}_\ell & = & r^{\rm CIB}_{143\times217} \sqrt{ A^{\rm CIB}_{143}A^{\rm CIB}_{217}} \left ( {\ell \over 3000} \right )^{\gamma^{\rm CIB}}, \label{CIB0c}
\end{eqnarray}
\endlet
where $A^{\rm CIB}_{143}$ and $A^{\rm CIB}_{143}$ are expressed in $\mu{\rm K}^2$.
As explained above, we define these amplitudes at the \planck\ CMB
frequencies of $143$ and $217\,$GHz and compute scalings to adjust
these amplitudes to the effective frequencies for a dust-like
spectrum for each experiment.
The scalings are
\begin{eqnarray}
&&
\hspace{-0.02\textwidth}
{\cal D}^{{\rm CIB}_{\nu_i\times \nu_j}}_\ell =    {\cal D}^{{\rm CIB}_{{\nu_{i0}\times \nu_{\! j0}}}}_{3000} \left ( {g(\nu_i) g(\nu_j)  \over
g(\nu_{i0}) g(\nu_{j0}) } \right)
\left(\frac{\nu_i\nu_j}{\nu_{i0} \nu_{j0}}\right)^{\beta_{\rm d}}
\frac{B_{\nu_i}(T_{\rm d})}{B_{\nu_{i0}}(T_{\rm d})}\frac{B_{\nu_j}(T_{\rm d})}{B_{\nu_{j0}}(T_{\rm d})},
\nonumber \\
&&  \label{CIB1}
\end{eqnarray}
where $B_{\nu}(T_{\rm d})$ is the Planck function at a frequency $\nu$,
\begin{equation}
g(\nu) = \left[\partial B_\nu(T)/ \partial T \right]^{-1}|_{ T_{\rm CMB}} \label{CIB2}
\end{equation}
converts antenna temperature to
thermodynamic temperature, $\nu_i$ and $\nu_j$ refer to the \planck/ACT/SPT dust effective
frequencies, and $\nu_{i0}$ and $\nu_{j0}$ refer to the corresponding reference CMB
\planck\ frequencies. In the analysis presented here, the parameters
of the CIB spectrum are fixed to $\beta_{\rm d} = 2.20$ and $T_{\rm d}=9.7\,$K,
as discussed in \citet{Addison:12}.  The model of Eq.~(\ref{CIB1}) then relates the \planck\ reference
amplitudes of Eqs.~ ({\ref{CIB0a}--{\ref{CIB0c}) to the
neighbouring \planck, ACT, and SPT effective frequencies, assuming
that the CIB is perfectly correlated over these small frequency
ranges.

    It has been common practice in recent CMB parameter studies to fix
    the slope of the CIB spectrum to $\gamma^{\rm CIB}=0.8$
    \citep[e.g.,][]{Story:12,Dunkley:13}. In fact, the shape
    of the CIB spectrum is  poorly constrained at frequencies
    below $353\,$GHz and we have decided to reflect this uncertainty
    by allowing the slope $\gamma^{\rm CIB}$ to vary. We adopt a
Gaussian prior on $\gamma^{\rm CIB}$ with a mean of $0.7$ and a dispersion
 of $0.2$. In reality, the CIB spectrum
    is likely to have some degree of curvature reflecting the
    transition between linear (two-halo) and non-linear (one-halo)
    clustering \citep[see e.g.,][]{Cooray02,planck2011-6.6,Amblard:11,Thacker:12}.
   However, a single
   power law is an adequate approximation within
   the restricted multipole range ($500 \la \ell \la 3000$) over which the
   CIB contributes significantly to the \Planck/ACT/SPT high-frequency
   spectra (as judged by the foreground-corrected power spectrum residuals
   shown in Figs.~\ref{PlanckandHighL}, \ref{SPT} and \ref{ACT} below). The 
   prior on $\gamma^{\rm CIB}$ is motivated, in part, by the map-based
   \planck\ CIB analysis discussed in ~\citet{planck2013-pip56}
   \citep[see also][]{planck2013-p13}. Appendix \ref{app:test}
   explores different parameterizations of the CIB power spectrum.

\begin{table*}[tmb] 
\begingroup 
\newdimen\tblskip \tblskip=5pt
\caption{Astrophysical parameters used to model foregrounds in our analysis,
plus instrumental calibration and beam parameters. We include the symbol
for each parameter, the prior range adopted for the MCMC analysis and a
summary definition (see text for details). Square brackets denote hard
priors, parentheses indicate Gaussian priors.  Note that the beam eigenmode
amplitudes require a correlation matrix to describe fully their joint prior,
and that all but $\beta^1_1$ are internally marginalized over rather than
sampled over for the main MCMC runs.  The bottom two blocks are only
used in the analysis including the ACT and SPT high-$\ell$ CMB data.}
\label{tab:fgparams}
\vskip -4mm
\footnotesize 
\setbox\tablebox=\vbox{
\newdimen\digitwidth
\setbox0=\hbox{\rm 0}
\digitwidth=\wd0
\catcode`*=\active
\def*{\kern\digitwidth}
\newdimen\signwidth
\setbox0=\hbox{+}
\signwidth=\wd0
\catcode`!=\active
\def!{\kern\signwidth}
\halign{%
\hbox to 0.9in{#\leaderfil}\tabskip=0.2cm&
            \hfil#\hfil\tabskip=0.2cm&
            #\hfil\tabskip=0pt\cr
\noalign{\doubleline}
\omit \hfil Parameter \hfil& Prior range& \hfil Definition \hfil\cr
\noalign{\vskip 3pt\hrule\vskip 3pt}
$A^{\mathrm{PS}}_{100}$&$[0,360]$&Contribution of Poisson point-source power to $\mathcal{D}^{100\times 100}_{3000}$ for \planck\ (in $\mu\mathrm{K}^2$)\cr
\noalign{\vskip 2pt}
$A^{\mathrm{PS}}_{143}$&$[0,270]$&As for $A^{\mathrm{PS}}_{100}$, but at $143\,$GHz\cr
\noalign{\vskip 2pt}
$A^{\mathrm{PS}}_{217}$&$[0,450]$&As for $A^{\mathrm{PS}}_{100}$, but at $217\,$GHz\cr
\noalign{\vskip 2pt}
$r^{\mathrm{PS}}_{143\times 217}$&$[0,1]$&Point-source correlation coefficient
for \planck\ between $143$ and $217\,$GHz\cr
\noalign{\vskip 2pt}
$A^{\mathrm{CIB}}_{143}$&$[0,20]$&Contribution of CIB power to $\mathcal{D}^{143\times 143}_{3000}$ at the \planck\ CMB frequency for $143\,$GHz (in $\mu\mathrm{K}^2$)\cr
\noalign{\vskip 2pt}
$A^{\mathrm{CIB}}_{217}$&$[0,80]$&As for $A^{\mathrm{CIB}}_{143}$, but for $217\,$GHz\cr
\noalign{\vskip 2pt}
$r^{\mathrm{CIB}}_{143\times 217}$&$[0,1]$&CIB correlation coefficient
between $143$ and $217\,$GHz\cr
\noalign{\vskip 2pt}
$\gamma^{\mathrm{CIB}}$&$[-2,2]\,(0.7\pm0.2)$&Spectral index of the CIB angular power ($\mathcal{D}_\ell \propto \ell^{\gamma^{\mathrm{CIB}}}$)\cr
\noalign{\vskip 2pt}
$A^{\mathrm{tSZ}}$&$[0,10]$&Contribution of tSZ to $\mathcal{D}_{3000}^{143\times 143}$ at $143\,$GHz (in $\mu\mathrm{K}^2$)\cr
\noalign{\vskip 2pt}
$A^{\mathrm{kSZ}}$&$[0,10]$&Contribution of kSZ to $\mathcal{D}_{3000}$ (in $\mu\mathrm{K}^2$)\cr
\noalign{\vskip 2pt}
$\xi^{\mbox{\scriptsize{tSZ$\times$CIB}}}$&$[0,1]$&Correlation coefficient between the CIB and tSZ (see text)\cr
\noalign{\vskip 3pt\hrule\vskip 3pt}
$c_{100}$&$[0.98,1.02]\,(1.0006\pm0.0004)$&Relative power spectrum calibration for \planck\ between
$100\,$GHz and $143\,$GHz\cr
\noalign{\vskip 2pt}
$c_{217}$&$[0.95,1.05]\,(0.9966\pm0.0015)$&Relative power spectrum calibration for \planck\ between
$217\,$GHz and $143\,$GHz\cr
\noalign{\vskip 2pt}
$\beta^i_j$&$(0\pm1)$&Amplitude of the $j$th beam eigenmode ($j=1$--5) for
the $i$th cross-spectrum ($i=1$--4)\cr 
\noalign{\vskip 3pt\hrule\vskip 3pt}
$A^{\mathrm{PS},\,\mathrm{ACT}}_{148}$&$[0,30]$&Contribution of Poisson point-source power to $\mathcal{D}^{148\times 148}_{3000}$ for ACT (in $\mu\mathrm{K}^2$)\cr
\noalign{\vskip 2pt}
$A^{\mathrm{PS},\,\mathrm{ACT}}_{218}$&$[0,200]$&As for $A^{\mathrm{PS},\,\mathrm{ACT}}_{148}$,
but at $218\,$GHz\cr
\noalign{\vskip 2pt}
$r^{\mathrm{PS}}_{150\times 220}$&$[0,1]$&Point-source correlation coefficient
between $150$ and $220\,$GHz (for ACT and SPT)\cr
\noalign{\vskip 2pt}
$A^{\mathrm{ACTe}}_{\mathrm{dust}}$&$[0,5]\,(0.8\pm0.2)$&Contribution from Galactic cirrus to
$\mathcal{D}_{3000}$ at $150\,$GHz for ACTe (in $\mu\mathrm{K}^2$)\cr
\noalign{\vskip 2pt}
$A^{\mathrm{ACTs}}_{\mathrm{dust}}$&$[0,5]\,(0.4\pm0.2)$&As $A^{\mathrm{ACTe}}_{\mathrm{dust}}$, but for
ACTs\cr
\noalign{\vskip 2pt}
$y^{\mathrm{ACTe}}_{148}$&$[0.8,1.3]$&Map-level calibration of ACTe at $148\,$GHz relative
to \planck\ $143\,$GHz\cr
\noalign{\vskip 2pt}
$y^{\mathrm{ACTe}}_{217}$&$[0.8,1.3]$&As $y^{\mathrm{ACTe}}_{148}$, but at $217\,$GHz\cr
\noalign{\vskip 2pt}
$y^{\mathrm{ACTs}}_{148}$&$[0.8,1.3]$&Map-level calibration of ACTs at $148\,$GHz relative
to \planck\ $143\,$GHz\cr
\noalign{\vskip 2pt}
$y^{\mathrm{ACTs}}_{217}$&$[0.8,1.3]$&As $y^{\mathrm{ACTs}}_{148}$, but at $217\,$GHz\cr
\noalign{\vskip 3pt\hrule\vskip 3pt}
$A^{\mathrm{PS},\,\mathrm{SPT}}_{95}$&$[0,30]$&Contribution of Poisson point-source power to $\mathcal{D}^{95\times 95}_{3000}$ for SPT (in $\mu\mathrm{K}^2$)\cr
\noalign{\vskip 2pt}
$A^{\mathrm{PS},\,\mathrm{SPT}}_{150}$&$[0,30]$&As for $A^{\mathrm{PS},\,\mathrm{SPT}}_{95}$,
but at $150\,$GHz\cr
\noalign{\vskip 2pt}
$A^{\mathrm{PS},\,\mathrm{SPT}}_{220}$&$[0,200]$&As for $A^{\mathrm{PS},\,\mathrm{SPT}}_{95}$,
but at $220\,$GHz\cr
\noalign{\vskip 2pt}
$r^{\mathrm{PS}}_{95\times 150}$&$[0,1]$&Point-source correlation coefficient
between $95$ and $150\,$GHz for SPT\cr
\noalign{\vskip 2pt}
$r^{\mathrm{PS}}_{95\times 220}$&$[0,1]$&As $r^{\mathrm{PS}}_{95\times 150}$, but between
$95$ and $220\,$GHz\cr
\noalign{\vskip 2pt}
$y^{\mathrm{SPT}}_{95}$&$[0.8,1.3]$&Map-level calibration of SPT at $95\,$GHz relative
to \planck\ $143\,$GHz\cr
\noalign{\vskip 2pt}
$y^{\mathrm{SPT}}_{150}$&$[0.8,1.3]$&As for $y^{\mathrm{SPT}}_{95}$, but at $150\,$GHz\cr
\noalign{\vskip 2pt}
$y^{\mathrm{SPT}}_{220}$&$[0.8,1.3]$&As for $y^{\mathrm{SPT}}_{95}$, but at $220\,$GHz\cr
\noalign{\vskip 3pt\hrule\vskip 3pt}}}
\endPlancktable 
\endgroup
\end{table*}

\paragraph{Thermal-SZ/CIB cross-correlation:} The cross-correlation between dust emission from CIB galaxies and SZ
emission from clusters (tSZ$\times$CIB) is expected to be
non-zero. Because of uncertainties in the modelling of the CIB, it is
difficult to compute this correlation with a high degree
of precision. \citet{Addison:12b} present a halo-model approach to
model this term and conclude that anti-correlations of around
10--20\% are plausible between the clustered CIB components
and the SZ at $150$ GHz.  The tSZ$\times$CIB correlation is therefore expected to make a minor contribution
to the unresolved foreground emission, but it is nevertheless worth
including to determine how it might interact with other sub-dominant
components, in particular the kSZ contribution.  We use the
\citet{Addison:12b} template spectrum in this paper and model
the frequency dependence of the power spectrum as follows:
\begin{eqnarray}
{\cal D}_\ell^{\tSZCIB_{\nu_i \times \nu_j}} &=&- \xi^{\tSZCIB}
{\cal D}_\ell^{\tSZCIB\,\mathrm{template}} \nonumber \\
&&\hspace{-0.015\textwidth}\times \left ( \sqrt {{\cal D}^{{\rm CIB}_{\nu_i\times\nu_i}}_{3000} {\cal D}^{{\rm tSZ}_{\nu_j\times \nu_j}}_{3000}}
  +  \sqrt {{\cal D}^{{\rm CIB}_{\nu_j\times \nu_j}}_{3000} {\cal D}^{{\rm tSZ}_{\nu_i\times \nu_i}}_{3000}}
\right ) , \label{CIB3}
\end{eqnarray}
where ${\cal D}_\ell^{\tSZCIB\,\mathrm{template}}$ is the \citet{Addison:12b}
template spectrum normalized to unity at $\ell=3000$ and
${\cal D}_\ell^{{\rm CIB}_{\nu_i\times \nu_i}}$ and ${\cal D}_\ell^{{\rm tSZ}_{\nu_i\times \nu_i}}$ are
given by Eqs.~(\ref{SPTF1}) and (\ref{CIB2}). The tSZ$\times$CIB contribution
is therefore characterized by the dimensionless cross-correlation coefficient
$\xi^{\tSZCIB}$. With the definition of Eq.~(\ref{CIB3}), a positive
value of $\xi^{\tSZCIB}$ corresponds to an anti-correlation 
between the CIB and the tSZ signals.

\paragraph{Galactic dust:} For the masks used in the \planck\
{\tt CamSpec} likelihood,  Galactic dust
makes a small contribution to ${\cal
  D}_{3000}$ of around $5\, {\mu}{\rm K}^2$ to the $217\times217$
power spectrum, $1.5\, {\mu}{\rm K}^2$ to the $143\times217$
spectrum, and around $0.5\, {\mu}{\rm K}^2$ to the $143\times143$ spectrum.  
We subtract the Galactic dust  contributions from these 
power spectra using a ``universal'' dust template
spectrum (at high multipoles this is accurately represented by a
power law $\mathcal{D}^{\rm dust}_{\ell} \propto \ell^{-0.6}$). The template
spectrum is  based on an
analysis of the $857\,$GHz \planck\ maps described in ~\citet{planck2013-p08}, which
uses mask-differenced power spectra to separate Galactic dust from
an isotropic extragalactic CIB contribution. This Galactic dust
correction is kept fixed with an amplitude determined by template
fitting the $217$ and $143\,$GHz \planck\ maps to the $857\,$GHz map, as
described in  ~\citet{planck2013-p08}. Galactic dust contamination is
ignored in the $100\times 100$ spectrum.\footnote{\referee{The contribution of Galactic emission in the $100\times 100\,{\rm GHz}$ spectrum used in the \camspec\
likelihood is undetectable at multipoles $\ell > 50$, either via cross-correlation with the $857$\,GHz
maps or via analysis of mask-differenced $100\times100$ spectra.}} The Galactic dust template spectrum
is actually a good fit to the dust contamination at low multipoles,
$\ell \ll 1000$; however, we limit the effects of any inaccuracies in dust
subtraction at low multipoles by truncating the $217\times217$ and
$143\times217$ spectra at a minimum multipole of
$\ell_{\rm min}=500$. (At multipoles $\ell \la 1000$, the \planck\ temperature
power spectra are signal dominated, so the $100\times100$ and $143\times143$
spectra contain essentially all of the information on cosmology.)

\begin{figure}
\centering
\includegraphics[width=8.8cm]{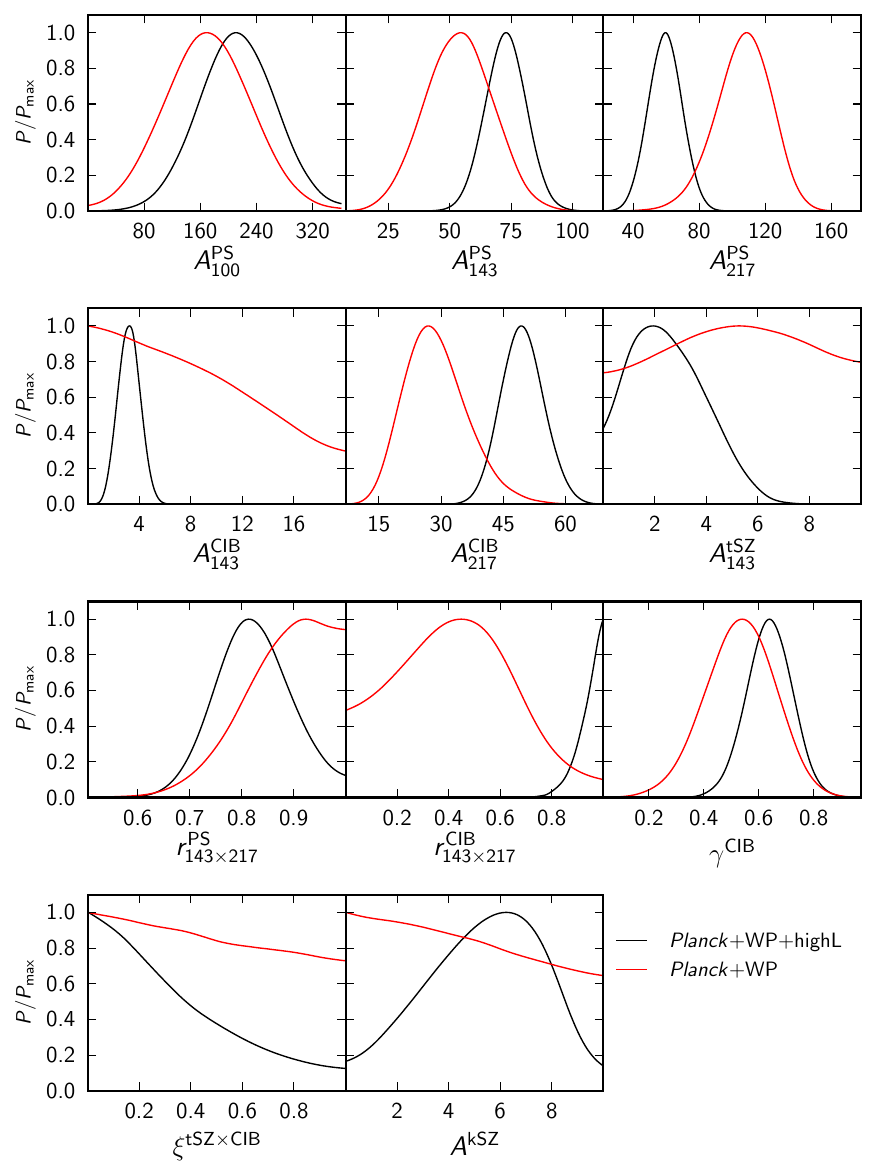}
\caption {Comparison of the posterior distributions of the foreground parameters for \planck+\WP\
(red) and \planck+\WP+\highL\ (black).
}
\label{fig:PlanckandHighLparams}
\end{figure}

Compared to the contribution of Poisson point sources and the CIB,
Galactic dust is a minor foreground component
at $217\,$GHz within our default mask, which retains 37\% of the sky.
However,  the contribution of Galactic dust
emission rises rapidly as more sky area is used.  Extending the sky mask
to  $65\%$ of the sky
\citep[using the sequence of masks described in []{planck2013-p08},
Galactic dust contributes to
${\cal D}_{3000}$ around $50\,\mu{\rm K}^2$ at $217\,$GHz 
(rising to around $200\,\mu{\rm K}^2$ on the scale of
the first acoustic peak) and becomes a {\it major}
foreground component, with an amplitude close to the 
net contribution of Poisson point sources and the clustered
CIB. There is therefore a trade-off between limiting the
signal-to-noise at $143$ and $217\,$GHz, by restricting the sky area, and
potential systematic errors associated with modelling Galactic dust
over a large area of sky (i.e., sensitivity to the assumption of a
``universal'' dust template spectrum).  We have chosen to be conservative 
in this first cosmological analysis of \planck\ by limiting the sky area
at $143$ and $217\,\mathrm{GHz}$ so that dust contamination is a minor foreground at high multipoles.
As a further test of the importance of
Galactic dust, we have analysed a \planck\ likelihood 
that retains only  $24.7\%$ of the sky \citep[see][]{planck2013-p08}
at $217\,$GHz. Within this mask the CIB dominates over Galactic dust
at multipoles $\ell \ga 500$. There is a signal-to-noise
penalty in using such a small area of sky at $217\,$GHz, but otherwise
the results from this likelihood are in good agreement with the 
results presented here. 
 With the conservative choices adopted in this paper, 
 Galactic dust has no significant impact on our
cosmological results.

We follow R12 and subtract a small-scale dust contribution of ${\cal
  D}_\ell^{\rm dust} = 2.19\,\mu\mathrm{K}^2 (\ell/3000)^{-1.2}$ from the
R12 $220\,$GHz spectrum.
This correction was  determined by cross-correlating the SPT data with
model 8 of \citet{Finkbeiner:99}.  For the ACT data we marginalize over a residual Galactic dust
component ${\cal D}^{\rm dust}_\ell =
A^{\mathrm{ACTe/s}}_{\mathrm{dust}} (\ell/3000)^{-0.7}$, with different
amplitudes for the southern and equatorial spectra, imposing Gaussian priors
and frequency scaling as described in \citet{Dunkley:13}.

Notice that the spectral index of the SPT dust correction is
significantly steeper than the dust correction applied to the \planck\
spectra.  In future analyses it would be useful to derive more
accurate dust corrections for the high-resolution CMB data by
cross-correlating the SPT and ACT  maps with the \planck\ 545 and $857\,$GHz
maps.  Since the dust corrections are relatively small  for the high-resolution
data used here, we adopt the correction described above in this paper.

In application of the likelihood to \planck\ data alone, the model for
unresolved foregrounds and relative calibrations contains 13
parameters. In addition, we can solve for up to 20 beam eigenmode
amplitudes (five amplitudes for each of the four spectra used in the
\planck\ likelihood; see ~\citealt{planck2013-p08}).  In practice, we
find that (usually) only the first beam eigenmode for the
$100\times100$ spectrum, $\beta^1_1$, has a posterior distribution that
differs perceptibly from the prior, and we obtain nearly identical results on both foreground and
cosmological parameters if we treat only the amplitude of this eigenmode as a parameter
and analytically marginalize over the rest. This is the default
adopted in this paper. (The analytic marginalization improves stability of the minimisation for best-fit
searches, and makes the \planck\ likelihood
less  cumbersome for the user.)

The addition of ACT and SPT data introduces 17 extra
parameters. We provide a summary of the 50 foreground and nuisance
parameters in Table~\ref{tab:fgparams}, including the prior ranges
adopted in our MCMC analysis\footnote{%
Note that the foreground, calibration and beam parameters are all ``fast'' parameters as regards the MCMC sampling, and their inclusion has only a small impact on the computational speed. Marginalising over 19 of the \planck\ beam parameters therefore leads only to $O(1)$ improvements in speed.}. The choice of priors for many of these
parameters is, to a large extent, subjective. They were chosen at an
early stage in the \planck\ analysis to reflect ``theoretically
plausible'' allowed ranges of the foreground parameters and to be broad
compared to the results from high-resolution CMB experiments (which
evolved over the course of this analysis as results from more ACT and SPT data were
published). The foreground parameters from ACT and SPT depend on the assumptions
of the underlying cosmology, and hence it is possible to introduce biases in the
solutions for extensions to the base \lcdm\ cosmology if overly restrictive
foreground priors are imposed on the \planck\ data. Using the 
priors summarized in Table~\ref{tab:fgparams},
the consistency between
the \planck-alone results and the solutions for \planck\ combined with ACT and SPT provides a crude
(but informative) measure of the sensitivity of cosmological results on 
the foreground model. Appendix \ref{app:test} discusses the effects
on extended \lcdm\ models of varying the priors on minor  foreground
components.

\subsection{The base \lcdm\ model}

\begin{table*}
\begin{center}
\caption{Best-fit values and $68\%$ confidence limits for the base \lcdm\ model.
Beam and calibration parameters, and additional nuisance parameters for
``highL'' data sets are not listed for brevity
but may be found in the Explanatory Supplement \citep{planck2013-p28}.
\label{LCDMForegroundparams}
}

\begingroup
\newdimen\tblskip \tblskip=5pt
\nointerlineskip
\vskip -3mm
\scriptsize
\setbox\tablebox=\vbox{
    \newdimen\digitwidth
    \setbox0=\hbox{\rm 0}
    \digitwidth=\wd0
    \catcode`"=\active
    \def"{\kern\digitwidth}
    \newdimen\signwidth
    \setbox0=\hbox{+}
    \signwidth=\wd0
    \catcode`!=\active
    \def!{\kern\signwidth}
\halign{
\hbox to 0.9in{$#$\leaderfil}\tabskip=1.5em&\hfil$#$\hfil\tabskip=0.5em&
\hfil$#$\hfil\tabskip=1.7em&
\hfil$#$\hfil\tabskip=0.5em&
\hfil$#$\hfil\tabskip=1.7em&
\hfil$#$\hfil\tabskip=0.5em&
\hfil$#$\hfil\tabskip=1.7em&
\hfil$#$\hfil\tabskip=0.5em&
\hfil$#$\hfil\tabskip=0pt\cr
\noalign{\doubleline}
\multispan1\hfil \hfil&\multispan2\hfil \planck+\WP\hfil&\multispan2\hfil \Planck+\WP+\HighL\hfil&\multispan2\hfil \Planck+\lensing+\WP+\HighL\hfil&\multispan2\hfil \Planck+\WP+\HighL+BAO\hfil\cr
\noalign{\vskip -3pt}
\omit&\multispan2\hrulefill&\multispan2\hrulefill&\multispan2\hrulefill&\multispan2\hrulefill\cr
\omit\hfil Parameter\hfil&\omit\hfil Best fit\hfil&\omit\hfil 68\% limits\hfil&\omit\hfil Best fit\hfil&\omit\hfil 68\% limits\hfil&\omit\hfil Best fit\hfil&\omit\hfil 68\% limits\hfil&\omit\hfil Best fit\hfil&\omit\hfil 68\% limits\hfil\cr
\noalign{\vskip 3pt\hrule\vskip 5pt}
\Omega_{\mathrm{b}} h^2&0.022032&0.02205\pm 0.00028&0.022069&0.02207\pm 0.00027&0.022199&0.02218\pm 0.00026&0.022161&0.02214\pm 0.00024\cr
\noalign{\vskip 3pt}
\Omega_{\mathrm{c}} h^2&0.12038&0.1199\pm 0.0027&0.12025&0.1198\pm 0.0026&0.11847&0.1186\pm 0.0022&0.11889&0.1187\pm 0.0017\cr
\noalign{\vskip 3pt}
100\theta_{\mathrm{MC}}&1.04119&1.04131\pm 0.00063&1.04130&1.04132\pm 0.00063&1.04146&1.04144\pm 0.00061&1.04148&1.04147\pm 0.00056\cr
\noalign{\vskip 3pt}
\tau&0.0925&0.089^{+0.012}_{-0.014}&0.0927&0.091^{+0.013}_{-0.014}&0.0943&0.090^{+0.013}_{-0.014}&0.0952&0.092\pm 0.013\cr
\noalign{\vskip 3pt}
n_\mathrm{s}&0.9619&0.9603\pm 0.0073&0.9582&0.9585\pm 0.0070&0.9624&0.9614\pm 0.0063&0.9611&0.9608\pm 0.0054\cr
\noalign{\vskip 3pt}
\ln(10^{10} A_\mathrm{s})&3.0980&3.089^{+0.024}_{-0.027}&3.0959&3.090\pm 0.025&3.0947&3.087\pm 0.024&3.0973&3.091\pm 0.025\cr
\noalign{\vskip 5pt\hrule\vskip 3pt}
A^{\mathrm{PS}}_{100}&152&171\pm 60&209&212\pm 50&204&213\pm 50&204&212\pm 50\cr
\noalign{\vskip 3pt}
A^{\mathrm{PS}}_{143}&63.3&54\pm 10&72.6&73\pm 8&72.2&72\pm 8&71.8&72.4\pm 8.0\cr
\noalign{\vskip 3pt}
A^{\mathrm{PS}}_{217}&117.0&107^{+20}_{-10}&59.5&59\pm 10&60.2&58\pm 10&59.4&59\pm 10\cr
\noalign{\vskip 3pt}
A^{\mathrm{CIB}}_{143}&0.0&< 10.7&3.57&3.24\pm 0.83&3.25&3.24\pm 0.83&3.30&3.25\pm 0.83\cr
\noalign{\vskip 3pt}
A^{\mathrm{CIB}}_{217}&27.2&29^{+6}_{-9}&53.9&49.6\pm 5.0&52.3&50.0\pm 4.9&53.0&49.7\pm 5.0\cr
\noalign{\vskip 3pt}
A^{\mathrm{tSZ}}_{143}&6.80&\dots&5.17&2.54^{+1.1}_{-1.9}&4.64&2.51^{+1.2}_{-1.8}&4.86&2.54^{+1.2}_{-1.8}\cr
\noalign{\vskip 3pt}
r^{\mathrm{PS}}_{143\times 217}&0.916&> 0.850&0.825&0.823^{+0.069}_{-0.077}&0.814&0.825\pm 0.071&0.824&0.823\pm 0.070\cr
\noalign{\vskip 3pt}
r^{\mathrm{CIB}}_{143\times 217}&0.406&0.42\pm 0.22&1.0000&> 0.930&1.0000&> 0.928&1.0000&> 0.930\cr
\noalign{\vskip 3pt}
\gamma^{\mathrm{CIB}}&0.601&0.53^{+0.13}_{-0.12}&0.674&0.638\pm 0.081&0.656&0.643\pm 0.080&0.667&0.639\pm 0.081\cr
\noalign{\vskip 3pt}
\xi^{\mathrm{tSZ}\times\mathrm{CIB}}&0.03&\dots&0.000&< 0.409&0.000&< 0.389&0.000&< 0.410\cr
\noalign{\vskip 3pt}
A^{\mathrm{kSZ}}&0.9&\dots&0.89&5.34^{+2.8}_{-1.9}&1.14&4.74^{+2.6}_{-2.1}&1.58&5.34^{+2.8}_{-2.0}\cr
\noalign{\vskip 5pt\hrule\vskip 3pt}
\Omega_\Lambda&0.6817&0.685^{+0.018}_{-0.016}&0.6830&0.685^{+0.017}_{-0.016}&0.6939&0.693\pm 0.013&0.6914&0.692\pm 0.010\cr
\noalign{\vskip 3pt}
\sigma_8&0.8347&0.829\pm 0.012&0.8322&0.828\pm 0.012&0.8271&0.8233\pm 0.0097&0.8288&0.826\pm 0.012\cr
\noalign{\vskip 3pt}
z_{\mathrm{re}}&11.37&11.1\pm 1.1&11.38&11.1\pm 1.1&11.42&11.1\pm 1.1&11.52&11.3\pm 1.1\cr
\noalign{\vskip 3pt}
H_0&67.04&67.3\pm 1.2&67.15&67.3\pm 1.2&67.94&67.9\pm 1.0&67.77&67.80\pm 0.77\cr
\noalign{\vskip 3pt}
\mathrm{Age}/\mathrm{Gyr}&13.8242&13.817\pm 0.048&13.8170&13.813\pm 0.047&13.7914&13.794\pm 0.044&13.7965&13.798\pm 0.037\cr
\noalign{\vskip 3pt}
100\theta_\ast&1.04136&1.04147\pm 0.00062&1.04146&1.04148\pm 0.00062&1.04161&1.04159\pm 0.00060&1.04163&1.04162\pm 0.00056\cr
\noalign{\vskip 3pt}
r_{\mathrm{drag}}&147.36&147.49\pm 0.59&147.35&147.47\pm 0.59&147.68&147.67\pm 0.50&147.611&147.68\pm 0.45\cr
\noalign{\vskip 5pt\hrule\vskip 3pt}
} 
} 
\endPlancktable
\endgroup

\end{center}
\end{table*}

Cosmological and foreground parameters for the base six-parameter \lcdm\
model are listed in Table~\ref{LCDMForegroundparams}, which
gives best-fit values and 68\% confidence limits.  The first two
columns list the parameters derived from the \Planck+\WP\ analysis
discussed in Sect.~\ref{sec:planck_lcdm}, and are repeated here for easy
reference.  The next two columns list the results of combining the
\planck+\WP\ likelihoods with the ACT and SPT likelihoods following
the model described above. We refer to this combination as
``\planck+\WP+\highL'' in this paper. The remaining columns list the
parameter constraints combining the \Planck+\WP+\highL\ likelihood with
the \planck\ lensing and BAO likelihoods (see Sect.~\ref{sec:datasets}). Table \ref{LCDMForegroundparams} lists the cosmological parameters
for the base \lcdm\ model and a selection of derived cosmological parameters.
These parameters are remarkably stable for such data combinations. We also
list the values of the parameters describing the \planck\ foregrounds.
A full list of all parameter
values, including nuisance parameters, is given in the Explanatory Supplement
\citep{planck2013-p28}.

A comparison of the foreground parameter constraints from \planck+\WP\ and
\Planck+\WP+\highL\ is shown in Fig.~\ref{fig:PlanckandHighLparams}; the
corresponding cosmological parameter constraints are shown in
Fig.~\ref{fig:margeLCDM}.

We can draw the following general conclusions.
\begin{itemize}
\item The cosmological parameters for the base \lcdm\ model are
extremely insensitive to the foreground model described in the
previous subsection. The addition of the ACT and SPT data causes the
posterior distributions of cosmological parameters to shift by much
less than one standard deviation.
\item With \planck\ data alone, the CIB amplitude at $217\,$GHz is
strongly degenerate with the $217\,$GHz Poisson point source
amplitude. This degeneracy is broken by the addition of the high-resolution CMB data. This degeneracy must be borne in mind when
interpreting \planck-only solutions for CIB parameters; the sum of
the Poisson point source and CIB contributions are well constrained by
\planck\ at $217\,$GHz (and in good agreement with the map-based CIB
\planck\ analysis reported in \citealt{planck2013-CIB}), whereas the
individual contributions are not. Another feature of the CIB
parameters is that we typically find smaller values of the CIB spectral index,
$\gamma^{\rm CIB}$, in \planck-alone solutions compared to
\planck+\highL\ solutions (which can be seen in
Fig.~\ref{fig:PlanckandHighLparams}). This provided additional
motivation to treat $\gamma^{\rm CIB}$ as a parameter in the \planck\
likelihood rather than fixing it to a particular value. There is
evidence from the \planck\ spectra (most clearly seen by differencing
the $217\times 217$ and $143\times 143$ spectra) that the CIB spectrum
at $217\,$GHz flattens in slope over the multipole range $500 \la \ell
\la 1000$. This will be explored in further detail in future papers
(see also Appendix \ref{app:test}).
\item The  addition of the ACT and SPT data constrains the
thermal SZ amplitude, which is poorly determined by \planck\ alone. In
the \planck-alone analysis, the tSZ amplitude is strongly degenerate
with the Poisson point source amplitude at $100\,$GHz. This degeneracy is
broken when the high-resolution CMB data are added to \planck.
\end{itemize}

The last two points are demonstrated clearly in Fig.~\ref{PlanckandHighL},
which shows the residuals of the \planck\ spectra
with respect to the best-fit cosmology for the \planck+\WP\ analysis
compared to the \planck+\WP+\highL\ fits.  The addition of high-resolution
CMB data also strongly
constrains the net contribution from the kSZ and tSZ$\times$CIB components
(dotted lines), though these components are degenerate with each other
(and tend to cancel). 

Although the foreground parameters for the \planck+\WP\ fits
can differ substantially from those for \planck+\WP+\highL,
the total foreground spectra are insensitive to the addition of the high-resolution CMB data. For example, for the $217\times 217$ spectrum, the differences in the total foreground solution are less than $10\,\mu\mathrm{K}^2$ at $\ell=2500$.
The net residuals after subtracting both the foregrounds and CMB spectrum  (shown in the lower panels of each sub-plot in Fig.~\ref{PlanckandHighL})
are  similarly insensitive to the addition of the high-resolution CMB data. 
The foreground model is sufficiently complex that it has a high ``absorptive capacity''
to any smoothly-varying frequency-dependent differences between spectra (including
beam errors).

\begin{figure*}
\centering
\hspace*{2mm}\includegraphics[width=57mm,angle=0]{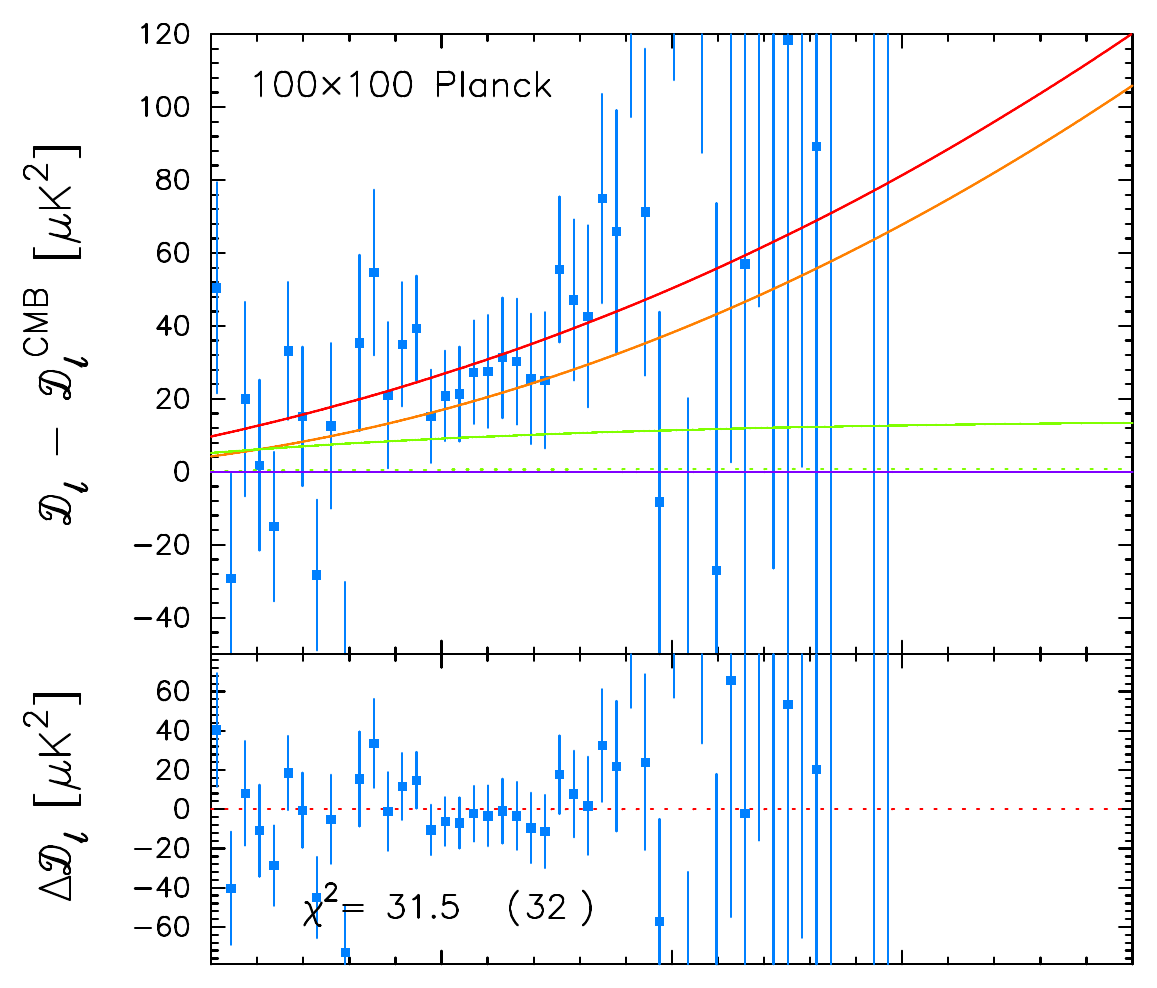}
\hspace*{2mm}\includegraphics[width=48mm,angle=0]{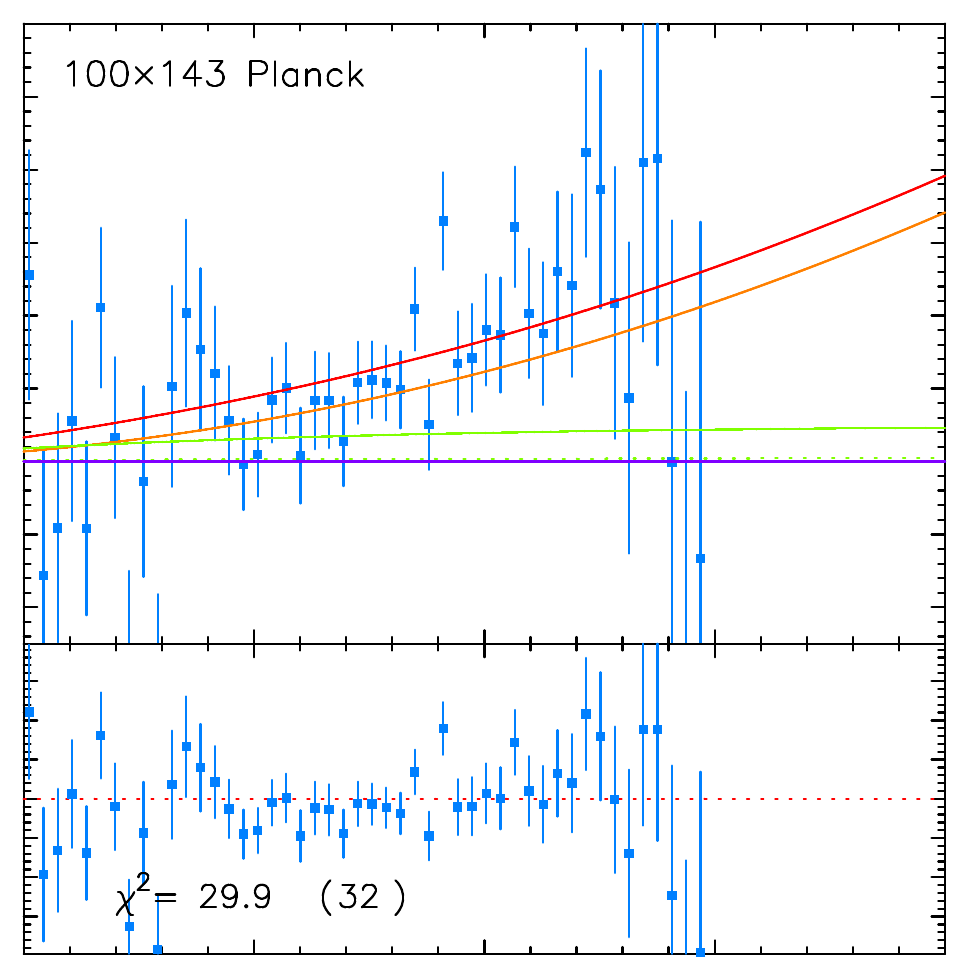}
\hspace*{2mm}\includegraphics[width=48mm,angle=0]{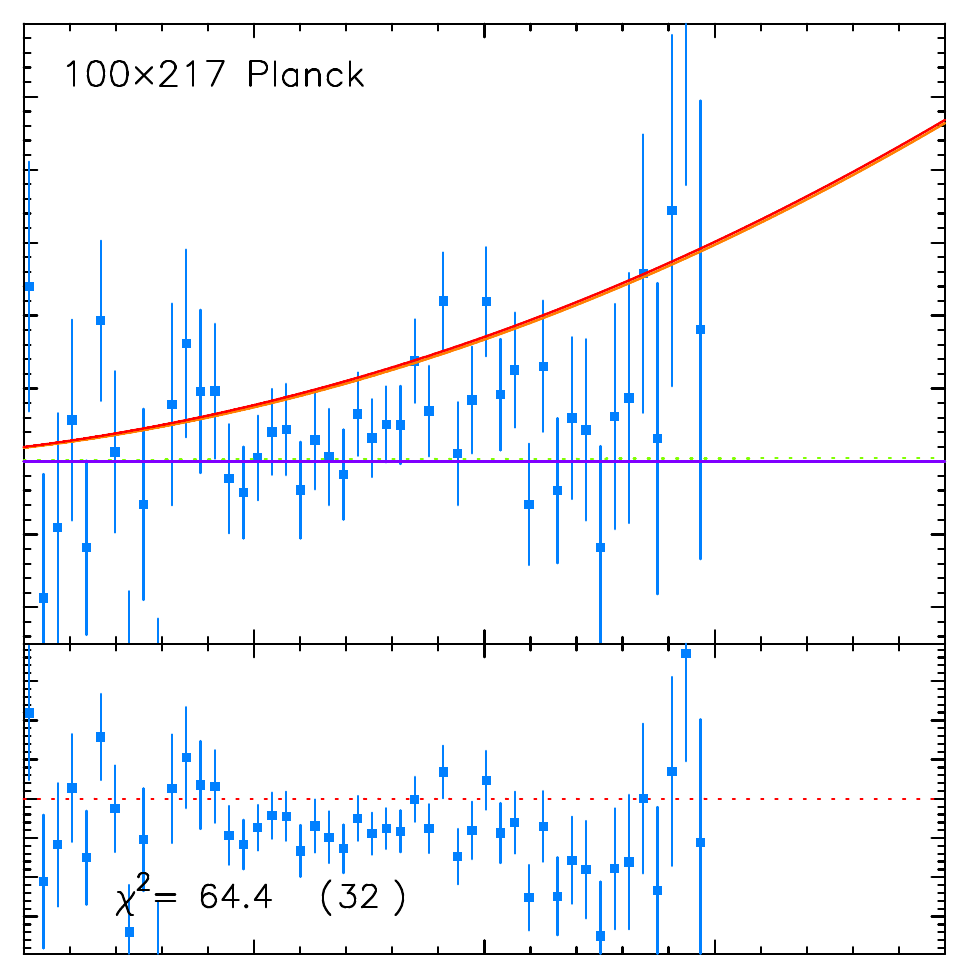}
\\
\hspace*{2mm}\includegraphics[width=57mm,angle=0]{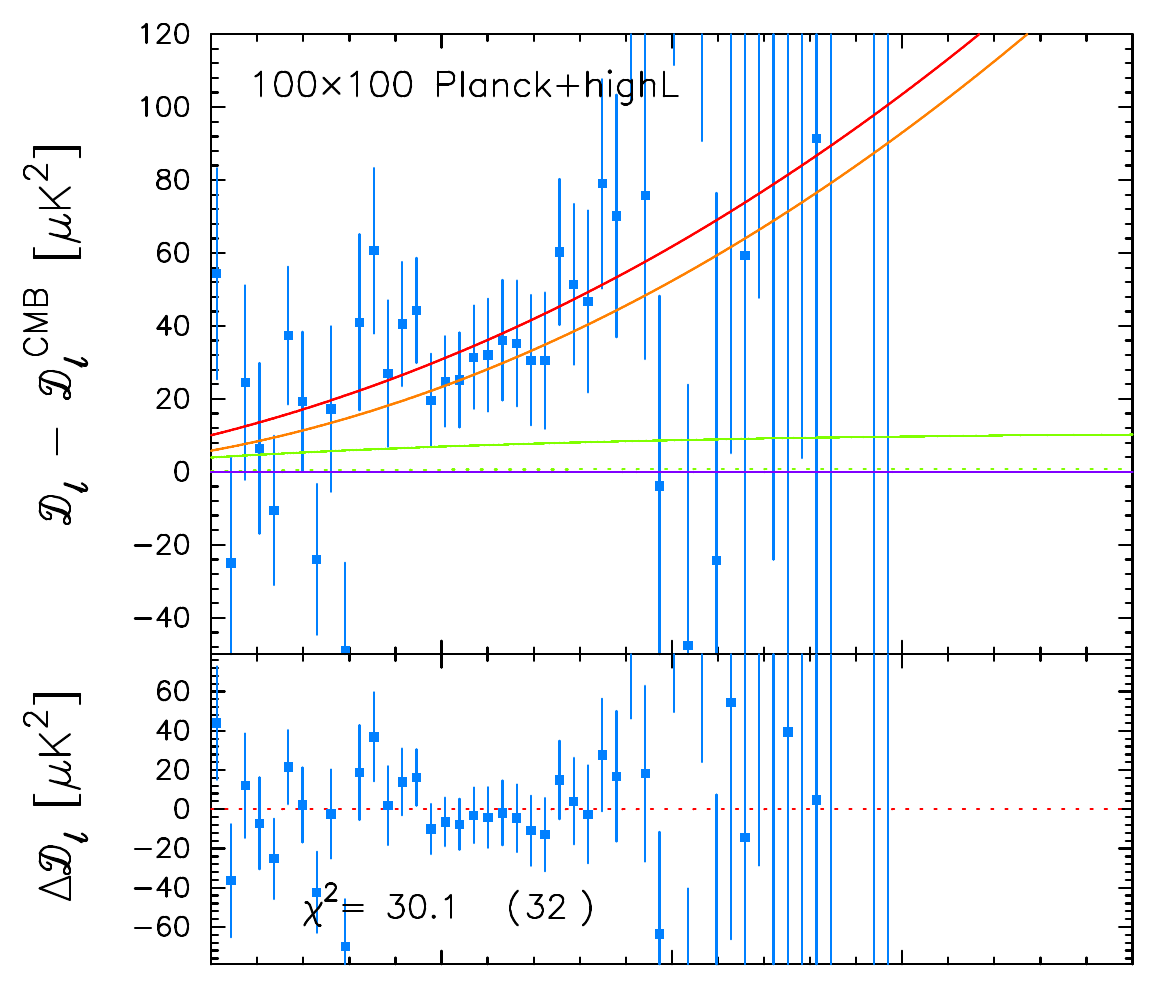}
\hspace*{2mm}\includegraphics[width=48mm,angle=0]{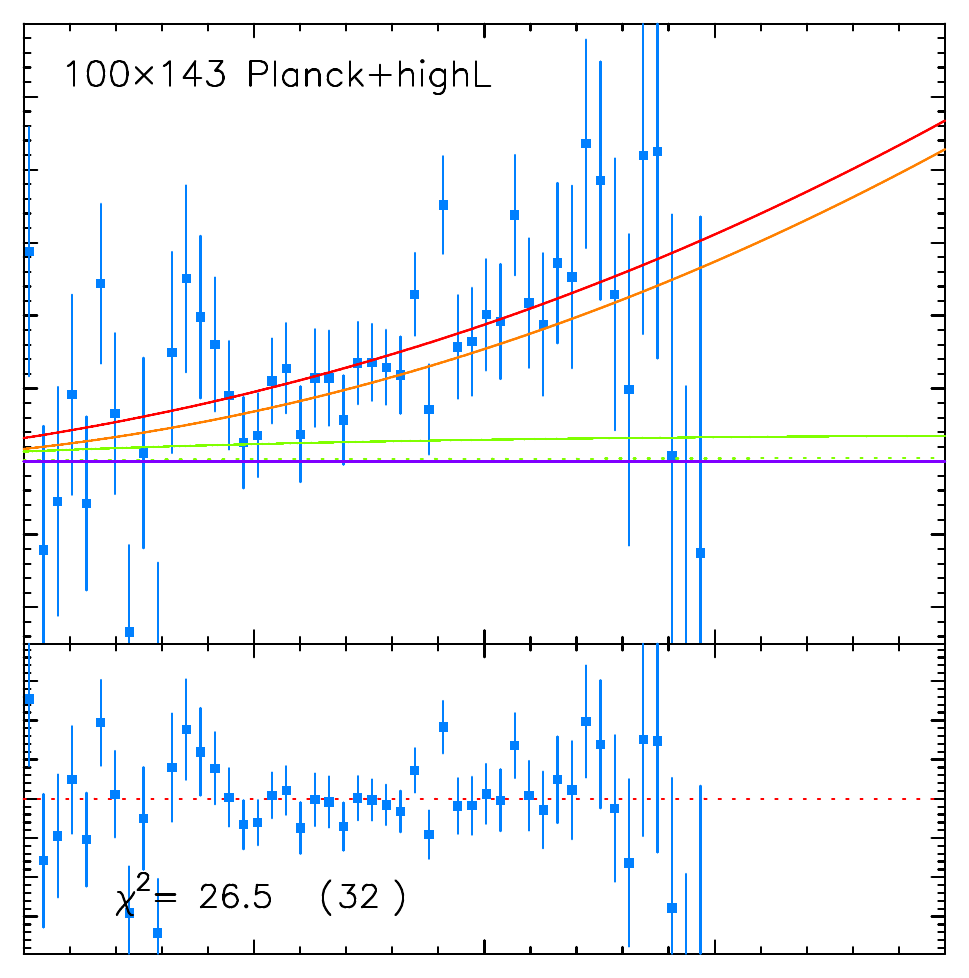}
\hspace*{2mm}\includegraphics[width=48mm,angle=0]{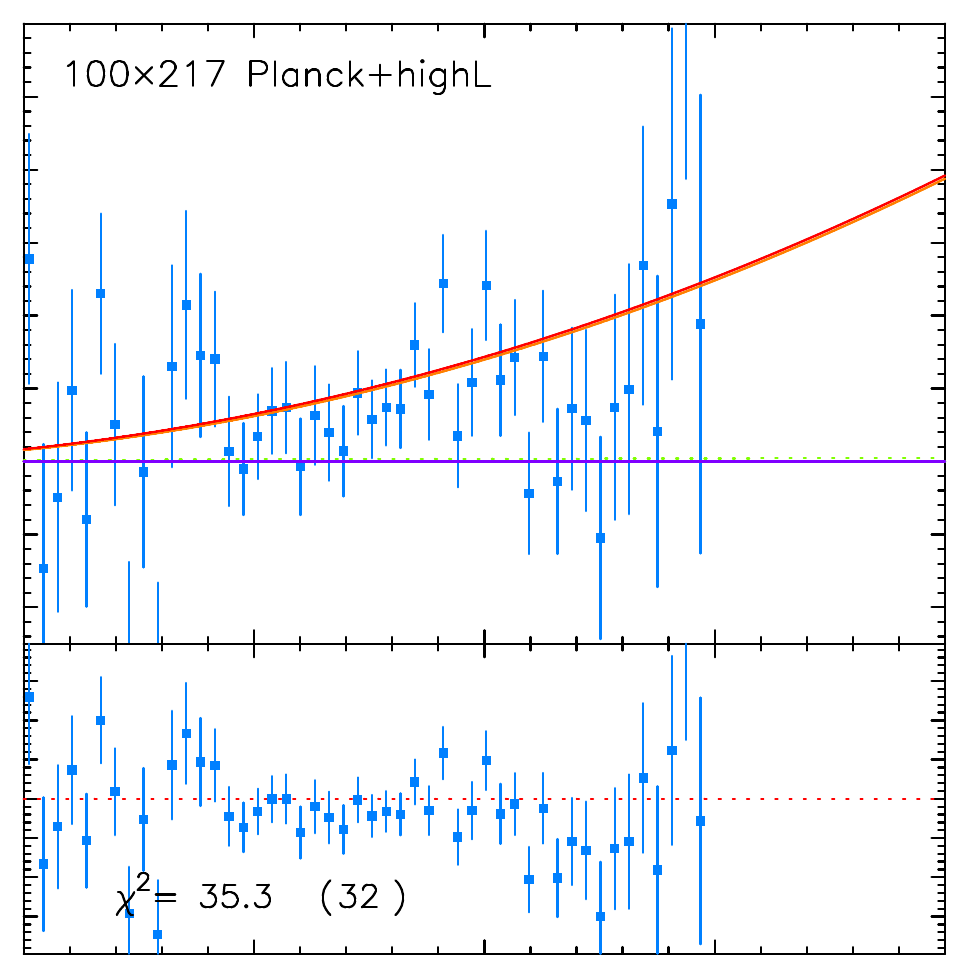}
\\
\hspace*{2mm}\includegraphics[width=57mm,angle=0]{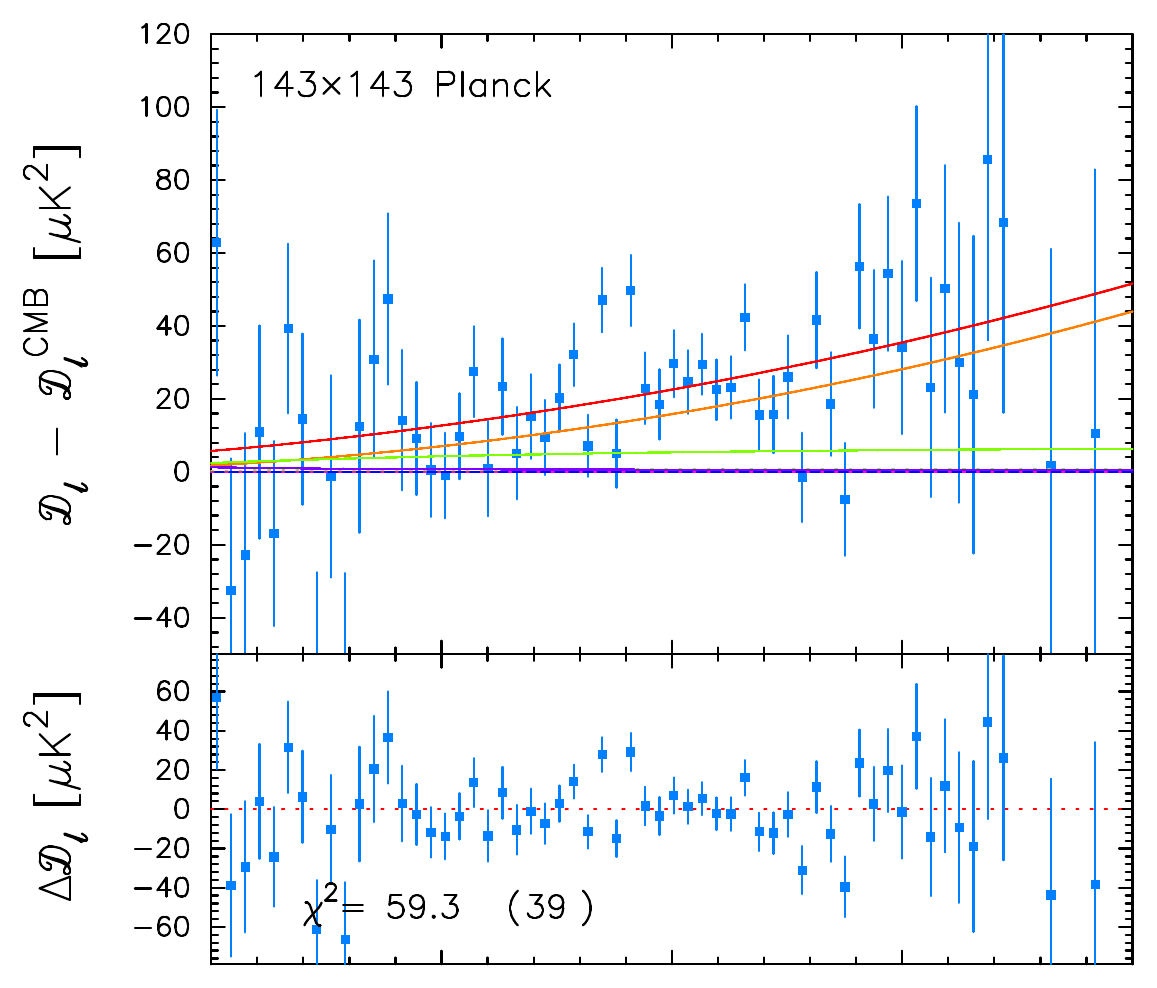}
\hspace*{2mm}\includegraphics[width=48mm,angle=0]{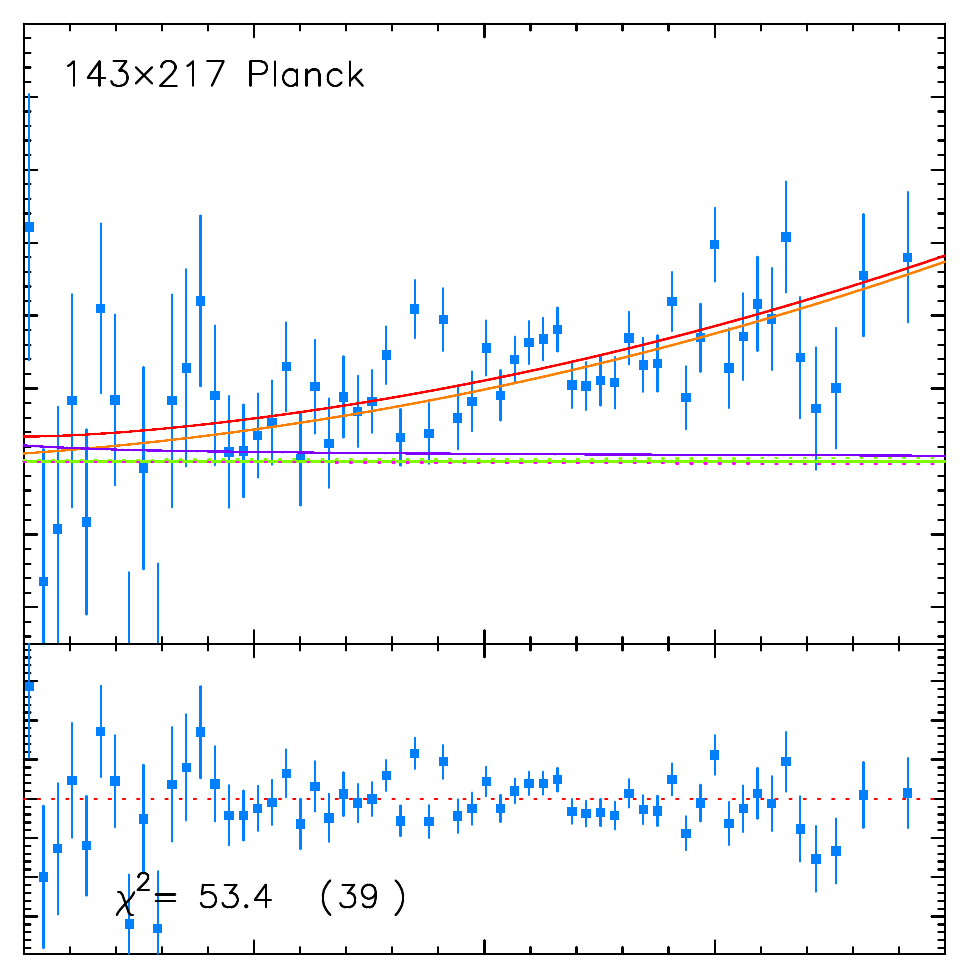}
\hspace*{2mm}\includegraphics[width=48mm,angle=0]{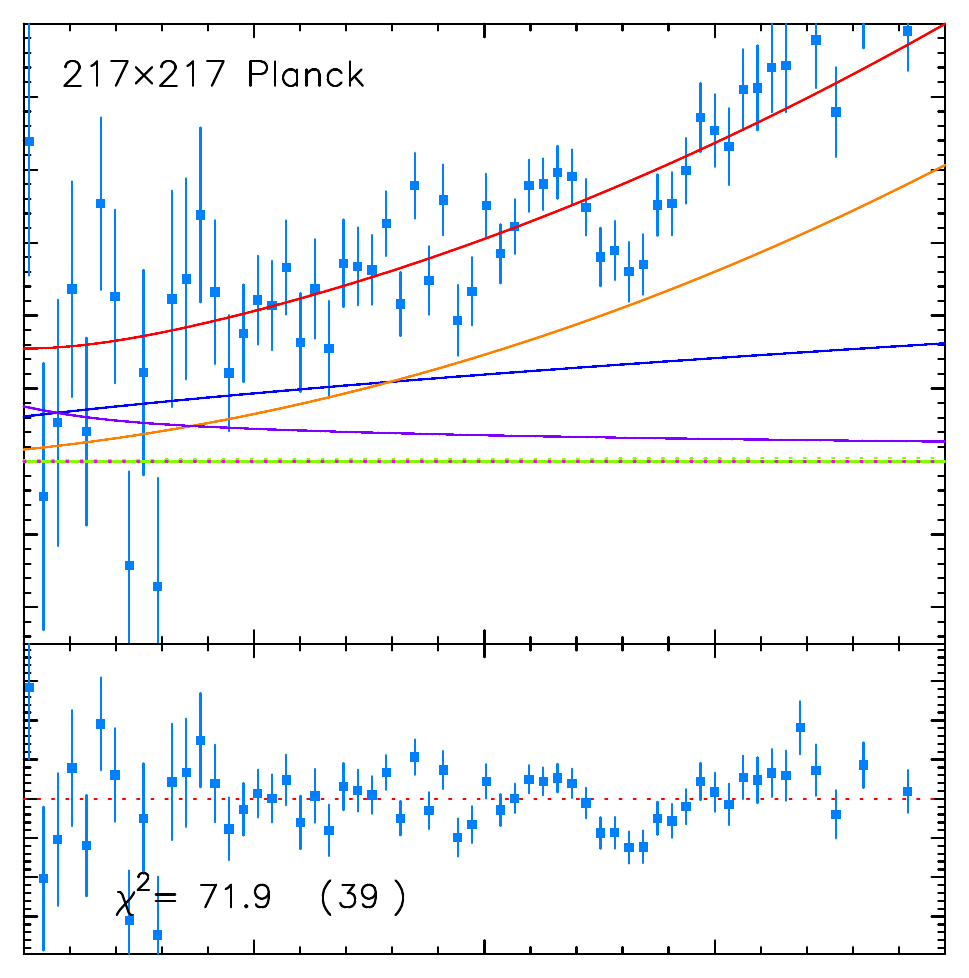}
\\
\hspace*{2.0mm} \includegraphics[width=59.5mm,angle=0]{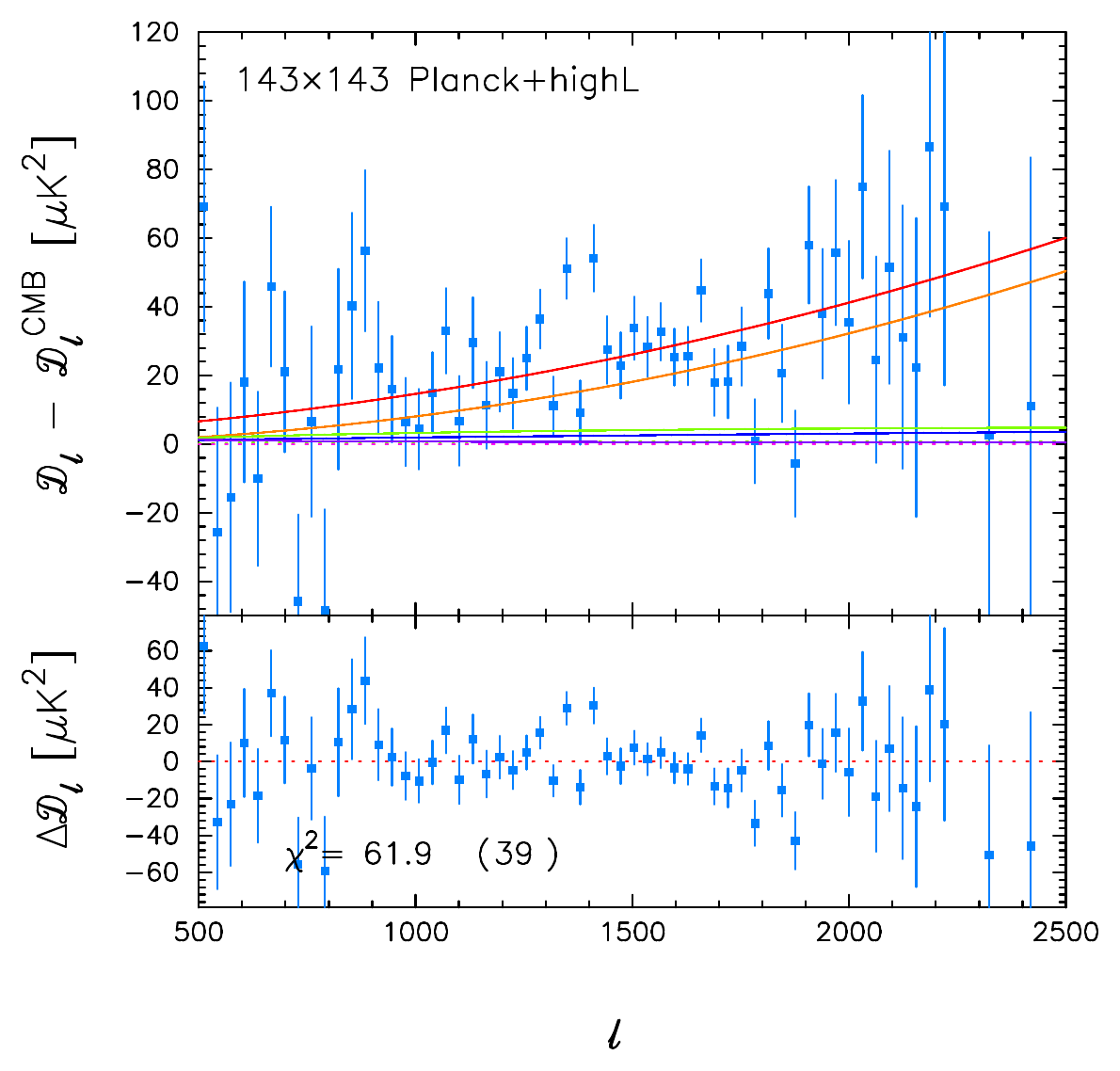}
\hspace*{-1.5mm}\includegraphics[width=51mm,angle=0]{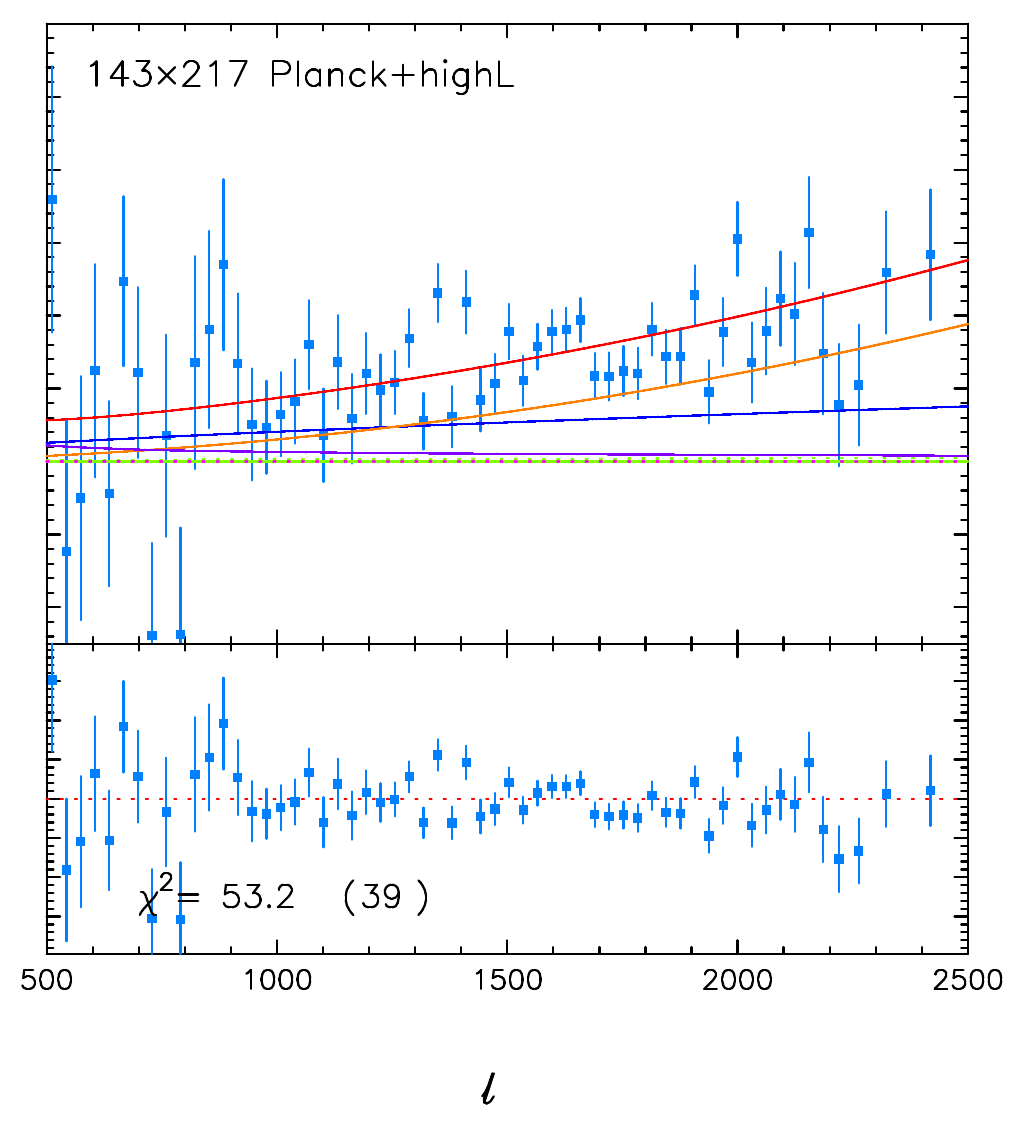}
\hspace*{-1.3mm}\includegraphics[width=51mm,angle=0]{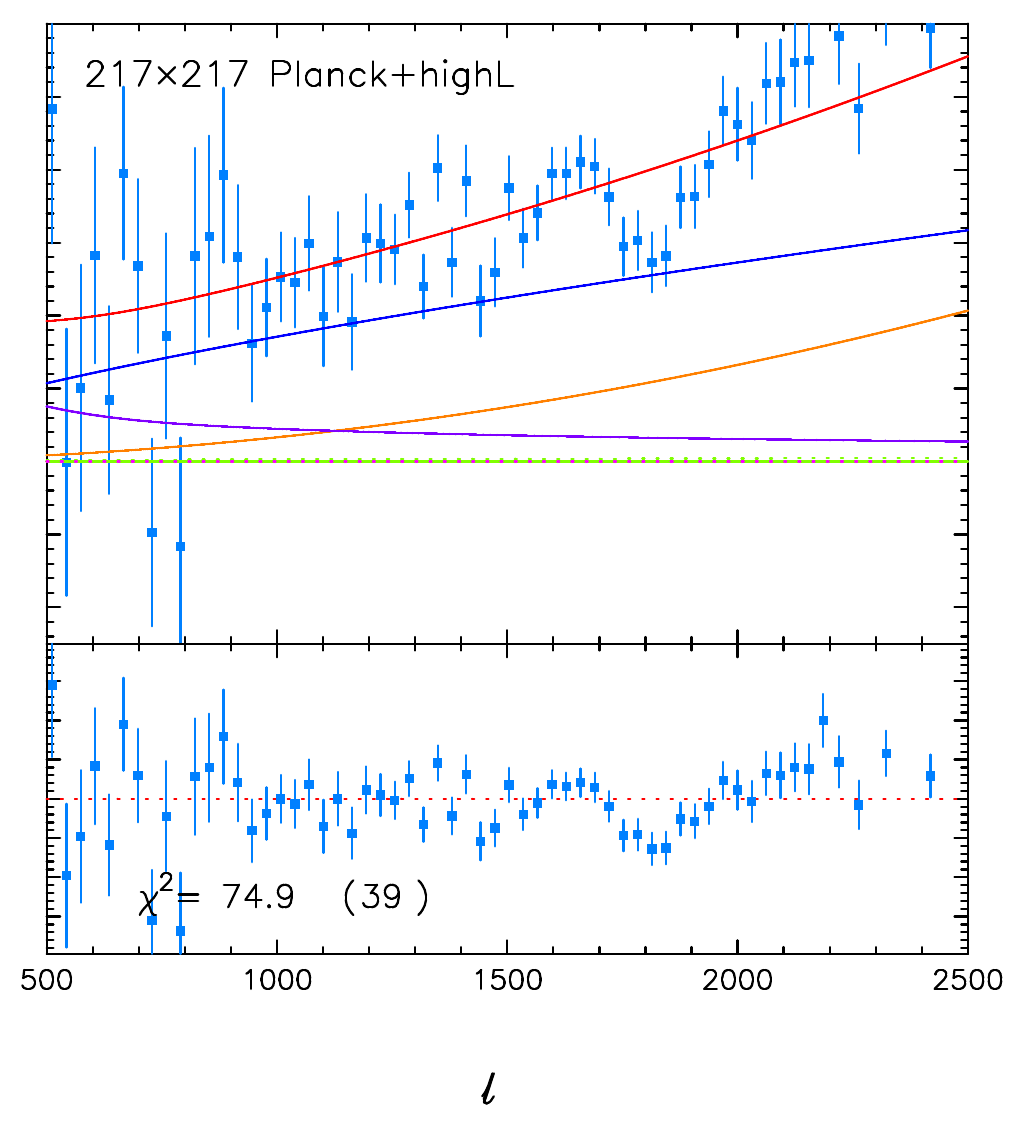}
\\
\caption {Power spectrum residual plots illustrating the accuracy of the
foreground modelling. For each cross-spectrum, there are two sub-figures. The upper sub-figures show the residuals with
  respect to the \planck+\WP\ best-fit solution
  (from Table~\ref{LCDMForegroundparams}). The lowers sub-figure show the residuals with respect to the  \planck+\WP+\highL\ solution
 The upper panel in each sub-figure shows the
  residual between the measured power spectrum and the best-fit
  (lensed) CMB power spectrum.  The lower panels show the residuals
  after further removing the best-fit foreground model. The lines in the upper
  panels show the various foreground components. Major foreground
  components are shown by the solid lines, colour coded as follows:
  total foreground spectrum (red); Poisson point sources (orange); clustered
  CIB (blue); thermal SZ (green); and Galactic dust (purple).  Minor
  foreground components are shown by the dotted lines colour coded as
  follows: kinetic SZ (green); tSZ$\times$CIB cross-correlation
  (purple). We also show residuals for the two spectra $100\times143$
  and $100\times217$ that are not used in the \planck\ likelihood.
  For these, we have assumed Poisson point-source correlation coefficients
  of unity. \referee{The $\chi^2$ values of the residuals, and the number of bandpowers, are listed in the lower panels.}}
\label{PlanckandHighL}
\end{figure*}

\begin{figure*}
\centering
\includegraphics[width=69mm,angle=00, clip]{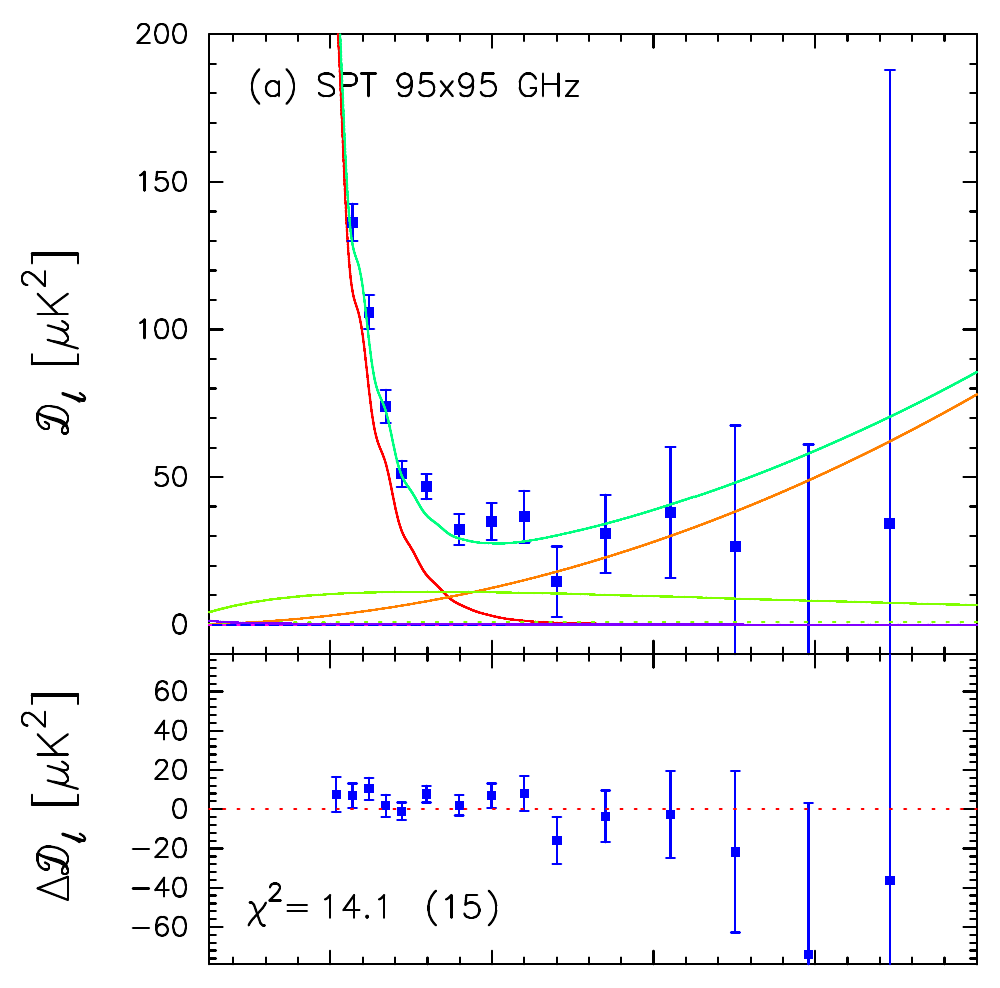} 
\includegraphics[width=69mm,angle=0]{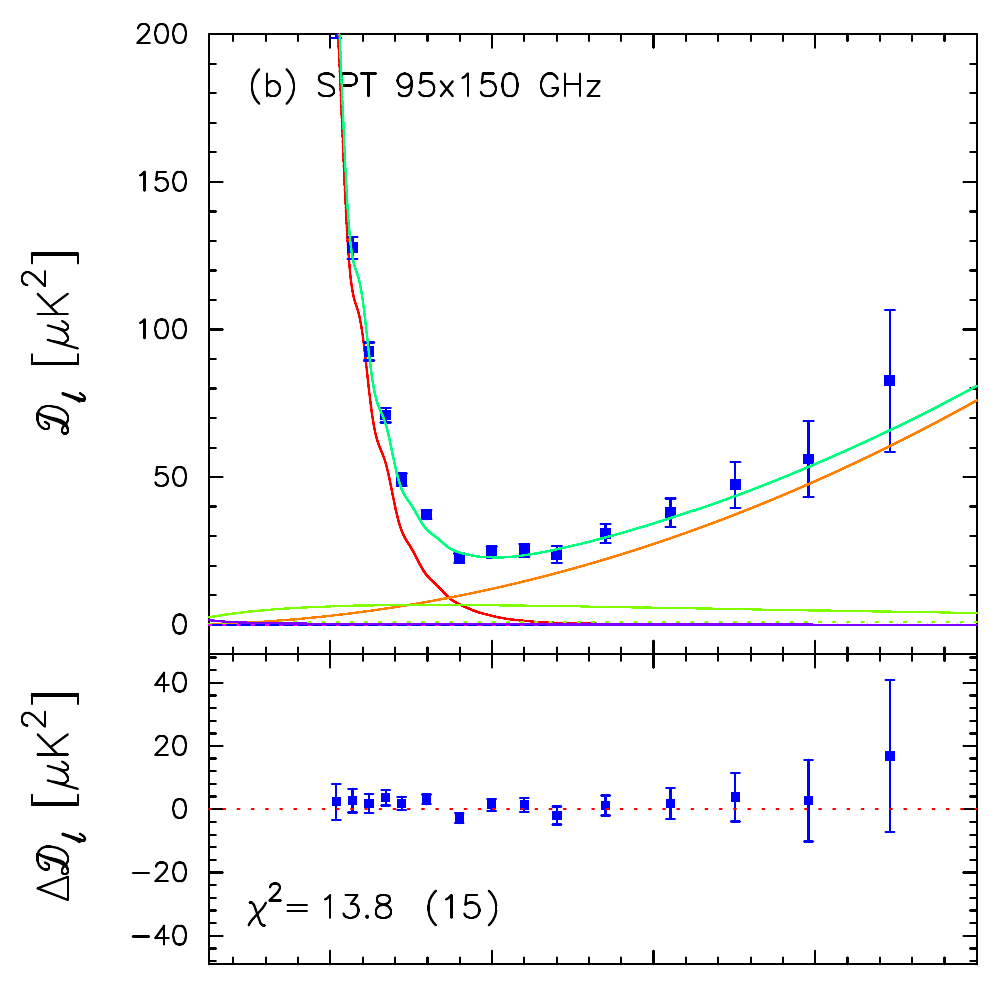}
\\
\includegraphics[width=69mm,angle=0]{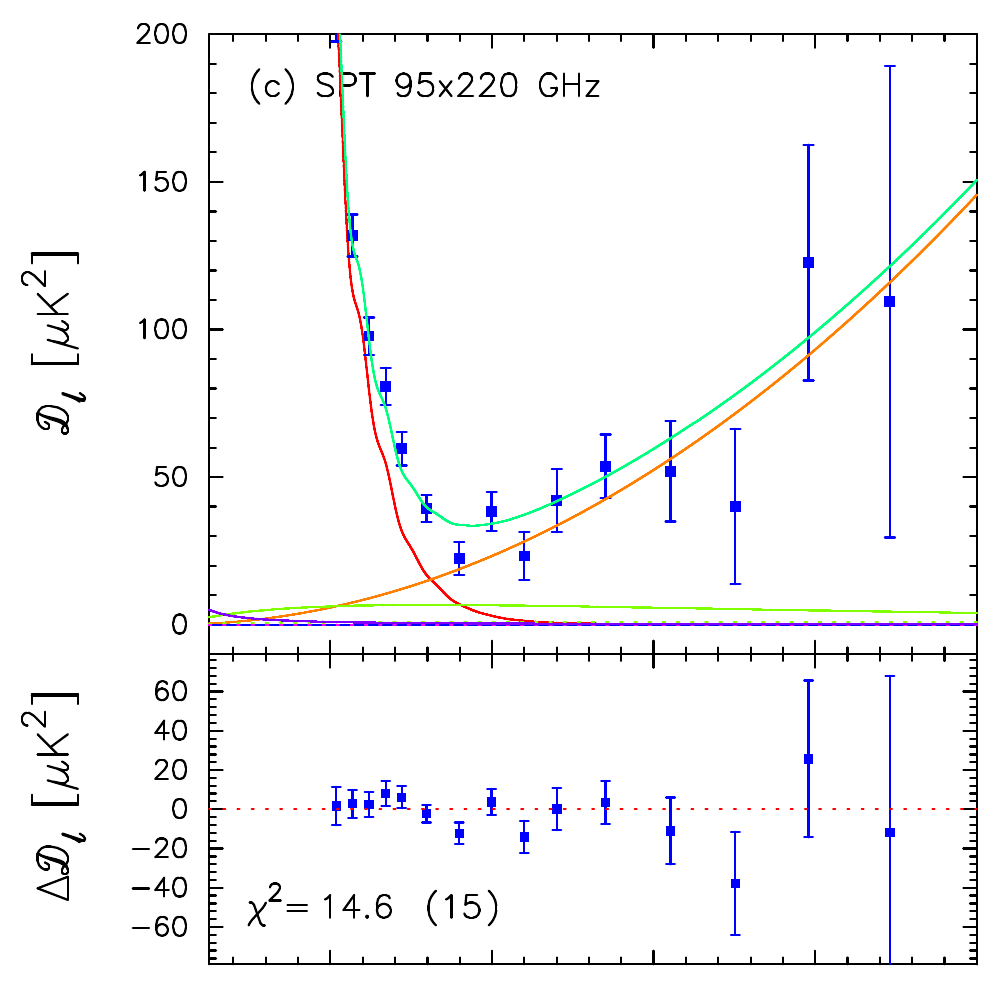} 
\includegraphics[width=69mm,angle=0]{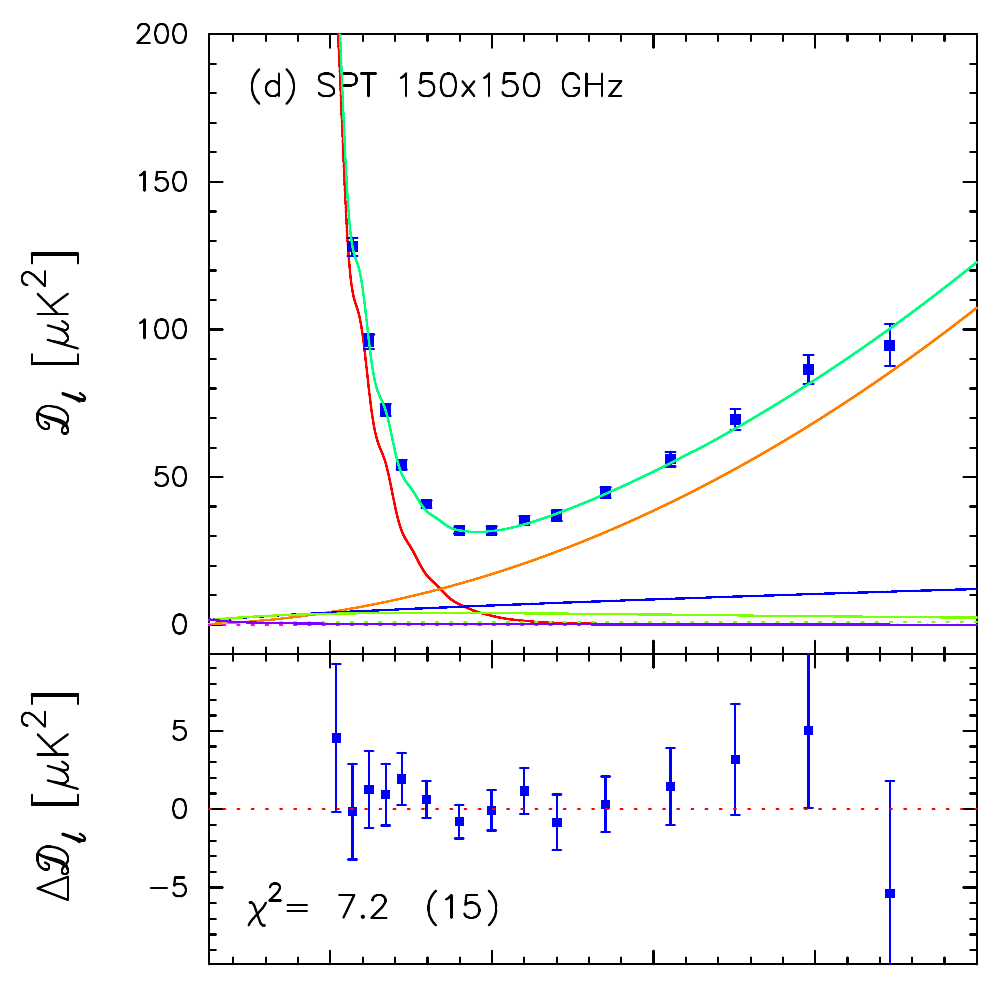}
\\
\hspace*{2mm}\includegraphics[width=71mm,angle=0]{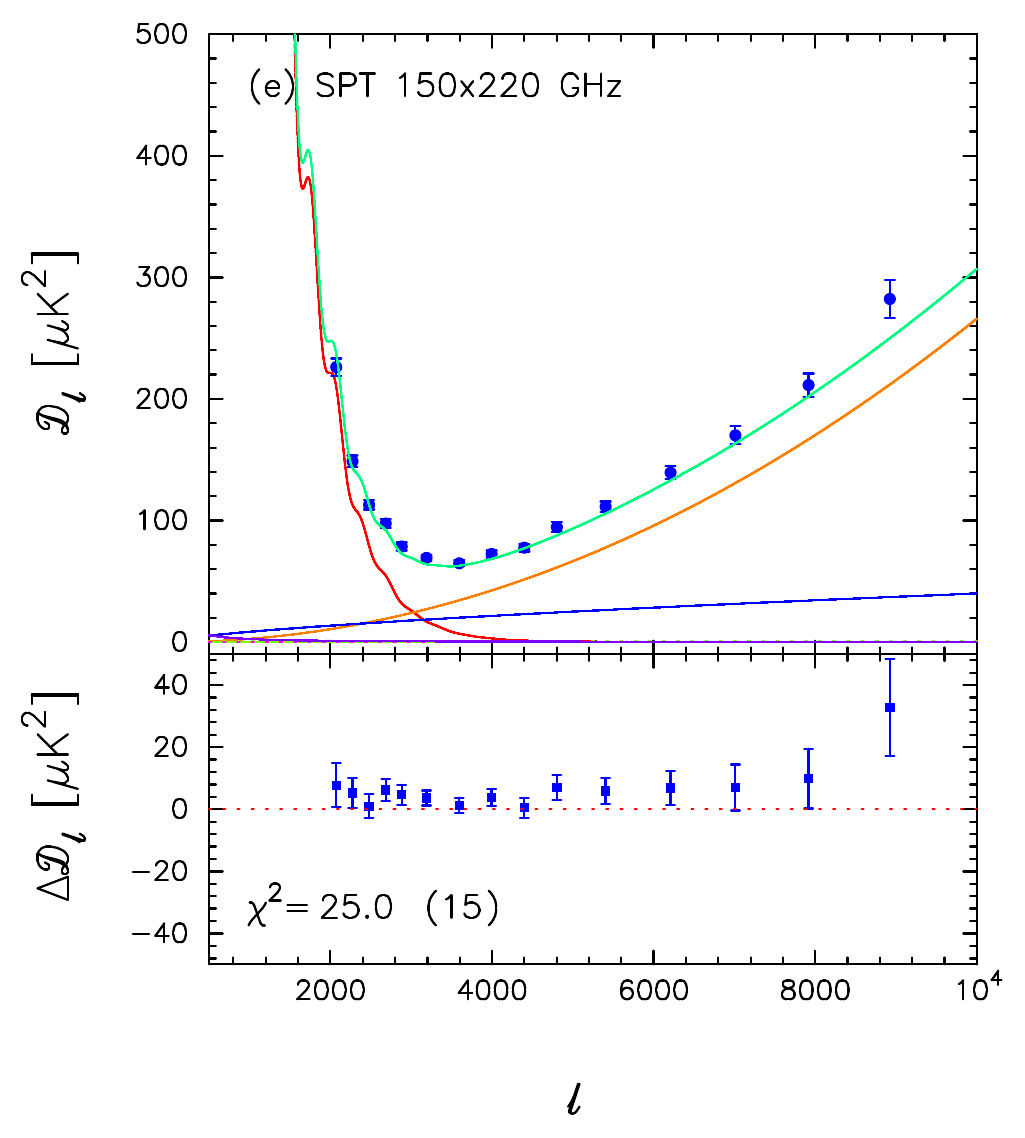} 
\hspace*{-2mm}\includegraphics[width=71mm,angle=0]{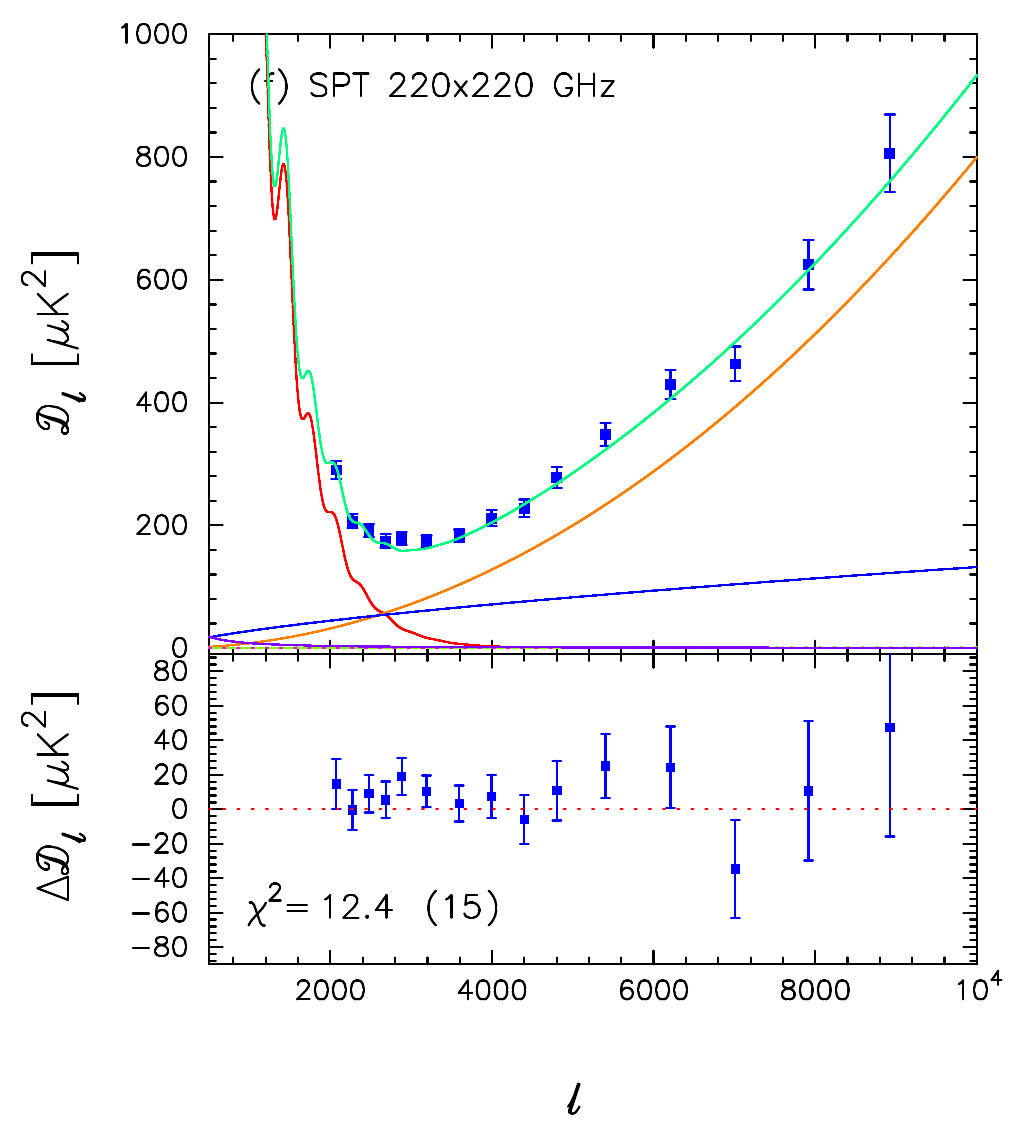}
\caption {SPT power spectra at high multipoles using the foreground model developed in this paper. The SPT R12 power spectra for each frequency combination are
  shown by the blue points, together with $1\,\sigma$ error bars. The
  foreground components, determined from the \planck+\WP+\highL\ analysis of \lcdm\ models, are shown in the upper panels using the same
  colour coding as in Fig.~\ref{PlanckandHighL}. Here, the spectrum of the best-fit CMB is shown in red and the total spectra are the upper green curves. The lower panel in
  each sub-figure shows the residuals with respect to the best-fit 
  base  \lcdm\ cosmology+foreground model. The $\chi^2$ values of the
  residuals, and the number of SPT bandpowers, are listed in the lower
  panels. }
\label{SPT}
\end{figure*}

\begin{figure*}
\centering
\includegraphics[width=69mm,angle=0]{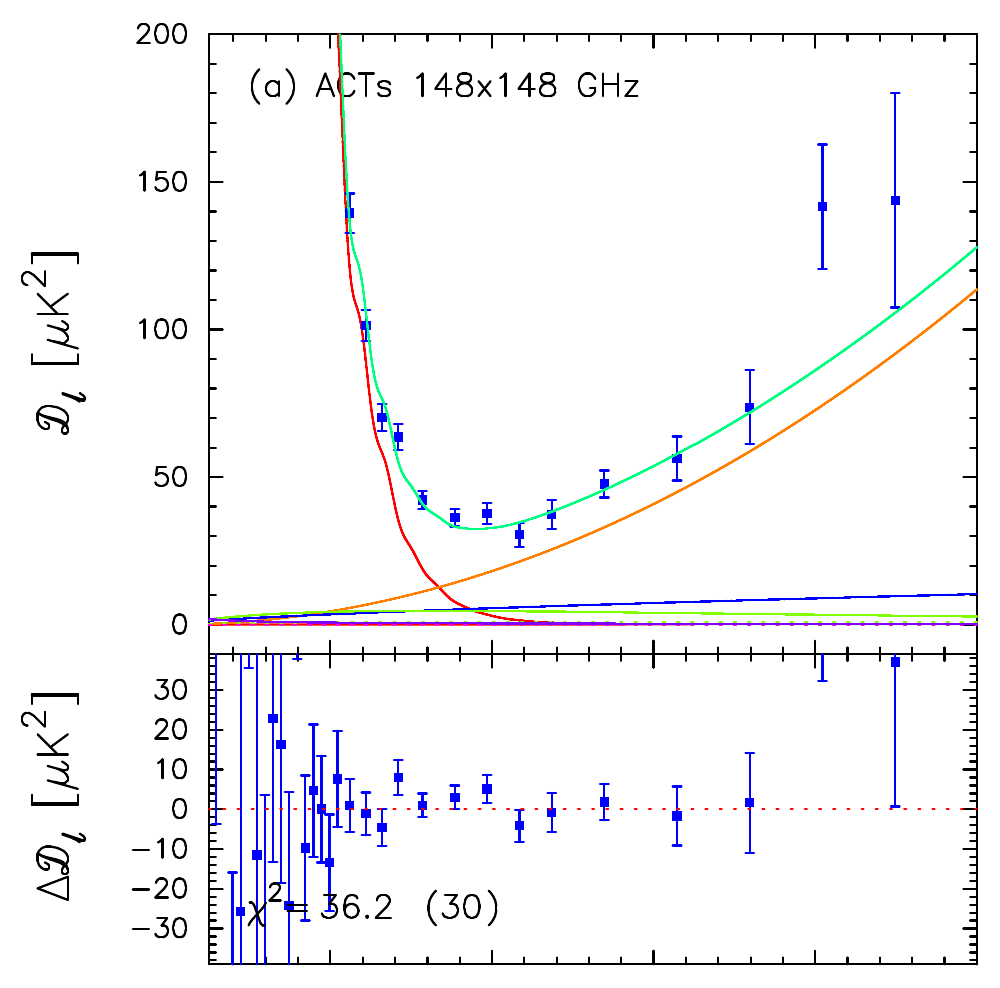}
\includegraphics[width=69mm,angle=0]{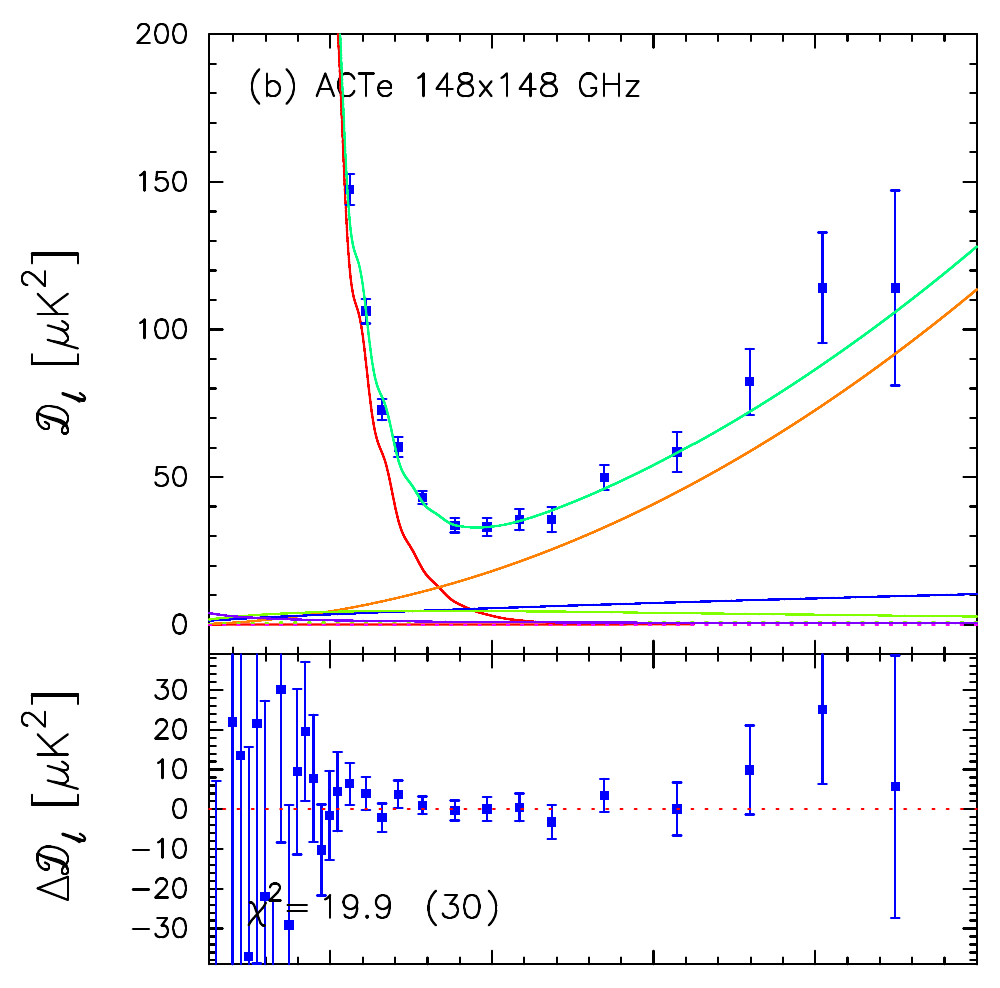}
\\
\includegraphics[width=69mm,angle=0]{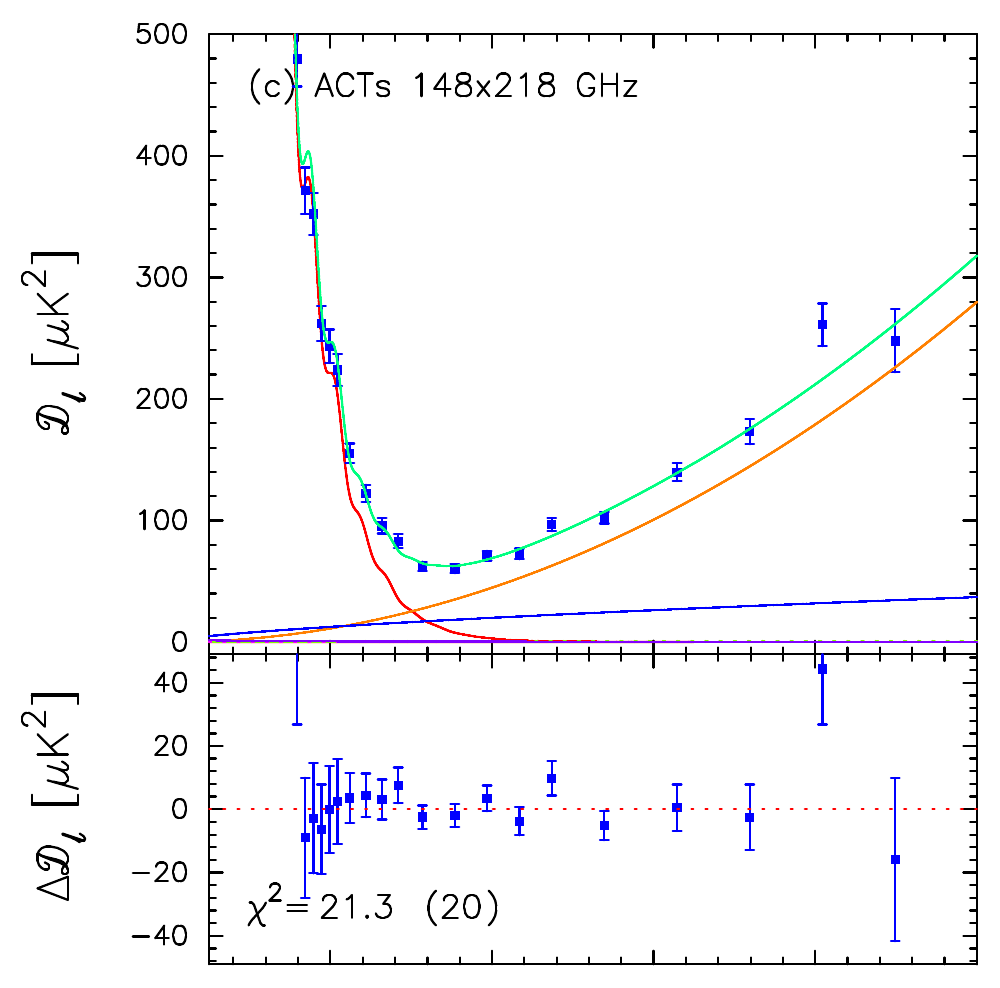}
\includegraphics[width=69mm,angle=0]{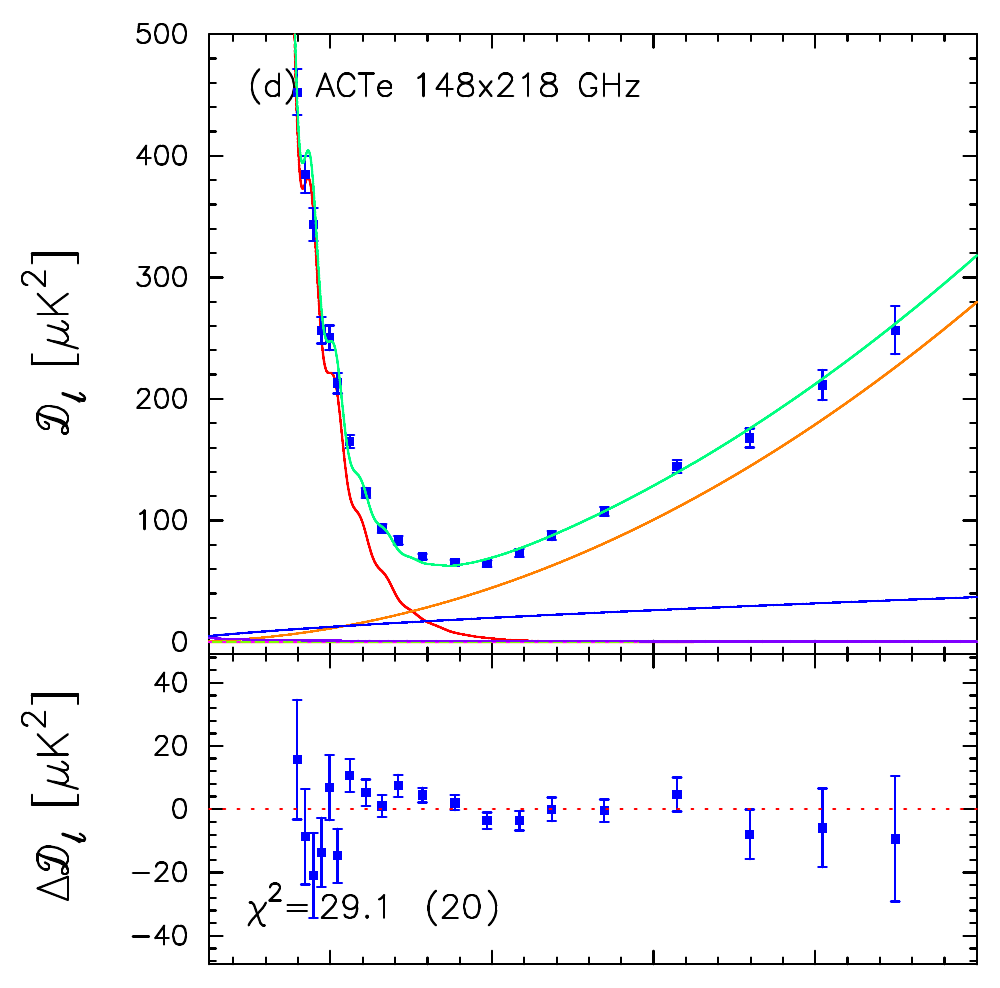}
\\
\hspace{2mm} \includegraphics[width=71mm,angle=0]{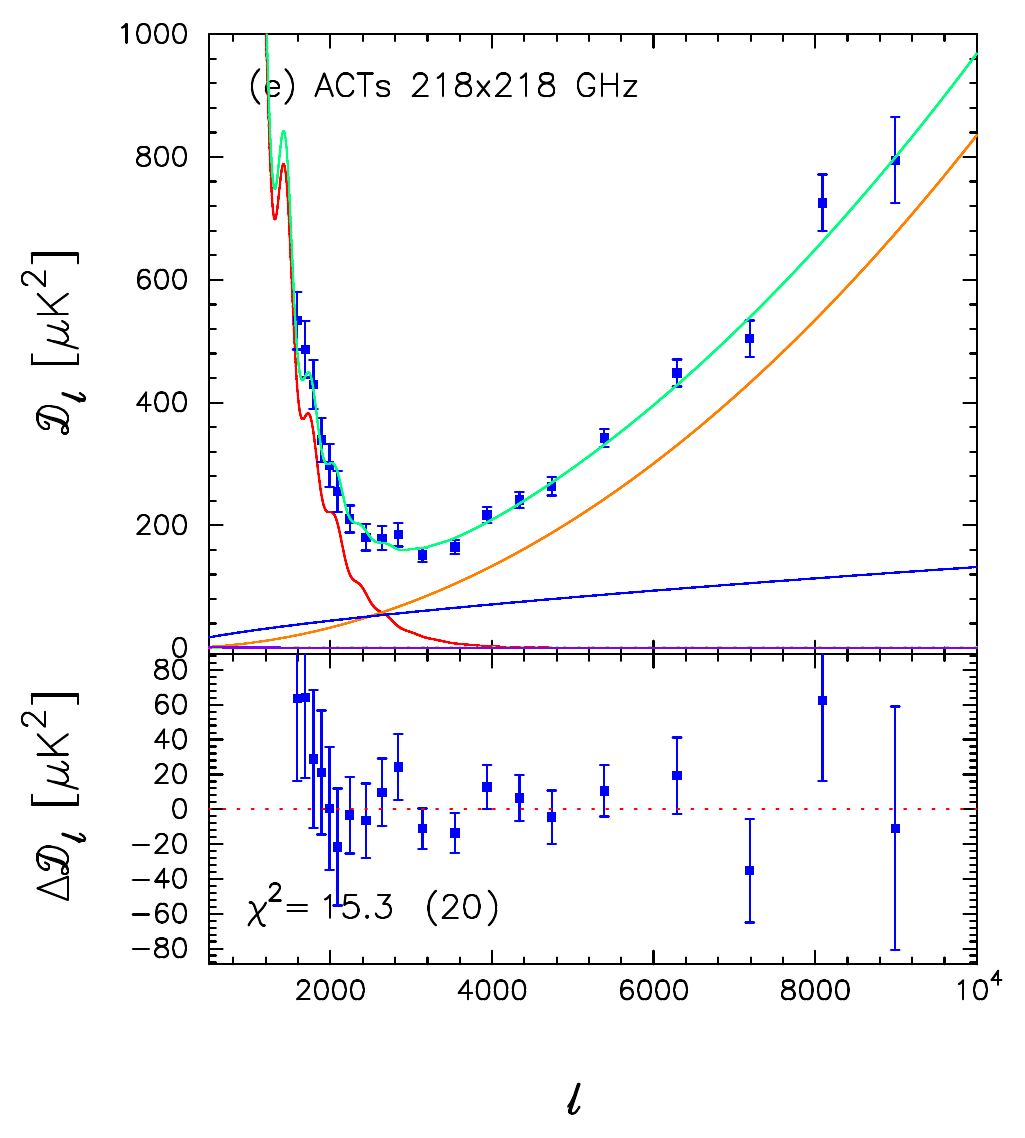}
\hspace*{-2mm}\includegraphics[width=71mm,angle=0]{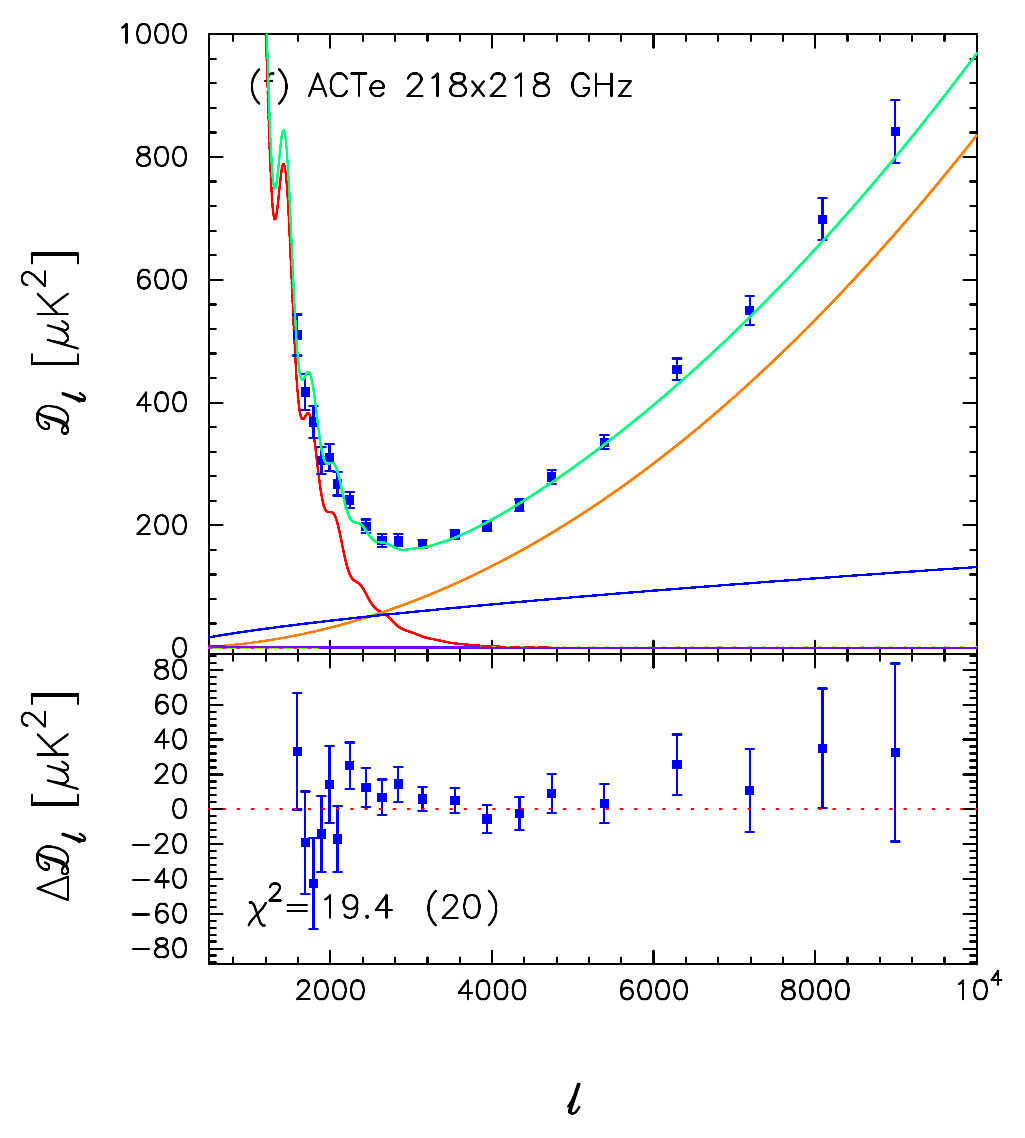}
\caption {As Fig.~\ref{SPT}, but for the ACT south and ACT
  equatorial power spectra.}
\label{ACT}
\end{figure*}

\begin{table}[tmb]                 
\begingroup
\newdimen\tblskip \tblskip=5pt
\caption{Goodness-of-fit tests for the \planck\ spectra. The quantity
$\Delta \chi^2=\chi^2-N_\ell$ is the difference in $\chi^2$ from the expected value if the
model is correct. \referee{The sixth column expresses $\Delta \chi^2$ in units of
the expected dispersion, $\sqrt{2N_\ell}$, and the last column lists the
probability to exceed (PTE) the tabulated value of $\chi^2$.}}
\label{planckchi2}         
\nointerlineskip
\vskip -3mm
\footnotesize
\setbox\tablebox=\vbox{
   \newdimen\digitwidth
   \setbox0=\hbox{\rm 0}
   \digitwidth=\wd0
   \catcode`*=\active
   \def*{\kern\digitwidth}
   \newdimen\signwidth
   \setbox0=\hbox{+}
   \signwidth=\wd0
   \catcode`!=\active
   \def!{\kern\signwidth}
\halign{\hfil#\hfil\tabskip=1em&
        \hfil#\hfil&
        \hfil#\hfil&
        \hfil#\hfil&
        \hfil#\hfil&
        \hfil#\hfil&
        \hfil#\hfil\tabskip=0pt\cr                
\noalign{\doubleline}
Spectrum&$\ell_{\rm min}$&$\ell_{\rm max}$&$\chi^2$&$\chi^2/N_\ell$&$\Delta\chi^2/\sqrt{2N_\ell}$&PTE\cr 
\noalign{\vskip 3pt\hrule\vskip 5pt}
$100\times100$&*50&1200&1158&1.01&!0.14&$44.4\%$\cr
$143\times143$&*50&2000&1883&0.97&$-1.09$& $86.2\%$\cr
$217\times217$&500&2500&2079&1.04&!1.23& $10.9\%$\cr
$143\times217$&500&2500&1930&0.96&$-1.13$& $87.1\%$\cr
All&50&2500&2564&1.05&!1.62 & !$5.3\%$\cr
\noalign{\vskip 3pt\hrule\vskip 4pt}}}
\endPlancktable                    
\endgroup
\end{table}                        

To quantify the consistency of the model fits shown in
Fig.~\ref{PlanckandHighL} for \planck\ we compute the $\chi^2$ statistic
\be
  \chi^2 =   \sum_{\ell \ell^{\prime}} (C_\ell^{\rm data} - C_\ell^{\rm CMB} - C_\ell^{\rm fg}) {\cal M}^{-1}_{\ell \ell^\prime} (C_{\ell^\prime} ^{\rm data} - C_{\ell^\prime}
^{\rm CMB} - C_{\ell^\prime}^{\rm fg}),  \label{GF1}
\ee
for each of the spectra, where the sums extend over the multipole ranges
$\ell_{\rm min}$ and $\ell_{\rm max}$ used in the likelihood, ${\cal M}_{\ell \ell^\prime}$ is the
covariance matrix for the spectrum $C^{\rm data}_\ell$ (including corrections
for beam eigenmodes and calibrations),  $C_\ell^{\rm CMB}$
is the best-fit primordial CMB spectrum and $C_\ell^{\rm fg}$ is the
best-fit foreground model appropriate to the data spectrum.
 We expect $\chi^2$
to be approximately Gaussian distributed with a mean of $N_{\ell} = \ell_{\rm max} - \ell_{\rm min} +1$ and
dispersion $\sqrt{2N_{\ell}}$. Results are summarized in Table \ref{planckchi2}
for the \planck+\WP+\highL\ best-fit parameters of Table \ref{LCDMForegroundparams}.
(The $\chi^2$ values for the \planck+\WP\  fit are almost identical.)
Each of the spectra gives an acceptable global fit to the model, quantifying the
high degree of consistency of these spectra
described in~\citet{planck2013-p08}. (Note
that \citealt{planck2013-p08} presents an alternative way of investigating
consistency between these spectra via  power spectrum differences.)

Figures~\ref{SPT} and \ref{ACT} show the fits and residuals with
respect to the best-fit \planck+\WP+\highL\ model of
Table~\ref{LCDMForegroundparams},
for each of the SPT and
ACT spectra.  The SPT and ACT spectra are reported
as band-powers, with associated window functions $[W^{\rm SPT}_b(\ell)/\ell]$
and $W^{\rm ACT}_b(\ell)$. The definitions of these window functions
differ between the two experiments.

For SPT, the contribution of the CMB and foreground spectra in each band is
\begin{equation}
{\cal D}_b = \sum_\ell [W_b^{\rm SPT}(\ell)/\ell] {\ell(\ell + 1/2) \over 2 \pi}
\left( C^{\rm CMB}_\ell + C^{\rm fg}_\ell \right) . \label{SPTF3}
\end{equation}
(Note that this differs from the equations given in R12 and S12.)

For ACT, the window functions operate on the power spectra:
\begin{equation}
 C_b = \sum_\ell W_b^{\rm ACT}(\ell) \left( C^{\rm CMB}_\ell + C^{\rm fg}_\ell \right).   \label{ACTW1}
\end{equation}
In Fig.~\ref{ACT}. we plot ${\cal D}_b =  \ell_b (\ell_b + 1)C_b /(2 \pi)$, where $\ell_b$ is
the effective multipole for band $b$.

The upper panels of each of the sub-plots in Figs.~\ref{SPT} and \ref{ACT} show the spectra of the best-fit CMB, and the total CMB+foreground, as well as the individual
contributions of the foreground components using the same colour
codings as in Fig.~\ref{PlanckandHighL}. The lower panel in each sub-plot
shows the residuals with respect to the best-fit
cosmology+foreground model. For each spectrum, we list the value of
$\chi^2$, neglecting correlations between the (broad) ACT and SPT
bands, together with the number of data points.
The quality of the fits is generally very good.  For SPT, the
residuals are very similar to those inferred from Fig.~3 of R12. The
SPT $150\times220$ spectrum has the largest $\chi^2$ (approximately a
$1.8\,\sigma$ excess). This spectrum shows systematic positive residuals
of a few ${\mu}{\rm K}^2$ over the entire multipole range.  For ACT,
the residuals and $\chi^2$ values are close to those plotted in Fig.~4 of \citet{Dunkley:13}. All of the ACT spectra plotted in
Fig.~\ref{ACT} are well fit by the model (except for some residuals
at multipoles $\ell \la 2000$, which are also seen by \citealt{Dunkley:13}).

\begin{figure*}
\centering
\includegraphics[angle=0,width=16cm]{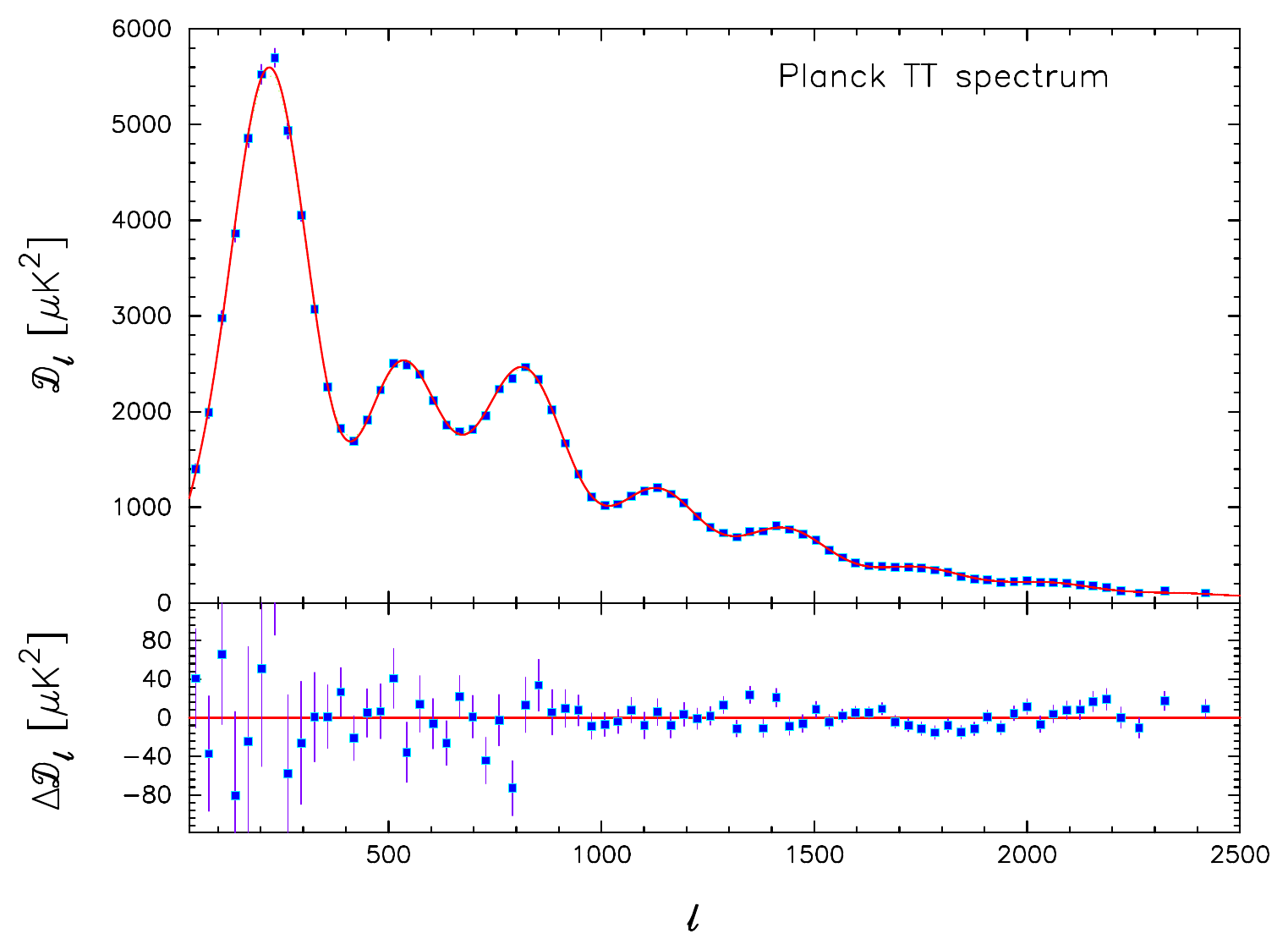}
\caption {\Planck\ $TT$ power spectrum.
The points in the upper panel show the maximum-likelihood
  estimates of the primary CMB spectrum 
  computed as described in the text
  for the best-fit foreground and nuisance
  parameters of the \planck+\WP+\highL\ fit listed in Table~\ref{LCDMForegroundparams}. The red line shows the best-fit base \lcdm\
  spectrum. The lower panel shows the residuals with respect to the
  theoretical model.  The error bars are computed from the full
  covariance matrix, appropriately weighted across each band (see
  Eqs.~\ref{PBF1a} and \ref{PBF1b}) and include beam uncertainties
  and uncertainties in the foreground model parameters.}
\label{Planckbestfitcl}
\end{figure*}

\begin{figure*}
\centering
\vspace{5mm}
\includegraphics[width=90mm,angle=0]{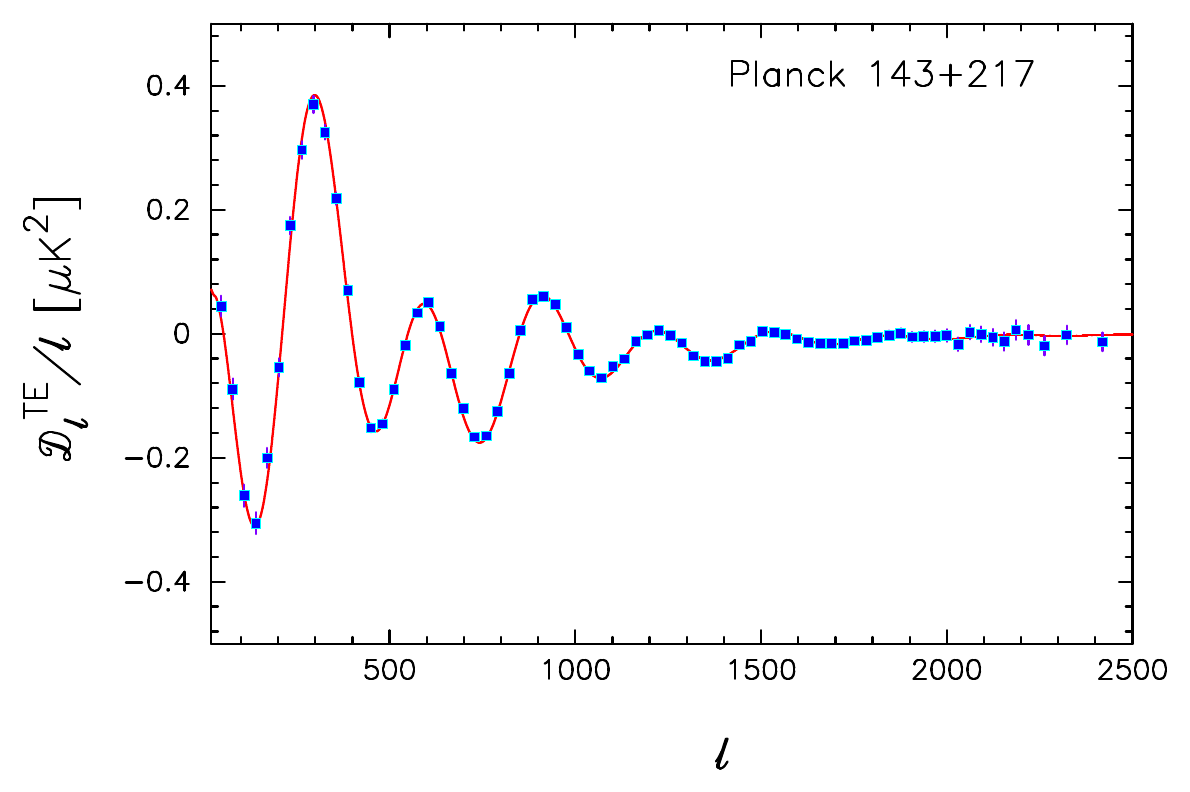}
\includegraphics[width=90mm,angle=0]{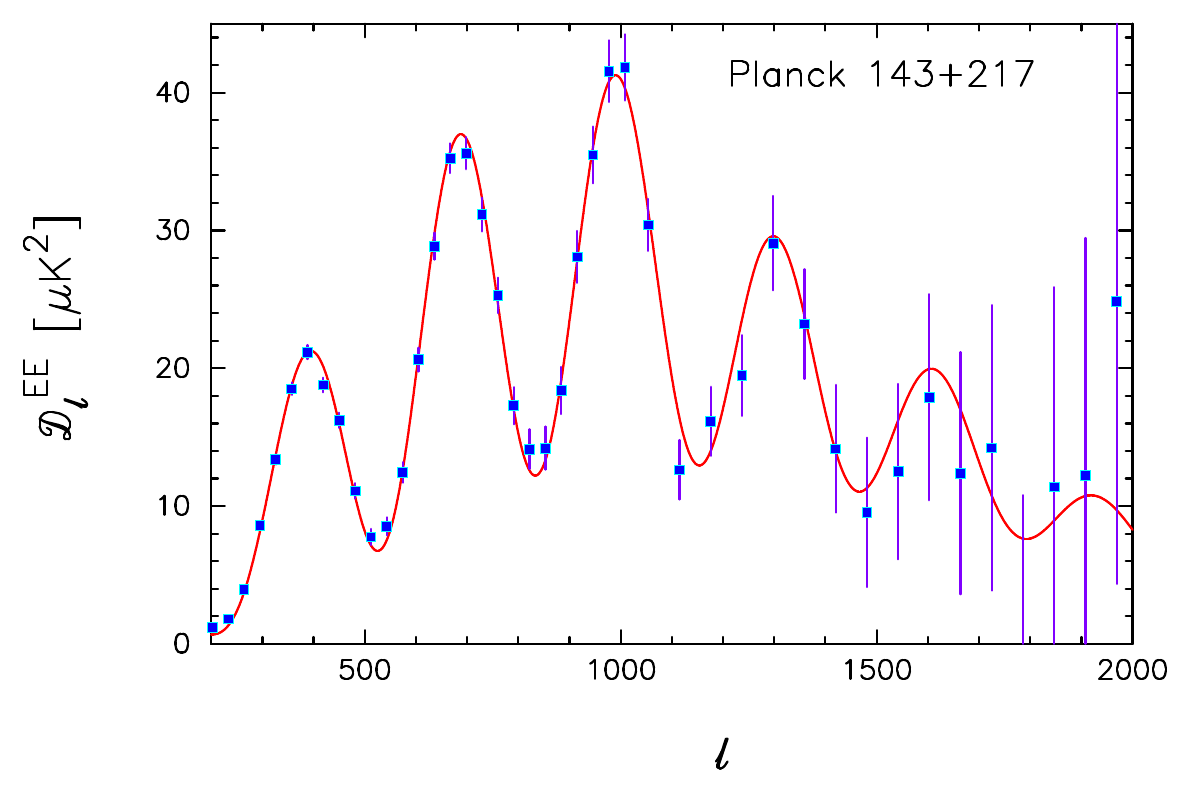}
\caption {\planck\ $TE$ (left) and $EE$ spectra (right) computed as
  described in the text.  The red lines show the polarization spectra
  from the base \lcdm\ \planck+\WP+\highL\ model, {\it which is
    fitted to the TT data only}.}
\label{Planckpolspec}
\vspace{5mm}
\end{figure*}

Having determined a solution for the best-fit foreground and other
``nuisance'' parameters, we can correct the four spectra used in the
\planck\ likelihood and combine them to reconstruct a ``best-fit''
primary CMB spectrum and covariance matrix as described in ~\citet{planck2013-p08}.  This best-fit \planck\ CMB spectrum
is plotted in the upper panels of Figs.~\ref{TTspec} and
\ref{Planckbestfitcl} for \planck+\WP+\highL\ foreground parameters.
The spectrum in Fig.~\ref{Planckbestfitcl} has been band-averaged in
bins of width $\Delta \ell \sim 31$ using a window function $W_b(l)$:
\beglet
\be \hat {\cal D}_b = \sum_\ell W_b(\ell) \hat{\cal
  D}_\ell, \label{PBF1a} \ee \be W_b(\ell) = \left\{ \begin{array}{ll}
  (\hat {\cal M}^{\cal D}_{\ell \ell})^{-1} / \sum_{\ell=\ell^b_{\rm
      min}}^{\ell^b_{\rm max}} (\hat {\cal M}^{\cal D}_{\ell
    \ell})^{-1}, & \ell^b_{\rm min} \le \ell < \ell^b_{\rm max}, \\ 0,
  & \mbox{otherwise}. \end{array} \right. 
\label{PBF1b} 
\ee \endlet
Here, $\ell^b_{\rm min}$ and $\ell^b_{\rm max}$ denote the
minimum and maximum multipole ranges of band $b$, and $\hat {\cal
  M}^{\cal D}_{\ell \ell^\prime}$ is the covariance matrix of the
best-fit spectrum $\hat {\cal D}_\ell$, computed as described in~\citet{planck2013-p08}, and to which we have added corrections for beam and foreground errors (using the curvature
matrix of the foreground model parameters from the MCMC chains).  The
solid lines in the upper panels of Figs.~\ref{TTspec} and
\ref{Planckbestfitcl} show the spectrum for the best-fit
\lcdm\ cosmology.  The residuals with respect to this cosmology are
plotted in the lower panel. To assess the goodness-of-fit, we compute
$\chi^2$:
\be \chi^2 = \sum_{\ell \ell^{\prime}} (\hat C_\ell^{\rm
  data} - C_\ell^{\rm CMB}) \hat {\cal M}^{-1}_{\ell \ell^\prime}
(\hat C_{\ell^\prime} ^{\rm data} - C_{\ell^\prime} ^{\rm
  CMB}), \label{GF2} 
\ee 
using the covariance matrix for the best-fit
data spectrum (including foreground and beam errors\footnote{Though the $\chi^2$ value is similar if
  foreground and beam errors are not included in the covariance
  matrix.}).  The results are given in the last line of Table
\ref{planckchi2} labelled ``All.''  The lower panel of
Fig.~\ref{Planckbestfitcl} shows the residuals with respect to the
best-fit cosmology (on an expanded scale compared to
Fig.~\ref{TTspec}).  There are some visually striking residuals in
this plot, particularly in the regions $\ell\sim 800$ and $\ell\sim
1300$--$1500$ (where we see ``oscillatory'' behaviour).  As discussed
in detail in \citet{planck2013-p08}, these residuals are reproducible to high
accuracy across \planck\ detectors and across \planck\ frequencies; see also Fig.~\ref{PlanckandHighL}. There is therefore strong evidence that the residuals at these
multipoles, which are in the largely signal dominated region of the
spectrum, are real features of the primordial CMB sky.  These features
are compatible with statistical fluctuations of a Gaussian
\lcdm\ model, and are described accurately by the covariance matrix
used in the \planck\ likelihood.  As judged by the $\chi^2$ statistic
listed in Table \ref{planckchi2}, the best fit reconstructed
\planck\ spectrum is compatible with the base \lcdm\ cosmology to within
$1.6\,\sigma$\footnote{\citet{planck2013-p17} describes a specific statistical
test designed to find features in the primordial power spectrum. This test 
responds to the extended ``dip'' in the \planck\ power spectrum centred
at about $\ell \sim 1800$, tentatively suggesting  $2.4$--$3.1\,\sigma$
evidence for a feature. \referee{As discussed in Sect.~\ref{sec:introduction},
after submission of the
\Planck\ 2013 papers, we found strong evidence that this feature is 
a small systematic in the $217\times217$ spectrum caused by
incomplete removal of 4\,K cooler lines.  This feature can be seen
in the residual plots in Fig.~\ref{PlanckandHighL} and contributes to the
high (almost $2\,\sigma$) values of $\chi^2$ in the $217\times217$ residual 
plots.}}

\emph{To the extremely high accuracy afforded by the \planck\ data, the
  power spectrum at high multipoles is compatible with the predictions
  of the base six parameter \lcdm\ cosmology.}  This is the main
result of this paper.  Fig.~\ref{TTspec} does, however, suggest that
the power spectrum of the best-fit base \lcdm\ cosmology has a higher
amplitude than the observed power spectrum at multipoles $\ell \la
30$. We will return to this point in Sect.~\ref{sec:conclusions}.

Finally, Fig.~\ref{Planckpolspec} shows examples of \planck\ $TE$ and $EE$
spectra. These are computed by performing a straight average of the
(scalar) beam-corrected $143\times143$, $143\times217$, and
$217\times217$ cross-spectra (ignoring auto-spectra). There are $32$ $TE$
and $ET$ cross-spectra contributing to the mean $TE$ spectrum plotted
in Fig.~\ref{Planckpolspec}, and six $EE$ spectra contributing to the
mean $EE$ spectrum. \planck\ polarization data, including LFI and
$353\,$GHz data not shown here, will be analysed in
detail, and incorporated into a \planck\ likelihood, following this
data release. The purpose of presenting these figures here
is twofold: first, to demonstrate the potential of \planck\ to deliver
high quality polarization maps and spectra, as described in the
\planck\ ``blue-book'' \citep{planck2005-bluebook}; and, second, to
show the consistency of these polarization spectra with the temperature
spectrum shown in Fig.~\ref{Planckbestfitcl}. As discussed in
\citet{planck2013-p03} and \citet{planck2013-p08}, at present, the HFI polarization spectra at low
multipoles ($\ell \la 200$) are affected by systematic errors
that cause biases. For the HFI channels used in Fig.~\ref{Planckpolspec},
there are two primary sources of systematic error
arising from non-linear gain-like variations, and residual bandpass
mismatches between detectors. However,  these systematics rapidly become
unimportant at higher multipoles\footnote{The main focus
of current work on \planck\ polarization is  to reduce the effects
of these systematics on the polarization maps at large angular scales.}.

The errors on the mean $TE$ and $EE$ spectra shown in Fig.
\ref{Planckpolspec} are computed from the analytic formulae given in
\citet{Efstathiou:06}, using an effective beam-width adjusted to
reproduce the observed scatter in the polarization spectra at high
multipoles. The spectra are then band-averaged as in Eq.~(\ref{GF2}). The error bars shown in Fig.~\ref{Planckpolspec} are
computed from the diagonal components of the band-averaged covariance
matrices.

The solid lines in the upper panels of Fig.~\ref{Planckpolspec}} show the theoretical $TE$ and $EE$
spectra expected in the best-fit \planck+\WP+\highL\ \lcdm\  model (i.e., the model used to compute the theory $TT$ spectrum plotted in
Fig.~\ref{Planckbestfitcl}). These theoretical spectra are determined
entirely from the $TT$ analysis and make no use of the \planck\
polarization data.  As with the $TT$ spectra, the \lcdm\ model
provides an extremely good match to the polarization spectra.
Furthermore, polarized foreground emission is expected to be
unimportant at high multipoles \citep[e.g.,][]{Tucci:12}
and so no foreground corrections have been made to the spectra in 
Fig.~\ref{Planckpolspec}. The
agreement between the polarization spectra and the theoretical
spectra therefore provides strong evidence that the best-fit
cosmological parameters listed in Table \ref{LCDMForegroundparams} are not
strongly affected by the modelling of unresolved foregrounds in the $TT$
analysis.

\section{Comparison of the Planck  base \lcdm\ model with other astrophysical
data sets}
\label{sec:datasets}

Unlike CMB data, traditional astrophysical data sets  -- e.g.,
measurements of the Hubble parameter, type Ia supernovae (SNe Ia), and
galaxy redshift surveys -- involve complex physical systems that are not
understood at a fundamental level. Astronomers are therefore reliant
on internal consistency tests and empirical calibrations to limit the
possible impact of systematic effects. Examples include calibrating
the metallicity dependence of the Cepheid period luminosity relation,
calibrating the colour-decline-rate-luminosity relation of Type Ia
supernovae, or quantifying the relationship between the spatial
distributions of galaxies and dark matter. In addition, there are more
mundane potential sources of error, which can affect certain types of
astrophysical observations (e.g., establishing consistent
photometric calibration systems).  We must be open to the possibility that
unknown, or poorly quantified, systematic errors may be present in the
astrophysical data, especially when used in combination with the
high precision data from \planck.

We have seen in the previous section that the base \lcdm\
 model provides an acceptable fit to the \planck\ $TT$ power
spectra (and  the \planck\ $TE$ and $EE$ spectra) and also to the ACT and SPT
temperature power spectra. The cosmological parameters of
this model are determined to high precision. We therefore review
whether these parameters provide acceptable fits to other
astrophysical data. If they do not, then we need to assess whether the
discrepancy is a pointer to new physics, or evidence of some type of
poorly understood systematic effect. Unless stated otherwise, we use
the \planck+\WP+\highL\ parameters listed in Table~\ref{LCDMForegroundparams}
as the default ``\planck'' parameters for the base \lcdm\  model.

\subsection{CMB lensing measured by \Planck}
\label{sec:lensing}

\begin{figure}
\begin{center}
\includegraphics[width=65mm,angle=-90]{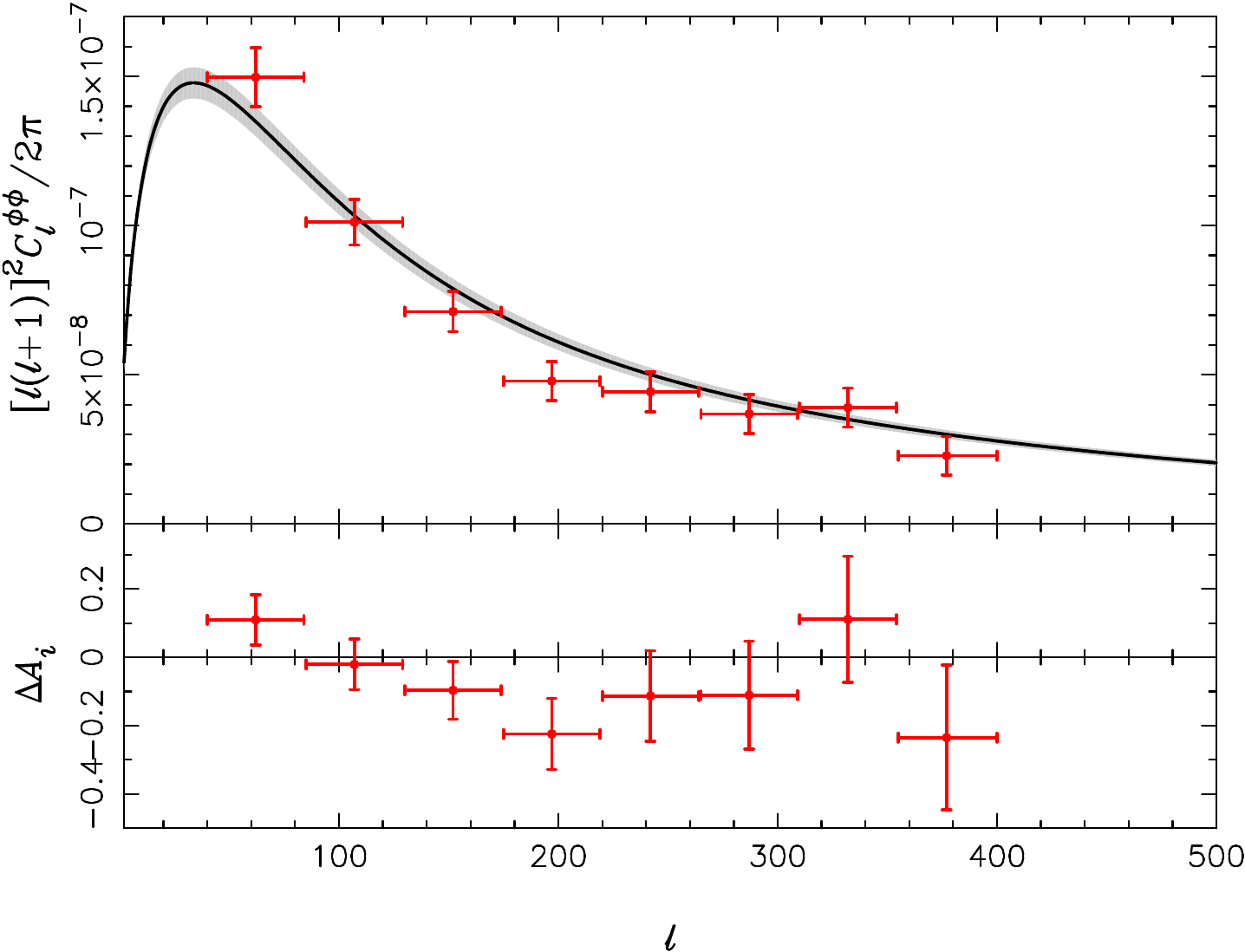}
\end{center}
\caption{\Planck\ measurements of the lensing power spectrum compared
  to the prediction for the best-fitting \Planck+\WP+\HighL\ \LCDM\ model
  parameters.  In the top panel, the data points are the measured bandpowers and
 $\pm 1\,\sigma$ error ranges from the diagonal of the
  covariance matrix.
The measured bandpowers are compared to the $C_\ell^{\phi\phi}$ in the best-fit model (black
  line). The grey region shows the $1\,\sigma$ range in
$C_\ell^{\phi\phi}$ due to \LCDM\ parameter uncertainties.
The lower panel shows the differences between the bandpower amplitudes
  $\hat{A}_i$ and the predictions for their
  expectation values in the best-fit model, $\Ailensbestfit$.}
\label{fig:CMBlensing_LCDM}
\end{figure}

Weak gravitational lensing by large-scale structure subtly alters
the statistics of the CMB anisotropies, encoding information about the
late-time Universe which is otherwise degenerate in the primary anisotropies
laid down at last-scattering (see~\citealt{2006PhR...429....1L} for a review).
The lensing deflections are given by the gradient of the lensing potential
$\phi(\hat{\vec{n}})$, which corresponds to an integrated measure of the
matter distribution along the line of sight with peak sensitivity to
structures around redshift 2. The rms deflection is expected to
be around $2.5\,\mathrm{arcmin}$ and to be coherent over several degrees.
We include
the effect of lensing on the temperature power spectrum in all our parameter
analysis, but for some results we also include the lensing information
encoded in the non-Gaussian trispectrum (connected 4-point function) of the CMB.
Lensing generates a non-zero trispectrum, which, at leading order, is
proportional to the power spectrum $C_\ell^{\phi\phi}$ of the lensing
potential~\citep{2001PhRvD..64h3005H}.

In~\citet{planck2013-p12}, we present a detailed
analysis of CMB lensing
with \Planck\ data, including estimation of
$C_\ell^{\phi\phi}$ from the trispectrum computed from \Planck's maps.
This paper also describes the construction of a lensing likelihood.
Briefly,
we first reconstruct an estimate of the lensing potential using near-optimal
quadratic estimators, following~\citet{2003PhRvD..67h3002O},
with various Galactic and point-source masks.
The empirical power spectrum of this reconstruction, after subtraction of
the Gaussian noise bias (i.e., the disconnected part of the 4-point function),
is then used to estimate
$C_\ell^{\phi\phi}$ in bandpowers. The associated bandpower errors are
estimated from simulations.
The lensing power
spectrum is estimated from channel-coadded \Planck\ maps at 100, 143
and 217\,GHz in the multipole range $\ell=10$--$1000$, and also from
a minimum-variance combination of the 143 and 217\,GHz maps. An empirical
correction for the shot-noise trispectrum of unresolved point sources is made
to each spectrum, based on the measured amplitude of a generalized kurtosis
of the appropriate maps. Additionally, the $N^{(1)}$ bias
of~\citet{2003PhRvD..67l3507K},
computed for a fiducial \LCDM\ spectrum determined from a pre-publication analysis of the \planck\ data,
is subtracted from each spectrum. This latter correction is proportional
to $C_\ell^{\phi\phi}$ and accounts for sub-dominant couplings of the
trispectrum, which mix lensing power over a range of scales into the
power spectrum estimates. Excellent internal consistency of the various
$C_\ell^{\phi\phi}$ estimates is found over the full multipole range.

The \Planck\ lensing likelihood is based on reconstructions from the
minimum-variance combination of the 143 and 217\,GHz maps with 30\% of the
sky masked. Conservatively, only multipoles in the range $\ell=40$--$400$
are included, with a bandpower width $\Delta \ell = 45$.
The range $\ell=40$--$400$ captures 90\% of the signal-to-noise
on a measurement of the amplitude of a fiducial $C_\ell^{\phi\phi}$, while
minimizing the impact of imperfections in modelling the effect
of survey anisotropies on the large-scale $\phi$ reconstruction (the
``mean-field'' of~\citealt{planck2013-p12}), and the large Gaussian noise bias
on small scales. Note, however, that by restricting the range of angular scales
we do lose some ability to distinguish between scale-dependent
modifications of $C_\ell^{\phi\phi}$, such as from massive neutrinos,
and almost scale-independent modifications, such as from changes in the
equation of state of unclustered dark energy or spatial curvature.
\referee{Correlated uncertainties in the
beam transfer functions, point-source corrections, and the cosmology
dependence of the $N^{(1)}$ bias give very broad-band correlations between the bandpowers. These are modelled as a sum of rank-one corrections to the covariance matrix and induce bandpower correlations that are small, less than $4\%$, but
very broad.} Bandpower correlations induced by masking are estimated to be less than
5\% for neighbouring bins and are neglected. The likelihood is modelled
as a Gaussian in the bandpowers with a fiducial (i.e., parameter-independent)
covariance. For verification of this approximation, see~\citet{Schmittfullinprep}.

The connected four-point function is related to the fully-reduced
trispectrum $\mathbb{T}^{\ell_1 \ell_2}_{\ell_3 \ell_4}(L)$ by
\begin{eqnarray}
\langle T_{\ell_1 m_1}  T_{\ell_2 m_2} T_{\ell_3 m_3} T_{\ell_4 m_4} \rangle_{\mathrm{c}} &=& \frac{1}{2} \sum_{LM} (-1)^M \left( \begin{array}{ccc} \ell_1 & \ell_2 & L \\
m_1 & m_2 & M\end{array}\right) \nonumber \\
&& \mbox{} \hspace{-0.03\textwidth} \times
\left( \begin{array}{ccc} \ell_3 & \ell_4 & L \\
m_3 & m_4 & -M\end{array}\right) \mathbb{T}^{\ell_1 \ell_2}_{\ell_3 \ell_4}(L)
+ \mathrm{perms}\, ,
\end{eqnarray}
\citep{2001PhRvD..64h3005H}.
In the context of lensing reconstruction, the CMB trispectrum due to lensing
takes the form
\begin{equation}
\mathbb{T}^{\ell_1 \ell_2}_{\ell_3 \ell_4}(L) \approx C_L^{\phi\phi}
C_{\ell_2}^{TT} C_{\ell_4}^{TT} F_{\ell_1 L \ell_2} F_{\ell_3 L \ell_4} \, ,
\label{eq:lensing_trispectrum}
\end{equation}
where $C_\ell^{TT}$ is the \emph{lensed} temperature power spectrum and
$F_{\ell_1 L \ell_2}$ is a geometric mode-coupling function~\citep{2001PhRvD..64h3005H,2011PhRvD..83d3005H}.
Our estimates of $C_\ell^{\phi\phi}$ derive from the measured trispectrum. They
are normalized using the fiducial lensed power spectrum to account for the factors of $C_{\ell}^{TT}$ in
Eq.~(\ref{eq:lensing_trispectrum}). In the likelihood, we renormalize the
parameter-dependent $C_\ell^{\phi\phi}$
to account for the mismatch between the parameter-dependent $C_\ell^{TT}$
and that in the fiducial model. Since the best-fit \LCDM\ model we consider in this
section has a lensed temperature power spectrum that is very close to that
of the fiducial model, the renormalisation factor differs from unity by less
than $0.25\%$.

The estimated lensing power spectrum $C_\ell^{\phi\phi}$ is not independent
of the measured temperature power spectrum $C_\ell^{TT}$, but the
dependence is very weak for \Planck,
and can be accurately ignored~\citep{Schmittfullinprep,planck2013-p12}.
As discussed in detail in~\citet{Schmittfullinprep}, there are
several effects to consider. First, the reconstruction noise in the estimated $\phi$
derives from chance correlations in the unlensed CMB. If, due to cosmic
variance, the unlensed CMB fluctuates high at some scale, the noise in the
reconstruction will generally increase over a broad range of scales. Over the
scales relevant for \Planck\ lensing reconstruction, the correlation between the
measured $C_\ell^{\phi\phi}$ and $C_{\ell'}^{TT}$ from this effect is
less than $0.2\%$ and, moreover, is removed by a
data-dependent Gaussian noise bias removal that we adopt
following~\citet{2011PhRvD..83d3005H} and~\cite{2012arXiv1209.0091N}. The second effect derives from cosmic variance of the
lenses. If a lens on a given scale fluctuates high, the estimated
$C_\ell^{\phi\phi}$ will fluctuate high at that scale. In tandem, there
will be more smoothing of the acoustic peaks in the measured
$C_{\ell'}^{TT}$,
giving broad-band correlations that are negative at acoustic peaks and
positive at troughs. The maximum correlation is around $0.05\%$. If we consider
estimating the amplitude of a fiducial lensing power spectrum independently
from the smoothing effect of $C_\ell^{TT}$ and the measured
$C_\ell^{\phi\phi}$ in the range $\ell=40$--$400$, the correlation between
these estimates due to the cosmic variance of the lenses is only $4\%$.
This amounts to a mis-estimation of the error on a lensing amplitude in
a joint analysis of $C_\ell^{\phi\phi}$ and $C_\ell^{TT}$, treated as
independent, of only $2\%$. For physical parameters, the mis-estimation
of the errors is even smaller:~\citet{Schmittfullinprep} estimate around
$0.5\%$ from a Fisher analysis.  A third negligible effect is due to the
$T$--$\phi$ correlation sourced by the late integrated Sachs-Wolfe effect
\citep[see][]{planck2013-p14}.
This produces only \emph{local} correlations between the measured
$C_\ell^{\phi\phi}$ and
$C_\ell^{TT}$ which are less than $0.5\%$ by $\ell=40$ and fall rapidly on
smaller scales. They produce a negligible correlation between lensing
amplitude estimates for the multipole ranges considered here.
The $T$--$\phi$ correlation is potentially a powerful probe of
dark energy dynamics~(e.g., \citealt{2002PhRvD..65d3007V}) and
modified theories of gravity~(e.g., \citealt{2004PhRvD..70b3515A}).
The power spectrum $C_\ell^{T\phi}$ can be measured from the \Planck\ data
using  the CMB 3-point function~\citep{planck2013-p09a} or, equivalently,
by cross-correlating the $\phi$ reconstruction with the large-angle temperature
anisotropies~\citep{planck2013-p14} although the detection significance is
only around $3\,\sigma$. The power-spectrum based analysis in this paper
discards the small amount of information in the $T$--$\phi$ correlation
from \Planck. In summary,
we can safely treat the measured temperature and lensing power spectra as
independent and simply multiply their respective likelihoods in a joint
analysis.

We note that ACT~\citep{2011PhRvL.107b1301D,2013arXiv1301.1037D} and SPT~\citep{vanEngelen:2012va} have both measured the
lensing power spectrum with significances of $4.6\,\sigma$ and $6.3\,\sigma$,
respectively, in the multipole ranges $\ell=75$--$2050$ and $\ell=100$--$1500$.
The \Planck\ measurements used here represent a
$26\,\sigma$ detection. We therefore do not expect the published
lensing measurements from these other experiments to carry much statistical
weight in a joint analysis with \Planck, despite the complementary range
of angular scales probed, and we choose not to include them in the
analyses in this paper.

In the lensing likelihood, we characterize the estimates of $C_\ell^{\phi\phi}$ with a set of eight (dimensionless) amplitudes $\hat{A}_i$, where
\begin{equation}
\hat{A}_i= \sum_\ell \mathcal{B}^\ell_i \hat{C}_\ell^{\phi\phi} \, .
\end{equation}
Here, $\mathcal{B}^\ell_i$ is a binning operation with
\begin{equation}
\mathcal{B}^\ell_i = \frac{C_\ell^{\phi\phi,\,\mathrm{fid}} V_\ell^{-1}}
{\sum_{\ell'=\ell_{\mathrm{min}}^i}^{\ell_{\mathrm{max}}^i} \left(C_{\ell'}^{\phi\phi,\,
\mathrm{fid}}\right)^2 V_{\ell'}^{-1}} \, ,
\end{equation}
for $\ell$ within the band defined by a minimum multipole
$\ell_{\mathrm{min}}^i$ and a maximum $\ell_{\mathrm{max}}^i$. The inverse of
the weighting function, $V_\ell$, is an approximation to the variance of
the measured $\hat{C}_\ell^{\phi\phi}$ and $C_\ell^{\phi\phi,\,\mathrm{fid}}$
is the lensing power spectrum of the fiducial model, which is used throughout the analysis.
The $\hat{A}_i$ are therefore near-optimal estimates of the amplitude
of the fiducial power spectrum within the appropriate multipole range,
normalized to unity in the fiducial model. Given some parameter-dependent
model $C_\ell^{\phi\phi}$, the expected values of the $\hat{A}_i$ are
\begin{equation}
\langle \hat{A}_i \rangle = A_i^{\mathrm{theory}} = \sum_\ell \mathcal{B}^\ell_i \left[1+\Delta^\phi(C_\ell^{TT})\right]^2
C_\ell^{\phi\phi} \, ,
\label{eq:meanA}
\end{equation}
where the term involving $\Delta^\phi(C_\ell^{TT})$, which depends on the
parameter-dependent $C_\ell^{TT}$, accounts for the renormalisation step
described above. The lensing amplitudes $\hat{A}_i$ are compared to the
$\Ailensbestfit$ for the best-fitting \LCDM\ model to the \planck+\WP+\highL\
data combination (i.e., not including the lensing likelihood) in
Table~\ref{tab:CMBlensing_LCDM}. The differences
between $\hat{A}_i$ and $\Ailensbestfit$ are plotted in the bottom
panel of Fig.~\ref{fig:CMBlensing_LCDM} while in the top panel
the bandpower estimates are compared to $C_\ell^{\phi\phi}$ in the best-fitting
model. The \Planck\ measurements of $C_\ell^{\phi\phi}$ are consistent
with the prediction from the best-fit \LCDM\ model to \Planck+\WP+\highL.
Using the full covariance matrix, we find $\chi^2 = 10.9$ with eight degrees of
freedom, giving an acceptable probability to exceed of approximately $21\%$.
It is worth recalling here that the parameters
of the \LCDM\ model are tightly constrained by the CMB 2-point function (as
probed by our \Planck+\WP+\highL\ data combination) which derives from physics
at $z\approx 1100$ seen in angular projection. \emph{It is a significant further
vindication of the \LCDM\ model that its predictions for the evolution of
structure and geometry at much lower redshifts (around $z=2$) fit so well
with \Planck's CMB lensing measurements.}

\begin{table}[tmb]                 
\begingroup
\newdimen\tblskip \tblskip=5pt
\caption{\Planck\ CMB lensing constraints. The $\Ailensbestfit$
are renormalized power spectrum amplitudes in the best-fit \LCDM\ model
to \Planck+\WP+\highL\ within the $i$th band (from $\ell_{\rm min}$ to $\ell_{\rm max}$).
The errors $\sigma(A_i)$ on the amplitudes are the
square root of the diagonals of the $\hat{A}_i$ covariance matrix.}                          
\label{tab:CMBlensing_LCDM}         
\nointerlineskip
\vskip -0mm
\footnotesize
\setbox\tablebox=\vbox{
   \newdimen\digitwidth
   \setbox0=\hbox{\rm 0}
   \digitwidth=\wd0
   \catcode`*=\active
   \def*{\kern\digitwidth}
   \newdimen\signwidth
   \setbox0=\hbox{+}
   \signwidth=\wd0
   \catcode`!=\active
   \def!{\kern\signwidth}
\halign{%
\hbox to 0.6in{#\leaderfil}\tabskip=1em&
        \hfil#\hfil&
        \hfil#\hfil&
        \hfil#\hfil&
        \hfil#\hfil&
        \hfil#\hfil\tabskip=0pt\cr                
\noalign{\doubleline}
\omit\hfil Band\hfil&$\ell_{\mathrm{min}}$&$\ell_{\mathrm{max}}$&$\hat{A}$&$\Alensbestfit$&$\sigma(A)$\cr 
\noalign{\vskip 3pt\hrule\vskip 5pt}
1&*40&*84&1.11&1.00&0.07\cr
2&*85&129&0.97&0.99&0.07\cr
3&130&174&0.90&0.98&0.08\cr
4&175&219&0.77&0.98&0.10\cr
5&220&264&0.88&0.98&0.13\cr
6&265&309&0.88&0.98&0.16\cr
7&310&354&1.10&0.98&0.18\cr
8&355&400&0.75&0.98&0.21\cr        
\noalign{\vskip 5pt\hrule\vskip 3pt}}}
\endPlancktable                    
\endgroup
\end{table}                        

The discussion above does not account for the small spread in the
$C_\ell^{\phi\phi}$ predictions across the \Planck+WP+highL \LCDM\ posterior
distribution.
To address this, we
introduce a parameter $\Aphiphi$ which, at any point in
parameter space, scales the lensing  trispectrum.
Note that $\Aphiphi$ does not alter the lensed temperature
power spectrum, so it can be used to assess directly how well the \LCDM\ 
predictions from $C_\ell^{TT}$ agree with the lensing measurements;
in \LCDM\ we have $\Aphiphi=1$.
The marginalized posterior distribution for $\Aphiphi$ in
a joint analysis of \Planck+\WP+\HighL\ and the \Planck\ lensing
likelihood is given in Fig.~\ref{fig:CMBlensing_Alensphiphi}. The agreement
with $\Aphiphi=1$ is excellent, with
\begin{equation}
\Aphiphi=0.99\pm 0.05 \quad (\mbox{68\%; \planck+\lensing+\WP+\HighL}).
\end{equation}
The significance of the detection of lensing using $\Aphiphi$ in \lcdm\ is a little less than the $26\,\sigma$ detection of lensing power reported in~\citet{planck2013-p12}, due to the small spread in $C_\ell^{\phi\phi}$ from \lcdm\ parameter uncertainties.

\begin{figure}
\centering
\includegraphics[width=88mm,angle=0]{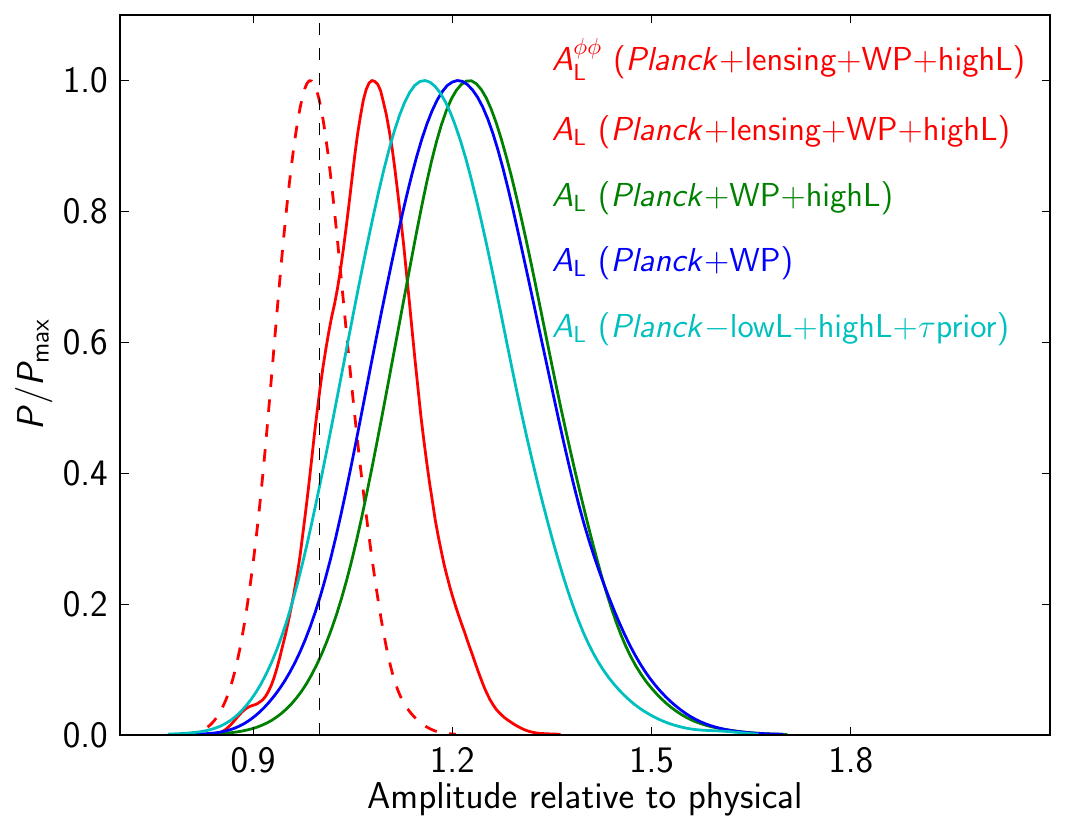}
\caption{Marginalized posterior distributions for $\Aphiphi$ (dashed) and $\Alens$ (solid). For $\Aphiphi$ we use the data combination \Planck+\lensing+WP+\highL. For $\Alens$ we consider \Planck+\lensing+WP+\highL\ (red), \Planck+WP+\highL\ (green), \Planck+\WP\ (blue) and \Planck$-$lowL+\highL+$\tau$prior (cyan; see text).}
\label{fig:CMBlensing_Alensphiphi}
\end{figure}

Lensing also affects the temperature power spectrum, primarily by smoothing
the acoustic peaks and troughs on the scales relevant for \Planck. The most
significant detection of the lensing effect in the power spectrum to date
is from SPT.  Introducing a parameter $\Alens$~\citep{2008PhRvD..77l3531C}
which takes $C_\ell^{\phi\phi} \rightarrow \Alens C_\ell^{\phi\phi}$ when
computing the lensed temperature power spectrum (we shall shortly extend the
action of this parameter to include the computation of the lensing
trispectrum), \citet{Story:12} report $\Alens = 0.86^{+0.15}_{-0.13}$
(68\%; SPT+\textit{WMAP}-7). Results for $\Alens$ from \Planck\ in
combination with \WMAP\ low-$\ell$ polarization and the high-$\ell$ power
spectra from ACT and SPT are also shown in
Fig.~\ref{fig:CMBlensing_Alensphiphi}. Where we include the \Planck\ lensing
measurements, we define $\Alens$ to scale the explicit $C_\ell^{\phi\phi}$
in Eq.~(\ref{eq:lensing_trispectrum}), as well as modulating the lensing
effect in the temperature power spectrum.
Figure~\ref{fig:CMBlensing_Alensphiphi} reveals a preference for
$\Alens>1$ from the \Planck\ temperature power spectrum (plus \WMAP\
polarization). This is most significant when combining with the high-$\ell$
experiments for which we find
\begin{equation}
\Alens = 1.23\pm 0.11 \quad \mbox{(68\%; \Planck+\WP+\highL)}, \label{Alens}
\end{equation}
i.e., a $2\,\sigma$ preference for $\Alens>1$. Including the lensing
measurements, the posterior narrows but shifts to lower $\Alens$, becoming
consistent with $\Alens=1$ at the $1\,\sigma$ level as expected from the
$\Aphiphi$ results.

\begin{figure*}
\centering
\includegraphics[width=88mm,angle=0]{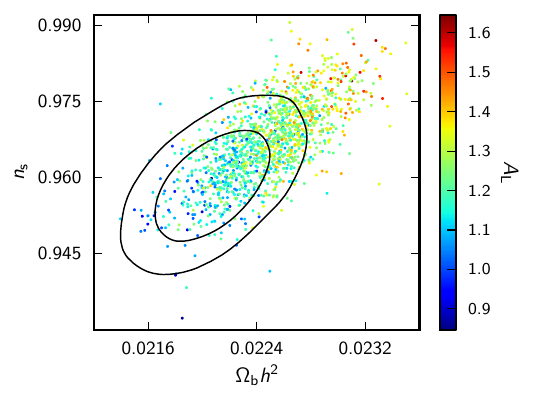}
\includegraphics[width=88mm,angle=0]{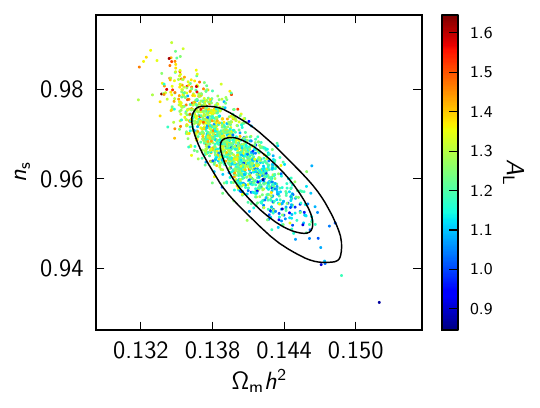}
\caption{Effect of allowing $\Alens$ to vary on the degeneracies between $\Omb h^2$ and $\ns$ (left) and $\Omm h^2$ and $\ns$ (right). In both panels the data combination is \Planck+\WP+\highL. The contours enclose the $68\%$ and $95\%$ confidence regions in the base \lcdm\ model with $\Alens=1$. The samples are from models with variable $\Alens$ and are colour-coded by the value of $\Alens$.}
\label{fig:ombc_ns_Alens}
\end{figure*}

We do not yet have a full understanding of what is driving the preference for high $\Alens$ in the temperature power spectrum. As discussed in Appendix~\ref{app:test}, the general preference is stable to assumptions about foreground modelling and cuts of the \planck\ data in the likelihood. To gain some insight, we consider the range of multipoles that drive the preference for $\Alens>1$. For our favoured data combination of \Planck+\WP+\highL, $\Delta \chi^2  = -5.2$ going from the best-fit $\Alens=1$ model to the best-fit model with variable $\Alens$. The improvement in fit comes only from the low-$\ell$ temperature power spectrum
($\Delta \chi^2 = -1.9$) and the ACT+SPT data ($\Delta\chi^2 = - 3.3$); for this data combination, there is no preference for high $\Alens$ from the \Planck\ temperature data at intermediate and high multipoles ($\Delta \chi^2 = + 0.2$).  The situation at low-$\ell$ is similar if we exclude the high-$\ell$ experiments, with $\Delta\chi^2 =-1.6$ there, but there is then a preference for the high $\Alens$ best-fit from the \planck\ data on intermediate and small scales ($\Delta\chi^2 = -3.4$). However, as discussed in Sect.~\ref{sec:highell}, there is more freedom in the foreground model when we exclude the high-$\ell$ data, and this can offset smooth differences in the CMB power spectra such as the transfer of power from large to small scales by lensing that is enhanced for $\Alens > 1$.

Since the low-$\ell$ temperature data seem to be partly responsible for pulling $\Alens$ high, we consider the effect of removing the low-$\ell$ likelihood from the analysis. In doing so, we also remove the \WMAP\ large-angle polarization which we compensate by introducing a simple prior on the optical depth; we use a Gaussian with mean $0.09$ and standard deviation $0.013$, similar to the constraint from \WMAP\ polarization~\citep{hinshaw2012}. We denote this data combination, including the high-$\ell$ experiments, by \Planck$-$lowL+\highL+$\tau$prior and show the posterior for $\Alens$ in Fig.~\ref{fig:CMBlensing_Alensphiphi}. 
As anticipated, the peak of the posterior moves to lower $\Alens$ giving $\Alens=1.17^{+0.11}_{-0.13}$ (68\% CL). The $\Delta\chi^2 = +1.1$ between the best-fit model (now at $\Alens=1.18$) and the $\Alens=1$ model for the \Planck\ data (i.e.\ no preference for the higher $\Alens$) while $\Delta \chi^2 = -3.6$ for the high-$\ell$ experiments.

Since varying $\Alens$ alone does not alter the power spectrum on large scales, why should the low-$\ell$ data prefer higher $\Alens$? The reason is due to a chain of parameter degeneracies that are illustrated in Fig.~\ref{fig:ombc_ns_Alens}, and the deficit of power in the measured $C_\ell$s on large scales compared to the
best-fit \lcdm\ model (see Fig.~\ref{TTspec} and Sect.~\ref{sec:conclusions}). In models with a power-law primordial spectrum, the temperature power spectrum on large scales can be reduced by increasing $\ns$. The effect of an increase in $\ns$ on the relative heights of the first few acoustic peaks can be compensated by increasing $\omb$ and reducing $\omm$, as shown by the contours in Fig.~\ref{fig:ombc_ns_Alens}. However, on smaller scales, corresponding to modes that entered the sound horizon well before matter-radiation equality, the effects of baryons on the mid-point of the acoustic oscillations (which modulates the relative heights of even and odd peaks) is diminished since the gravitational potentials have pressure-damped away during the oscillations in the radiation-dominated phase (e.g., \citealt{1996ApJ...471...30H,1997ApJ...479..568H}).
Moreover, on such scales the radiation-driving at the onset of the oscillations that amplifies their amplitude happens early enough to be unaffected by small changes in the matter density. The net effect is that, in models with $\Alens=1$, the extent of the degeneracy involving $\ns$, $\omb$ and $\omm$ is limited by the higher-order acoustic peaks, and there is little freedom to lower the large-scale temperature power spectrum by increasing $\ns$ while preserving the good fit at intermediate and small scales. Allowing $\Alens$ to vary changes this picture, letting the degeneracy extend to higher $\ns$, as shown by the samples in Fig.~\ref{fig:ombc_ns_Alens}. The additional smoothing of the acoustic peaks due to an increase  in $\Alens$ can mitigate the effect of increasing $\ns$ around the fifth peak, where the signal-to-noise for \Planck\ is still high.\footnote{Since models with high $\Alens$ that fit the \Planck\ data have lower $\omm$, the additional smoothing of the acoustic peaks at high $\Alens$ is typically a few percent less than is suggested by $\Alens$ alone.} This allows one to decrease the spectrum at low $\ell$, while leaving it essentially unchanged on those smaller scales where \Planck\ still has good sensitivity. Above $\ell \sim 2000$, the best-fit $\Alens$ model has a little more power than the base model (around $3\,\mu\mathrm{K}^2$ at $\ell=2000$), while the \planck, ACT, and SPT data have excess power over the best-fit $\Alens=1$ \lcdm+foreground model at the level of a few $\mu\mathrm{K}^2$ (see Sect.~\ref{sec:highell}). It is plausible that this may drive the preference for high $\Alens$ in the $\chi^2$ of the high-$\ell$ experiments. We note that a similar $2\,\sigma$  preference for $\Alens > 1$ is also found combining ACT and \WMAP\ data~\citep{Sievers:13} and, as we find here, this tension is reduced when the lensing power spectrum is included in the fit.

To summarize, there is no preference in the \planck\ lensing power spectrum for $\Alens > 1$. The general preference for high $\Alens$ from the CMB power spectra in our favoured data combination (\planck+\WP+\highL) is mostly driven by two effects:
the difficulty that \lcdm\ models have in fitting the low-$\ell$ spectrum when calibrated from the smaller-scale spectrum; and, plausibly, from excess residuals at the $\mu\mathrm{K}^2$ level in the high-$\ell$ spectra relative to the best-fit $\Alens=1$ \lcdm+foregrounds model on scales where extragalactic foreground modelling is critical.

\subsection {Baryon acoustic oscillations}
\label{sec:BAO}

Baryon acoustic oscillations (BAO) in the matter power spectrum were first
detected in analyses of  the 2dF Galaxy Redshift Survey \citep{Cole:05} and the
SDSS redshift survey \citep{Eisenstein:05}. Since then, accurate BAO
measurements have been made using a number of different galaxy
redshift surveys, providing constraints on the distance luminosity
relation spanning the redshift range $0.1 \la z \la 0.7$\footnote{Detections of a BAO feature have recently been reported
in the three-dimensional correlation function of the Ly$\alpha$ forest
in large samples of quasars at a mean redshift of $z\approx 2.3$
\citep{Busca:12,  Slosar:13}. These remarkable results, probing
cosmology well into the matter-dominated regime, are based on new
techniques that are less mature than galaxy BAO measurements. For
this reason, we do not include Ly$\alpha$ BAO measurements as
supplementary data to \planck. For the models considered here and in
Sect. \ref{sec:grid}, the galaxy BAO results give significantly tighter
constraints than the Ly$\alpha$ results.}.
 Here we use the results from four redshift
surveys: the SDSS DR7 BAO measurements at effective redshifts $z_{\rm
  eff}=0.2$ and $z_{\rm eff}=0.35$, analysed by \citet{Percival:10};
the $z=0.35$ SDSS DR7 measurement at $z_{\rm eff} = 0.35$
reanalyzed by \citet{Padmanabhan:2012hf}; the WiggleZ measurements
at $z_{\rm eff}=0.44$, $0.60$ and $0.73$ analysed by \citet{Blake:11};
the BOSS DR9 measurement at $z_{\rm eff}=0.57$ analyzed by
\citet{Anderson:2012sa}; and the 6dF Galaxy Survey measurement at
$z=0.1$ discussed  by \citet{Beutler:11}.

BAO surveys measure the distance ratio
\begin{equation}
d_{z} = { r_{\rm s}(z_{\rm drag}) \over D_{\rm V}(z)}, 
\end{equation}
where $ r_{\rm s}(z_{\rm drag})$ is the comoving sound horizon at the baryon
drag epoch (when baryons became dynamically decoupled from the
photons) and $D_{\rm V}(z)$  is a combination of
the angular-diameter distance, $D_{\rm A}(z)$, and the Hubble parameter,
$H(z)$,   appropriate for the analysis of spherically-averaged two-point statistics:
\begin{equation}
 D_{\rm V}(z) = \left [ (1+z)^2 D^2_{\rm A}(z) {cz \over H(z)} \right ]^{1/3}.  \label{BAO2}
\end{equation}
In the \LCDM\ cosmology (allowing for spatial curvature), the angular diameter distance to redshift $z$ is
\begin{eqnarray}
D_{\rm A}(z) &=& {c \over H_0} \hat D_{\rm A}.  \nonumber \\
   & =&
{ c \over H_0} {1
\over \vert \Omega_{K} \vert^{1/2}(1+z)} {\rm sin}_{K} 
 \left[ \vert \Omega_{K} \vert^{1/2} x(z, \Omega_{\rm m},
 \Omega_\Lambda)\right],  \label{BAO1}
\end{eqnarray}
where
\begin{equation}
x (z, \Omega_{\rm m}, \Omega_\Lambda) = \int_0^z {dz^\prime \over
[ \Omega_{\rm m} (1+z^\prime)^3 + \Omega_{K} (1+z^\prime)^2 + \Omega_\Lambda]^{1/2}},  
\label{eq:BAOdist}
\end{equation}
and ${\rm sin}_{K} = {\rm sinh}$ for $\Omega_{K} >0$ and ${\rm sin}_{K} = \sin$ for $\Omega_{K} < 0$. (The small effects of the $0.06\,{\rm eV}$ massive neutrino in our base cosmology are ignored in Eq.~\ref{eq:BAOdist}.)
Note that the luminosity distance, $D_{\rm L}$, relevant for the
analysis of Type Ia supernovae (see Sect.~ \ref{subsec:datasetsSNe}) is related to the
angular diameter distance via $D_{\rm L} = (c/H_0)\hat D_{\rm L} = D_{\rm A}(1+z)^2$.

\begin{figure}
\begin{center}
\includegraphics[width=85mm,angle=0]{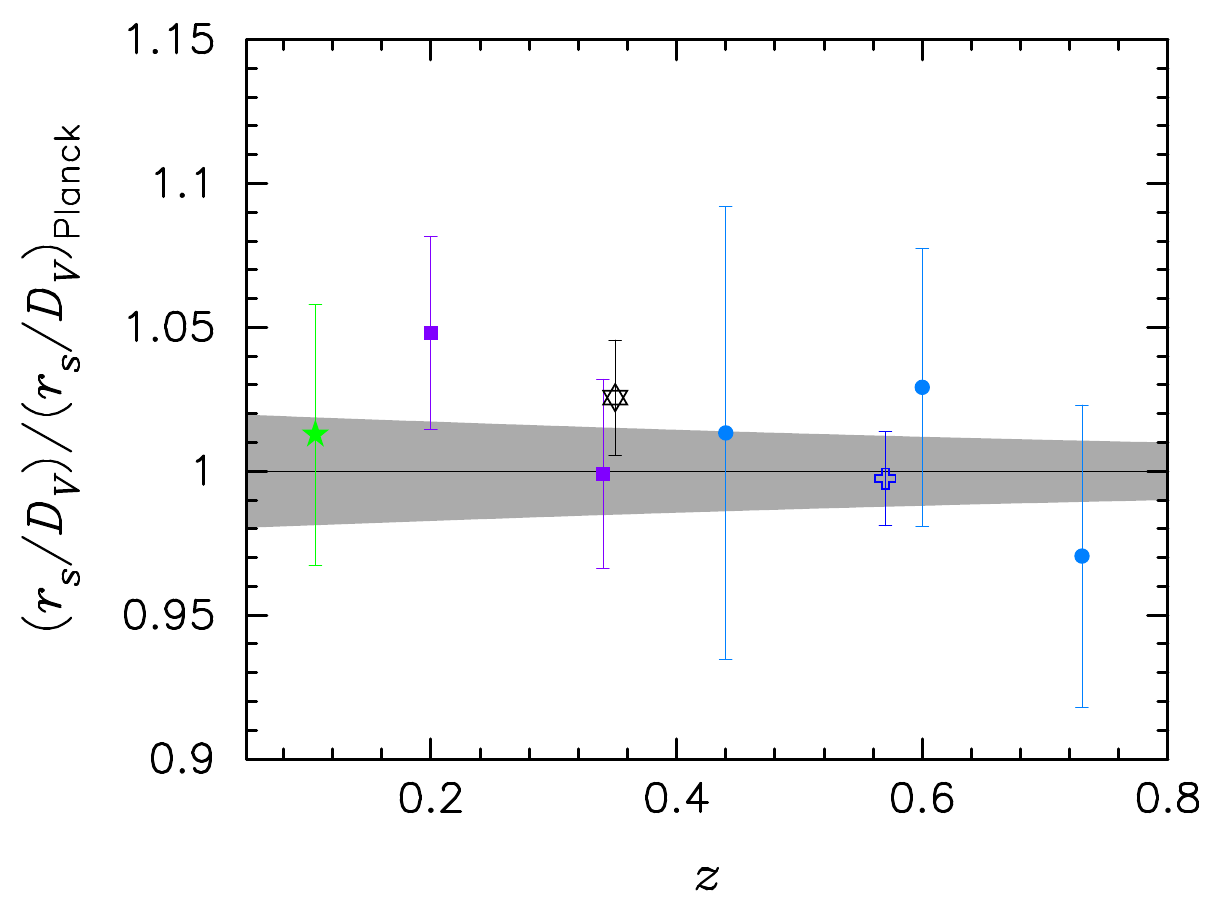}
\end{center}
\caption {Acoustic-scale distance ratio $r_{\rm s}/D_{\rm V}(z)$
  divided by the distance ratio of the \planck\ base \LCDM\ model.
  The points are colour-coded as follows: green star
  (6dF); purple squares \citep[SDSS DR7 as analyzed by][]{Percival:10};
  black star \citep[SDSS DR7 as analyzed by][]{Padmanabhan:2012hf}; 
blue cross (BOSS DR9); and blue circles (WiggleZ). 
  The grey band shows the approximate
$\pm 1\,\sigma$ range allowed by \planck\ (computed from the \COSMOMC\ chains).}
\label{BAO}
\end{figure}

Different groups fit and characterize BAO features in different
ways. For example, the WiggleZ team encode some shape information on
the power spectrum to measure the acoustic parameter $A(z)$, introduced
by \citet{Eisenstein:05},
\be
A(z) = { D_{\rm V}(z)\sqrt{\Omm H_0^2} \over cz}, \label{BAO3} 
\ee
which is almost independent of $\omega_{\rm m}$.  To
simplify the presentation, Fig.~\ref{BAO} shows estimates of
$r_{\rm s}/D_{\rm V}(z)$ and $1\,\sigma$ errors,  as quoted by each of the
  experimental groups,  divided by the expected relation for the 
\planck\  base \LCDM\  parameters. Note that the experimental groups use the
approximate formulae of \citet{Eisenstein:98} to compute $z_{\rm
  drag}$ and $r_{\rm s}(z_{\rm drag})$, though they fit power spectra
computed with Boltzmann codes, such as \CAMB, generated for a set of
fiducial-model parameters. The measurements have now become so precise
that the small difference between the \citet{Eisenstein:98}
approximations and the accurate values of $z_{\rm drag}$ and $r_{\rm drag} =
r_{\rm s}(z_{\rm drag})$ returned by \CAMB\ need to be taken into account. In
\COSMOMC\ we multiply the accurate numerical value of $r_{\rm s}(z_{\rm
  drag})$ by a constant factor of $1.0275$ to match the Eisenstein-Hu
approximation in the fiducial model.  This
correction is sufficiently accurate over the range of $\omega_{\rm m}$ and
$\omega_{\rm b}$ allowed by the  CMB in the base \lcdm\ cosmology \citep[see e.g.][]{Mehta:12}
and also for the extended \lcdm\ models  discussed
in Sect. \ref{sec:grid}.

The \citet{Padmanabhan:2012hf} result plotted in Fig.~\ref{BAO} is a
reanalysis of the $z_{\rm eff} = 0.35$ SDSS DR7 sample discussed by
\citet{Percival:10}. \citet{Padmanabhan:2012hf}
achieve a higher precision than \citet{Percival:10} by employing a
reconstruction technique \citep{Eisenstein:07} to correct (partially)
the baryon oscillations for the smearing caused by galaxy peculiar
velocities.  The \citet{Padmanabhan:2012hf} results are therefore strongly
correlated with those of \citet{Percival:10}.  
We refer to the \citet{Padmanabhan:2012hf}
``reconstruction-corrected'' results as SDSS(R). A similar
reconstruction technique was applied to the BOSS survey by
\citet{Anderson:2012sa} to achieve $1.6\%$ precision in $D_{\rm V}(z=0.57)/r_{\rm s}$, the
most precise determination of the acoustic oscillation scale to date.

All of the BAO measurements are compatible with the base \LCDM\ parameters
from \planck. The grey band in Fig. \ref{BAO} shows the $\pm 1\,\sigma$
range in the acoustic-scale distance ratio computed from the 
\planck+\WP+\highL\  \COSMOMC\ chains for the base \lcdm\ model.
 To get a qualitative feel for how the BAO measurements constrain
parameters in the base \lcdm\ model, we form
 $\chi^2$,
\begin{equation}
\chi_{\rm BAO}^2  =  (\vx- \vx^{\Lambda{\rm CDM}})^T \tens{C}_{\rm BAO}^{-1} (\vx -\vx^{\Lambda{\rm CDM}}),
\label{BAO5}
\end{equation}
where $\vx$ is the data vector, $\vx^{\Lambda{\rm CDM}}$ denotes the theoretical
prediction for the \LCDM\ model and $\tens{C}_{\rm BAO}^{-1}$ is the inverse
covariance matrix for the data vector $\vx$.  The data vector is as
follows: $D_{\rm V}(0.106) = (457 \pm 27) \,{\rm Mpc}$ (6dF); 
$r_{\rm s}/D_{\rm V}(0.20)
= 0.1905 \pm 0.0061$, $r_{\rm s}/D_{\rm V}(0.35) = 0.1097 \pm 0.0036$ (SDSS);
$A(0.44) = 0.474 \pm 0.034$, $A(0.60) = 0.442 \pm 0.020$, $A(0.73) =
0.424 \pm 0.021$ (WiggleZ); $D_{\rm V}(0.35)/r_{\rm s}=8.88 \pm 0.17$ (SDSS(R));
and $D_{\rm V}(0.57)/r_{\rm s} = 13.67 \pm 0.22$, (BOSS). The off-diagonal components
of $\tens{C}_{\rm  BAO}^{-1}$ for the SDSS and WiggleZ results are given in
\citet{Percival:10} and \citet{Blake:11}. We ignore any covariances
between surveys. Since the SDSS and SDSS(R) results are based on the
same survey, we include either one set of results or the
other in the analysis described below, but not both together.

 The Eisenstein-Hu values of $r_{\rm s}$ for the \planck\ and \mbox{\textit{WMAP}-9}
base \LCDM\ parameters differ by only $0.9\%$, significantly smaller than the
errors in the BAO measurements. We can obtain an {\it approximate\/}
idea of the complementary information provided by BAO measurements
by minimizing Eq.~(\ref{BAO5}) with
respect to either $\Omega_{\rm m}$ or $H_0$, fixing $\omega_{\rm m}$ and
 $\omega_{\rm b}$
to the CMB best-fit parameters. (We use the \planck+\WP+\highL\ parameters from Table~\ref{LCDMForegroundparams}.) The results are listed in Table~\ref{BAOTable}\footnote{As an indication of the accuracy of Table \ref{BAOTable}, the full likelihood results for the \planck+\WP+6dF+SDSS(R)+BOSS BAO data sets give $\Omega_{\rm m}=0.308 \pm 0.010$ and $H_0=67.8 \pm 0.8\, {\rm km}\, {\rm s}^{-1}\, {\rm Mpc}^{-1}$,  for the base \LCDM\ model.}.

\begin{table}[tmb]                 
\begingroup
\newdimen\tblskip \tblskip=5pt
\caption{Approximate constraints with 68\%\ errors on $\Omega_{\rm m}$ and $H_0$ (in units of ${\rm km}\, {\rm s}^{-1}\,  {\rm Mpc}^{-1}$) from BAO, with $\omega_{\rm m}$ and $\omega_{\rm b}$ fixed to the best-fit \planck+\WP+\highL\ values for the base \lcdm\
cosmology.}                          
\label{BAOTable}         
\nointerlineskip
\vskip -3mm
\footnotesize
\setbox\tablebox=\vbox{
   \newdimen\digitwidth
   \setbox0=\hbox{\rm 0}
   \digitwidth=\wd0
   \catcode`*=\active
   \def*{\kern\digitwidth}
   \newdimen\signwidth
   \setbox0=\hbox{+}
   \signwidth=\wd0
   \catcode`!=\active
   \def!{\kern\signwidth}
\halign{\hbox to 2.0in{#\leaderfil}\tabskip 2.2em&
        \hfil#\hfil&
       \hfil#\hfil\tabskip=0pt\cr                
\noalign{\doubleline}
\omit\hfil Sample\hfil&$\Omega_{\rm m}$&$H_0$\cr
\noalign{\vskip 4pt\hrule\vskip 6pt}
6dF&$0.305^{+0.032}_{-0.026}$&$68.3^{+3.2}_{-3.2}$\cr
\noalign{\vskip 3pt}
SDSS&$0.295^{+0.019}_{-0.017}$&$69.5^{+2.2}_{-2.1}$\cr
\noalign{\vskip 3pt}
SDSS(R)&$0.293^{+0.015}_{-0.013}$&$69.6^{+1.7}_{-1.5}$\cr
\noalign{\vskip 3pt}
WiggleZ&$0.309^{+0.041}_{-0.035}$&$67.8^{+4.1}_{-2.8}$\cr
\noalign{\vskip 3pt}
BOSS&$0.315^{+0.015}_{-0.015}$&$67.2^{+1.6}_{-1.5}$\cr
\noalign{\vskip 3pt}
6dF+SDSS+BOSS+WiggleZ&$0.307^{+0.010}_{-0.011}$&$68.1^{+1.1}_{-1.1}$\cr
\noalign{\vskip 3pt}
6dF+SDSS(R)+BOSS&$0.305^{+0.009}_{-0.010}$&$68.4^{+1.0}_{-1.0}$\cr
\noalign{\vskip 3pt}
6dF+SDSS(R)+BOSS+WiggleZ&$0.305^{+0.009}_{-0.008}$&$68.4^{+1.0}_{-1.0}$\cr
\noalign{\vskip 3pt\hrule\vskip 4pt}}}
\endPlancktable                    
\endgroup
\end{table}                        

As can be seen, the results are very stable from survey to survey
and  are in excellent agreement with the base \LCDM\ 
parameters listed in Tables \ref{LCDMparams} and
\ref{LCDMForegroundparams}. The values of $\chi^2_{\rm BAO}$ are also
reasonable. For example, for the six data points of the
6dF+SDSS(R)+BOSS+WiggleZ combination, we find $\chi^2_{\rm BAO}=4.3$,
evaluated for the \planck+\WP+\highL\ best-fit \LCDM\ parameters.

The high value of $\Omm$ is consistent with the parameter analysis
described by \citet{Blake:11} and with the ``tension'' discussed by
\citet{Anderson:2012sa} between BAO distance measurements and direct
determinations of $H_0$ \citep{Riess:2011yx, Freedman:12}.
Furthermore, if the errors on the BAO measurements are accurate, the
constraints on $\Omm$ and $H_0$ (for fixed $\omega_{\rm m}$ and
$\omega_{\rm b}$) are of comparable accuracy to those from \planck.

The results of this section show that  BAO measurements are an extremely
valuable complementary data set to \planck. The measurements are
basically geometrical and free from complex systematic effects that
plague many other types of astrophysical measurements. The results are
consistent from survey to survey and are of comparable precision to
\Planck. In addition, BAO measurements can be used to break parameter
degeneracies that limit  analyses based purely on CMB data. For example, from
the excellent agreement with the base \LCDM\ model evident in Fig.
\ref{BAO}, we can infer that the combination of \planck\ and BAO
measurements will lead to tight constraints favouring $\Omega_{K}=0$
(Sect. \ref{subsec:params_early}) and a dark energy equation-of-state
parameter, $w=-1$ (Sect. \ref{subsec:darkenergy}).
\referee{Since the BAO measurements are primarily geometrical, they are used in
preference to more complex astrophysical datasets to break CMB parameter
degeneracies in this paper.}

Finally, we note that we choose to use the 6dF+SDSS(R)+BOSS data combination
in the likelihood analysis of Sect.~\ref{sec:grid}.  This choice includes the
two most accurate BAO measurements and, since the effective redshifts
of these samples are widely separated, it should be a very good
approximation to neglect correlations between the surveys.

\subsection{The Hubble constant}
\label{sec:hubble}

\begin{figure}
\begin{center}
\includegraphics[width=80mm,angle=0]{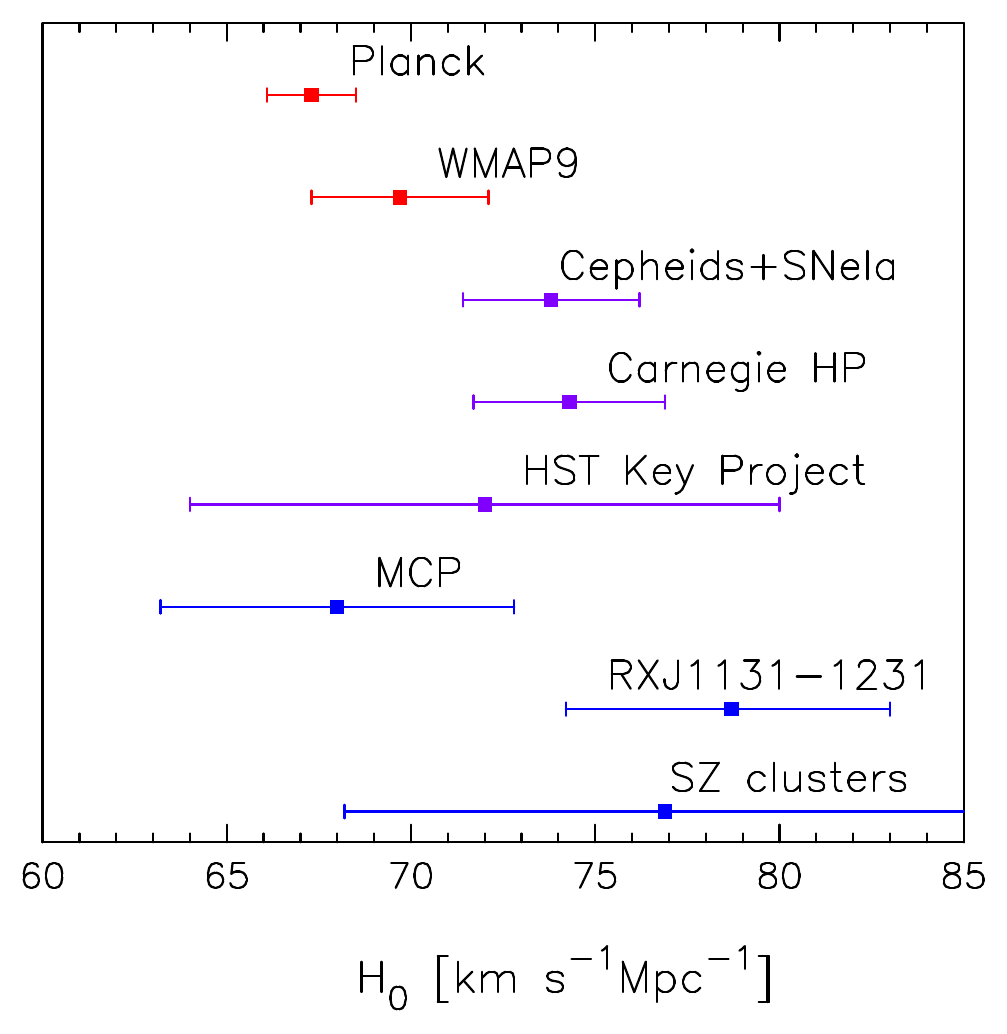}
\end{center}
\caption {Comparison of $H_0$ measurements, with estimates of $\pm 1\,\sigma$
  errors, from a number of techniques (see text
  for details).
  These are compared with the spatially-flat \LCDM\ model constraints
  from \planck\ and \WMAP-9.}
\label{H0}
\vspace{-3mm}
\end{figure}

A  striking result from the fits of the base \LCDM\ model
to \planck\ power spectra is the low value of the Hubble constant,
which is tightly constrained by CMB data alone in this model.
From the \planck+\WP+\highL\  analysis we find
\begin{equation}
H_0 =  (67.3\pm 1.2)  \, {\rm km}\, {\rm s}^{-1}\, {\rm Mpc}^{-1}  \quad \mbox{(68\%; \planck+\WP+\highL)}. \label{H01}
\end{equation}

A low value of $H_0$ has been found in other CMB experiments, most
notably from the recent \textit{WMAP}-9 analysis. Fitting the base \LCDM\ model,
\citet{hinshaw2012} find\footnote{The quoted \WMAP-9 result does not include the $0.06\,{\rm eV}$ neutrino mass of our base \lcdm\ model. Including this mass, we find $H_0 =  (69.7\pm  2.2) \, {\rm km}\,{\rm s}^{-1}\,{\rm Mpc}^{-1} $ from the \WMAP-9 likelihood.}
\begin{equation}
H_0 =  (70.0\pm  2.2) \, {\rm km}\,{\rm s}^{-1}\,{\rm Mpc}^{-1}  \quad
\mbox{(68\%; \ \textit{WMAP}-9}), \label{H01a}
\end{equation}
consistent with Eq.~(\ref{H01}) to within  $1\,\sigma$. We emphasize
here that the CMB estimates are \emph{highly
  model dependent}.  It is important therefore to compare with
astrophysical measurements of $H_0$, since any discrepancies could be a
pointer to new physics.

\begin{figure*}[!]
\begin{centering}
\includegraphics[width=0.4\paperwidth]{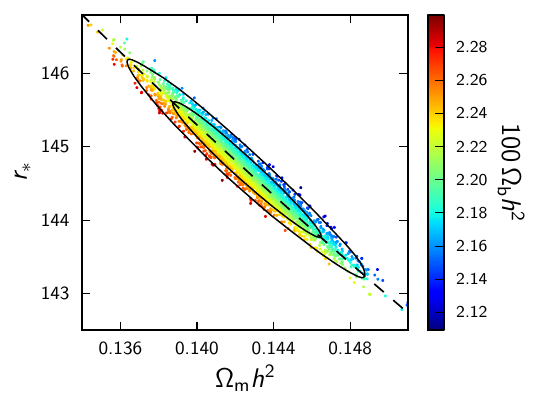}
\includegraphics[width=0.4\paperwidth]{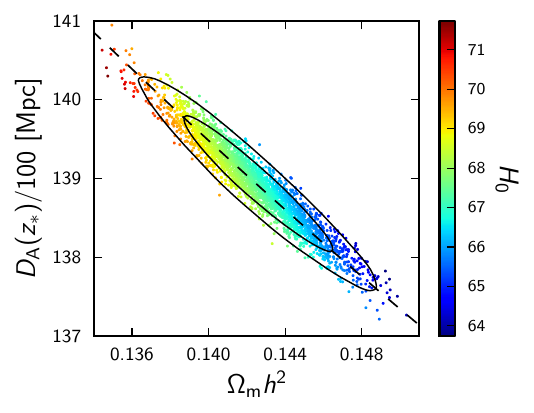}
\par\end{centering}
\caption {MCMC samples and contours in the $r_\ast$--$\Omm h^2$ plane (left) and the $D_{\rm A}(z_\ast)$--$\Omm h^2$ plane (right) for \lcdm\ models analysed with \planck+\WP+\highL. The lines in these
plots show the expected  degeneracy directions in the base \lcdm\ cosmology.
Samples are colour-coded by the values of $\Omb h^2$ (left) and $H_0$ (right).}
\label{rsdafig}
\end{figure*}

There have been remarkable improvements in the precision of the
cosmic distance scale in the last decade or so.  The final results of
the {\it Hubble Space Telescope\/} ({\it HST\/})
Key Project~\citep{Freedman:01}, which used Cepheid
calibrations of secondary distance indicators, resulted in a Hubble
constant of $H_0 = (72 \pm 8) \,{\rm km}\,{\rm s}^{-1}\,{\rm Mpc}^{-1}$
(where the error includes estimates of both $1\,\sigma$ random and systematic
errors). This estimate has been used widely in combination with CMB
observations and other cosmological data sets to constrain cosmological
parameters \citep[e.g.,][]{Spergel:03, spergel2007}.  It
has also been recognized that an accurate
measurement of $H_0$ with around 1\%\ precision, when combined with CMB and other cosmological
data, has the potential to reveal exotic new physics, for example, a
time-varying dark energy equation of state, additional relativistic
particles, or neutrino masses \citep[see e.g.,][and references therein]{Suyu:12b}.  
Establishing a more accurate cosmic distance
scale is, of course, an important problem in its own right. The
possibility of uncovering new fundamental physics provides an
additional incentive.

Two recent analyses have greatly improved the precision of the cosmic
distance scale. \citet{Riess:2011yx} use {\it HST\/}
observations of Cepheid variables in the host galaxies of eight SNe Ia to
calibrate the supernova magnitude-redshift relation.  Their ``best
estimate'' of the Hubble constant, from fitting the calibrated SNe
magnitude-redshift relation,  is
\begin{equation}
H_0 = (73.8 \pm 2.4)  \, {\rm km}\, {\rm s}^{-1}\,{\rm Mpc}^{-1}  \quad \mbox{(Cepheids+SNe Ia)}, \label{H02}
\end{equation}
where the error is $1\,\sigma$ and includes known sources of systematic
errors. At face value, this measurement is discrepant with
the \planck\  estimate in Eq.~(\ref{H01}) at about the $2.5\,\sigma$ level.

\citet{Freedman:12}, as part of the {\it Carnegie Hubble Program},
use {\it Spitzer Space Telescope\/} mid-infrared observations to
recalibrate secondary distance methods used in the {\it HST\/} Key
Project. These authors find
\begin{eqnarray}
H_0 &=&  [74.3 \pm 1.5 \,\, \mbox{(statistical)} \pm 2.1 \,\,
\mbox{(systematic)}] \, {\rm km}\, {\rm s}^{-1}\, {\rm Mpc}^{-1}   \nonumber \\
    & & \hspace{0.2\textwidth} \mbox{(Carnegie HP)}. \label{H02a}
\end{eqnarray}
We have added the two sources of error in quadrature in the
error range shown in Fig.~\ref{H0}. This estimate agrees well with
Eq.~(\ref{H02}) and is also discordant with the \planck\ value
(Eq.~\ref{H0}) at about the $2.5\,\sigma$ level.
The error analysis in Eq.~(\ref{H02a}) does not include a number of known
sources of systematic error and is very likely an underestimate.
For this reason, and because of the relatively good agreement between
Eqs.~(\ref{H02}) and (\ref{H02a}), we do not use the
estimate in Eq.~(\ref{H02a}) in the
likelihood analyses described in Sect.~\ref{sec:grid}.

The dominant source of error in the estimate in Eq.~(\ref{H02}) comes from
the first rung in the distance ladder.  Using the megamaser-based
distance to NGC4258, \citet{Riess:2011yx} find
$(74.8 \pm 3.1) \, {\rm km}\,{\rm s}^{-1}\,{\rm Mpc}^{-1}$.\footnote{\referee{As
noted in Sect.~\ref{sec:introduction},
after the submission of this paper \citet{Humphreys:2013} reported 
a new geometric maser distance to NGC4258 that leads to a reduction of
the \citet{Riess:2011yx} NGC4258 value of $H_0$ from
$(74.8\pm 3.1) \, {\rm km}\,{\rm s}^{-1}\,{\rm Mpc}^{-1}$ to 
$H_0 = (72.0 \pm 3.0) \, {\rm km}\,{\rm s}^{-1}\,{\rm Mpc}^{-1}$.}}
Using parallax measurements for 10
Milky Way Cepheids, they find
$(75.7 \pm 2.6) \, {\rm km}\,{\rm s}^{-1}\,{\rm Mpc}^{-1}$,
and using Cepheid observations and a revised distance to
the Large Magellanic Cloud, they find
$(71.3 \pm 3.8) \, {\rm km}\,{\rm s}^{-1}\,{\rm Mpc}^{-1}$.
These estimates are consistent with each
other, and the combined estimate in Eq.~(\ref{H02}) uses all three
calibrations. The fact that the error budget of measurement
(\ref{H02}) is dominated by the ``first-rung'' calibrators is a point of
concern. A mild underestimate of the distance errors to these
calibrators could eliminate the tension with \Planck.

Figure~\ref{H0} includes three estimates of $H_0$
based on ``geometrical'' methods.\footnote{Note that each of these
estimates is weakly dependent on the assumed background
cosmology.} \referee{The estimate labelled ``MCP'' shows the result $H_0 =
(68.0 \pm 4.8) \, {\rm km}\,{\rm s}^{-1}\,{\rm Mpc}^{-1}$ from the Megamaser Cosmology Project
\citep{Braatz:13} based on observations of megamasers in UGC 3789, NGC 6264 and Mrk 1419
\citep[see also][for a detailed analysis of UGC 3789]{Reid:12}.}
 The point labelled ``RXJ1131-1231'' shows the estimate
$H_0 = 78.7^{+4.3}_{-4.5} \, {\rm km}\,{\rm s}^{-1}\,{\rm Mpc}^{-1}$ derived
from gravitational lensing time delay measurements of
the system RXJ1131-1231, observed as part of the ``COSmological MOnitoring of
GRAvitational Lenses''  (COSMOGRAIL) project (\citealt{Suyu:13},  \referee{see also \citealt{Courbin:11, Tewes:13}}).
Finally, the point labelled SZ clusters shows the value $H_0 =
76.9^{+10.7}_{-8.7} \, {\rm km}\,{\rm s}^{-1}\,{\rm Mpc}^{-1}$
\citep{Bonamente:06}, derived by combining tSZ and X-ray measurements
of rich clusters of galaxies \citep[see][and references therein]{Carlstrom:02}. 
These geometrical methods bypass the need for local
distance calibrators, but each has its own sources of systematic
error that need to be controlled. The geometrical methods are
consistent with the Cepheid-based methods, but at present, the
errors on these methods are quite large. The COSMOGRAIL measurement
(which involved a ``blind'' analysis to prevent experimenter bias)
is discrepant at about $2.5\,\sigma$ with the \planck\ value
in Eq.~(\ref{H01}). We note here a number of other direct measurements 
of $H_0$ \citep{Jones:05,Sandage:06,Oguri:07,Tammann:2013} that give lower values than the measurements summarized in Fig.~\ref{H0}.

The tension between the CMB-based estimates and the astrophysical
measurements of $H_0$ is intriguing and merits further discussion.  In
the base \LCDM\ model, the sound horizon depends primarily on $\Omm h^2$
(with a weaker dependence on $\Omb h^2$). This is illustrated by the
left-hand panel of Fig.~\ref{rsdafig}, which shows samples from the
\planck+\WP+\highL\ MCMC chains in the $r_\ast$--$\Omega_{\rm m}h^2$ plane colour coded according to $\Omega_{\rm b}h^2$. The acoustic scale parameter $\theta_*$ is
tightly constrained by the CMB power spectrum, and so a change in
$r_\ast$ must be matched by a corresponding shift in the angular diameter
distance to the last scattering surface $D_{\rm A}(z_\ast)$. In the base \LCDM\ model,
$D_{\rm A}$ depends on $H_0$ and $\Omega_{\rm m}h^2$, as shown in the right-hand panel
of Fig.~\ref{rsdafig}.  The $2.7 \, {\rm km}\,{\rm s}^{-1}\,{\rm
Mpc}^{-1}$ shift in $H_0$ between \planck\ and \textit{WMAP}-9 is primarily a
consequence of the slightly higher matter density determined by
\planck\ ($\Omega_{\rm m}h^2 = 0.143 \pm 0.003$) compared to \textit{WMAP}-9 
($\Omega_{\rm m}h^2 =
0.136 \pm 0.004$). A shift of around $7 \,{\rm km}\,{\rm s}^{-1}\,{\rm
Mpc}^{-1}$, necessary to match the astrophysical measurements of $H_0$
would require an even larger change in $\Omega_{\rm m}h^2$, which is
disfavoured by the \planck\ data. The tension between \planck\ and the
direct measurements of $H_0$ cannot be easily resolved by varying the
parameters of the base \LCDM\ model. Section \ref{sec:grid} will explore whether
there are any extensions to the base \LCDM\ model that can relieve
this tension. In that section, results
labelled ``$H_0$'' include a Gaussian prior on $H_0$ based on the
\citet{Riess:2011yx} measurement given in Eq.~(\ref{H02}).

\subsection{Type Ia supernovae}
\label{subsec:datasetsSNe}

\begin{figure*}[ht]
\centering
\hspace{-4mm}\includegraphics[width=0.41\paperwidth,angle=0]{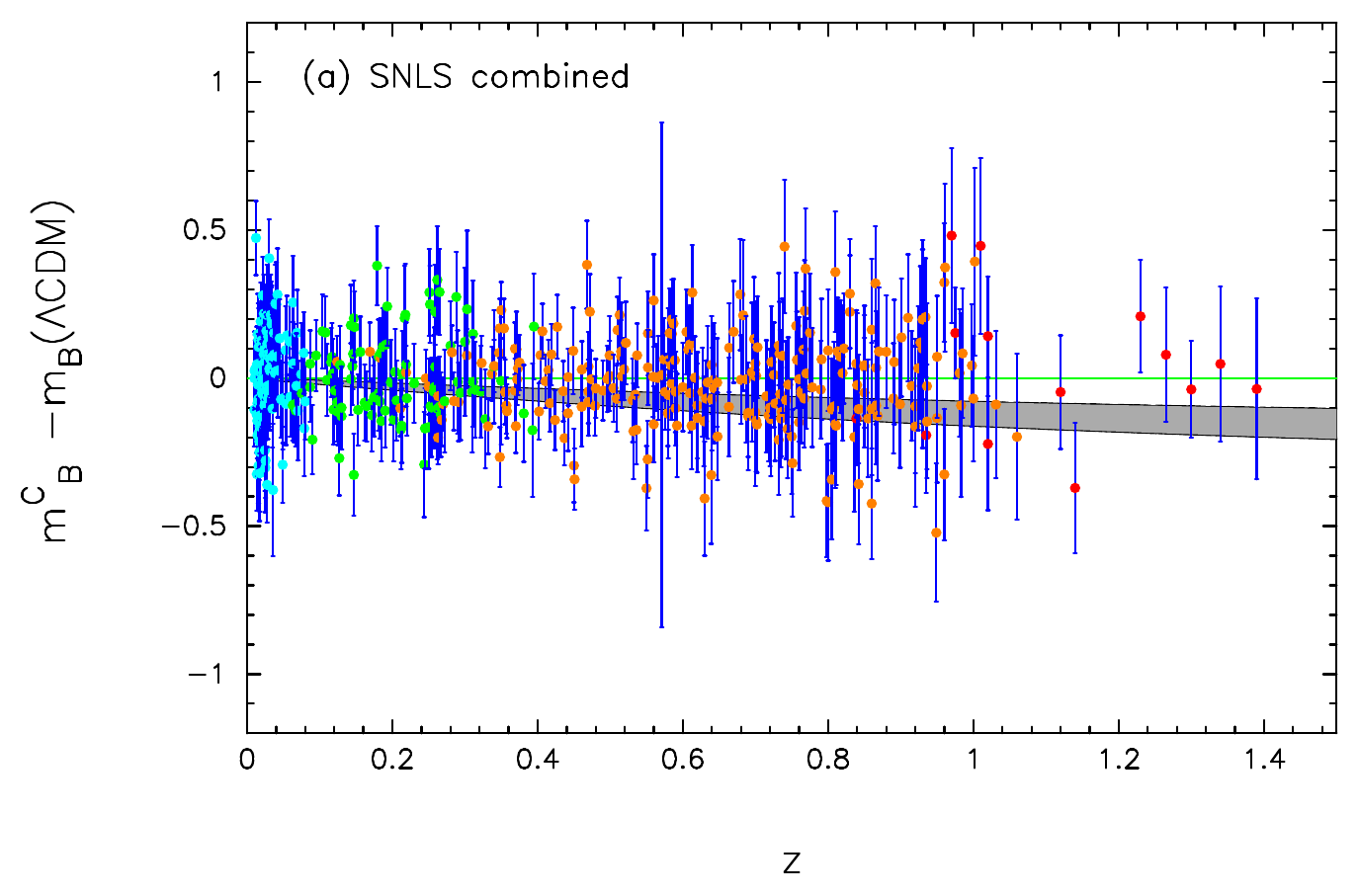}
\hspace{3mm}\includegraphics[width=0.41\paperwidth,angle=0]{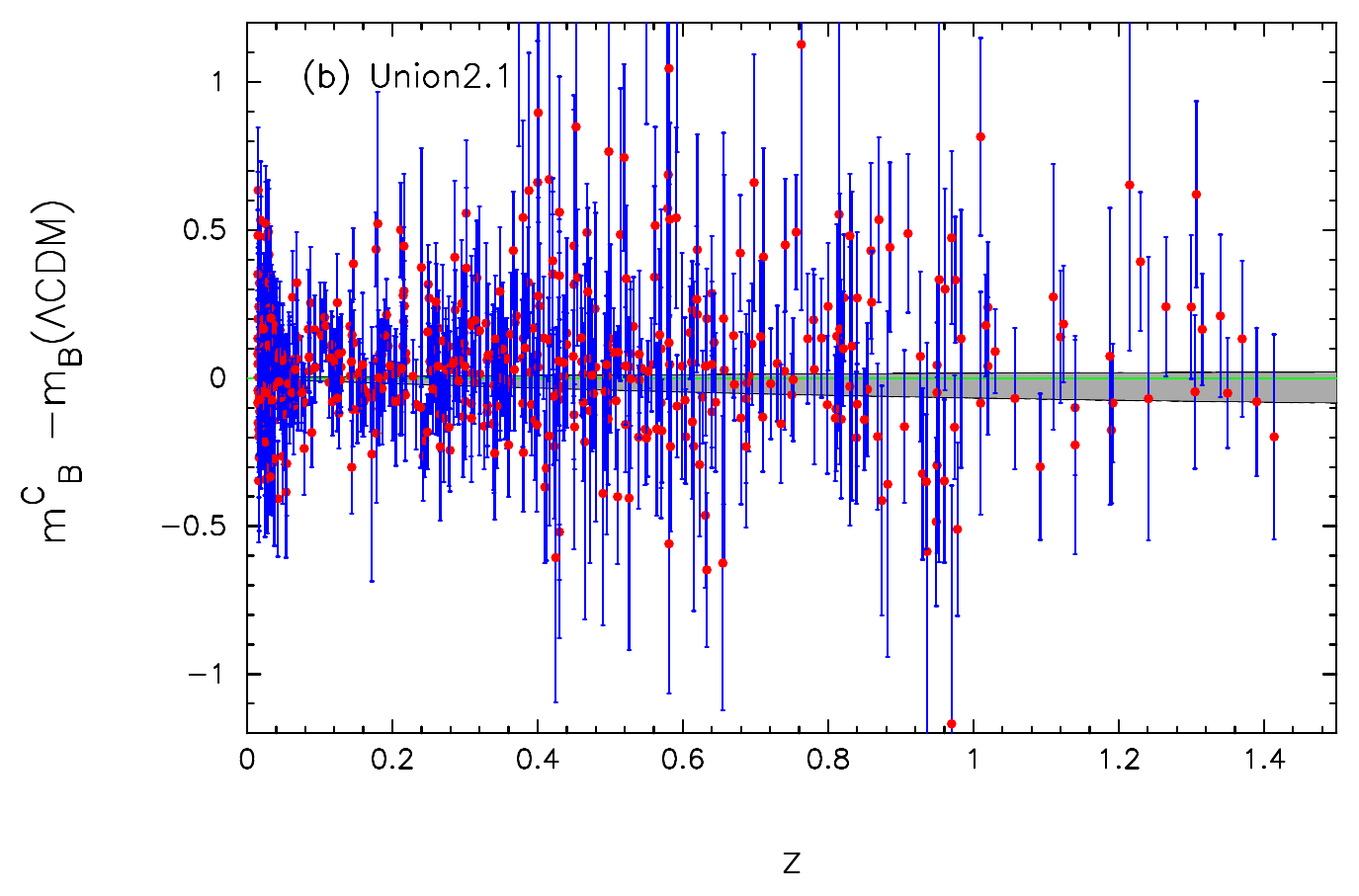}
\caption {Magnitude residuals relative to the base \lcdm\ model that best fits
 the SNLS combined sample (left) and the Union2.1 sample (right).
 The error bars show the $1\,\sigma$ (diagonal) errors on $m_B$.  The
 filled grey regions show the residuals between the expected  magnitudes and
 the best-fit to the SNe sample as $\Omm$ varies across the $\pm 2\,\sigma$
 range allowed by \planck+\WP+\highL\ in the base \lcdm\ cosmology.
 The colour coding of the SNLS samples are as follows: low redshift
 (blue points); SDSS (green points); SNLS three-year sample (orange points);
 and {\it HST\/} high redshift (red points).}
\label{SN}
\end{figure*}

\begin{table*}[tmb]                 
\begingroup
\newdimen\tblskip \tblskip=5pt
\caption{Best-fit parameters for the SNLS compilations.}                          
\label{SNLSTable}         
\nointerlineskip
\vskip -3mm
\footnotesize
\setbox\tablebox=\vbox{
   \newdimen\digitwidth
   \setbox0=\hbox{\rm 0}
   \digitwidth=\wd0
   \catcode`*=\active
   \def*{\kern\digitwidth}
   \newdimen\signwidth
   \setbox0=\hbox{+}
   \signwidth=\wd0
   \catcode`!=\active
   \def!{\kern\signwidth}
\halign{\hbox to 2.0in{#\leaderfil}\tabskip 2.2em&
        \hfil#\hfil&
        \hfil#\hfil&
        \hfil#\hfil&
        \hfil#\hfil&
        \hfil#\hfil&
        \hfil#\hfil&
        \hfil#\hfil\tabskip=0pt\cr                
\noalign{\doubleline}
\omit\hfil Data set\hfil&$N_{\rm SNe}$&$M^1_B$&$M^2_B$&$\alpha$&$\beta$&$\Omega_{\rm m}$ &$\chi^2$\cr
\noalign{\vskip 4pt\hrule\vskip 6pt}
SNLS combined&472&$-19.16$&$-19.21$&1.425&3.256&0.227&407.8\cr
SNLS SiFTO&468&$-19.15$&$-19.20$&1.352&3.375&0.223&414.9\cr
SNLS SALT2&473&$-19.15$&$-19.20$&1.698&3.289&0.247&376.7\cr
SNLS combined (CMB $\Omega_{\rm m}$)&472&$-19.12$&$-19.18$&1.417&3.244&0.317&412.5\cr
SNLS SiFTO (CMB $\Omega_{\rm m}$)&468&$-19.12$&$-19.18$&1.339&3.351&0.317&420.1\cr
SNLS SALT2 (CMB $\Omega_{\rm m}$)&473&$-19.12$&$-19.18$&1.691&3.302&0.317&378.9\cr
\noalign{\vskip 3pt\hrule\vskip 4pt}}}
\endPlancktablewide                 
\endgroup
\end{table*}                        

In this subsection, we analyse two SNe Ia samples: the sample of $473$
SNe as reprocessed by \citet{Conley:11}, which we will refer to as the
``SNLS'' compilation; and the updated Union2.1 compilation of 580 SNe 
described  by \citet{Suzuki:12}.

\subsubsection{The SNLS compilation}

The SNLS ``combined'' compilation consists of $123$ SNe Ia at low
redshifts, $242$ SNe Ia from the three-year Supernova Legacy Survey
(SNLS; see \citealt{Regnault:09,Guy:10,Conley:11}),
93 intermediate redshift SNe Ia from the Sloan Digital Sky Survey
(SDSS; \citealt{Holtzman:08,Kessler:09}) and 14 objects at
high redshift observed with the {\it Hubble Space Telescope\/} ({\it HST\/};
\citealt{Riess:07}).

The ``combined'' sample of \citet{Conley:11} combines the results of two
light-curve fitting codes, SiFTO~\citep{Conley:08} and SALT2~\citep{Guy:07}, to produce a peak apparent $B$-band magnitude, $m_B$,
stretch parameter $s$ and colour ${\cal C}$ for each supernova. To
explore the impact of light-curve fitting, we also analyse separately the
SiFTO and SALT2 parameters. The SiFTO and SALT2 samples differ by a few
SNe from the combined sample because of colour and stretch
constraints imposed on the samples. We also use ancillary data, such as
estimates of the stellar masses of the host galaxies and associated
covariance matrices, as reported by
\citet{Conley:11}\footnote{\url{https://tspace.library.utoronto.ca/handle/1807/25390}. We use the module supplied with \COSMOMC.}.

In this section, we focus exclusively on the base \lcdm\ model
(i.e., $w=-1$ and $\Omega_K=0$). For a flat Universe, the expected apparent magnitudes are then given by
\begin{equation}
m^{\Lambda{\rm CDM}}_{B} = 5 {\rm log}_{10} \hat D_{\rm L}(z_{\rm hel}, z_{\rm CMB}, \Omm) - \alpha (s-1) + \beta{\cal C} + 
{\cal M}_B, \label{SN1}
\end{equation}
where $\hat{D}_{\rm L}$ is the dimensionless luminosity distance\footnote{Note
  that the luminosity distance depends on both the heliocentric,
  $z_{\mathrm{hel}}$, and CMB frame, $z_{\mathrm{CMB}}$, redshifts of the SNe. This distinction is
  important for low-redshift objects.} and ${\cal M}_{B}$ absorbs the
Hubble constant. As in \citet{Sullivan:11}, we express values
of the parameter(s) ${\cal M}_{B}$ in terms of an effective absolute
magnitude
\begin{equation}
M_{B} = {\cal M}_{B} - 5 {\rm log}_{10} \left  ({c \over H_0} \right  ) - 25, \label{SN2}
\end{equation}
for a value of $H_0 = 70 \, {\rm km}\,{\rm s}^{-1}\, {\rm Mpc}^{-1}$.

The likelihood for this sample is then constructed as in \citet{Conley:11} and
\citet{Sullivan:11}:
\begin{equation}
  \chi_{\rm SNe}^2  =   (\vM_{B} - \vM^{\Lambda{\rm CDM}}_{B})^T 
\tens{C}_{\rm SNe}^{-1} (\vM_{B} - \vM^{\Lambda{\rm CDM}}_{B}),  
 \label{SN3a} 
\end{equation}
where $\vM_B$ is the vector of effective absolute magnitudes and $\tens{C}_{\rm SNe}$
is the sum of the non-sparse covariance matrices of
\citet{Conley:11} quantifying statistical and systematic errors.  As in
\citet{Sullivan:11}, we divide the sample according to the
estimated stellar mass of the host galaxy and solve for two parameters,
$M^1_{B}$ for $M_{\rm host} < 10^{10}{\rm M}_\odot$ and $M^2_{B}$ for
$M_{\rm host} \ge 10^{10}{\rm M}_\odot$. We adopt the estimates of the
``intrinsic'' scatter in $m_{B}$ for each SNe sample given in Table 4 of
\citet{Conley:11}.

Fits to the SNLS combined sample are shown in the left-hand panel of
Fig.~\ref{SN}. The best-fit parameters for the combined, SiFTO and
SALT2 samples are given in Table~\ref{SNLSTable}. In the base \lcdm\
model, the SNe data provide a constraint on $\Omm$, independent of the
CMB. As can be seen from Table \ref{SNLSTable} (and also in the
analyses of \citealt{Conley:11} and \citealt{Sullivan:11}), the SNLS
combined compilation favours a lower value of $\Omm$ than we find from
the CMB. The key question, of course, is whether the SNe data are
statistically compatible with the Planck data. The last three rows of
Table~\ref{SNLSTable} give the best-fit SNe parameters constraining
$\Omm$ to the \planck+\WP+\highL\ best-fit value $\Omm=0.317$. The grey
bands in Fig. \ref{SN} show the magnitude residuals expected for a
$\pm 2\,\sigma$ variation in the value of $\Omm$ allowed by the CMB
data.  The CMB band lies systematically low by about $0.1$ magnitude
over most of the redshift range shown in Fig. \ref{SN}a.

Table \ref{SNLSTable} also lists the $\chi^2$ values for the
$\Omm=0.317$ fits.\footnote{We caution the reader that, generally, the $\chi_{\rm SNe}^2$ obtained from Eq.~(\ref{SN3a}) will differ from that quoted in the online parameter tables in cases where the SNLS data is importance sampled. For importance sampling, we modified the SNLS likelihood to marginalize numerically over the $\alpha$ and $\beta$ parameters.}  The likelihood ratio for the SiFTO fits is
\begin{equation}
{{\cal L}_{\rm SNe} \over {\cal L}_{\rm SNe+CMB \; \Omm}} = {\rm exp} \left ( {1 \over 2} (\chi^2_{\rm SNe} - 
\chi^2_{\rm SNe+CMB \; \Omm}) \right ) \approx 0.074. \label{SN4}
\end{equation}
This is almost a $2\,\sigma$ discrepancy. (The discrepancy would appear
to be much more significant if only the diagonal statistical errors
were included in the covariance matrix in Eq.~\ref{SN3a}). The
likelihood ratio for the combined sample is slightly larger ($0.095$)
and is larger still for the SALT2 sample ($0.33$). In summary, there
is some tension between the SNLS compilations and the base \lcdm\
value of $\Omm$ derived from \planck. The degree of tension
depends on the light-curve fitter and is stronger for the SiFTO and
combined SNLS compilations.\footnote{\referee{As noted in Sect.~\ref{sec:introduction},
recent revisions to the photometric calibrations between the SDSS and SNLS 
observations relieve some of the tensions discussed in this paper
between the SNe data and the \planck\ base \LCDM\ cosmology.}}

\subsubsection{The Union2.1 compilation}

The Union2.1 compilation \citep{Suzuki:12} is the latest application
of a scheme  for combining multiple SNe data sets
described by \citet{Kowalski:08}.  The Union2.1 compilation contains
19 data sets and includes early high-redshift SNe data \citep[e.g.,][]
{Riess:98, Perlmutter:99} as well as recent data from
the {\it HST\/} Cluster Supernova Survey \citep{Amanullah:10,
Suzuki:12}. The SNLS and Union2.1 compilations contain
$256$  SNe in common and are therefore not independent.

The SALT2 model \citep{Guy:07} is used to fit the light curves
returning a $B$-band magnitude at maximum light, a light-curve shape
parameter and a colour correction. \referee{(Note that the version of SALT2 used in the Union2.1 analysis is
not exactly the same as that used in the SNLS analysis.)} As in Eq.~(\ref{SN1}), the
theoretically-predicted magnitudes include nuisance parameters
$\alpha$ and $\beta$ multiplying the shape and colour corrections, and
an additional nuisance parameter $\delta$ describing the variation of
SNe luminosity with host galaxy mass \citep[see Eq.~3 of][]
{Suzuki:12}.  The \COSMOMC\ module associated with the Union2.1
sample\footnote{\url{http://supernova.lbl.gov/Union}.} holds the nuisance
parameters fixed ($\alpha=0.1218$, $\beta=2.4657$, and $\delta=-0.03634$)
and computes a $\chi^2$ via Eq.~(\ref{SN3a}) using a fixed
covariance matrix that includes a model for systematic errors.  An
analysis of the base \lcdm\  model then requires minimization with
respect to only two parameters, $\Omm$ and ${\cal M}_{B}$ (or
equivalently, $M_{B}$).

Maximizing the Union2.1 likelihood, we find best-fit parameters of
$\Omm=0.296$ and $M_B=-19.272$ (defined as in Eq.~\ref{SN2} for
a value of $H_0 = 70 \, {\rm km}\,{\rm s}^{-1} \,{\rm Mpc}^{-1}$) and
$\chi^2_{\rm Union2.1} = 545.11$ (580 SNe).  The magnitude residuals
with respect to this fit are shown in the right-hand panel of Fig.
\ref{SN}. Notice that the scatter in this plot is significantly larger
than the scatter of the SNLS compilation (left-hand panel)
reflecting the more diverse range of data and the lower precision of
some of the earlier SNe data used in the Union2.1 compilation.
Nevertheless, the Union2.1 best-fit is close to (and clearly
compatible with) the \Planck\ base \lcdm\ value of $\Omm$.

\subsubsection{SNe: Summary}

\begin{figure}
\begin{center}
\includegraphics[angle=0,width=8.8cm]{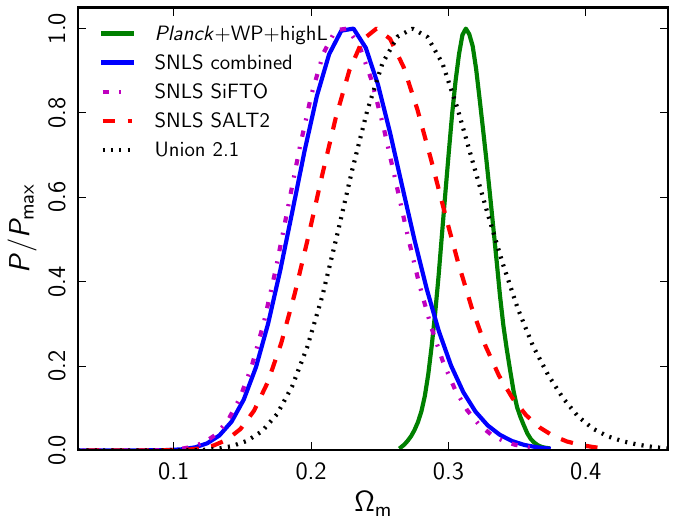}
\end{center}
\caption {Posterior distributions for $\Omega_{\rm m}$ (assuming a flat
  cosmology) for the SNe compilations described in the text. The
  posterior distribution for $\Omega_{\rm m}$ from the \planck+\WP+\highL\
  fits to the base \lcdm\ model is shown by the solid green line.}
\label{SNposteriors}
\end{figure}

The results of this subsection are summarized in Fig.~\ref{SNposteriors}. This shows the posterior distributions for $\Omm$ in the base
\lcdm\ cosmology, marginalized over nuisance parameters, for each of the
SNe samples. These distributions are broad (with the Union2.1
distribution somewhat broader than the SNLS distributions) and show
substantial overlap. There is no obvious inconsistency between the SNe
samples. The posterior distribution for $\Omm$ in the base \lcdm\
model fit to \planck+\WP+\highL\ is shown by the narrow green curve. This is consistent with
 the Union2.1 and SNLS SALT2 results, but is in some
tension with the distributions from the SNLS combined and SNLS SiFTO
samples. As we will see in Sect. \ref{sec:grid}, \planck\ combined with
\planck\ lensing and BAO measurements overwhelm SNe data for most of
the extensions of the \lcdm\ model considered in this paper. 
 However, the results presented here
suggest that there could  be residual systematic errors in the SNe
data that are not properly accounted for in the covariance matrices.
Hints of new physics based on combining CMB and SNe data should 
therefore be treated with caution.

\subsection{Additional data}
\label{subsec:additional}

In this subsection we review a number of other astrophysical
data sets that have sometimes been combined with CMB data.
These data sets are not used with \planck\ in this paper, either
because they are statistically less powerful than the data reviewed
in previous subsections and/or they involve complex physics (such as
the behaviour of intra-cluster gas  in rich clusters of galaxies) which is not
yet well understood. 

\subsubsection{Shape information on the galaxy/matter power spectrum}

\begin{figure}
\begin{center}
\includegraphics[width=86mm,angle=0]{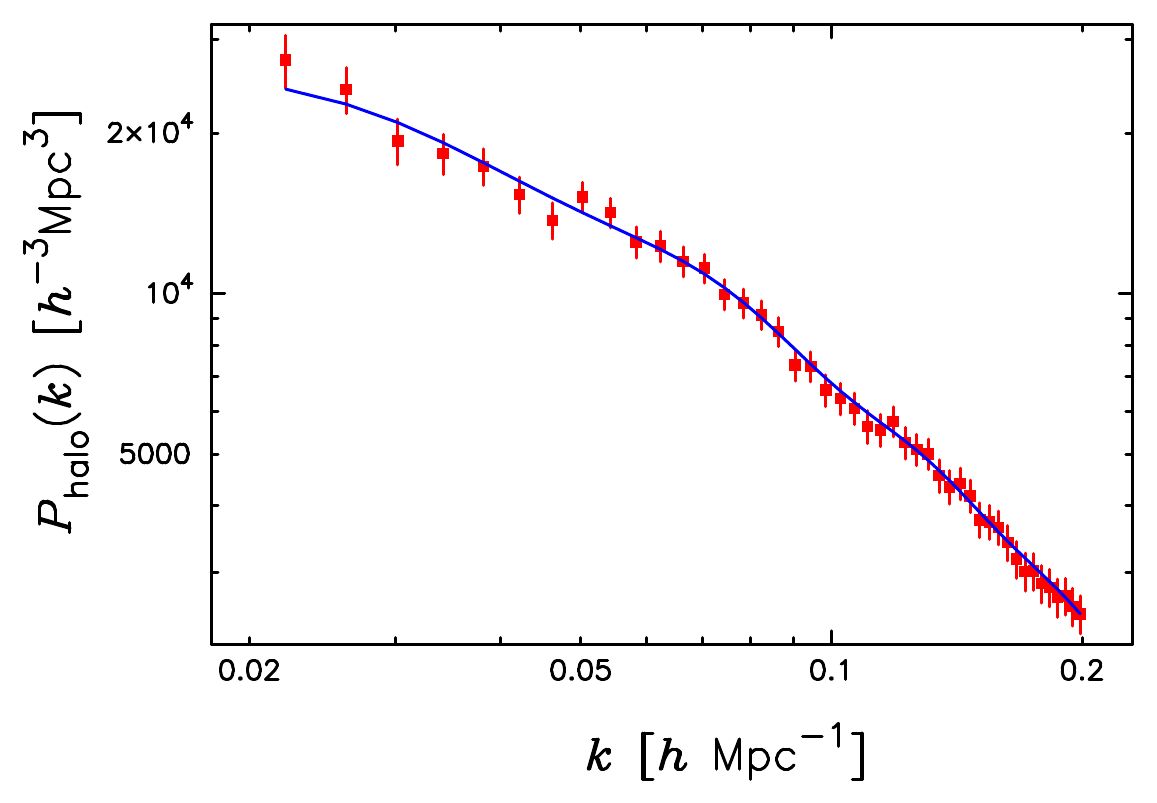}
\end{center}
\caption {Band-power estimates of the halo power spectrum, $P_{\rm halo}(k)$,
from \citet{Reid:10} together with $1\,\sigma$ errors. (Note that these
data points are strongly correlated.) The line shows the predicted
spectrum  for the best-fit \planck+\WP+\highL\ base \lcdm\ parameters.}
\label{LRG}
\end{figure}

\citet{Reid:10} present an estimate of the dark matter halo power spectrum, $P_{\rm halo}(k)$, derived from $110{,}756$ luminous red galaxies (LRGs) from
the SDSS 7th data release \citep{SDSS_DR7}. The sample extends to
redshifts $z \approx 0.5$, and is processed to identify LRGs occupying the
same dark matter halo, reducing the impact of redshift-space distortions
and recovering an approximation to the halo density field. The power spectrum
$P_{\rm halo}(k)$ is reported in $45$ bands, covering the wavenumber
range $0.02\, h\,{\rm Mpc}^{-1} < k < 0.2\,h\,{\rm Mpc}^{-1}$. The window functions, covariance
matrix and \COSMOMC\ likelihood module are available on the 
{\it NASA} LAMBDA web site\footnote{\url{http://lambda.gsfc.nasa.gov/toolbox/lrgdr}.}.

The halo power spectrum is plotted in Fig. \ref{LRG}. The blue line shows
the predicted halo power spectrum from our best-fit base \lcdm\ parameters
convolved with the \citet{Reid:10} window functions. Here we show the
predicted halo power spectrum for the best-fit values of the ``nuisance''
parameters $b_0$ (halo bias), $a_1$, and $a_2$ (defined in Eq.~15
of \citealt{Reid:10}) which relate the halo power spectrum to the dark matter
power spectrum (computed using \CAMB). \referee{The \planck\ model gives
 $\chi^2_{\rm LRG}=40.4$, very close to the value $\chi^2_{\rm LRG}=40.0$
of the best-fit  model of
\citet{Reid:10}.}

Figure~\ref{LRG} shows that the \planck\ parameters provide a good match to
the shape of the halo power spectrum. However, we do not use these data
(in this form) in conjunction with \planck. The BAO scale derived from these
and other data is used with \planck, as summarized in Sect.~\ref{sec:BAO}. As discussed
by \citet[][see their figure~5]{Reid:10} there is little
additional information on cosmology once the BAO features are filtered from the
spectrum, and hence little to be gained by adding this information
to \planck. The corrections for nonlinear evolution, though small
in the wavenumber range $0.1$--$0.2\,h\,{\rm Mpc}^{-1}$, add to the 
complexity of using shape information from the halo power spectrum.

\subsubsection{Cosmic shear}
\label{subsec:cosmicshear}

{Another key cosmological observable is the distortion of distant galaxy images
by the gravitational lensing of large-scale structure, often called cosmic
shear.  The shear probes the (non-linear) matter density projected along the
line of sight with a broad kernel.
It is thus sensitive to the geometry of the Universe and the growth of
large-scale structure, with a strong sensitivity to the amplitude of the
matter power spectrum.

\referee{The most recent, and largest, cosmic shear data sets are provided by the
CFHTLenS survey \citep{Heymans:12,Erben:12}, which
covers\footnote{Approximately 61\% of the survey is fit for cosmic shear
science.} $154\,{\rm deg}^2$ in five optical bands with accurate shear
measurements and photometric redshifts.
The CFHTLenS team has released several cosmic shear results that are relevant
to this paper.  \citet{Benjamin:13} present results from a two-bin tomographic
analysis and \citet{Heymans:13} from a finely binned tomographic analysis.
\citet{Kilbinger:13} present constraints from a 2D analysis.
The constraints from all of the analyses show a high degree of consistency.

\citet{Heymans:13} estimate shear correlation functions associated with six
redshift bins.  Assuming a flat, $\Lambda$CDM model, from the weak lensing
data alone they find
$\sigma_8\left(\Omega_{\rm m}/0.27\right)^{0.46\pm0.02}=0.774\pm0.04$
(68\% errors)
which is consistent with the constraint found by \citet{Benjamin:13}.
For comparison, we find
\begin{eqnarray}
&&\hspace{-0.015\textwidth}
\sigma_8\left(\Omega_{\rm m}/0.27\right)^{0.46} = 0.89 \pm 0.03
\quad \mbox{(68\%; \planck+\WP+\highL)}, \nonumber \\
&& \hspace{0.2\textwidth}
\end{eqnarray}
which is discrepant at about the $2\,\sigma$ level.
Combining the tomographic lensing data with CMB constraints from \WMAP-7,
\citet{Heymans:13} are able to constrain the individual parameters of the flat,
$\Lambda$CDM model to be $\Omega_{\rm m}=0.255\pm0.014$ and $h=0.717\pm0.016$.
The best-fit \planck\ value of $\Omega_{\rm m}$ is $4\,\sigma$ away from this
value, while $h$ is discrepant at nearly $3\,\sigma$.
As might be expected, given the good agreement between the \planck\ and BAO
distance scales, the best-fit CFHTLenS $\Lambda$CDM cosmology is also
discrepant with the BOSS data, predicting a distance ratio to $z=0.57$ which
is 5\% lower than measured by BOSS \citep{Anderson:2012sa}.
This is discrepant at approximately the  $3\,\sigma$ level, comparable to the
discrepancy with the \planck\ values.
The source of the discrepancies between \planck\ and the CFHTLenS tomographic
analyses is at present unclear, and further work will be needed to resolve them.

\citet{Kilbinger:13} give a tight constraint in the $\sigma_8$--$\Omega_{\rm m}$
plane for flat $\Lambda$CDM models from their 2D (i.e., non-tomographic)
analysis.
They find $\sigma_8\left(\Omega_{\rm m}/0.27\right)^{0.6}=0.79\pm0.03$, which,
when combined with \WMAP-7, gives
$\Omega_{\rm m}=0.283\pm 0.010$ and $h=0.69\pm 0.01$.
These results are still discrepant with the \planck\ best-fit, but with
lower significance than the results reported by \citet{Heymans:13}.

It is also worth noting that a recent analysis of galaxy-galaxy lensing in the
SDSS survey \citep{Mandelbaum:13} leads to the constraint
$\sigma_8\left(\Omega_{\rm m}/0.25\right)^{0.57}=0.80\pm0.05$ for the base
\LCDM\ cosmology. This is about $2.4\,\sigma$ lower than expected from \Planck. 

}

\subsubsection{Counts of rich clusters}

For the base \lcdm\ model we find $\sigma_8=0.828\pm0.012$ from
\planck+\WP+\highL. This value
is in excellent agreement with the \textit{WMAP}-9 value of $\sigma_8 = 0.821
\pm 0.023$ \citep{hinshaw2012}.  There are other ways to probe the
power spectrum normalization, in addition to the cosmic shear
measurements discussed above. For example, the abundances of rich clusters of
galaxies are particularly sensitive to the normalization \citep[see
  e.g.,][]{Komatsu:02}.  Recently, a number of studies have used
tSZ-cluster mass scaling relations to constrain combinations of
$\sigma_8$ and $\Omm$ \citep[e.g.,][]{Benson:13, Reichardt:13,
  Hasselfield:13} including an analysis of a sample of \planck\ tSZ
clusters \citep[see][]{planck2013-p05,planck2013-p05a}
 reported in this series of papers \citep{planck2013-p15}\footnote{There is additionally a study of the statistical properties of
the \Planck-derived Compton-$y$ map \citep{planck2013-p05b} from which
other parameter estimates can be obtained.}.

The \planck\ analysis uses a relation between cluster mass and tSZ
signal based on comparisons with X-ray mass measurements.
\referee{To take into account departures from hydrostatic equilibrium,
X-ray temperature calibration, modelling of the selection function,
uncertainties in scaling relations and analysis uncertainties,
\citet{planck2013-p15} assume a ``bias'' between the X-ray derived
masses and the true cluster  masses.
If the mass bias, $(1-b)$, is allowed to vary uniformly between $0.7$
and $1.0$, \citet{planck2013-p15} find
$\sigma_8(\Omm/0.27)^{0.3} = 0.76 \pm 0.03$ for the base \lcdm\ model.
In comparison, for the same model we find
\begin{eqnarray}
&&\hspace{-0.015\textwidth}
\sigma_8\left(\Omega_{\rm m}/0.27\right)^{0.3} = 0.87 \pm 0.02
\quad \mbox{(68\%; \planck+\WP+\highL)}, \nonumber \\
&& \hspace{0.2\textwidth} \label{SZclus1}
\end{eqnarray}
which is a significant (around $3\,\sigma$) discrepancy that remains
unexplained.
}
Qualitatively similar results are found from analyses of SPT clusters
[$\sigma_8(\Omm/0.27)^{0.3} = 0.77 \pm 0.04$]. Key difficulties with
this type of measurement, as discussed in \citet{planck2013-p15}, 
include adequately modelling selection biases
and calibrating cluster masses. These effects are discussed  in the
analysis of ACT clusters by \citet{Hasselfield:13}, who adopt
a number of approaches, including folding in dynamical mass
measurements, to calibrate  biases in clusters mass estimates.  Some of these
approaches give joint $\sigma_8$--$\Omm$ constraints consistent
with the base \lcdm\ parameters reported here.

 At this stage of our understanding of the biases and scatter in the
 cluster mass calibrations, we believe that for the purposes of this
 paper it is premature to use cluster counts together with CMB
 measurements to search for new physics. \citet{planck2013-p15}
 explore a number of possibilities for reducing the tension between
 \planck\  CMB measurements and tSZ cluster counts, including non-zero
neutrino masses.

\section{Extensions to the base  \LCDM\ model}

\label{sec:grid}

\subsection{ Grid of models}

\begin{figure*}
\centering
\includegraphics[width=17cm]{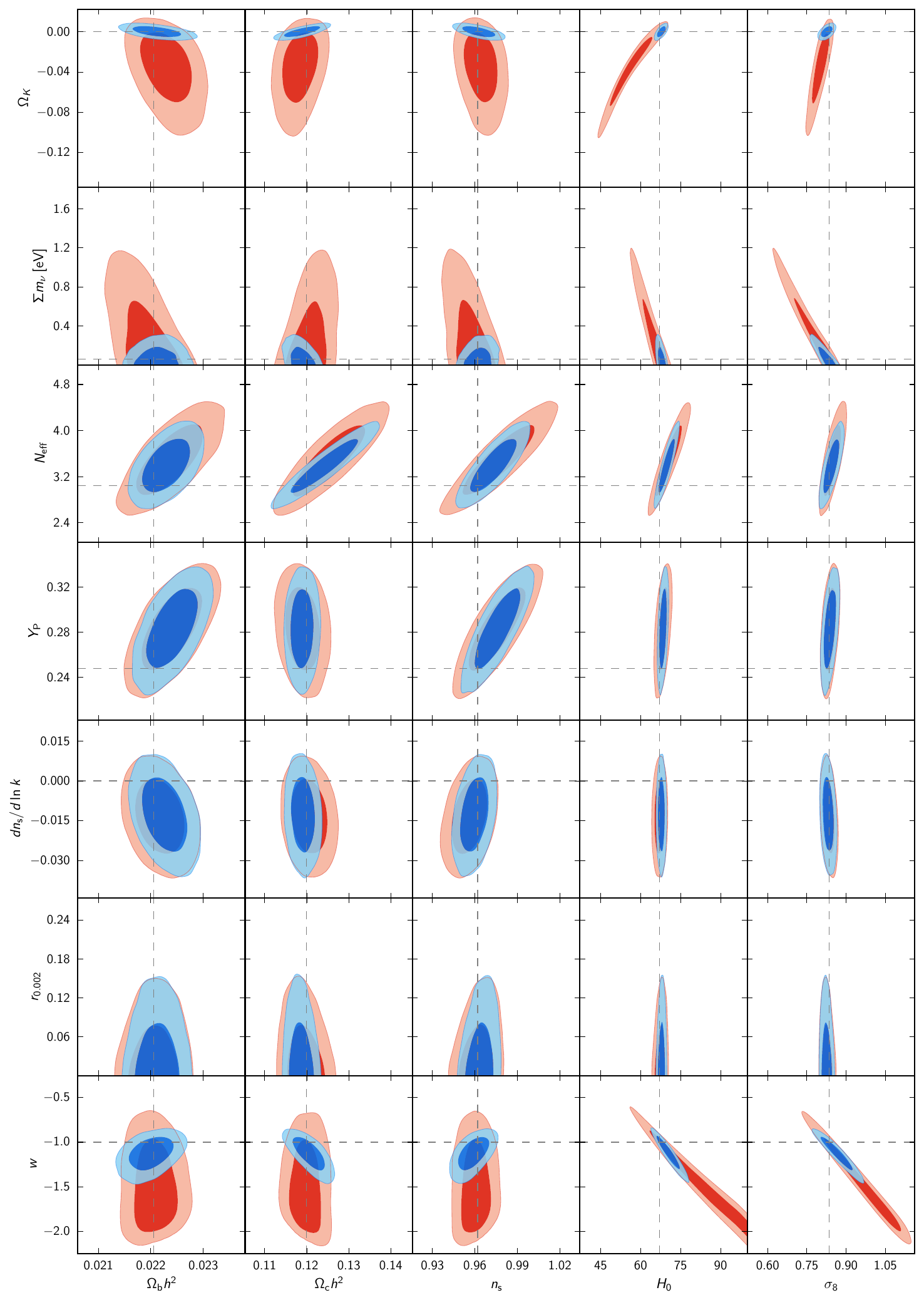}
\caption {68\% and 95\% confidence regions on one-parameter extensions of
the base \LCDM\ model for \planck+\WP\ (red) and \planck+\WP+BAO (blue).
Horizontal dashed lines correspond to the fixed base model parameter value, and
vertical dashed lines show the mean posterior value in the base model for \planck+\WP.}
\label{fig:grid_1paramext}
\end{figure*}



\begin{table*}
\begin{center}
\caption{Constraints on one-parameter extensions to the base \lcdm\ model. Data combinations all include
\Planck\ combined with \WMAP\ polarization, and results are shown for combinations with high-$\ell$ CMB data and BAO. Note that we quote 95\% limits here.
\label{tab:grid_1paramext}
}
\begingroup
\newdimen\tblskip \tblskip=5pt
\nointerlineskip
\vskip -3mm
\footnotesize
\setbox\tablebox=\vbox{
    \newdimen\digitwidth
    \setbox0=\hbox{\rm 0}
    \digitwidth=\wd0
    \catcode`"=\active
    \def"{\kern\digitwidth}
    \newdimen\signwidth
    \setbox0=\hbox{+}
    \signwidth=\wd0
    \catcode`!=\active
    \def!{\kern\signwidth}
\halign{
\hbox to 0.9in{$#$\leaderfil}\tabskip=1.5em&\hfil$#$\hfil\tabskip=0.5em&
\hfil$#$\hfil\tabskip=1.7em&
\hfil$#$\hfil\tabskip=0.5em&
\hfil$#$\hfil\tabskip=1.7em&
\hfil$#$\hfil\tabskip=0.5em&
\hfil$#$\hfil\tabskip=1.7em&
\hfil$#$\hfil\tabskip=0.5em&
\hfil$#$\hfil\tabskip=0pt\cr
\noalign{\doubleline}
 \multispan1\hfil \hfil& \multispan2\hfil \Planck+\WP\hfil& \multispan2\hfil \Planck+\WP+BAO\hfil& \multispan2\hfil \Planck+\WP+\highL\hfil& \multispan2\hfil \Planck+\WP+\HighL+BAO\hfil\cr
\noalign{\vskip -3pt}
\omit&\multispan2\hrulefill&\multispan2\hrulefill&\multispan2\hrulefill&\multispan2\hrulefill\cr
 \omit\hfil Parameter\hfil & \omit\hfil Best fit\hfil & \omit\hfil 95\% limits\hfil & \omit\hfil Best fit\hfil & \omit\hfil 95\% limits\hfil & \omit\hfil Best fit\hfil & \omit\hfil 95\% limits\hfil & \omit\hfil Best fit\hfil & \omit\hfil 95\% limits\hfil\cr
\noalign{\vskip 3pt\hrule\vskip 5pt}
%
\Omega_K & -0.0326 & -0.037^{+0.043}_{-0.049} & 0.0006 & 0.0000^{+0.0066}_{-0.0067} & -0.0389 & -0.042^{+0.043}_{-0.048} & -0.0003 & -0.0005^{+0.0065}_{-0.0066}\cr
\noalign{\vskip 3pt}
\Sigma m_\nu\,[\mathrm{eV}] & 0.002 & < 0.933 & 0.000 & < 0.247 & 0.000 & < 0.663 & 0.001 & < 0.230\cr
\noalign{\vskip 3pt}
N_{\mathrm{eff}} & 3.25 & 3.51^{+0.80}_{-0.74} & 3.32 & 3.40^{+0.59}_{-0.57} & 3.38 & 3.36^{+0.68}_{-0.64} & 3.33 & 3.30^{+0.54}_{-0.51}\cr
\noalign{\vskip 3pt}
Y_{\mathrm{P}} & 0.2896 & 0.283^{+0.045}_{-0.048} & 0.2889 & 0.283^{+0.043}_{-0.045} & 0.2652 & 0.266^{+0.040}_{-0.042} & 0.2701 & 0.267^{+0.038}_{-0.040}\cr
\noalign{\vskip 3pt}
dn_{\mathrm{s}}/d\ln k & -0.0125 & -0.013^{+0.018}_{-0.018} & -0.0097 & -0.013^{+0.018}_{-0.018} & -0.0146 & -0.015^{+0.017}_{-0.017} & -0.0143 & -0.014^{+0.016}_{-0.017}\cr
\noalign{\vskip 3pt}
r_{0.002} & 0.000 & < 0.120 & 0.000 & < 0.122 & 0.000 & < 0.108 & 0.000 & < 0.111\cr
\noalign{\vskip 3pt}
w & -1.94 & -1.49^{+0.65}_{-0.57} & -1.106 & -1.13^{+0.24}_{-0.25} & -1.94 & -1.51^{+0.62}_{-0.53} & -1.113 & -1.13^{+0.23}_{-0.25}\cr
\noalign{\vskip 5pt\hrule\vskip 3pt}
} 
} 
\endPlancktable
\endgroup
\end{center}
\end{table*}

To explore possible deviations from \LCDM\ we have analysed an
extensive grid of models that covers many well-motivated extensions of
\LCDM. As in the exploration of  the base \LCDM\ cosmology,  we have also considered a variety of data
combinations for each model. For models involving more than one
additional parameter we restrict ourselves to \planck+\WP\ combinations
in order to obtain tighter constraints by leveraging the relative
amplitude of the power spectrum at very low $\ell$ and high $\ell$.  Most
models are run with \planck, \planck+\WP, and \planck+\WP+\highL;
additionally all are
importance sampled with \planck\ lensing (Sect.~\ref{sec:lensing}),
BAO (Sect.~\ref{sec:BAO}),
SNe (Sect.~\ref{subsec:datasetsSNe}), and the~\citet{Riess:2011yx} direct $H_0$ measurement (Sect.~\ref{sec:hubble}). For models where the non-CMB data
give a large reduction in parameter volume (e.g. $\Omega_K$ models),
we run separate chains instead of importance sampling.

These runs provide no compelling evidence for deviations from the
base \lcdm\ model, and indeed, as shown in Table~\ref{tab:grid_1paramext}
and Fig.~\ref{fig:grid_1paramext}, the posteriors for individual extra
parameters generally overlap the fiducial model within one standard deviation.
The inclusion of BAO data shrinks further the allowed scope for
deviation. The parameters of the base  \LCDM\ model are relatively
robust to inclusion of additional parameters, but the errors on some do broaden
significantly when additional degeneracies open up, as can be seen in 
Fig.~\ref{fig:grid_1paramext}

The full grid results are available online\footnote{\url{http://www.sciops.esa.int/index.php?project=planck&page=Planck_Legacy_Archive}}.
Here we summarize some of the key results, and also consider a few additional extensions.

\subsection{Early-Universe physics}
\label{subsec:params_early}

Inflationary cosmology offers elegant explanations of key features of
our Universe, such as its large size and near spatially flat
geometry. Within this scenario, the Universe underwent a brief period
of accelerated expansion
\citep{starobinsky:1979,starobinsky:1982,kazanas:1980,guth:1981,sato:1981,linde:1982,albrecht/steinhardt:1982}
during which quantum fluctuations were inflated in scale to become the
classical fluctuations that we see today. In the simplest inflationary
models, the primordial fluctuations are predicted to be adiabatic,
nearly scale-invariant and Gaussian
\citep{mukhanov/chibisov:1981,hawking:1982,starobinsky:1982,guth/pi:1982,bardeen/steinhardt/turner:1983},
in good agreement with CMB observations and other probes of
large-scale structure.

Despite this success, the fundamental physics behind inflation is not
yet understood and there is no convincing evidence that rules out
alternative scenarios for the early Universe. A large number of
phenomenological models of inflation, some inspired by string theory,
have been discussed in the literature \citep[see][for
reviews]{liddle/lyth:CIALSS,bassett/tsujikawa/wands:2006,linde:2008},
as well as alternatives to inflation including  pre-big bang
scenarios
\citep[e.g.,][]{Gasperini:93,khoury/etal:2001,boyle/steinhardt/turok:2004,creminelli/senatore:2007,
Brandenberger:12}.
Many of these models lead to distinctive signatures, such as
departures from Gaussianity, isocurvature perturbations, or oscillatory
features in the power spectrum, that are potentially observable. The
detection of such signatures would offer valuable information on the
physics of the early Universe and is one of the main science goals of
\planck.

In this section we discuss basic aspects of the primordial power
spectrum, such as the spectral index, departures from a pure
power law, limits on tensor modes  etc., and discuss the
implications for inflationary cosmology.  Tests of more complex
models, such as multi-field inflation, are discussed in a separate
paper \citep{planck2013-p17}. In \citet{planck2013-p09a}, the \planck\ maps are used to
constrain possible deviations from Gaussianity via measurements of the
bispectrum and trispectrum. \citet{planck2013-p09} considers departures from
statistical isotropy and additional tests of non-Gaussianity.

\subsubsection{Scale dependence of primordial fluctuations}
\label{subsubsec:index}

The primordial fluctuations in the base \LCDM\ model are parameterized as a
pure power law with a spectral index $\ns$ (Eq.~\ref{PS1}).
Prior to  \planck, CMB observations have favoured a power
law index with slope $\ns<1$, which is expected in simple single-field
slow-roll inflationary models (see e.g.,
 \citealt{Mukhanov:07} and Eq.~\ref{GE1a} below). The final \WMAP\ nine-year data give
$\ns=0.972\pm0.013$ at 68\% confidence
\citep{hinshaw2012}. Combining this with damping-tail
measurements from ACT and SPT data gives $\ns=0.968 \pm 0.009$,
indicating a departure from scale invariance at the 3$\,\sigma$
level. The addition of BAO data has resulted in a stronger preference
for $\ns<1$
\citep{Anderson:2012sa,hinshaw2012,Story:12,Sievers:13}.
These
constraints assume the basic six-parameter \LCDM\ cosmological model. Any new physics that affects
the damping tail of the CMB spectrum, such as additional relativistic particles,
can alter these constraints substantially and still allow a precisely scale-invariant
spectrum.

With \planck, a robust detection of the deviation from scale invariance
can now be made from {\it  a single set of CMB observations spanning three decades in scale  from $\ell=2$ to $\ell=2500$}. We find
\be
\ns=0.959\pm0.007 \quad\mbox{(68\%; \planck+\WP+\highL)}, \label{GE0}
\ee
for the base \LCDM\ model, a roughly $6\,\sigma$ departure from scale invariance. This
is consistent with the results from previous CMB experiments cited above. The statistical
significance of this result is high enough that the difference between a purely scale
invariant spectrum can be seen easily in a
plot of the power spectrum. Figure \ref{fig:degen} shows the \planck\ spectrum
of Fig. \ref{Planckbestfitcl} plotted as $\ell^2 {\cal D_\ell}$ compared to
the base \lcdm\ fit with $n_{\rm s}=0.96$ (red dashed line) and to the 
best-fit base \lcdm\ cosmology with $n_{\rm s}=1$. The $\ns=1$ model has more power
at small scales and is strongly excluded by the \planck\ data.

\begin{figure}
\centering
\includegraphics[width=88mm]{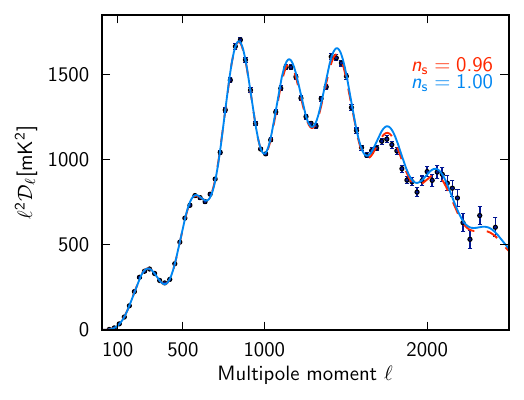}

\caption {The \planck\ power spectrum of Fig.~\ref{Planckbestfitcl} plotted as
$\ell^2 {\cal D_\ell}$ against multipole,  compared to the best-fit base
\lcdm\ model with $n_{\rm s}=0.96$ (red dashed line). The best-fit base
\lcdm\ model with $n_{\rm s}$ constrained
to unity is shown by the blue line.}
\label{fig:degen}
\end{figure}

{\it The unique contribution of \planck, compared to previous experiments,
is that we are able to show that the departure from scale invariance is
robust to changes in the underlying theoretical model.}
For example, Figs. \ref{fig:grid_1paramext} and  \ref{fig:ns_corr}
show that the departure from scale invariance is not sensitive
to the parameterization of the primordial
fluctuations. Even if we allow a
 possible running of the spectral index (the parameter $\nrun$
defined in equation  \ref{PS1}) and/or  a component of tensor
fluctuations, the \planck\ data favour a tilted spectrum at
a high significance level.

Our extensive grid of models allows us to investigate correlations of
the spectral index with a number of cosmological parameters beyond
those of the base \lcdm\ model (see Figs. \ref{fig:grid_1paramext} and
\ref{fig:ns_neff}). As expected, $\ns$ is uncorrelated with parameters
describing late-time physics, including the neutrino mass, geometry,
and the equation of state of dark energy.  The remaining correlations
are with parameters that affect the evolution of the early Universe,
including the number of relativistic species, or the helium
fraction. This is illustrated in Fig. \ref{fig:ns_neff}: modifying
the standard model by increasing the number of neutrinos species, or
the helium fraction, has the effect of damping the small-scale power
spectrum.   This can be partially
compensated by an increase in the spectral index. However, an increase
in the neutrino species must be accompanied by an increased matter density
to maintain the peak positions. A measurement of the matter density
from the BAO measurements helps to  break this degeneracy. This is
clearly seen in the upper panel of Fig. \ref{fig:ns_neff}, which shows
the improvement in the constraints when BAO measurements are added to
the \planck+\WP+\highL\ likelihood. With the addition of BAO measurements
we find more than a 3$\,\sigma$ deviation from $\ns=1$ even in this extended
model, with a best-fit value of $\ns=0.969\pm0.010$ for varying
relativistic species. As discussed in Sect.  \ref{sec:neutrino},
we see no evidence from the \planck\ data for non-standard neutrino physics.

\begin{figure}
\centering
\includegraphics[width=88mm]{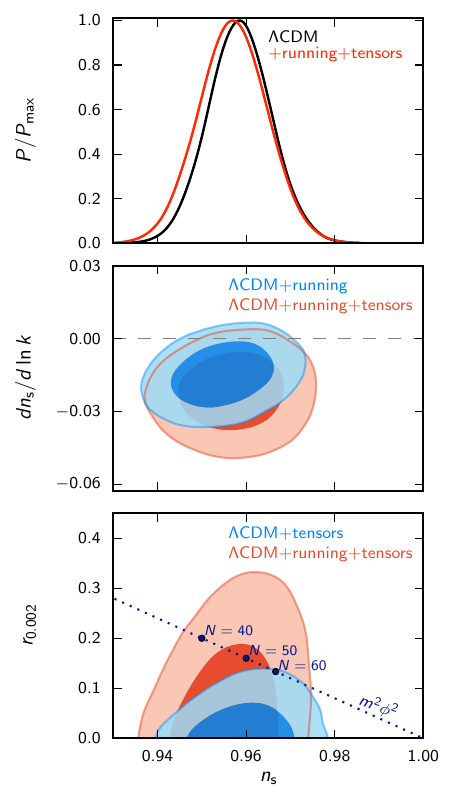}
\caption {\textit{Upper}: Posterior distribution for
  $n_{\rm s}$ for the base \lcdm\ model (black) compared to the
  posterior when a tensor component and running scalar spectral
  index are added to the model (red)
  \textit{Middle}: Constraints (68\% and 95\%) in the $\ns$--$\nrun$ plane for \lcdm\ models with running (blue) and additionally with tensors (red).
  \textit{Lower}: Constraints (68\% and 95\%) on $\ns$ and the tensor-to-scalar ratio $\rzerotwo$ for \lcdm\ models with tensors (blue) and additionally with running of the spectral index (red). The dotted line
  show the expected relation between $r$ and $n_{\rm s}$ for a $V(\phi)
  \propto \phi^2$ inflationary potential (Eqs.~\ref{GE1a} and
  \ref{GE1b}); here $N$ is the number of inflationary e-foldings as
  defined in the text. The dotted line should be compared to the blue 
contours, since this model predicts negligible running.  All of these results use the
  \planck+\WP+\highL\ data combination.}

\label{fig:ns_corr}
\end{figure}

\begin{figure}
\centering
\includegraphics[width=88mm]{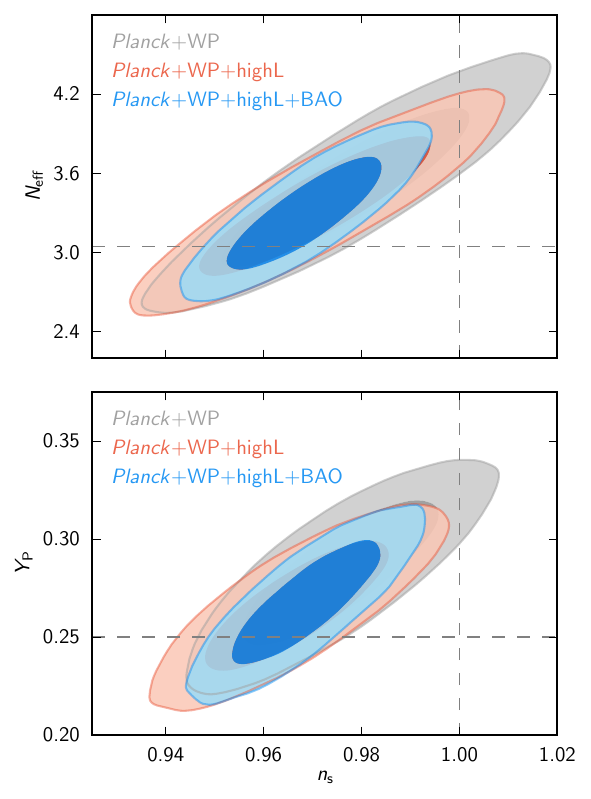}
\caption {Constraints on $n_{\rm s}$ for \lcdm\ models with non-standard
relativistic species, $\neff$, (upper) and helium fraction, $\yhe$,
(lower). We show 68\% and 95\% contours for various data combinations. Note the tightening of the constraints with the addition of BAO data.
}
\label{fig:ns_neff}
\end{figure}

The simplest single-field inflationary models predict that the running
of the spectral index should be of second order in inflationary
slow-roll parameters and therefore small [$\nrun \sim (\ns-1)^2$],
typically about an order of magnitude below the sensitivity limit of
\planck\ \citep[see e.g.,][]{kosowsky/turner:1995,baumann/etal:2009}.
Nevertheless, it is easy to construct inflationary models that have
a larger scale dependence (e.g.,  by adjusting the third derivative
of the inflaton potential) and so it is instructive to use the \planck\
data to constrain $\nrun$. A test for $\nrun$  is of particularly interest
given the results from previous CMB experiments.

Early results from \WMAP\ suggested a preference for a negative running
at the 1--2$\,\sigma$ level. In the final 9-year \WMAP\ analysis no
significant running was seen using \WMAP\ data alone, with
$\nrun=-0.019\pm0.025$ (68\% confidence; \citealt{hinshaw2012}. Combining
\WMAP\ data with the first data releases from ACT and SPT,
\citet{hinshaw2012} found a negative running at nearly the $2\,\sigma$
level with $\nrun=-0.022\pm0.012$ (see also
\citealt{dunkley11} and \citealt{keisler11} for analysis of ACT and
SPT with earlier data from \WMAP). The ACT 3-year release, which
incorporated a new region of sky, gave $\nrun=-0.003\pm0.013$
\citep{Sievers:13} when combined with \WMAP\ 7 year data.  With the wide
field SPT data at 150\,GHz, a negative running was seen at just over
the $2\,\sigma$ level, $\nrun = -0.024\pm0.011$ \citep{hou12}.

The
picture from previous CMB experiments is therefore mixed. The latest
\WMAP\ data show a $1\,\sigma$ trend for a running, but when combined with
the S12  SPT data, this trend is amplified to give a
potentially interesting result. The latest ACT data go in the other
direction, giving no support for a running spectral index when
combined with \WMAP\footnote{The differences between the \planck\
results reported here and the \WMAP-7+SPT results \citep{hou12} are
discussed in Appendix~\ref{app:spt}.}.

The results from \planck\ data are as follows
(see Figs.~\ref{fig:grid_1paramext} and \ref{fig:ns_corr}):
\beglet
\ba
\nrun &=& \hspace{-1mm} -0.013\pm0.009 \; \mbox{(68\%;
  \Planck+\WP)}; \label{nrunplanck} \\ \nrun &=& \hspace{-1mm}
-0.015\pm0.009 \; \mbox{(68\%;
  \Planck+\WP+\HighL)}; \label{nrunplanckhighL} \\
\nrun
&=& \hspace{-1mm} -0.011\pm0.008 \; \mbox{(68\%; \Planck+lensing}
\nonumber \\ & & \hspace{41mm}
\mbox{+\WP+\HighL)}. \label{nrunplanckhighL+lensing}
\ea
\endlet
The consistency between (\ref{nrunplanck}) and (\ref{nrunplanckhighL})
shows that these results are insensitive to modelling of unresolved
foregrounds. The preferred solutions have a small negative running,
but not at a high level of statistical significance. Closer inspection
of the best-fits shows that the change in $\chi^2$ when $\nrun$
is included as a parameter comes almost entirely from the low
multipole temperature likelihood. (The fits to the high
multipole \planck\ likelihood have a $\Delta \chi^2 = -0.4$ when
$\nrun$ is included.) The slight preference for a negative running is
therefore driven by the spectrum at low multipoles $\ell \la 50$.
The tendency for negative running is partly mitigated by including
the \planck\ lensing likelihood (Eq. \ref{nrunplanckhighL+lensing}).

The constraints on $\nrun$ are broadly
similar if tensor fluctuations are allowed in addition to a running of the
spectrum (Fig.~\ref{fig:ns_corr}) . Adding tensor fluctuations, the marginalized posterior distributions
for $\nrun$ give
\beglet
\ba
\nrun &=& \hspace{-1mm}-0.021\pm0.011 \;\mbox{(68\%; \Planck+\WP)}, \\
\nrun &=& \hspace{-1mm}-0.022\pm0.010 \;\mbox{(68\%;  \Planck+\WP+\HighL)}, \\
\nrun
&=& \hspace{-1mm} -0.019\pm0.010 \; \mbox{(68\%; \Planck+lensing}
\nonumber \\ & & \hspace{41mm}
\mbox{+\WP+\HighL)}.
\ea
\endlet
As with Eqs.~(\ref{nrunplanck})--(\ref{nrunplanckhighL+lensing}) the tendency to
favour negative running is driven by the low multipole 
component of the temperature likelihood {\it not by the \planck\ spectrum
at high multipoles}.

This is one of several examples discussed in this section where marginal
evidence for extensions to the base \lcdm\ model are favoured by the
$TT$ spectrum at low multipoles. (The low multipole spectrum is also
largely responsible for the pull of the lensing amplitude,  $A_{\rm L}$,
to values greater than unity discussed in Sect. \ref{sec:lensing}).
The mismatch between the best-fit base \lcdm\ model and the
$TT$ spectrum at multipoles $\ell \la 30$ is clearly visible
in Fig. \ref{TTspec}. The implications of this mismatch are
discussed further in Sect. \ref{sec:conclusions}.

 Beyond a simple running, various extended parameterizations have been
 developed by e.g., \citet{Bridle:03}, \citet{shafieloo/souradeep:2008},
   \citet{verde/peiris:2008}, and \citet{hlozek/etal:2012},
 to test for deviations from a
 power-law spectrum of fluctuations. Similar techniques are applied to
 the \planck\ data in ~\citet{planck2013-p17}.

\subsubsection{Tensor fluctuations}
\label{subsubsec:tensors}

In the base \LCDM\ model, the fluctuations are assumed to be
purely scalar modes. Primordial
tensor fluctuations could also contribute to the temperature and
polarization power spectra
\citep[e.g.,][]{grishchuk:1975,starobinsky:1979,basko/polnarev:1980,Crittenden:93, Crittenden:95}.
The most direct way of testing for a tensor contribution is to search
for a magnetic-type parity signature via a large-scale $B$-mode pattern
in CMB polarization \citep{Seljak:97,Zaldarriaga:97,Kamionkowski:97}.
Direct $B$-mode measurements are challenging as the expected signal is
small; upper limits measured by BICEP and QUIET give $95\%$ upper limits
of $r_{0.002}<0.73$ and $r_{0.002}<2.8$ respectively
\citep{chiang/etal:2010,araujo/etal:2012}\footnote{As discussed in \citet{planck2013-p02} and \citet{planck2013-p03}, residual
low-level polarization systematics in both the LFI and HFI data preclude
a \planck\ $B$-mode polarization analysis at this stage.}.

Measurements of the temperature power spectrum can also be used to
constrain the amplitude of tensor modes. Although such limits can
appear to be much tighter than the limits from $B$-mode measurements,
it should be borne in mind that such limits are
indirect because they are derived within the context of a
particular theoretical model. In the rest of this subsection, we will
review temperature based limits on tensor modes and then present the
results from \planck.

Adding a tensor component to the base \LCDM\ model, the \WMAP\ 9-year
results constrain $r_{0.002}<0.38$ at 95\% confidence
\citep{hinshaw2012}.  Including small-scale ACT and SPT data this
improves to $r_{0.002} <0.17$, and to $r_{0.002}<0.12$ with the
addition of BAO data. These limits are degraded substantially,
however, in models which allow running of the scalar spectral index in
addition to tensors. For such models, the \WMAP\ data give
$r_{0.002}<0.50$, and this limit is not significantly improved by
adding high resolution CMB and BAO data.

The precise determination of the fourth, fifth and sixth acoustic
peaks by \planck\ now largely breaks the degeneracy between the primordial
fluctuation parameters. For the \planck+\WP+\highL\ likelihood we find
\beglet
 \ba
r_{0.002}&<&0.11 \quad \mbox{(95\%;  no running)},   \label{rnorun}\\
r_{0.002}&<&0.26 \quad \mbox{(95\%; including running)}.  \label{rrun}
\ea
\endlet
As shown in Figs. \ref{fig:grid_1paramext} and  \ref{fig:ns_corr}, the tensor amplitude is weakly
correlated with the scalar spectral index; an increase in $\ns$  that
could match the first three peaks cannot fit the fourth and higher
acoustic peak in the \planck\ spectrum. Likewise,
the shape constraints from the fourth and higher acoustic
peaks give a reduction in the correlations between a tensor mode and a
running in the spectral index, leading to significantly tighter limits than from
previous CMB experiments.  These numbers in Eqs.~(\ref{rnorun}) and (\ref{rrun})
are driven by the temperature spectrum and change very little if we add non-CMB data
such as BAO measurements. The \planck\ limits are largely decoupled from
assumptions about the late-time evolution of the Universe and are close
to the tightest possible limits achievable from the temperature power spectrum
alone  \citep{Knox:94,knox:1995}.

These limits on a tensor mode have profound implications for
inflationary cosmology.  The limits translate directly to an upper
limit on the energy scale of inflation,
\be V_* = (1.94 \times 10^{16}
\ {\rm GeV})^4 (r_{0.002}/0.12) \ee
\citep{linde:1983,Lyth:84}, and to
the parameters of ``large-field'' inflation models.
Slow-roll inflation driven by a power law potential $V(\phi) \propto
\phi^\alpha$ offers a simple example of large-field inflation. The
field values in such a model must necessarily exceed the Planck scale
$m_{\rm Pl}$ , and lead to a scalar spectral index and tensor amplitude of 
\beglet
\ba
1-\ns & \approx & (\alpha+2)/2N,  \label{GE1a}\\
r &\approx & 4\alpha/N, \label{GE1b} \ea
\endlet
where $N$ is the number of e-foldings between the end of inflation and the
time that our present day Hubble scale crossed the inflationary horizon
\citep[see e.g.,][]{lyth/riotto:1999}. The 95\% confidence limits from the
\planck\ data
are now close to the predictions of  $\alpha=2$ models for $N\approx50$--60
e-folds (see Fig.~\ref{fig:ns_corr}). Large-field
models with quartic potentials \citep[e.g.,][]{linde:1982} are now firmly excluded by
CMB data. \planck\ constraints on power-law and on broader classes of inflationary models are discussed
in detail in \citet{planck2013-p09a}. Improved limits on $B$-modes will be required to further constrain high field
models of inflation.

\begin{figure*}[t]
\centering
\includegraphics[width=9cm]{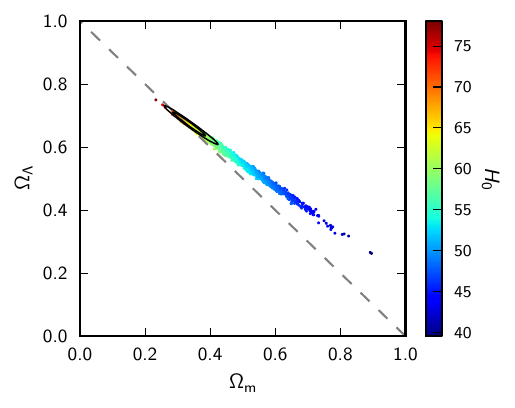}
\includegraphics[width=9cm]{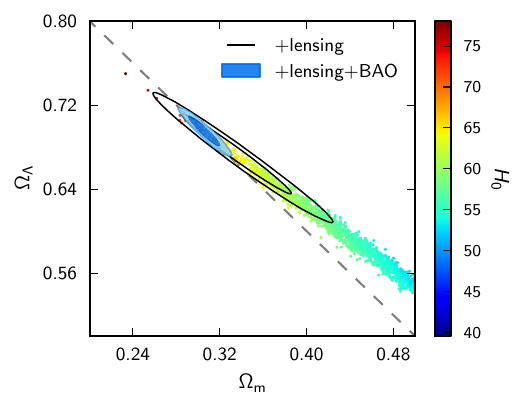}
\caption {The \planck+\WP+\highL\ data combination (samples; colour-coded by the value of $H_0$) partially breaks the geometric degeneracy between $\Omega_{\rm m}$ and $\Omega_{\Lambda}$ due to the effect of lensing in the temperature power spectrum.
These limits are significantly improved by the inclusion of the \Planck\ lensing reconstruction (black contours). Combining also with BAO (right; solid blue contours) tightly constrains the geometry to be nearly flat.}
\label{fig:geom}
\end{figure*}

\subsubsection{Curvature}
\label{subsubsec:curv}

An explanation of the near flatness of our observed Universe
was one of the primary  motivations for inflationary cosmology.
Inflationary models that allow  a large number of
e-foldings predict that our Universe should be very accurately
spatially flat\footnote{The effective curvature within our Hubble
  radius should then be of the order of the amplitude of the curvature
  fluctuations generated during inflation, $\Omk\sim
  O(10^{-5})$.}. Nevertheless, by introducing fine tunings it is possible
to construct inflation models with observationally interesting open
geometries \citep[e.g.,][]{Gott:82,Linde:95,Bucher:95, Linde:99} or
closed geometries \citep{Linde:03}. Even more
speculatively, there has been interest in models with open geometries
from considerations of tunnelling events between metastable vacua
within a ``string landscape'' \citep{Frievogel:06}. Observational
limits on spatial curvature therefore offer important additional
constraints on inflationary models and fundamental physics.

CMB temperature power spectrum measurements suffer from a well-known
``geometrical degeneracy'' \citep{Bond:97, Zaldarriaga_cosmo:97}.
Models with identical primordial spectra, physical  matter densities
and angular diameter distance to the last scattering surface, will have
almost identical CMB temperature power spectra. This is a near perfect
degeneracy (see Fig. \ref{fig:geom}) and is broken only via the
integrated Sachs-Wolfe (ISW)  effect on large angular scales and
gravitational lensing of the CMB spectrum \citep{Stompor:99}.
The geometrical degeneracy can also be
broken with the addition of probes of late time
physics, including BAO, Type
Ia supernova, and measurement of the Hubble constant
\citep[e.g.,][]{spergel2007}.

Recently, the detection of the gravitational lensing of the CMB by ACT
and SPT has been used to break the geometrical degeneracy, by measuring the
integrated matter potential distribution. ACT constrained
$\Omega_\Lambda=0.61\pm0.29$ (68\% CL)  in \citet{sherwin/etal:2011}, with the
updated analysis in \citet{2013arXiv1301.1037D} giving
$\Omk=-0.031\pm0.026$ (68\% CL) \citep{Sievers:13}.
The SPT lensing measurements combined with seven year \WMAP\ temperature
spectrum improved this
limit to $\Omk=-0.0014\pm0.017$ (68 \% CL) \citep{vanEngelen:2012va}.

With \planck\ we detect gravitational lensing at about $26\,\sigma$ through the 4-point function
\citep[Sect. \ref{sec:lensing} and][]{planck2013-p12}. This strong detection
of gravitational lensing allows us to
constrain the curvature to percent level precision using observations of the
CMB alone:
\beglet
\ba
100\Omk &=& -4.2^{+4.3}_{-4.8}  \quad\mbox{(95\%; \planck+\WP+\highL}); \label{GE5a}\\
100\Omk &=& -1.0^{+1.8}_{-1.9} \quad\mbox{(95\%; \planck+lensing} \nonumber \\
            &  & \hspace{44mm} \mbox{+ \WP+\highL)}. \label{GE5b}
\ea
\endlet

These constraints are improved substantially  by the addition of BAO data. We then
find
\beglet
\ba
100\Omk &=& -0.05^{+0.65}_{-0.66}  \ \ \mbox{(95\%; \planck+\WP+\highL+BAO}), \label{GE6a}\\
100\Omk &=& -0.10^{+0.62}_{-0.65}  \quad\mbox{(95\%; \planck+lensing+\WP} \nonumber \\
            &  & \qquad \qquad\qquad\qquad\qquad\quad\quad\mbox{+\highL+BAO)}. \label{GE6b}
\ea
\endlet
These limits are consistent with (and slightly tighter than)  the results reported by \citet{hinshaw2012}
from combining the nine-year \WMAP\ data with high resolution CMB measurements  and BAO data. We find broadly similar
results to Eqs.~(\ref{GE6a}) and (\ref{GE6b}) if the \citet{Riess:2011yx} $H_0$ measurement, or either of the SNe
compilations discussed in Sect. \ref{subsec:datasetsSNe},  are used in place of the BAO measurements.

In summary, there is no evidence from \planck\ for any departure from a spatially flat
geometry. The results of Eqs.~(\ref{GE6a}) and (\ref{GE6b}) suggest that our Universe is
spatially flat to an accuracy of better than a percent.

\subsection{Neutrino physics and constraints on relativistic components}
\label{sec:neutrino}

A striking illustration of the interplay between cosmology and particle
physics is the potential of CMB observations to constrain the properties of
relic neutrinos, and possibly of additional light relic particles in the
Universe~(see e.g., \citealt{Dodelson:1995es,Hu:1995fqa,bashinsky04,Ichikawa:2004zi,Lesgourgues:2006nd,Hannestad:2010kz}).
In the following subsections, we present \Planck\ constraints on the mass of
ordinary (active) neutrinos assuming no extra relics, on the density of light
relics assuming they all have negligible masses, and finally on models with
both light massive and massless relics.

\subsubsection{Constraints on the total mass of active neutrinos}

The detection of solar and atmospheric neutrino oscillations proves that
neutrinos are massive, with at least two species being non-relativistic today.
The measurement of the absolute neutrino mass scale is a challenge for both
experimental particle physics and observational cosmology. The combination of
CMB, large-scale structure and distance measurements already excludes a large
range of masses compared to beta-decay experiments. Current limits on the
total neutrino mass $\sumnu$ (summed over the three neutrino families) from
cosmology are rather model dependent and vary strongly with the data
combination adopted. The tightest constraints for flat models with three
families of neutrinos are typically around $0.3\,\mathrm{eV}$
(95\% CL; e.g.,~\citealt{2012ApJ...761...12D}). Since $\sumnu$ must be greater
than approximately $0.06\,\mathrm{eV}$ in the normal hierarchy scenario and
$0.1\,\mathrm{eV}$ in the degenerate hierarchy~\citep{GonzalezGarcia:2012sz},
the allowed neutrino mass window is already quite tight and could be closed
further by current or forthcoming
observations~\citep{Jimenez:2010ev, Lesgourgues13}.

Cosmological models, with and without neutrino mass, have different primary CMB power spectra. For observationally-relevant masses, neutrinos are still relativistic at recombination and the unique effects of masses in the primary power spectra are small. The main effect is around the first acoustic peak and is due to the early integrated Sachs-Wolfe (ISW) effect; neutrino masses have an impact here even for a fixed redshift of matter--radiation equality~\citep{Lesgourgues:2012uu,2012MNRAS.425.1170H,hou12,Lesgourgues13}. To date, this effect has been the dominant one in constraining the neutrino mass from CMB data, as demonstrated in \citet{hou12}.
As we shall see here, the \Planck\ data move us into a new regime where the dominant effect is from gravitational lensing.
Increasing neutrino mass, while adjusting other parameters to remain in a high-probability region of parameter space, increases the expansion rate at $z \ga 1$ and so suppresses clustering on scales smaller than the horizon size at the non-relativistic transition~\citep{Kaplinghat:2003bh,Lesgourgues:2005yv}. The net effect for lensing is a suppression of the CMB lensing potential and, for orientation, by $\ell=1000$ the suppression is around $10\%$ in power for $\sumnu = 0.66\,\mathrm{eV}$.

Here we report constraints assuming three species of degenerate
massive neutrinos. At the level of sensitivity of \Planck, the effect of mass splittings
is negligible, and the degenerate model can be assumed without loss of generality.

\begin{figure}
\centering
\includegraphics[width=\hsize]{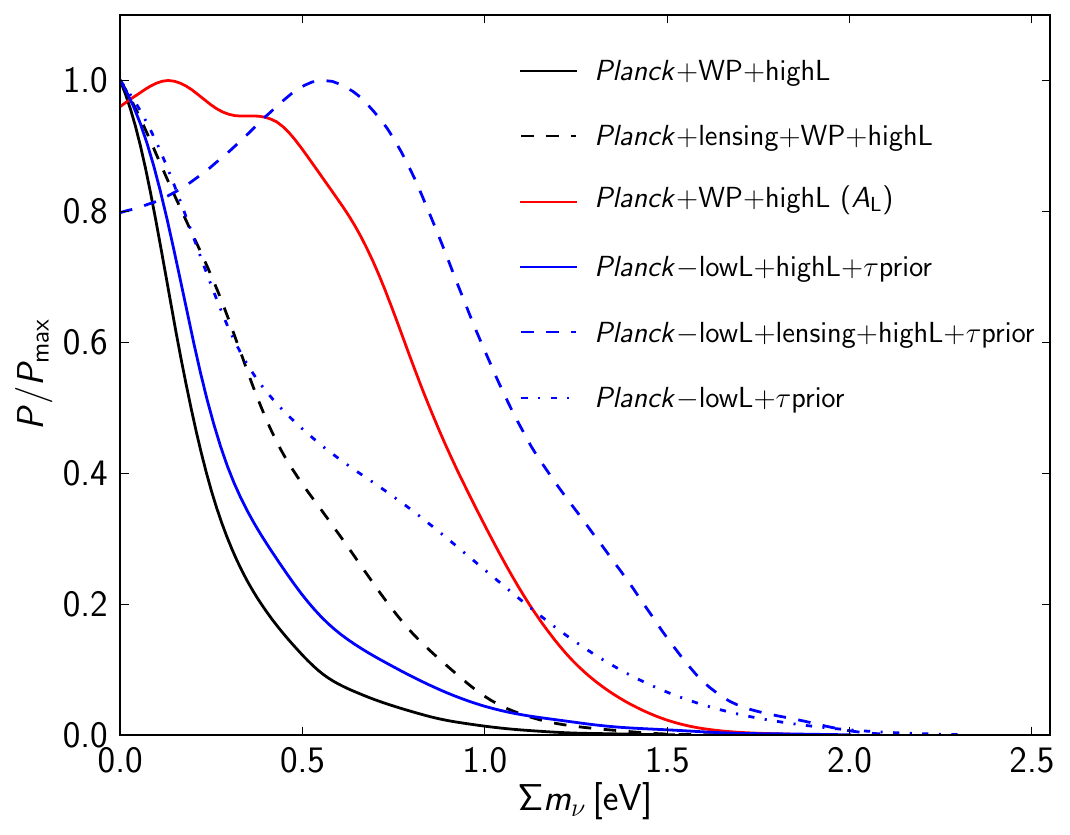}
\caption{Marginalized posterior distributions for $\sum m_\nu$ in flat models from CMB data.
We show results for \planck+\WP+\highL\ without (solid black) and with (red) marginalization over $\Alens$, showing how the posterior is significantly broadened by removing the lensing information from the temperature anisotropy power spectrum.
The effect of replacing the low-$\ell$ temperature and (\WMAP) polarization data with a $\tau$ prior is shown in solid blue (\planck$-$lowL+\highL+$\tau$prior) and of further removing the high-$\ell$ data in dot-dashed blue (\planck$-$lowL+$\tau$prior). We also show  the result of including the lensing likelihood with \planck+\WP+\highL\ (dashed black) and \planck$-$lowL+\highL+$\tau$prior (dashed blue).}
\label{fig:mnu}
\end{figure}

Combining the \Planck+\WP+\highL\ data, we obtain an upper limit on the summed neutrino mass of
\begin{equation}
\sum m_\nu < 0.66\, \mathrm{eV}\quad \mbox{(95\%; \planck+\WP+\highL)} .
\end{equation}
The posterior distribution is shown by the solid black curve in Fig.~\ref{fig:mnu}. To demonstrate
that the dominant effect leading to the constraint is gravitational lensing, we remove the lensing information by marginalizing over $\Alens$\footnote{%
The power spectrum of the temperature anisotropies is predominantly sensitive to changes in only one mode of the lensing potential power spectrum~\citep{2006PhRvD..74l3002S}. It follows that marginalizing over the single parameter $\Alens$ is nearly equivalent to marginalizing over the full amplitude and shape information in the lensing power spectrum as regards constraints from the temperature power spectrum.}.
We see that the posterior broadens considerably (see the red curve in Fig.~\ref{fig:mnu}) to give
\begin{equation}
\sum m_\nu < 1.08\, \mathrm{eV}\quad \mbox{[95\%; \planck+\WP+\highL\ ($\Alens$)]} ,
\end{equation}
taking us back close to the value of $1.3\,\mathrm{eV}$ (for $\Alens=1$) from the nine-year \WMAP\ data~\citep{hinshaw2012}, corresponding to the limit above which neutrinos become non-relativistic before recombination. (The resolution of \WMAP\ gives very little sensitivity to lensing effects.)

As discussed in Sect.~\ref{sec:lensing}, the \Planck+\WP+\highL\ data combination has a preference for high $\Alens$. Since massive neutrinos suppress the lensing power (like a low $\Alens$) there is a concern that the same tensions which drive $\Alens$ high may give artificially tight constraints on $\mnu$. We can investigate this issue by replacing the low-$\ell$ data with a prior on the optical depth (as in Sect.~\ref{sec:lensing}) and removing the high-$\ell$ data. Posterior distributions with the $\tau$ prior, and additionally without the high-$\ell$ data, are shown in Fig.~\ref{fig:mnu} by the solid blue and dot-dashed blue curves, respectively. The constraint on $\mnu$ does not degrade much by replacing the low-$\ell$ data with the $\tau$ prior only, but the degradation is more severe when the high-$\ell$ data are also removed: $\mnu < 1.31\, \mathrm{eV}$ (95\% CL).

Including the lensing likelihood (see Sect.~\ref{sec:lensing}) has a significant, but surprising, effect on our results. Adding the lensing likelihood to the \planck+\WP+\highL\ data combination \emph{weakens} the limit on $\mnu$,
\begin{equation}
\sum m_\nu < 0.85\,\mathrm{eV}\quad \mbox{(95\%; \planck+\lensing+\WP+\highL)} ,
\end{equation}
as shown by the dashed black curve in Fig.~\ref{fig:mnu}. This is representative of a general trend that the \planck\ lensing likelihood favours larger $\mnu$ than the temperature power spectrum. Indeed, if we use the data combination \planck$-$lowL+\highL+$\tau$prior, which gives a weaker constraint from the temperature power spectrum, adding lensing gives a best-fit away from zero ($\mnu  = 0.46\,\mathrm{eV}$; dashed blue curve in Fig.~\ref{fig:mnu}). However, the total $\chi^2$ at the best-fit is very close to that for the best-fitting base model (which, recall, has one massive neutrino of mass $0.06\,{\rm eV}$), with the improved fit to the lensing data ($\Delta \chi^2 = -2.35$) being cancelled by the poorer fit to high-$\ell$ CMB data ($\Delta \chi^2 = -2.15$).
There are rather large shifts in other cosmological parameters between these best-fit solutions corresponding to shifts along the acoustic-scale degeneracy direction for the temperature power spectrum. Note that, as well as the change in $H_0$ (which falls to compensate the increase in $\mnu$ at fixed acoustic scale), $\ns$, $\omb$ and $\omc$ change significantly keeping the \emph{lensed} temperature spectrum almost constant. These latter shifts are similar to those discussed for $\Alens$ in Sect.~\ref{sec:lensing}, with non-zero $\mnu$ acting like $\Alens < 1$.
The lensing power spectrum $C_\ell^{\phi\phi}$ is lower by $5.4\%$ for the higher-mass best fit at $\ell=400$ and larger below $\ell \approx 45$ (e.g.\ by $0.6\%$ at $\ell=40$), which is a similar trend to the residuals from the best-fit minimal-mass model shown in the bottom panel of Fig.~\ref{fig:CMBlensing_LCDM}. \cite{planck2013-p12} explores the robustness of the $C_\ell^{\phi\phi}$ estimates to various data cuts and foreground-cleaning methods. The first ($\ell=40$--$85$) bandpower is the least stable to these choices, although the variations are not statistically significant. We have checked that excluding this bandpower does not change the posterior for $\mnu$ significantly, as expected since most of the constraining power on $\mnu$ comes from the bandpowers on smaller scales. At this stage, it is unclear what to make of this mild preference for high masses from the 4-point function compared to the 2-point function. As noted in~\citet{planck2013-p12}, the lensing measurements from ACT~\citep{2013arXiv1301.1037D} and SPT~\citep{vanEngelen:2012va} show similar trends to those from \planck\ where they overlap in scale. With further \planck\ data (including polarization), and forthcoming measurements from the full $2500\,\mathrm{deg}^2$ SPT temperature survey, we can expect more definitive results on this issue in the near future.


Apart from its impact on the early-ISW effect and lensing potential, the total neutrino mass affects the angular-diameter distance to last scattering, and can be constrained through the angular scale of the first acoustic peak. However, this effect is degenerate with $\Oml$ (and so the derived $H_0$) in flat models and with other late-time parameters such as $\Omk$ and $w$ in more general models~\citep{Howlett:2012mh}.
Late-time geometric measurements help in reducing this ``geometric'' degeneracy.  Increasing the neutrino masses at fixed $\theta_\ast$ increases the angular-diameter distance for $0 \leq z \leq \zstar$ and reduces the expansion rate at low redshift ($z \la 1$) but increases it at higher redshift. The spherically-averaged BAO distance $D_{\rm V}(z)$ therefore increases with increasing neutrino mass at fixed $\theta_\ast$, and the Hubble constant falls; see Fig.~8 of~\citet{hou12}. With the BAO data of Sect.~\ref{sec:BAO}, we find a significantly lower bound on the neutrino mass:
\begin{equation}
\sum m_\nu < 0.23\, \mathrm{eV}\quad \mbox{(95\%; \planck+\WP+\highL+BAO)} . \label{Neutrinomass1}
\end{equation}
\referee{Following the philosophy of this paper, namely to give higher weight to the BAO data compared to more complex astrophysical data, we quote the result of
Eq.~(\ref{Neutrinomass1})
in the abstract as our most reliable limit on the neutrino mass.}
The \lcdm\ model with minimal neutrino masses was shown in
Sect.~\ref{sec:hubble} to be in tension with recent direct measurements of $H_0$ which favour higher values. Increasing the neutrino mass will only make this tension worse and drive us to artificially tight constraints on $\sumnu$.
If we relax spatial flatness, the CMB geometric degeneracy becomes three-dimensional in models with massive neutrinos and the constraints on $\sumnu$ weaken considerably to
\begin{equation}
\sum m_\nu < \left\{
\begin{tabular}{ll}$0.98\,\mathrm{eV}$ & \mbox{(95\%; \planck+\WP+\highL)} \\
$0.32\,\mathrm{eV}$ & \mbox{(95\%; \planck+\WP+\highL+BAO}).
\end{tabular}
\right.
\end{equation}

\subsubsection{Constraints on $N_{\rm eff}$}

As discussed in Sect.~\ref{sec:model}, the density of radiation in the Universe (besides photons) is usually parameterized by the effective neutrino number $N_{\rm eff}$. This parameter specifies the energy density when the species are relativistic in terms of the neutrino temperature assuming exactly three flavours and instantaneous decoupling. In the Standard Model, $\neff=3.046$, due
to non-instantaneous decoupling corrections~\citep{Mangano:2005cc}.

However, there has been some mild preference for $N_{\rm eff} > 3.046$
from recent CMB anisotropy measurements \citep{komatsu11, dunkley11, keisler11, Archidiacono:2011gq, hinshaw2012, hou12}. This is potentially interesting, since
an excess could be caused by a neutrino/anti-neutrino asymmetry,
sterile neutrinos, and/or any other light relics in the Universe. In this subsection we discuss the constraints on $\neff$ from \planck\ in scenarios where the extra relativistic degrees of freedom are effectively massless.

The physics of how $N_{\rm eff}$ is constrained by CMB anisotropies is explained in \citet{bashinsky04},~\citet{hou11} and~\cite{Lesgourgues13}. The main effect is that increasing the radiation density at fixed $\theta_\ast$ (to preserve the angular scales of the acoustic peaks) and fixed $\zeq$ (to preserve the early-ISW effect and so first-peak height) increases the expansion rate before recombination and 
reduces the age of the Universe at recombination. Since the diffusion length scales approximately as the square root of the age, while the sound horizon varies proportionately with the age, the angular scale of the photon diffusion length, $\theta_{\rm D}$, increases, thereby reducing power in the damping tail at a given multipole.
Combining \planck, \WMAP\ polarization and the high-$\ell$ experiments gives
\begin{equation}
\neff = 3.36_{-0.64}^{+0.68}\quad \mbox{(95\%; \planck+\WP+\highL)} . \label{Neutrino2}
\end{equation}
The marginalized posterior distribution is given in Fig.~\ref{fig:neff} (black curve). \referee{The result in Eq.~(\ref{Neutrino2}) is consistent with the value of $\neff = 3.046$ of the Standard Model, but it is important to aknowledge that it is
difficult to constrain $\neff$ accurately using CMB temperature 
measurements alone. Evidently, the nominal mission data from \Planck\ do not strongly rule out a value as high as $\neff = 4$.  }

\begin{figure}
\centering
    \includegraphics[width=8.8cm]{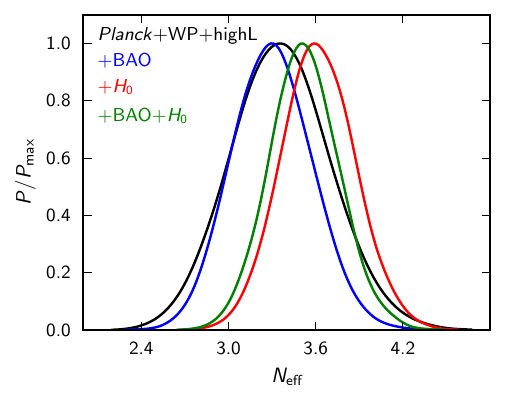}
    \caption{Marginalized posterior distribution of $\neff$ for \planck+\WP+\highL\ (black) and additionally BAO (blue), the $H_0$ measurement (red), and both BAO and $H_0$ (green).}
\label{fig:neff}
\end{figure}

Increasing $\neff$ at fixed $\theta_\ast$ and $\zeq$ necessarily raises the expansion rate at low redshifts too. Combining CMB with distance measurements can therefore improve constraints (see Fig.~\ref{fig:neff}) although for the BAO observable $\rdrag/D_{\rm V}(z)$ the reduction in both $\rdrag$ and $D_{\rm V}(z)$ with increasing $\neff$ partly cancel.
With the BAO data of Sect.~\ref{sec:BAO}, the $N_{\rm eff}$ constraint is tightened to
\begin{equation}
N_{\rm eff} = 3.30_{-0.51}^{+0.54}\quad \mbox{(95\%; \planck+\WP+\highL+BAO)} .
\end{equation}
Our constraints from CMB alone and CMB+BAO are compatible with the standard value $\neff= 3.046$ at the $1\,\sigma$ level, giving no evidence for extra relativistic degrees of freedom.

Since $\neff$ is positively correlated with $H_0$, the tension between the \planck\ data and direct measurements of $H_0$ in the base \lcdm\ model (Sect.~\ref{sec:hubble}) can be reduced at the expense of high $\neff$. The marginalized constraint is
\begin{equation}
N_{\rm eff} = 3.62_{-0.48}^{+0.50}\quad \mbox{(95\%; \planck+\WP+\highL+$H_0$)} .
\end{equation}
For this data combination, the $\chi^2$ for the best-fitting model allowing $\neff$ to vary is lower by $5.3$ than for the base $\neff=3.046$ model.
The $H_0$ fit is much better, with $\Delta \chi^2
= -4.4$, but there is no strong preference either way from the CMB. The low-$\ell$ temperature power spectrum does weakly favour the high $\neff$ model ($\Delta \chi^2 = -1.4$) -- since $\neff$ is positively correlated with $\ns$ (see Fig.~\ref{fig:ns_neff}) and increasing $\ns$ reduces power on large scales -- as does the rest of the \planck\ power spectrum ($\Delta \chi^2 = -1.8$). The high-$\ell$ experiments mildly disfavour high $\neff$ in our fits ($\Delta \chi^2 = 1.9$). Further including the BAO data pulls the central value downwards by around $0.5\,\sigma$ (see Fig.~\ref{fig:neff}):
\begin{equation}
N_{\rm eff} = 3.52_{-0.45}^{+0.48}\quad \mbox{(95\%; \planck+\WP+\highL+$H_0$+BAO)} .
\end{equation}
The $\chi^2$ at the best-fit for this data combination ($\neff = 3.48$) is lower by $4.2$ than the best-fitting $\neff=3.046$ model. While the high $\neff$ best-fit is preferred by \planck+\WP\ ($\Delta \chi^2 = -3.1$) and the $H_0$ data ($\Delta \chi^2 = -3.3$ giving an acceptable $\chi^2 = 1.8$ for this data point), it is disfavoured by the high-$\ell$ CMB data ($\Delta \chi^2 = 2.0$) and slightly by BAO ($\Delta \chi^2 = 0.5$). We conclude that the tension between direct $H_0$ measurements and the CMB and BAO data in the base \lcdm\ can be relieved at the cost of additional neutrino-like physics, but there is no strong preference for this extension from the CMB damping tail.

Throughout this subsection, we have assumed that all the relativistic components parameterized by $N_{\rm eff}$ consist of ordinary free-streaming relativistic particles. Extra radiation components with a different sound speed or viscosity parameter~\citep{Hu:1998kj} can provide a good fit to pre-\planck\ CMB data~\citep{Archidiacono:2013lva}, but are not investigated in this paper.

\subsubsection{Simultaneous constraints on $N_{\rm eff}$ and either $\sumnu$ or $\mnusterile$}

It is interesting to investigate simultaneous contraints on $N_{\rm
  eff}$ and $\sumnu$, since extra relics could coexist with neutrinos
of sizeable mass, or could themselves have a mass in the eV
range. Joint constraints on $N_{\rm eff}$ and $\sumnu$ have been
explored several times in the literature. These two parameters are
known to be partially degenerate when large-scale structure data are
used~\citep{Hannestad:2003ye,Crotty:2004gm}, but their impact in the CMB is different and does not
lead to significant correlations.

Joint constraints on $N_{\rm eff}$ and $\sumnu$ are always model-dependent: they vary strongly with assumptions about how the total mass is split between
different species (and they would also be different for models in which
massive species have chemical potentials or a non-thermal
phase-space distribution). We present here
\Planck\ constraints for two different models and describe the
scenarios that motivate them.

First, as in the previous subsection we assume that the three active
neutrinos share a mass of $\sumnu/3$, and may coexist
  with extra massless species contributing to $N_{\rm eff}$. In
  this model, when $N_{\rm eff}$ is greater than 3.046, $\Delta N_{\rm
    eff} = N_{\rm eff}-3.046$ gives the density of extra massless relics
  with arbitrary phase-space distribution. When $\neff < 3.046$,
 the temperature of the three active neutrinos is reduced accordingly, and no additional relativistic species are assumed. 
In this case, the CMB constraint is
\begin{equation}
\left.
\begin{array}{c}
N_{\rm eff} = 3.29_{-0.64}^{+0.67} \\
\sumnu< 0.60\, {\rm eV}
\end{array}
\right\} \quad\mbox{(95\%; \planck+\WP+\highL)}.
\end{equation}
These bounds tighten somewhat with the inclusion of BAO data, as illustrated in Fig.~\ref{fig:neff_mnu}; we find
\begin{equation}
\left.
\begin{array}{c}
N_{\rm eff} = 3.32_{-0.52}^{+0.54} \\
\sumnu< 0.28\, {\rm eV}
\end{array}
\right\} \quad\mbox{(95\%; \planck+\WP+\highL+BAO)}.
\end{equation}
We see that the joint constraints do not differ very much from the bounds
obtained when introducing these parameters separately. The physical effects of
neutrino masses and extra relativistic relics are sufficiently different to be
resolved separately at the level of accuracy of \planck.

\begin{figure*}
\centering
    \includegraphics[height=7cm]{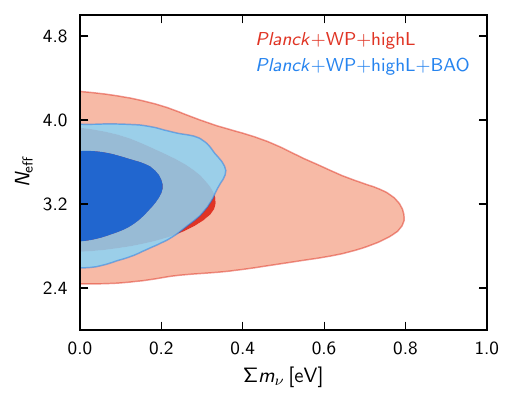}
    \includegraphics[height=7cm]{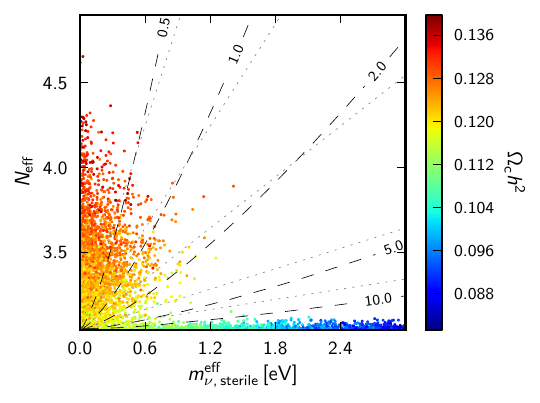}
    \caption{\emph{Left}: 2D joint posterior distribution between $\neff$ and
$\sumnu$ (the summed mass of the three active neutrinos) in models with extra
massless neutrino-like species. \emph{Right}: Samples in the
$\neff$--$\mnusterile$ plane, colour-coded by $\Omega_{\rm c}h^2$, in models
with one massive sterile neutrino family, with effective mass $\mnusterile$,
and the three active neutrinos as in the base \lcdm\ model. The physical mass
of the sterile neutrino in the thermal scenario,
$m_{\rm sterile}^{\rm thermal}$, is constant along the grey dashed lines, with
the indicated mass in $\mathrm{eV}$. The physical mass in the Dodelson-Widrow
scenario, $m_{\rm sterile}^{\rm DW}$, is constant along the dotted lines
(with the value indicated on the adjacent dashed lines).
\referee{Note the pile up of
points at low values of $\neff$, caused because  the sterile neutrino component
behaves like cold dark matter there, introducing a strong degeneracy between
the two components, as described in the text.}
}
\label{fig:neff_mnu}
\end{figure*}

In the second model, we assume the existence of one massive sterile neutrino,
in addition to the two massless and one massive active neutrino of the base
model. The active neutrino mass is kept fixed at $0.06\,\mathrm{eV}$.
In particle physics, this assumption can be motivated
in several ways. For example, there has recently been renewed interest in models
with one light sterile neutrino in order to explain the MiniBoone
anomaly reported in~\citet{AguilarArevalo:2012va}, as well as reactor
and Gallium anomalies~\citep{Giunti:2012bc}. The statistical significance of
these results is marginal and they should not be over-interpreted. However,
they do motivate investigating a model with three active neutrinos and one
heavier sterile neutrino with mass $m_{\rm sterile}$. If the sterile neutrino
were to thermalize with the same temperature as active neutrinos, this model
would have $N_{\rm eff} \approx 4$.

Since we wish to be more general, we assume that the extra eigenstate is
either: (i) thermally distributed with an arbitrary temperature $\Tsterile$;
or (ii) distributed proportionally to active neutrinos with an arbitrary
scaling factor $\chi_{\rm s}$ in which the scaling factor is a function of the
active--sterile neutrino mixing angle. This second case corresponds the
Dodelson-Widrow scenario~\citep{1994PhRvL..72...17D}.
The two cases are in fact
equivalent for cosmological observables and do not require separate
analyses~\citep{Colombi:1995ze,
Lesgourgues13}. Sampling the posterior with flat priors on $N_{\rm eff}$ and
$m_{\rm  sterile}$ would not be efficient, since in the limit of small
temperature $\Tsterile$, or small scaling factor $\chi_{\rm s}$, the mass would
be unbounded. Hence we adopt a flat prior on the ``effective sterile neutrino
mass'' defined as $\meffsterile\equiv (94.1 \omega_{\nu,\,\mathrm{sterile}})\,
\mathrm{eV}$\footnote{The factor of $94.1\,\mathrm{eV}$ here is the usual one
in the relation between physical mass and energy density for non-relativistic
neutrinos with physical temperature $T_\nu$.}. In the case of a
thermally-distributed sterile neutrino, this parameter is related to the true
mass via
\begin{equation}
\meffsterile = (\Tsterile/T_\nu)^3 m_{\rm sterile}^{\rm thermal} = (\Delta N_{\rm eff})^{3/4}  m_{\rm sterile}^{\rm thermal}  \,.
\end{equation}
Here, recall that $T_\nu =(4/11)^{1/3}T_\gamma$ is the active neutrino temperature in the instantaneous-decoupling limit and that the effective number is defined via the energy density, $\Delta N_{\rm eff}=(\Tsterile/T_\nu)^4$.
In the Dodelson-Widrow case the relation is given by
\begin{equation}
\meffsterile = \chi_{\rm s} m_{\rm sterile}^{\rm DW}\,,
\end{equation}
with $\Delta N_{\rm eff}=\chi_{\rm s}$. For a thermalized sterile neutrino with temperature $T_\nu$ (i.e., the temperature the active neutrinos would have if there were no heating at electron-positron annihilation), corresponding to $\Delta \neff  = 1$, the three masses are equal to each other.

Assuming flat priors on $ N_{\rm eff}$ and $\meffsterile$ with $\meffsterile<3\, \mathrm{eV}$, we find the results shown in Fig.~\ref{fig:neff_mnu}. The physical mass, $m_{\rm sterile}^{\rm thermal}$ in the thermal scenario is constant along the dashed lines in the figure and takes the indicated value in $\mathrm{eV}$. The physical mass, $m_{\rm sterile}^{\rm DW}$, in the Dodelson-Widrow scenario is constant on the dotted lines.
For low $\nnu$ the physical mass of the neutrinos becomes very large, so that they become non-relativistic well before recombination. In the limit in which the neutrinos become non-relativistic well before any relevant scales enter the horizon, they will behave exactly like cold dark matter, and hence are completely unconstrained within the overall total constraint on the dark matter density. For intermediate cases where the neutrinos become non-relativistic  well before recombination they behave like warm dark matter. The approach to the massive limit gives the tail of allowed models with large $\meffsterile$ and low $\nnu$ shown in Fig.~\ref{fig:neff_mnu}, with increasing $\meffsterile$ being compensated by decreased $\Omc h^2$ to maintain the total level required to give the correct shape to the CMB power spectrum.

For low $\meffsterile$ and $\Delta N_{\rm eff}$ away from zero the physical neutrino mass is very light, and the constraint becomes similar to the massless case. The different limits are continuously connected, and given the complicated shape seen in Fig.~\ref{fig:neff_mnu} it is clearly not appropriate to quote fully marginalized parameter constraints that would depend strongly on the assumed upper limit on $\meffsterile$. Instead we restrict attention to the case where the physical mass is $m_{\rm  sterile}^{\rm thermal}< 10\, \mathrm{eV}$, which roughly defines the region where (for the CMB) the particles are distinct from cold or warm dark matter. Using the \Planck+\WP+\highL\ (abbreviated to CMB below) data combination, this gives the marginalized one-parameter constraints
\begin{equation}
\left.
\begin{array}{c}
\nnu <3.91 \\
\meffsterile < 0.59\, \mathrm{eV} 
\end{array} \right\}
\quad \mbox{(95\%; CMB for $m_{\rm  sterile}^{\rm thermal}< 10\, \mathrm{eV}$)} \, .
\end{equation}
Combining further with BAO these tighten to 
\enlargethispage{10pt}
\begin{eqnarray}
&&\hspace{-0.025\textwidth}\left.
\begin{array}{c}
\nnu <3.80 \\
\meffsterile < 0.42\, \mathrm{eV} 
\end{array} \right\}
\hspace{2mm} \mbox{(95\%; CMB+BAO for $m_{\rm  sterile}^{\rm thermal}< 10\, \mathrm{eV}$)} \, .
\nonumber \\
&&
\end{eqnarray}
%
These bounds are only marginally compatible with a fully thermalized sterile neutrino ($N_{\rm eff}\approx 4$) with sub-eV mass $m_{\rm sterile}^{\rm thermal} \approx\meffsterile <0.5\, \mathrm{eV}$ that could explain the oscillation anomalies. 
The above contraints are also appropriate for the Dodelson-Widrow scenario, but for a physical mass cut of $m_{\rm  sterile}^{\rm DW}< 20\, \mathrm{eV}$.

The thermal and Dodelson-Widrow scenarios considered here are representative of a large number of possible models that have recently been investigated in the literature~\citep{Hamann:2011ge,Diamanti:2012tg,Archidiacono:2012gv,Hannestad:2012ky}.

\subsection{ Big bang nucleosynthesis}
\label{subsec:BBN}

Observations of light elements abundances created during big bang
nucleosynthesis (BBN) provided one of the earliest precision tests of
cosmology and were critical in establishing the existence of a hot big
bang.  Up-to-date accounts of nucleosynthesis are given by
\cite{Iocco:2008va} and \cite{Steigman:2012ve}. In the standard BBN model,
the abundance of light elements (parameterized by
$Y_{\rm P}^{\rm BBN}\equiv4n_{\mathrm{He}}/n_{\rm b}$ for helium-4 and
$y_{\rm DP}^{\rm BBN}\equiv10^5n_\mathrm{\rm D}/n_\mathrm{\rm H}$ for deuterium,
where $n_i$ is the number density of species $i$)\footnote{Observations of the primordial abundances are usually reported in terms of these number ratios. For helium, $Y_{\rm P}^{\rm BBN}$ differs from the mass fraction $Y_{\rm P}$, used elsewhere in this paper, by 0.5\% due to the binding energy of helium. Since the CMB is only sensitive to $Y_{\rm P}$ at the 10\% level, the distinction between definitions based on the mass or number fraction is ignored when comparing helium constraints from the CMB with those from observational data on primordial abundances.} can be
predicted as a function of the baryon density $\omega_{\rm b}$, the number
of relativistic degrees of freedom parameterized by $N_\mathrm{eff}$,
and of the lepton asymmetry in the electron neutrino
sector. Throughout this subsection, we assume for simplicity that
lepton asymmetry is too small to play a role at BBN. This is a
reasonable assumption, since \planck\ data cannot improve existing
constraints on the asymmetry\footnote{A primordial lepton asymmetry
  could modify the outcome of BBN only if it were very large (of the
  order of $10^{-3}$ or bigger). Such a large asymmetry
  is not motivated by particle physics, and is strongly constrained by
  BBN. Indeed, by taking into account neutrino oscillations in the
  early Universe, which tend to equalize the distribution function of
  three neutrino species, \cite{Mangano:2011ip} derived strong
  bounds on the lepton asymmetry. CMB data cannot improve these
  bounds, as shown by~\cite{Castorina:2012md}; an exquisite
  sensitivity to $N_\mathrm{eff}$ would be required. Note that the
  results of \cite{Mangano:2011ip} assume that $N_\mathrm{eff}$
  departs from the standard value only due to the lepton asymmetry. A
  model with both a large lepton asymmetry and extra relativistic
  relics could be constrained by CMB data. However, we will not
  consider such a contrived scenario in this paper.}. We also assume
that there is no significant entropy increase between BBN and the
present day, so that our CMB constraints on the baryon-to-photon ratio
can be used to compute primordial abundances.

To calculate the dependence of $Y_{\rm P}^{\rm BBN}$ and $y_{\rm DP}^{\rm BBN}$ on the
parameters $\omega_{\rm b}$ and  $N_\mathrm{eff}$, we use the accurate public
code {\tt PArthENoPE}~\citep{Pisanti:2007hk}, which incorporates values of
nuclear reaction rates, particle masses and fundamental constants, and
an updated estimate of the neutron lifetime ($\tau_{\rm n}=880.1$\,s; 
\citealt{Beringer:1900zz}). Experimental uncertainties on each of these
quantities lead to a theoretical error for $Y_{\rm P}^{\rm BBN}(\omega_{\rm b},
N_\mathrm{eff})$ and $y_{\rm DP}^{\rm BBN}(\omega_{\rm b}, N_\mathrm{eff})$. For
helium, the error is dominated by the uncertainty in the neutron
lifetime, leading to\footnote{\cite{Serpico:2004gx} quotes
$\sigma(Y_{\rm P}^{\rm BBN})=0.0002$, but since that work, the uncertainty on
the neutron lifetime has been re-evaluated, from
$\sigma(\tau_{\rm n})=0.8$~s to $\sigma(\tau_{\rm n})=1.1$~s
\citep{Beringer:1900zz}.} $\sigma(Y_{\rm P}^{\rm BBN})=0.0003$. For deuterium,
the error is dominated by  uncertainties in several nuclear rates, and is
estimated to be $\sigma(y_{\rm DP}^{\rm BBN})=0.04$ \citep{Serpico:2004gx}.

These predictions for the light elements can be confronted with measurements of their abundances, and also with CMB data (which is sensitive to
$\omega_{\rm b}$, $N_\mathrm{eff}$, and $Y_{\rm P}$). We shall see below that for
the base cosmological model with $N_\mathrm{eff}=3.046$ (or even
for an extended scenario with free $N_\mathrm{eff}$) the CMB data
predict the primordial abundances, under the assumption of standard BBN,
with smaller uncertainties than those estimated for the measured abundances.
Furthermore, the CMB predictions are consistent with direct abundance measurements.

\subsubsection{Observational data on primordial abundances}

The observational constraint on the primordial helium-4 fraction used
in this paper is $Y_{\rm P}^{\rm BBN}= 0.2534\pm0.0083$ (68\% CL) from the
recent data compilation of \citet{Aver:2011bw}, based on spectroscopic
observations of the chemical abundances in metal-poor \ion{H}{ii}
regions. The error on this measurement is dominated by systematic
effects that will be difficult to resolve in the near future. It is
reassuring that the independent and conservative method presented in
\cite{Mangano:2011ar} leads to an upper bound for $Y_{\rm P}^{\rm BBN}$ that is
consistent with the above estimate. The recent measurement of the
proto-Solar helium abundance by \cite{serenelli10} provides an even
more conservative upper bound, $Y_{\rm P}^{\rm BBN} < 0.294$ at the $2\,\sigma$ level.

For the primordial abundance of deuterium, data points show excess
scatter above the statistical errors, indicative of systematic
errors. The compilation presented in \cite{Iocco:2008va}, based
on data accumulated over several years, gives $y_{\rm DP}^{\rm BBN} =
2.87\pm0.22$ (68\% CL). 
\cite{Pettini:2012ph}
 report an accurate deuterium abundance measurement in the $z=3.04984$
low-metallicity damped Ly$\,\alpha$ system in the spectrum of QSO SDSS J1419+0829, which they
argue is particularly well suited to deuterium abundance measurements. These authors find
$y_{\rm DP}^{\rm BBN} = 2.535\pm0.05$ (68\% CL), a significantly tighter constraint
than that from the \cite{Iocco:2008va} compilation. The Pettini-Cooke measurement
is,  however, a single data point, and it is important to acquire more observations of
 similar systems to assess whether their error estimate is
consistent with possible sources of systematic error. We adopt a
conservative position in this paper and compare both the
\cite{Iocco:2008va} and the \cite{Pettini:2012ph} measurements
to the CMB predictions

We consider only the $^4$He and D abundances in this paper.
We do not discuss measurements of $^3$He abundances since these
provide  only an upper bound on the true primordial $^3$He
fraction.  Likewise, we do not discuss lithium. There has been a long standing
discrepancy between  the low lithium abundances measured in metal-poor stars in our Galaxy
and the predictions of BBN. At present it is not clear whether this discrepancy
is caused by systematic errors in the abundance measurements, or has an
``astrophysical'' solution  (e.g.,  destruction of primordial lithium) or is caused by new physics
\citep[see][for a recent review]{Fields:11}.

\subsubsection{\planck\ predictions of primordial abundances in standard BBN}

\begin{figure}
\centering
\includegraphics[width=8.8cm]{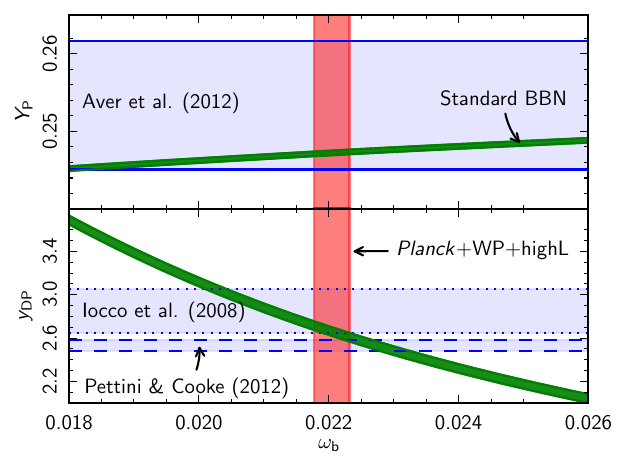}
\vspace{2mm}
\caption{
Predictions of standard BBN for the
primordial abundance of $^4$He (top) and deuterium (bottom), as a
function of the baryon density. The width of the green stripes corresponds
to 68\% uncertainties on nuclear reaction rates. The horizontal
bands show observational bounds on primordial element abundances
compiled by various authors, and the red  vertical band shows the
\planck+\WP+\highL\ bounds on $\omega_{\rm b}$ (all with 68\% errors). BBN
predictions and CMB results assume $N_\mathrm{eff}=3.046$ and no
significant lepton asymmetry.}
\label{fig:bbn_a}
\end{figure}

We first restrict ourselves to the base cosmological model, with no extra
relativistic degrees of freedom beyond ordinary neutrinos (and a
negligible lepton asymmetry), leading to $N_\mathrm{eff}=3.046$
\citep{Mangano:2005cc}. Assuming that standard BBN holds, and that
there is no entropy release after BBN, we can compute the spectrum of
CMB anisotropies using the relation $Y_{\rm P}(\omega_{\rm b})$ given by
{\tt PArthENoPE}. This relation is used as the default in the grid of models
discussed in this paper; we use the \COSMOMC\ implementation
developed by \citet{Hamann:2007sb}.
The \planck+\WP+\highL\  fits to the base \lcdm\  model
gives the following estimate of the baryon density,
\begin{equation}
\omega_{\rm b}=0.02207\pm0.00027 \quad \mbox{(68\%; \planck+\WP+\highL)},
\label{eq:bbn_omegab}
\end{equation}
as listed in Table~\ref{LCDMForegroundparams}. In
Fig.~\ref{fig:bbn_a}, we show this bound together with theoretical
BBN predictions for $Y_{\rm P}^{\rm BBN}(\omega_{\rm b})$ and $y_{\rm DP}^{\rm BBN}(\omega_{\rm b})$.
The bound of Eq.~(\ref{eq:bbn_omegab}) leads to the predictions
\beglet
\begin{eqnarray}
Y_{\rm P}^{\rm BBN}(\omega_{\rm b})&=&0.24725\pm0.00032, \\
y_{\rm DP}^{\rm BBN}(\omega_{\rm b})&=&2.656\pm0.067 ,
\end{eqnarray}
\endlet
where the errors here are 68\% and include theoretical errors that are added in quadrature to those arising from uncertainties in $\omega_{\rm b}$. (The theoretical error dominates the total error in the case of $Y_{\rm P}$.)\footnote{Note that, throughout this paper, our quoted CMB constraints on all parameters do not include the theoretical uncertainty in the BBN relation (where used).} For helium, this
prediction is in very good agreement with the data compilation of 
\cite{Aver:2011bw}, with an error that is  26 times smaller. For
deuterium, the CMB+BBN prediction lies midway between the best-fit
values of \cite{Iocco:2008va} and~\cite{Pettini:2012ph}, but agrees
with both at approximately the $1\,\sigma$ level. These results strongly support
standard BBN and show that within the framework of the base \lcdm\ model, 
 \planck\  observations lead to extremely precise predictions
of primordial abundances.

\subsubsection{Estimating the helium abundance directly from \planck\ data}

In the CMB analysis, instead of fixing $Y_{\rm P}$ to the BBN
prediction, $Y_{\rm P}^{\rm BBN}(\omega_{\rm b})$, we can relax any BBN prior and
let this parameter vary freely. The primordial helium fraction has an
influence on the recombination history and affects CMB anisotropies mainly
through the redshift of last scattering and the diffusion damping scale
\citep{Hu:1995fqa,Trotta:2003xg,Ichikawa:2006dt,Hamann:2007sb}.
Extending the base $\Lambda$CDM model by adding $Y_{\rm P}$ as a 
free parameter with a flat prior in the range $[0.1,0.5]$,
we find
\begin{equation}
Y_{\rm P}=0.266\pm0.021\quad \mbox{(68\%; \planck+\WP+\highL)}.
\end{equation}
Constraints in the $Y_{\rm P}$--$\omb$ plane are shown in Fig.~\ref{fig:bbn_b}.  This figure
shows that the CMB data have some sensitivity to the helium
abundance. In fact, the error on the CMB estimate of $Y_{\rm P}$
is only  $2.7$ times larger than the  direct measurements of the primordial
helium abundance by \cite{Aver:2011bw}. The CMB estimate of
$Y_{\rm P}$ is consistent with the observational measurements adding further support in favour of standard BBN.

\begin{figure}
\centering
\includegraphics[width=8.8cm]{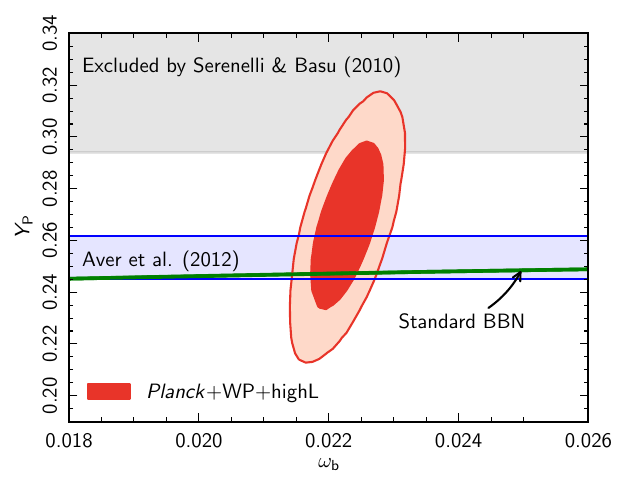}
\vspace{2mm}
\caption{Constraints in the $\omb$--$Y_{\rm P}$ plane from CMB and abundance measurements. The CMB constraints are for \planck+\WP+\highL\ (red 68\% and 95\%  contours) in \lcdm\ models with $Y_{\rm P}$ allowed to vary freely. The horizontal band shows observational bounds on $^4$He compiled by~\cite{Aver:2011bw} with 68\% errors, while the grey region at the top of the figure delineates the conservative 95\% upper bound inferred from Solar helium
  abundance by~\citet{serenelli10}. The green stripe shows the predictions of standard BBN for the
  primordial abundance of $^4$He as a function of the baryon density
  (with 68\% errors on nuclear reaction rates). Both BBN
  predictions and CMB results assume $N_\mathrm{eff}=3.046$ and no
  significant lepton asymmetry.}
\label{fig:bbn_b}
\end{figure}

\subsubsection{Extension to the case with extra relativistic relics}

We now consider the effects of additional relativistic degrees of freedom
on photons and ordinary neutrinos (obeying
the standard model of neutrino decoupling) by adding
$N_\mathrm{eff}$ as a free parameter. In the absence of
lepton asymmetry, we can predict the BBN
primordial abundances as a function of the two parameters $\omb$ and
$N_\mathrm{eff}$.

\begin{figure}[]
\centering
\includegraphics[width=8.8cm]{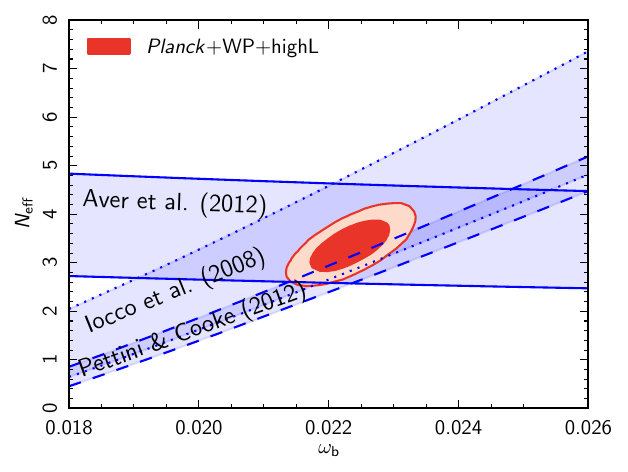}
\caption{Constraints in the $\omb$--$\neff$ plane from the CMB and abundance measurements. The blue stripes shows the 68\% confidence regions from
measurements of primordial
element abundances assuming standard BBN: $^4$He bounds compiled by \cite{Aver:2011bw}; and
deuterium bounds complied by \cite{Iocco:2008va} or measured by
\cite{Pettini:2012ph}.
We show for comparison the 68\% and 95\%
contours inferred from \planck+\WP+\highL, when $N_\mathrm{eff}$ is
left as a free parameter in the CMB analysis (and $Y_{\rm P}$ is fixed as
a function of $\omega_{\rm b}$ and $N_\mathrm{eff}$ according to BBN
predictions). These constraints assume no significant
lepton asymmetry.}
\label{fig:bbn_c}
\end{figure}

Figure~\ref{fig:bbn_c} shows the regions in  the $\omega_{\rm b}$--$N_\mathrm{eff}$ plane
preferred by primordial abundance measurements, and by the  CMB data
if the standard BBN picture is correct. The regions allowed by the abundance measurements
are defined by the  $\chi^2$ statistic
\begin{equation}
\chi^2(\omega_{\rm b}, N_\mathrm{eff}) \equiv \frac{\left[y(\omega_{\rm b}, N_\mathrm{eff})-y_\mathrm{obs}\right]^2}{\sigma_\mathrm{obs}^2+\sigma_\mathrm{theory}^2}~,
\end{equation}
where
$y(\omega_{\rm b}, N_\mathrm{eff})$ is the BBN prediction
for either $Y^{\rm BBN}_{\rm P}$ or $y_{\rm DP}^{\rm BBN}$, the quantity $y_\mathrm{obs}$
is the observed abundance, and the two errors in the denominator are the
observational and theoretical uncertainties.
Figure~\ref{fig:bbn_c} shows the edges of the 68\% preferred regions in
the $\omega_{\rm b}$--$N_\mathrm{eff}$ plane, given by
$\chi^2=\chi^2_\mathrm{min}+2.3$.

For the CMB data,  we fit a cosmological model with seven free parameters 
(the six parameters of the base \lcdm\ model,
plus $N_\mathrm{eff}$) to the \planck+\WP+\highL\ data, 
assuming that the primordial helium fraction is fixed by the
standard BBN prediction $Y_{\rm P}^{\rm BBN}(\omega_{\rm b}, N_\mathrm{eff})$.  Figure~\ref{fig:bbn_c} shows the
joint 68\% and 95\% confidence contours in the $\omega_{\rm b}$--$N_\mathrm{eff}$ plane.
The preferred regions in this plane from abundance measurements
and the CMB agree remarkably well. The CMB gives approximately
three times smaller error bars than primordial abundance data on both
parameters.

We can derive constraints on $N_\mathrm{eff}$ from primordial
element abundances \emph{and} CMB data together by combining their likelihoods.
The CMB-only confidence interval for $N_\mathrm{eff}$ is
\begin{equation}
N_\mathrm{eff} = 3.36\pm0.34 \quad \mbox{(68\%; \planck+\WP+\highL)} .
\end{equation}
When combined with the data reported respectively by
\cite{Aver:2011bw}, \cite{Iocco:2008va}, and~\cite{Pettini:2012ph}, the 68\% confidence
interval becomes
\begin{equation}
N_\mathrm{eff} = \left\{
\begin{tabular}{ll}
$3.41\pm0.30$, & $Y_{\rm P}$ (Aver et al.),\\
$3.43\pm0.34$, & $y_{\rm DP}$ (Iocco et al.),\\
$3.02\pm0.27$, & $y_{\rm DP}$ (Pettini and Cooke).
\end{tabular}
\right.
\end{equation}
Since there is no significant tension between CMB and primordial element results, all these bounds are in agreement with the CMB-only analysis. The small error bar derived from combining the CMB with the \cite{Pettini:2012ph} data point shows that further deuterium observations combined with \planck\ data have the potential to pin down the value of $N_\mathrm{eff}$ to high precision.

\subsubsection{Simultaneous constraints on both $N_{\rm eff}$ and $Y_{\rm P}$}

In this subsection, we discuss simultaneous constraints on both
$N_{\rm eff}$ and $Y_{\rm P}$ by adding them to the six parameters of
the base \lcdm\ model. Both $N_{\rm eff}$ and $Y_{\rm P}$ have an impact on the damping tail of the CMB power spectrum by altering the ratio $k^{-1}_{\rm D}/r_\ast$,
where $k^{-1}_{\rm D}$ is the photon diffusion length at last scattering and $r_\ast$
is the sound horizon there.  There is thus an approximate
degeneracy between these two parameters along which the ratio is nearly constant.
The extent of the degeneracy is limited
by the characteristic phase shift of the acoustic oscillations that arises due to the
free streaming of the neutrinos \citep{bashinsky04}.  As discussed by \citet{hou11}, the early ISW effect also partly breaks the degeneracy, but this is less important than the effect of the phase shifts.

\begin{figure}
\centering
\includegraphics[width=8.8cm]{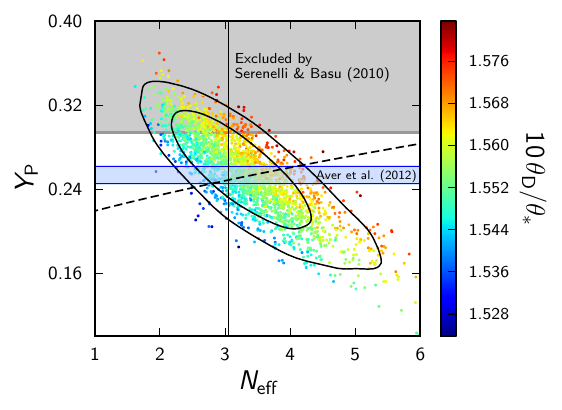}
\caption{
2D joint posterior distribution for $N_{\rm eff}$ and $Y_{\rm P}$
with both parameters varying freely, determined from \planck+\WP+\highL\ data.
Samples are colour-coded by the value of the angular ratio $\theta_{\rm D} / \theta_\ast$, which is constant along the degeneracy direction.
The $N_{\rm eff}$--$Y_{\rm P}$ relation from
BBN theory is shown by the dashed curve.  The vertical line shows
the standard value $N_{\rm eff} = 3.046$.  The region with $Y_{\rm P}>0.294$ is highlighted
in grey, delineating the region that exceeds the $2\,\sigma$ upper limit of the
recent measurement of initial Solar helium
abundance \citep{serenelli10}, and the blue horizontal region is the 68\% confidence region from the~\citet{Aver:2011bw} compilation of $^4$He measurements.
}
\label{fig:neff_yp}
\end{figure}

The joint posterior distribution for $N_{\rm eff}$ and $Y_{\rm P}$ from
the \planck+\WP+\highL\  likelihood is shown in Fig.~\ref{fig:neff_yp}, with
each MCMC sample colour-coded by the value of the observationally-relevant angular ratio $\theta_{\rm D}/\theta_\ast \propto
(k_{\rm D} r_\ast)^{-1}$. The main constraint on $N_{\rm eff}$ and $Y_{\rm P}$ comes from
the precise measurement of this ratio by the CMB, leaving the degeneracy along
the constant $\theta_{\rm D}/\theta_\ast$ direction.  The relation between
$N_{\rm eff}$ and $Y_{\rm P}$ from BBN theory is shown in the figure by the dashed
curve\footnote{For constant $N_{\rm eff}$, the variation due to the
uncertainty in the baryon density is too small to be visible, given the
thickness of the curve.}.  The standard BBN prediction with $N_{\rm
eff}=3.046$ is contained within the 68\% confidence region.  The grey
region is for $Y_{\rm P} > 0.294$ and is the $2\,\sigma$ conservative upper
bound on the primordial helium abundance from
\citet{serenelli10}. Most of the samples are consistent with this
bound.  The inferred estimates of $N_{\rm eff}$ and $Y_{\rm P}$ from
the \planck+\WP+\highL\ data are
\beglet
\begin{eqnarray}
    N_{\rm eff} &=& 3.33_{-0.83}^{+0.59} \quad\hspace{0.2mm} \mbox{(68\%; \planck+\WP+\highL)},\\
    Y_{\rm P} &=& 0.254_{-0.033}^{+0.041} \quad \mbox{(68\%; \planck+\WP+\highL)}.
\end{eqnarray}
\endlet

\begin{figure}
\centering
\includegraphics[width=8.8cm]{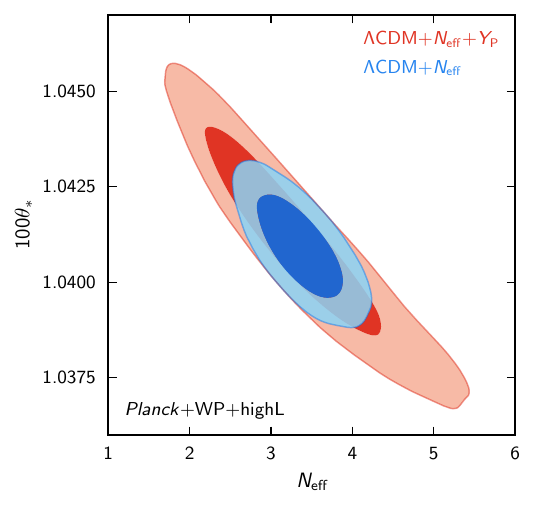}
\caption{2D joint posterior distribution between $N_{\rm eff}$ and $\theta_\ast$ for \lcdm\ models with variable $\neff$ (blue) and variable $\neff$ and $Y_{\rm P}$ (red).
Both cases are for \planck+\WP+\highL\  data.}
\label{fig:neff_thetas}
\end{figure}

With $Y_{\rm P}$ allowed to vary, $N_{\rm eff}$ is no longer tightly
constrained by the value of $\theta_{\rm D}/\theta_\ast$.  Instead, it is
constrained, at least in part, by the impact that varying $\neff$
has on the phase shifts of the acoustic oscillations. As discussed in
\citet{hou12}, this effect explains the observed correlation between
$\neff$ and $\theta_\ast$, which
is shown in Fig.~\ref{fig:neff_thetas}.  The
correlation in the \lcdm+$N_{\rm eff}$ model is also plotted in
the figure showing that the $N_{\rm eff}$--$Y_{\rm P}$ degeneracy combines with the phase shifts to generate a larger dispersion in $\theta_\ast$ in such models.

\subsection{Dark energy}
\label{subsec:darkenergy}

A major challenge for cosmology is to elucidate the nature of the dark energy
driving the accelerated expansion of the Universe.
Perhaps the most straightforward explanation is that dark energy is a cosmological constant.
An alternative is dynamical dark energy 
\citep{wetterich_1988,ratra_peebles_1988, caldwell:98},
usually based on a scalar field.  In the simplest models,  the field is very
light, has a canonical kinetic energy term and is minimally coupled to gravity.
In such models the dark energy sound speed equals the speed of light and
it has zero anisotropic stress.  It thus contributes very little to clustering.
We shall only consider such models in this subsection.

A cosmological constant has an equation of state $w\equiv p/\rho =  -1$,
 while scalar field models typically
have time varying $w$ with $w\ge -1$.
The analysis performed here is based on the ``parameterized post-Friedmann''
(PPF) framework of \citet{Hu:2007pj} and \citet{Hu:2008zd} as implemented in
\CAMB\  \citep{Fang:2008kc,Fang:2008sn} and discussed earlier in
Sect.~\ref{sec:model}. This allows us to investigate both regions of parameter space in which $w<-1$
(sometimes referred to as the ``phantom'' domain) and models in which $w$ changes with time.

\begin{figure}
\centering
\includegraphics[width=8.8cm]{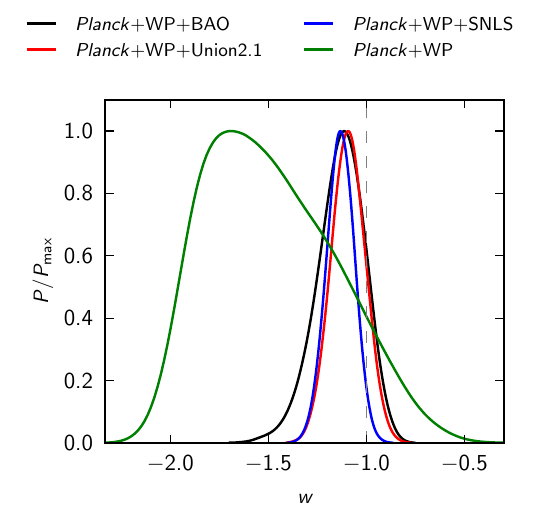}
\caption{Marginalized posterior distributions for the dark energy
  equation of state parameter $w$ (assumed constant), for \planck+\WP\ alone (green) and in combination with SNe data (SNSL in blue and the Union2.1 compilation in red) or BAO data (black). A flat prior on $w$ from $-3$
  to $-0.3$ was assumed and, importantly for the CMB-only constraints, the prior
$[20,100]\,{\rm km}\,{\rm s}^{-1}\,{\rm Mpc}^{-1}$ on $H_0$.
The dashed grey line indicates the
  cosmological constant solution, $w=-1$.}
\label{fig:wpost}
\end{figure}

Figure~\ref{fig:wpost} shows the marginalized posterior
distributions for $w$  for an extension of the base \lcdm\ cosmology to
models with constant $w$. 
We present results for \planck+\WP\ and in combination with SNe or BAO data.
(Note that adding in the high-$\ell$ data from ACT and SPT results in little
change to the posteriors shown in Fig. \ref{fig:wpost}.)
As expected, the CMB alone does not strongly constrain $w$, due to the two-dimensional geometric degeneracy in these models.
We can break this degeneracy by combining the CMB data with lower redshift distance measures. Adding in BAO data tightens the constraints
substantially, giving
\begin{equation}
w = -1.13^{+0.24}_{-0.25} \quad \mbox{(95\%;  \planck+\WP+BAO)},  \label{DE1}
\end{equation}
in good agreement with a cosmological constant ($w=-1$).
Using supernovae data leads to the constraints 
\beglet
\begin{eqnarray}
w &=& -1.09 \pm 0.17 \quad \mbox{(95\%;  \planck+\WP+Union2.1)}, \label{DE2a}\\
w &=& -1.13^{+0.13}_{-0.14} \quad \quad\mbox{(95\%; \planck+\WP+SNLS)}, \label{DE2b}
\end{eqnarray}
\endlet
The combination with SNLS data favours the phantom domain ($w < -1$) at $2\,\sigma$, while the Union2.1 compilation is more consistent with a cosmological constant.

If instead we combine \Planck+\WP\ with the \citet{Riess:2011yx}  measurement of $H_0$, 
we find
\begin{equation}
w=-1.24^{+0.18}_{-0.19}  \quad \mbox{(95\%; \planck+\WP+$H_0$)},  \label{DE3}
\end{equation}
which is in tension with $w=-1$ at more than the $2\,\sigma$ level.

The results in Eqs.\ (\ref{DE1}--{\ref{DE3}) reflect the tensions between 
the supplementary data sets and the \planck\ base \lcdm\ cosmology discussed
in Sect. \ref{sec:datasets}. The BAO data are in excellent agreement with the
\planck\ base \lcdm\ model, so there is no significant preference for $w\neq -1$ when combining BAO with \planck. In contrast,
the addition of the $H_0$ measurement,  or SNLS SNe data,  to the CMB data favours
models with exotic physics in the dark energy sector. These trends form a consistent
theme throughout this section. The SNLS data favours a lower $\Omega_{\rm m}$ in the \lcdm\ model than \planck, and hence larger dark energy density today. The tension can be relieved by making the dark energy fall away faster in the past than for a cosmological constant, i.e., $w< -1$.

The constant $w$ models are of limited physical interest. 
If $w\ne -1$ then it is likely to change with time.  To investigate
this we consider the simple linear relation in Eq.~(\ref{DE0}), 
$w(a)=w_0+w_a(1-a)$, which has often been used in the literature
\citep{chevallier_polarski_2001, linder_2003}.
This parameterization approximately captures the low-redshift behaviour of light,
slowly-rolling minimally-coupled scalar fields (as long
as they do not contribute significantly to the total energy density at
early times) and avoids the complexity of scanning a large number of possible
potential shapes and initial conditions.
The dynamical evolution of $w(a)$ can lead to distinctive imprints in the
CMB \citep{Caldwell:1997ii} which would show up in the \planck\ data.

\begin{figure}
\centering
\includegraphics[width=8.8cm]{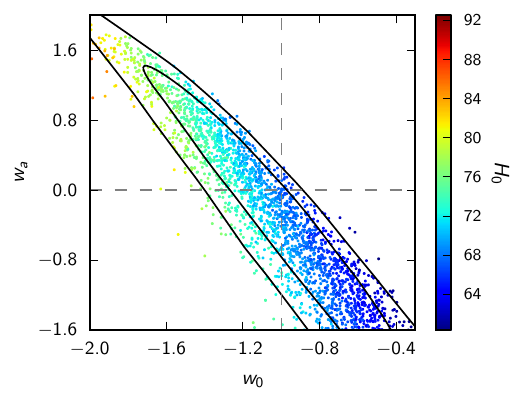}
\caption{2D marginalized posterior distribution for $w_0$ and $w_a$
  for \planck+\WP+BAO data. 
  The contours are 68\% and 95\%, and the samples are colour-coded according to the value of $H_0$. Independent flat priors of $-3< w_0<-0.3$
and $-2<w_a<2$ are assumed.  Dashed grey lines show the
  cosmological constant solution $w_0=-1$ and $w_a = 0$.}
\label{fig:wwaH0post}
\end{figure}

Figure~\ref{fig:wwaH0post} shows contours of the joint posterior
distribution  in the $w_0$--$w_a$ plane using \planck+\WP+BAO data
(colour-coded according to the value of $H_0$).
The points are coloured by the value of $H_0$, which shows a clear variation
with $w_0$ and $w_a$ revealing the three-dimensional nature of the geometric degeneracy in such models.
The cosmological constant point $(w_0,w_a)=(-1,0)$ lies within the
68\% contour and the marginalized posteriors for $w_0$ and $w_a$ are
\beglet
\begin{eqnarray}
w_0&=&-1.04^{+0.72}_{-0.69} \quad \mbox{(95\%; \planck+\WP+BAO)}, \label{DE4a}\\
w_a&<& 1.32 \hspace{12mm}  \mbox{(95\%; \planck+\WP+BAO)}. \label{DE4b}
\end{eqnarray}
\endlet
Including the $H_0$ measurement in place of the BAO data  moves $(w_0,w_a)$  away
from the cosmological constant solution towards negative $w_a$ at just under the 
$2\,\sigma$ level.

Figure~\ref{fig:wwapost} shows likelihood contours for $(w_0, w_a)$, now
adding SNe data to \planck. As discussed in detail in
Sect.~\ref{sec:datasets}, there is a dependence of the base
\lcdm\ parameters on the choice of SNe data set, and this is reflected
in Fig.~\ref{fig:wwapost}.  The results from the \planck+\WP+Union2.1
data combination are in better agreement with a cosmological constant
than those from the \planck+\WP+SNLS combination. For the latter data
combination, the cosmological constant solution lies on the $2\,\sigma$
boundary of the $(w_0, w_a)$ distribution.

\begin{figure}
\centering
\includegraphics[width=8.8cm]{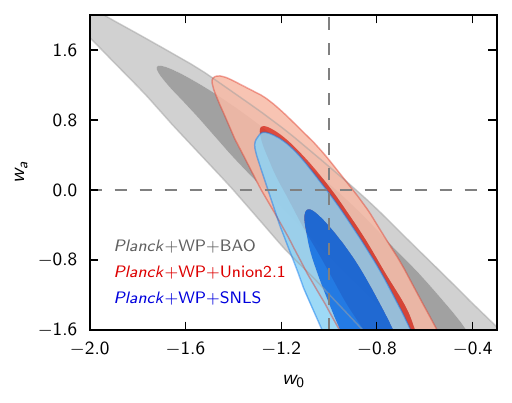}
\caption{2D marginalized posterior distributions for $w_0$ and $w_a$,
  for the data combinations \planck+\WP+BAO (grey), \planck+\WP+Union2.1 (red) and \planck+\WP+SNLS (blue). The contours are 68\% and 95\%, and dashed grey lines show the cosmological constant solution.}
\label{fig:wwapost}
\end{figure}

Dynamical dark energy models might also give a non-negligible contribution
to the energy density of the Universe at early times.
Such early dark energy \citep[EDE;][]{CWP} models may be very close to
$\Lambda$CDM recently, but have a non-zero dark energy density fraction,
$\Omega_{\rm e}$, at early times.
Such models complement the $(w_0,w_a)$ analysis by investigating how much
dark energy can be present at high redshifts.  
EDE has two main effects: it reduces structure growth in the period after
last scattering; and it changes the position and height of the peaks in the
CMB spectrum.

The model we adopt here is that of \citet{doran_robbers_2006}:
\begin{eqnarray}\label{eq:wde1}
\Omega_{\rm de}(a) = \frac{\Omega_{\rm de}^{0}-\Omega_{\rm e}
  (1-a^{-3 w_0})}{\Omega_{\rm de}^{0}+\Omega_{\rm m}^0 a^{3 w_0}}
  +\Omega_{\rm e}(1-a^{-3 w_0})  \,\, .
\end{eqnarray}
It requires two additional parameters to those of the base \lcdm\ model:
 $\Omega_{\rm e}$, the dark energy density
relative to the critical density at early times
(assumed constant in this treatment);
and the present-day dark energy equation of state parameter $w_0$.
Here $\Omega_{\rm m}^{0}$ is the present matter density and
$\Omega_{\rm de}^{0}=1-\Omega_{\rm m}^{0}$ is the present dark energy
abundance (for a flat Universe).
Note that the model of Eq.~(\ref{eq:wde1}) has dark energy present over
a large range of redshifts;
the bounds on $\Omega_{\rm e}$ can be substantially weaker if dark energy
is only present over a limited range of redshifts
\citep{pettorino_amendola_wetterich_2013}.
The presence or absence of dark energy at the epoch of last scattering
is the dominant effect on the CMB anisotropies and hence the constraints
are insensitive to the addition of low redshift supplementary data such as
BAO.

The most precise bounds on EDE arise from the analysis of CMB anisotropies
\citep{Doran_etal_2001, caldwell_etal_2003, calabrese_etal_2011, reichardt_etal_2012, Sievers:13, hou12, pettorino_amendola_wetterich_2013}. Using \planck+\WP+\highL,
we find
\begin{equation}
\Omega_{\rm e}   < 0.009  \quad \mbox{(95\%; \planck+\WP+\highL)}. \label{EDE1b}
\end{equation}
(The limit for \planck+\WP\ is very similar: $\Omega_{\rm e}   < 0.010$.)
These bounds are  consistent with and improve the recent ones of \citet{hou12},  who give $\Omega_e < 0.013$ at 95\% CL, and
\citet{Sievers:13}, who find $\Omega_e < 0.025$ at 95\% CL.

In summary, the results on dynamical dark energy (except for those on early dark energy
discussed above) are dependent on
exactly what supplementary data are used in conjunction with the CMB data.
(\planck\ lensing does not significantly improve the constraints on the models
discussed here.) Using the direct measurement of $H_0$, or the SNLS SNe
sample, together with \planck\ we see preferences for dynamical dark energy
at about the $2\,\sigma$ level reflecting the tensions between these data sets and
\planck\ in the \lcdm\ model. 
In contrast, the BAO measurements together
with \planck\ give tight constraints which are consistent with a cosmological
constant. Our inclination is to give greater weight to the BAO measurements
and to conclude that there is no strong evidence that the dark energy is
anything other than a cosmological constant.

\subsection{Dark matter annihilation}

Energy injection from dark matter (DM) annihilation can change the recombination history 
and affect the shape of the angular CMB spectra \citep{Chen:2003gz,Padmanabhan:2005es,Zhang:2006fr,Mapelli:2006ej}.
As recently shown in several papers \citep[see e.g.,][]{Galli:2009zc,Galli:2011rz,Giesen:2012rp,Hutsi:2011vx,Natarajan:2012ry,Evoli:2012}
CMB anisotropies offer an  opportunity to constrain DM annihilation models.

High-energy particles injected in the high-redshift thermal gas by DM
annihilation are typically cooled down to the keV scale by high energy
processes; once the shower has reached this energy scale, the
secondary particles produced can ionize, excite or heat the thermal gas
\citep{Shull:1985,Valdes:2009cq}; the first two processes modify the evolution
of the free electron fraction $x_{\rm e}$, while the third affects the
temperature of the baryons.
 
The rate of energy release, $dE/dt$, per unit volume
by a relic annihilating DM particle is given by
\begin{equation}
\label{enrateselfDM}
\frac{dE}{dt}(z)=  2\, g\, \rho^2_{\rm c} c^2 \Omega^2_{\rm c} (1+z)^6 p_{\rm ann}(z), \, \, \,
\end{equation}
where $p_{\rm ann}$ is, in principle, a function of redshift $z$,  defined as
\begin{equation}
\label{pann}
p_{\rm ann} (z) \equiv f(z) \frac{\langle\sigma v\rangle}{m_\chi}, 
\end{equation}
where $\langle\sigma v\rangle$ is the thermally averaged
annihilation cross-section, $m_\chi$ is the mass of the DM particle,
$\rho_{\rm c}$ is the critical density of the Universe today, $g$ is a
degeneracy factor equal to $1/2$ for Majorana particles and $1/4$
for Dirac particles (in the following, constraints will refer to
Majorana particles), and the parameter $f(z)$ indicates the fraction of
energy which is absorbed {\it overall} by the gas at redshift $z$. 

In Eq.~(\ref{pann}), the  factor $f(z)$ depends on the details of the
annihilation process, such as the mass of the DM particle and the
annihilation channel \cite[see e.g.,][]{Slatyer:2009yq}. The
functional shape of $f(z)$ can be taken into account using
generalized parameterizations
\citep{Finkbeiner:2011dx,Hutsi:2011vx}. However, as shown in
\cite{Galli:2011rz}, \cite{Giesen:2012rp}, and \cite{Finkbeiner:2011dx}
it is possible to
neglect the redshift dependence of $f(z)$ to first approximation,
since current data shows very little sensitivity to variations of this
function.  The effects of DM annihilation can therefore be
well parameterized by a single constant parameter, $p_{\rm ann}$, that
encodes the dependence on the properties of the DM particles.

We compute here the theoretical angular power in the presence of DM
annihilations, by modifying the {\tt RECFAST} routine in the {\tt camb}
code as in
\cite{Galli:2011rz} and by making use of the package {\tt CosmoMC}
for Monte Carlo parameter estimation.  We checked that we obtain the
same results by using the {\tt CLASS} Boltzmann
code~\citep{Lesgourgues:2011re} and the {\tt Monte Python} package
\citep{Audren:2012wb}, with DM annihilation effects calculated either
by {\tt RECFAST} or {\tt HyRec} \citep{AliHaimoud:2010dx}, as detailed in
\cite{Giesen:2012rp}. Besides $p_{\rm ann}$, we sample the parameters of the 
base \lcdm\ model and the foreground/nuisance parameters
described in Sect. \ref{sec:highell}. 

From \Planck+\WP\ we find
\begin{eqnarray}
p_{\rm ann} < 5.4 \times 10^{-6}\, \mathrm{m}^3\, \mathrm{s}^{-1}\,\mathrm{kg}^{-1}\quad \mbox{(95;  \planck+\WP)}. \label{DM1}
\end{eqnarray}
This constraint is weaker than that found from the
full {\it WMAP}9 temperature and polarization likelihood, $p_{\rm
  ann} < 1.2 \times 10^{-6}\, \mathrm{m^3s^{-1}kg^{-1}}$ because the
  \planck\ likelihood does not yet include polarization information at
  intermediate and high multipoles.  In fact, the damping effect of DM
  annihilation on the CMB temperature power spectrum is highly
  degenerate with other cosmological parameters, in particular with
  the scalar spectral index and the scalar amplitude, as first shown
  by \cite{Padmanabhan:2005es}. As a consequence, the constraint on
  the scalar spectral index is significantly weakened when $p_{\rm
    ann}$ is allowed to vary, $n_{\rm s}=0.984^{+0.012}_{-0.026}$, to be
  compared to the constraint listed in Table \ref{LCDMparams} for the
  base \lcdm\ cosmology, $n_{\rm s}=0.9603\pm 0.0073$.

These degeneracies can be broken by polarization data. The
effect of DM annihilation on polarization is in fact an overall
enhancement of the amplitude at large and intermediate scales, and a
damping at small scales (see e.g., Fig.~1 in \citealt{Galli:2009zc} or
Fig.~3 in \citealt{Giesen:2012rp}). We thus expect the constraint to
improve significantly with the forthcoming \Planck\ polarization data
release.  We verified that adding BAO, {\it HST\/} or \highL\ data to 
\Planck+\WP\ improves the constraints only marginally,  as these datasets are
not able to  break the degeneracy between $p_{\rm ann}$ and
$n_{\rm s}$.

On the other hand, we observe a substantial improvement in the constraints when we combine the 
\planck+\WP\ data with the \planck\ lensing likelihood data. For this data combination we find an upper
limit of 
\begin{equation}
p_{\rm ann} < 3.1 \times 10^{-6}\, \mathrm{m}^3\,\mathrm{s}^{-1}\,\mathrm{kg}^{-1}
\quad  \mbox{(95\%; \planck+\lensing+\WP}).
\end{equation}
The improvement over Eq.~(\ref{DM1}) comes from the constraining power of the lensing likelihood on $A_{\rm s}$ and $n_{\rm s}$, that 
partially breaks the degeneracy with $p_{\rm ann}$.

Our results are consistent with previous work and show no evidence for
DM annihilation. Future release of \Planck\ polarization data will
help to break the degeneracies which currently limit the accuracy of the
constraints presented here.

\subsection{Constraints on a stochastic background of primordial magnetic fields}

Large-scale magnetic fields of the order of a few $\mu$G observed in
galaxies and galaxy clusters may be the product of the amplification
during structure formation, of primordial magnetic seeds
\citep{Ryu:2011hu}. Several models of the early Universe predict the
generation of primordial magnetic fields (hereafter PMF), either
during inflation or during later phase transitions (see
\citealt{Widrow:2002ud} and \citealt{Widrow:2011hs} for reviews).

PMF have an impact on cosmological perturbations and in particular on
CMB anisotropy angular power spectra \citep{Subramanian:2006xs}, 
that can be used to constrain the PMF amplitude.  In this
section we will derive the constraints  from {\Planck} data on a
stochastic background of PMF.  We are mainly interested in constraints
from CMB temperature anisotropies. Therefore, we will not
consider the effect of Faraday rotation on CMB polarization
anisotropies \citep{Kosowsky:1996yc, Kosowsky:2004zh} nor non-Gaussianities
associated with PMF
\citep{Brown:2005kr,Caprini:2009vk, Seshadri:2009sy, Trivedi:2010gi}.  We will
restrict the analysis reported here to the non-helical case.

A stochastic background of PMF is modelled as a fully inhomogeneous
component whose energy-momentum tensor is quadratic in the fields.  We
assume the usual magnetohydrodynamics limit, in which PMF are frozen
and the time evolution is simply given by the dilution with
cosmological expansion, $B(k,\eta)=B(k)/a(\eta)^2$.  We model the PMF
with a simple power-law power spectrum: $P_B(k)=A k^{n_B}$, with a
sharp cut off at the damping scale $k_{\rm D}$,  as computed in
\citet{Jedamzik:1996wp} and \citet{Subramanian:1997gi},  to model the
suppression of PMF on small scales.

It is customary to specify the amplitude of the PMF power spectrum with $B_{\lambda}$, the root-mean-square of the field
smoothed over length scale $\lambda$, defined such that 
\begin{equation}
B^2_\lambda = \int_0^{\infty} \frac{d{k \, k^2}}{2 \pi^2} e^{-k^2 \lambda^2}
P_B (k).  
\label{gaussian}
\end{equation}
Given our assumed model and conventions,  PMF are fully described by 
two parameters: the smoothed amplitude
 $B_\lambda$; and the spectral index $n_B$.  Here, we
set $\lambda = 1\,$Mpc and hence use $B_{1\,\mathrm{Mpc}}$ as the parameter.

The components of the energy momentum tensor of PMF source all types
of linear cosmological perturbations, i.e., scalar, vector, and tensor.
 In particular, the source terms are given
by the magnetic energy density and anisotropic stress for scalar
magnetized perturbations,  whereas vector and tensor modes are sourced
only by the magnetic anisotropic stress.  In addition, both scalar and
vector perturbations are affected by the Lorentz force; PMF induce a
Lorentz force on baryons modifying their evolution and in particular
their velocity, but during the tight-coupling regime between matter
and radiation the Lorentz force also has an indirect effect on
photons.

For the computation of magnetized angular power spectra, we use the analytic approximations for
the PMF energy-momentum tensor components given in \citet{Paoletti:2010rx}.
We consider here the regular mode for magnetic scalar perturbations, with the initial conditions 
of \citet{Paoletti:2008ck} (see \citealt{Giovannini:2004aw} for earlier calculations) and  \citet{Shaw:2009nf} (which describes
the singular passive mode, depending on the generation time of PMF).

Previous analyses show that the main impact of PMF on the CMB anisotropy angular power spectrum is 
at small angular scales, well into  the Silk damping regime. 
The dominant mode is the magnetic vector mode which peaks at $\ell\sim
2000$--3000 \citep{Mack:2001gc, Lewis:2004kg}.
The scalar magnetic mode is the dominant PMF contribution on large and
intermediate angular scales \citep{Giovannini:2007ks,Giovannini:2007qn,  Finelli:2008xh}.
The tensor contribution is always subdominant with respect
to the other two and it is negligible for the purposes of this analysis.

We include the scalar and vector magnetized contributions to the
angular power spectrum within the MCMC analysis  to derive the
constraints on the PMF amplitude and spectral index using \Planck\
$TT$ data.  We vary the magnetic parameters
$B_{1\,\mathrm{Mpc}}/{\rm nG}$ and $n_B$, in addition to the other
cosmological parameters of the base \lcdm\ cosmology (this analysis
assumes massless neutrinos,  rather than the default value of a single
eigenstate of mass $0.06\ {\rm eV}$ used in the rest of this
paper).  We adopt as prior ranges for the parameters $\left[0 \,, 10
\right]$ for $B_{1\,\mathrm{Mpc}}/{\rm nG}$ and $\left[ -2.99 \,, 3
\right]$ for the spectral index $n_B$.  The lower bound $
  n_B> -3$ is necessary to avoid infrared divergences in the PMF
energy momentum tensor correlators.
 
\begin{figure}
\includegraphics[width=8.8cm]{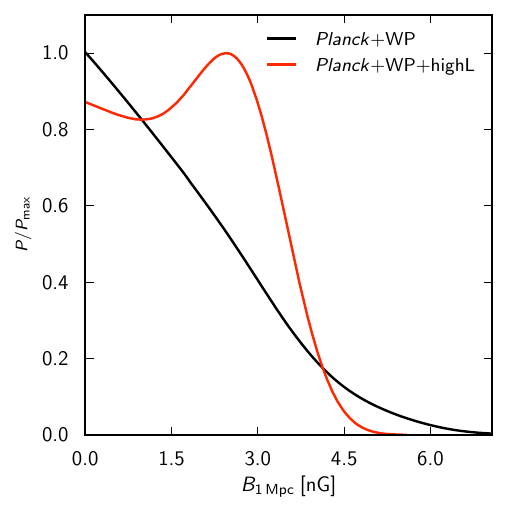}
\caption{Constraints on the root-mean-square amplitude of the primordial magnetic field (for a smoothing scale of $1\,{\rm Mpc}$)
obtained with \Planck+\WP\ (black) and \Planck+\WP+highL (red). }
\label{PlanckACTSPT}
\end{figure}

We perform analyses with \Planck+\WP\ and \Planck+\WP+\highL\
likelihood combinations. Results are shown in Fig.~\ref{PlanckACTSPT}.
We find that the cosmological parameters are in agreement with those
estimated assuming no PMF, confirming that the magnetic parameters are
not degenerate with the cosmological parameters of the base \lcdm\ model.  
The constraints on PMF
with the \Planck+\WP\ likelihood are $B_{1\,\mathrm{Mpc}}<4.1\,$nG, with a
preference for negative spectral indices at the $95\%$ confidence
level. These limits are improved using \Planck+\WP+\highL\
to $B_{1\,\mathrm{Mpc}}<3.4\,$nG with $n_B < 0$ preferred at
the $95\%$ confidence level. The new constraints are consistent with, and
slightly tighter, than previous limits based on combining \textit{WMAP}-7
data with high-resolution CMB data \citep[see e.g.][]{Paoletti:2010rx, 
Shaw:2010ea, Paoletti:2012bb}.



\subsection{Constraints on variation of the fine-structure constant}
\label{subsec:alpha}

\begin{table*}[tmb]                 
\begingroup
\newdimen\tblskip \tblskip=5pt
\caption{Constraints on the cosmological parameters of the base \lcdm\ model with the addition
of a varying fine-structure constant. We quote $\pm 1\,\sigma$ errors.
 Note that for \textit{WMAP}  there is a strong degeneracy
between $H_0$ and $\alpha$, which is why the error on $\alpha/\alpha_0$ is 
much larger than for
\Planck.}                          
\label{table:alpha-tab1}                            
\nointerlineskip
\vskip -3mm
\footnotesize
\setbox\tablebox=\vbox{
   \newdimen\digitwidth 
   \setbox0=\hbox{\rm 0} 
   \digitwidth=\wd0 
   \catcode`*=\active 
   \def*{\kern\digitwidth}
   \newdimen\signwidth 
   \setbox0=\hbox{+} 
   \signwidth=\wd0 
   \catcode`!=\active 
   \def!{\kern\signwidth}
\halign{\hbox to 0.9in{#\leaderfil}\tabskip=1.5em&
 \hfil#\hfil \tabskip 1em&\hfil#\hfil
 \tabskip 1em&\hfil#\hfil\tabskip 0pt\cr                
\noalign{\doubleline}
\omit& \Planck+\WP& \Planck+\WP+BAO& \textit{WMAP}-9\cr
\noalign{\vskip 3pt\hrule\vskip 3pt}
$\Omega_{\rm b}h^2$& $0.02206\pm0.00028$& $0.02220\pm0.00025$&
 $0.02309\pm0.00130$\cr
$\Omega_{\rm c}h^2$& $0.1174\pm0.0030$& $0.1161\pm0.0028$&
 $0.1148\pm0.0048$\cr
$\tau$& $0.095\pm0.014$& $0.097\pm0.014$& $0.089\pm0.014$\cr
$H_0$& $65.2\pm1.8*$& $66.7\pm1.1*$& $74\pm11$\cr
$n_{\rm s}$& $0.974\pm0.012$& $0.975\pm0.012$& $0.973\pm0.014$\cr
$\log(10^{10} A_{\rm s})$& $3.106\pm0.029$& $3.100\pm0.029$&
 $3.090\pm0.039$\cr
$\alpha/\alpha_0$& $0.9936\pm0.0043$& $0.9989\pm0.0037$& $1.008\pm0.020$\cr
\noalign{\vskip 3pt\hrule\vskip 5pt}
}
}
\endPlancktablewide                 
\endgroup
\end{table*}                        

\begin{figure*}[htb]
\begin{center}
\includegraphics[scale=0.36]{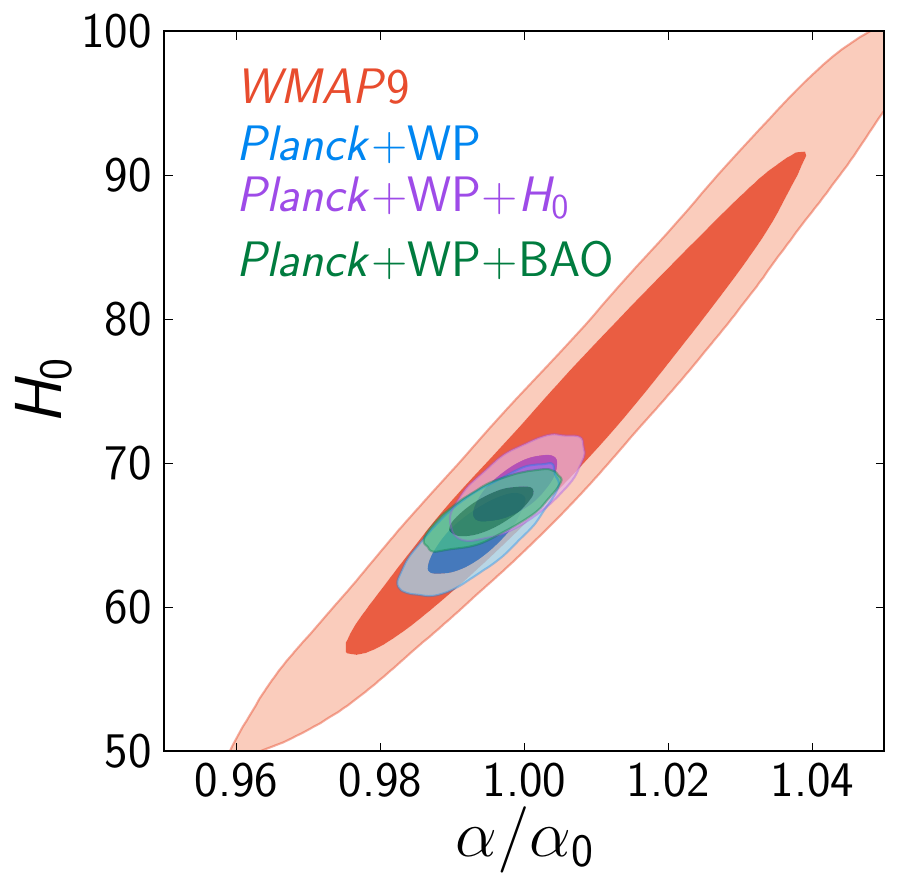}\includegraphics[scale=0.36]{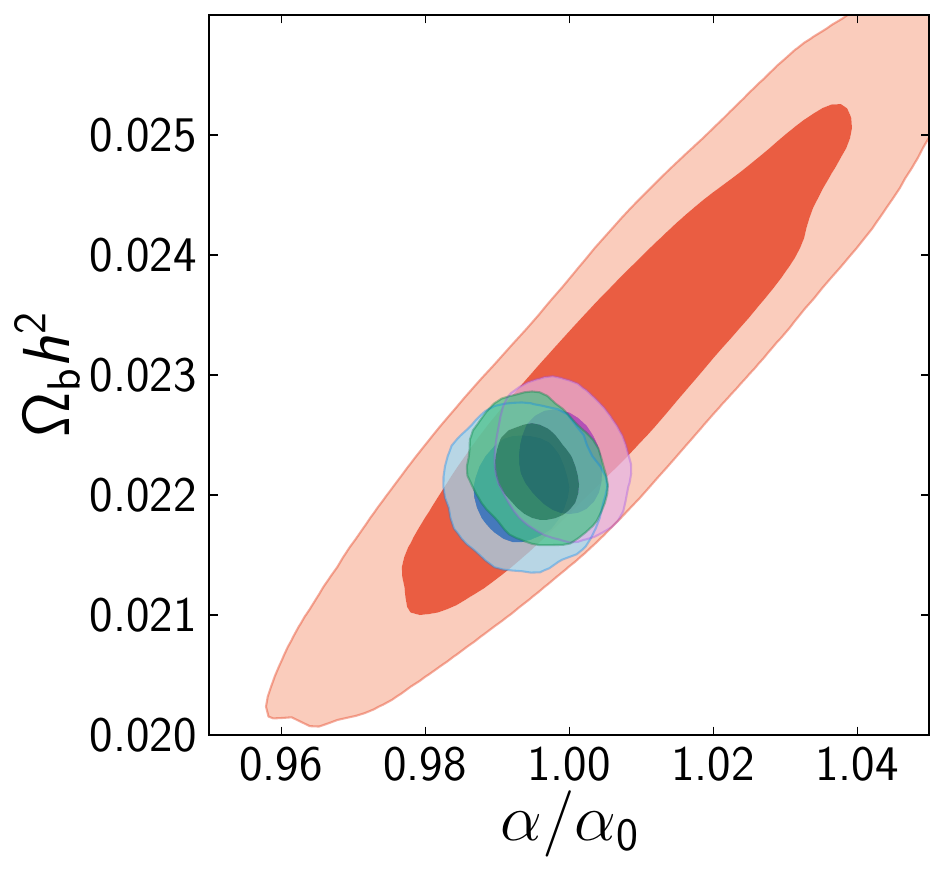}\includegraphics[scale=0.36]{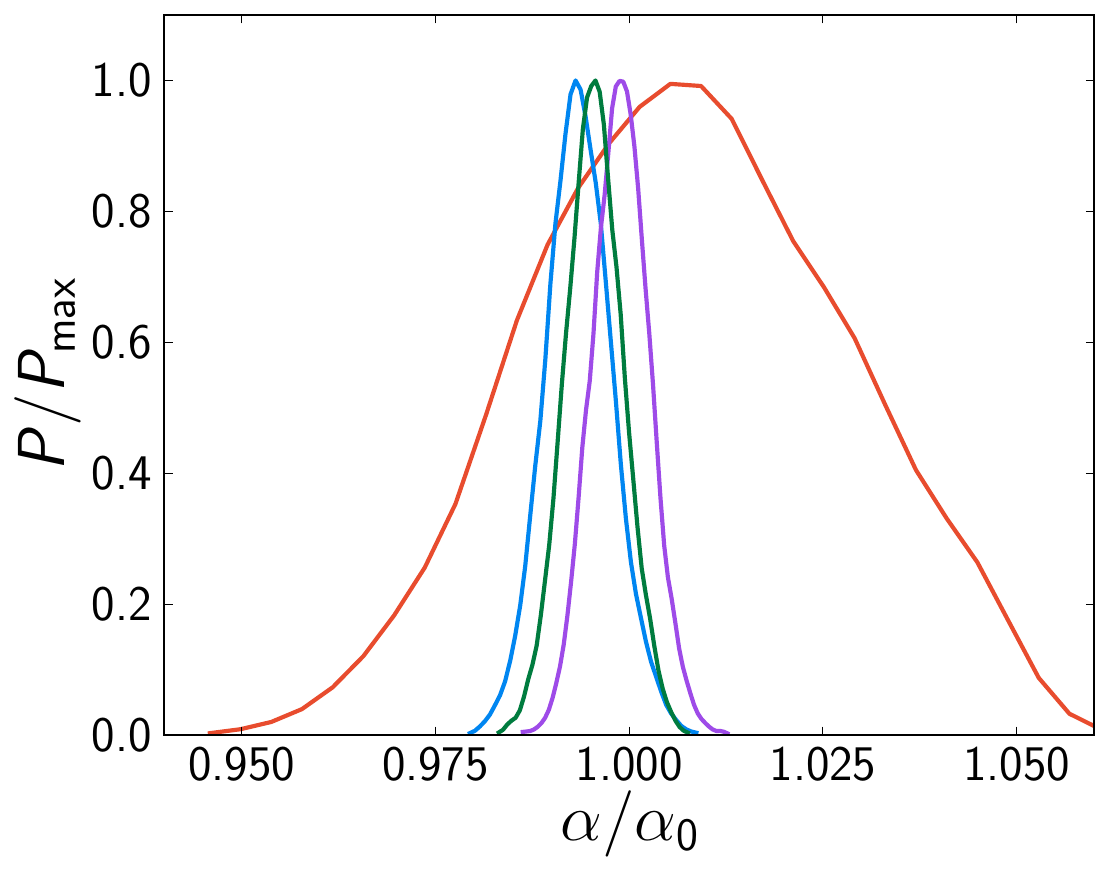}
\caption{\emph{Left}: Likelihood contours (68\% and 95\%) in the $\alpha/\alpha_0$--$H_0$ plane for
the \textit{WMAP}-9 (red), \Planck+\WP\ (blue), \Planck+\WP+$H_0$ (purple),
and \Planck+\WP+BAO (green) data combinations. \emph{Middle}: As left, but in the
$\alpha/\alpha_0$--$\Omega_{\rm b}h^{2}$ plane.
\emph{Right}: Marginalized posterior distributions
of $\alpha/\alpha_0$ for these data combinations.}
\label{ceff-h0}
\end{center}
\end{figure*}

The $\Lambda$CDM model
assumes the validity of General Relativity on cosmological scales, as
well as the physics of the standard model of particle physics.  One
possible extension, which may have motivations in fundamental physics, is
to consider variations of dimensionless constants.  Such variations can
be constrained through tests on astrophysical scales
\citep{jpu-revue,jpu-llr}.

A number of physical systems have been used, spanning different time
scales, to set constraints on  variations of the fundamental
constants. These range  from atomic 
clocks in the laboratory at a redshift $z=0$ to BBN
at $z\sim10^8$.  However, apart from the claims of varying $\alpha$ based on
high resolution
quasar absorption-line spectra ~\citep{webb01, murphyalpha}\footnote{
See however \cite{vlt1,vlt2}.}, there is no
other evidence for time-variable fundamental constants.

CMB temperature anisotropies have
been used extensively to constrain the variation of fundamental constants
over cosmic timescales. The temperature power spectrum is
sensitive to the variation of the fine-structure constant $\alpha$,
the electron-to-proton mass ratio $\mu$, and
the gravitational constant $\alpha_{\rm g}\equiv Gm_{\rm p}^2/\hbar c$.
A variation of $G$ can affect the Friedmann equation, and also raises the
issue of consistency in the overall theory of gravity.  
However, a variation of 
the non-gravitational constants ($\alpha$ and $m_{\rm e}$) is more
straightforward to analyse, mostly inducing a modification of the interaction between
light and atoms (shifts in the energy levels and binding energy of hydrogen
and helium). This induces a modification of the ionization history of the
Universe. In particular, a variation of $\alpha$  modifies the redshift
of recombination through the shift in the energy levels and  the Thomson
scattering cross-section. An increase in $\alpha$ induces a
shift of the position of the first acoustic peak, which
is inversely proportional to the sound horizon at last scattering.
The larger redshift of last scattering also produces
a larger early ISW effect, and hence a higher
amplitude of the first acoustic peak. 
Finally, an increase in $\alpha$ 
decreases diffusive damping at high multipoles. For earlier studies
of varying constants using the CMB \citep[see e.g.,][]{cmb-kap,cmb-avelino00, 
cmb-rocha1,cmb1-rocha2,wmap-alpha, naka00, Menegoni1, scoccola3}.

The analysis presented here 
focuses solely on the time variation of the fine-structure constant $\alpha$,
in addition to the parameters of the base \lcdm\ model,  using a 
modified form of the {\tt RECFAST} recombination code~\citep{cmb-han,cmb-rocha1,cmb1-rocha2}.
Selected results  are given in Table~\ref{table:alpha-tab1}, which
compares parameter
constraints from \Planck\ and from our own analysis of
the full \textit{WMAP}-9 $TT$, $TE$ and $EE$ likelihood.
From CMB data alone, \Planck\ improves the
constraints from a $2\%$ variation in $\alpha$ to about $0.4\%$.
\Planck\ thus improves the limit by a factor of around five, while the
constraints on the parameters of the base \lcdm\ model
 change very little with the addition of a time-varying  $\alpha$.  These results are in 
good agreement with earlier forecasts \citep{cmb1-rocha2}.

Given the apparent tension between the base \lcdm\ parameters
from \planck\ and direct measurements of  $H_0$  discussed in Sect.~\ref{sec:hubble}), we include further
information from the $H_0$ prior~ and BAO data (see Sect.~\ref{sec:BAO}).
Figure~\ref{ceff-h0}  compares the constraints in the 
($\alpha/\alpha_0,H_0$) and ($\alpha/\alpha_0,\Omega_{\rm b}h^{2}$) planes
and also shows the marginalized posterior distribution of
$\alpha/\alpha_0$ for the various data combinations.

The constraint on $\alpha$ is slightly improved by including the BAO
data (via a tightening of the parameters of the base \lcdm\ model).
Note that the central value of the prior on $H_0$ is outside the
95\% confidence region, even for the \Planck+\WP+$H_0$ combination.
Adding a varying $\alpha$ does not resolve the tension between 
direct measurements of $H_0$ and the value determined from the CMB.

In summary, \Planck\ data  improve the constraints on $\alpha/\alpha_0$,
with respect to those from \textit{WMAP}-9 by a factor of about five.
Our analysis of \Planck\ data limits any variation in the fine-structure constant from
$z\sim 10^3$ to the present day to be less than approximately $0.4\%$.

\section{Discussion and conclusions\footnote{Unless otherwise stated,  we quote 68\% confidence
   limits in this section for the \planck+WP+\highL\  data
   combination.}} \label{sec:conclusions}

The most important conclusion from this paper is the excellent
agreement between the \planck\ temperature power spectrum at high multipoles
with the predictions of the base \lcdm\ model. The base \lcdm\ model
also provides a good match to the \planck\ power spectrum of the
lensing potential, $C^{\phi\phi}_\ell$, and to the $TE$ and $EE$ power
spectra at high multipoles.

 The high statistical significance of the \planck\ detection of
 gravitational lensing of the CMB leads to some interesting science
 conclusions using \planck\ data alone. For example, gravitational
 lensing breaks the ``geometrical degeneracy'' and we find that the
 geometry of the Universe is consistent with spatial flatness to
 percent-level precision {\it using CMB data alone}. The
 \planck\ lensing power spectrum also leads to an interesting
 constraint on the reionization optical depth of $\tau = 0.089\pm
 0.032$, independent of CMB polarization measurements at low
 multipoles.

 The parameters of the base \lcdm\ model are determined to extremely
 high precision by the \planck\ data. For example, the scalar spectral
 index is determined as $\ns=0.9585 \pm 0.0070$, a $6\,\sigma$ deviation
 from exact scale invariance. Even in the base \lcdm\ model, we find
 quite large changes in some parameters compared to previous CMB
 experiments\footnote{The tension between the \planck\ and  SPT S12 results is
   discussed in detail in Appendix~\ref{app:spt}.}. In particular,
 from \planck\ we find a low value of the Hubble constant, $H_0 = (67.3
 \pm 1.2)\, {\rm km}\,{\rm s}^{-1}\,{\rm Mpc}^{-1}$, and a high matter
 density, $\Omega_{\rm m}=0.315 \pm 0.016$. If we accept that the base
 \lcdm\ model is the correct cosmology, then as discussed in Sect.
 \ref{sec:datasets} \planck\ is in tension with direct measurements of
 the Hubble constant (at about the $2.5\,\sigma$ level) and in mild
 tension with the SNLS Type Ia supernova compilation (at about the
 $2\,\sigma$ level). For the base \lcdm\ model, we also find a high
 amplitude for the present-day matter fluctuations, $\sigma_8 = 0.828
 \pm 0.012$, in agreement with previous CMB experiments. This value is
 higher than that inferred from counts of rich clusters of galaxies,
 including our own analysis of \planck\ cluster counts
 \citep{planck2013-p15}, \referee{and in tension with the cosmic shear measurements
discussed in Sect.~\ref{subsec:cosmicshear}.}

 One possible interpretation of these tensions is that 
 systematic errors are not completely understood in some astrophysical
 measurements. The fact that the \planck\ results for the base \lcdm\
 model are in such good agreement with BAO data, which are based on a
 simple geometrical measurement, lends support to this view. An
 alternative explanation is that the base \lcdm\ model is incorrect. In
 summary, at high multipoles, the base \lcdm\ cosmology provides an
 excellent fit to the spectra from \planck, ACT and SPT (for all
 frequency combinations), as illustrated in
 Figs. \ref{PlanckandHighL}, \ref{SPT} and~\ref{ACT}, but the
 parameters derived from the CMB apparently conflict with some types of
 astrophysical measurement.

\begin{figure*}
\centering
\hspace{-3mm}\includegraphics[width=95mm,angle=0]{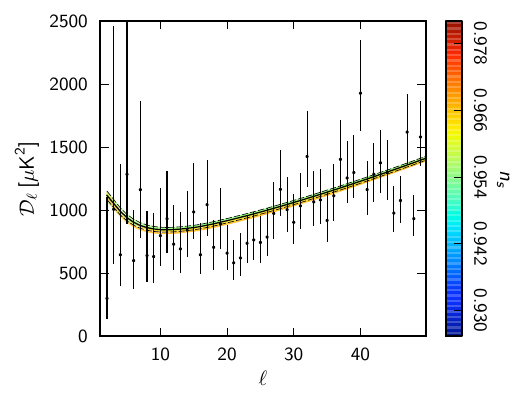}
\hspace{0mm}\raisebox{3.5mm}{\includegraphics[width=84mm,angle=0]{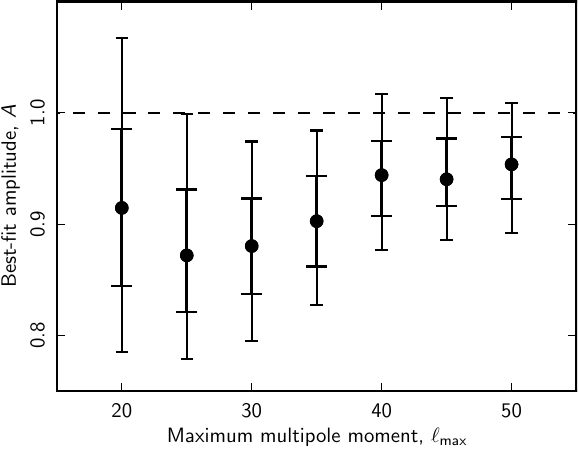}}
\caption {\emph{Left}: \planck\ $TT$ spectrum at
low multipoles with $68\%$ ranges on the posteriors.  The ``rainbow'' band show the
best fits to the entire \planck+\WP+\highL\ likelihood for the base \lcdm\
cosmology, colour-coded according to the value of the 
scalar spectral index $n_{\rm s}$. \emph{Right}: Limits (68\% and 95\%) on the relative amplitude of the base \lcdm\ 
fits to the \planck+\WP\ likelihood 
{\it fitted only to the \planck\ $TT$ likelihood} over the multipole range
$2 \le \ell \le \ell_{\rm max}$.}
\label{lowellfunnies}
\end{figure*}

Before summarizing our results on extensions to the base \lcdm\ model,
it is worth making some remarks on foreground modelling and the impact
of this modelling on our error estimates.  The addition of CMB data at
high multipoles helps to constrain the model of unresolved
foregrounds, in particular, the contribution from ``minor''
components, such as the kinetic SZ, which are poorly constrained
from \planck\ alone. For the base \lcdm\ model, the cosmological
parameters are not limited by foreground modelling\footnote{Even in the restricted case of the
  base \lcdm\ model, parameters can shift as a result of small changes
  to the theoretical assumptions. An example is given in Sect.
  \ref{subsec:LCDM_hubble}, where we show that changing from our
  default assumption of $\sum m_\nu = 0.06\, {\rm eV}$ to $\sum m_\nu
  = 0$, causes an upward shift of $0.4\,\sigma$ in the value of
  $H_0$.}, as
illustrated in Fig.~\ref{fig:margeLCDM}.
As discussed in Appendix \ref{app:test},
foreground modelling becomes more important in analysing
extended CDM models, particularly those that have strong parameter
degeneracies that are broken only via precision measurements of the
damping tail in the CMB spectrum.  As a crude measure of the
importance of foreground modelling, we can compare parameter values
with and without inclusion of the ACT and SPT data at high
multipoles. A large shift in parameter values indicates a possible
sensitivity to foreground modelling, and so any such  result should be
treated with caution.  We have thus normally adopted the
\planck+\WP+\highL\ likelihood combination as offering the most
reliable results for extensions to the base \lcdm\ cosmology.

{\it From an analysis of an extensive grid of models, we find no
  strong evidence to favour any extension to the base \lcdm\
  cosmology, either from the CMB temperature power spectrum alone, or
  in combination with the \planck\ lensing power spectrum and other
  astrophysical data sets.}

We find the following notable results using CMB data alone:
\begin{itemize}
\item The deviation of the scalar spectral index  from unity is robust
to the addition of tensor modes and to changes in the
 matter content of the Universe. For example, adding a tensor component
we find $\ns = 0.9600 \pm 0.0072$, a $5.5\,\sigma$ 
departure from $\ns=1$. 
\item A  95\% upper limit
on the tensor-to-scalar ratio of $r_{0.002} < 0.11$. The combined contraints
on $\ns$ and $r_{0.002}$ are on the borderline of compatibility with single-field inflation with  a quadratic potential (Fig.~\ref{fig:ns_corr}).
\item A 95\% upper limit
on the summed neutrino mass of $\sum m_\nu < 0.66 \, {\rm eV}$.
\item A determination of the effective number of neutrino-like
relativistic degrees of freedom of $N_{\rm eff} = 3.36 \pm 0.34$,
compatible with the standard value of $3.046$. 
\item The results from \planck\ are consistent with the results
of standard big bang nucleosynthesis. In fact,  combining the CMB data
with the most recent results on the deuterium abundance,
leads to the constraint $N_{\rm eff} = 3.02 \pm 0.27$, again compatible
with the standard value of $3.046$. 
\item New limits on a possible variation of the fine-structure constant, dark matter
annihilation and primordial magnetic fields.
\end{itemize}

We also find a number of marginal (around $2\,\sigma$) results, perhaps
indicative of internal tension within the \planck\ data. Examples
include the preference of the (phenomenological) lensing parameter for
values greater than unity ($\Alens = 1.23 \pm 0.11$; Eq.\
\ref{Alens}) and for negative running ($\nrun=-0.015 \pm 0.09$;
Eq.\ \ref{nrunplanckhighL}). In \citet{planck2013-p17}, the
\planck\ data indicate a preference for anti-correlated isocurvature
modes and for models with a truncated power spectrum on large
scales. None of these results have a decisive level of statistical
significance, but they can all be traced to an unusual aspect of the
temperature power spectrum at low multipoles. As can be seen in
Fig. \ref{TTspec}, and on an expanded scale in the left-hand panel of
Fig. \ref{lowellfunnies}, the measured power spectrum shows a dip
relative to the best-fit base \lcdm\ cosmology in the multipole range
$20 \la \ell \la 30$ and an excess at $\ell=40$.  The existence
of ``glitches'' in the power spectrum at low multipoles was noted by the
\WMAP\ team in the first-year papers \citep{Hinshaw:03, Spergel:03}
and acted as motivation to fit an inflation model with a step-like
feature in the potential \citep{Peiris:03}.  Similar investigations have been carried
out by a number of authors, \citep[see e.g.,][and references
therein]{Mortonson:09}.  At these low multipoles, the \planck\
spectrum is in excellent agreement with the \WMAP\ nine-year spectrum
\citep{planck2013-p08}, so it is unlikely that any of the features
such as the low quadrupole or ``dip'' in the multipole range $20$--$30$
are caused by instrumental effects or Galactic foregrounds. {\it These
  are real features of the CMB anisotropies.}

The \planck\ data, however, constrain the parameters of the base
\lcdm\ model to such high precision that there is little
remaining flexibility to fit the low-multipole part of the spectrum.
To illustrate this point, the right-hand panel of
Fig. \ref{lowellfunnies} shows the 68\% and 95\% limits on the
relative amplitude of the base \lcdm\ model (sampling the chains
constrained by the full likelihood) fitted only to the \planck\ $TT$
likelihood over the multipole range $2 \le \ell \le \ell_{\rm max}$.
From multipoles $\ell_{\rm max} \approx 25$ to multipoles $\ell_{\rm max}
 \approx 35$, we
see more than a $2\,\sigma$ departure from values of unity. (The maximum
deviation from unity is $2.7\,\sigma$ at $\ell = 30$.)  It is difficult
to know what to make of this result, and we present it here as a
``curiosity'' that needs further investigation.  The \planck\
temperature data are remarkably consistent with the predictions of the
base \lcdm\ model at high multipoles, but it is also conceivable that
the \lcdm\ cosmology fails at low multipoles.  There are other
indications, from both \WMAP\ and \planck\ data for ``anomalies'' at
low multipoles \citep{planck2013-p09}, that may be indicative of new
physics operating on the largest scales in our Universe.
Interpretation of large-scale anomalies (including the results shown
in Fig. \ref{lowellfunnies}) is difficult in the absence of a
theoretical framework.  The problem here is assessing the role of {\it
  a posteriori} choices, i.e., that inconsistencies attract our
attention and influence our choice of statistical test.  Nevertheless,
we know so little about the physics of the early Universe that we
should be open to the possibility that there is new physics beyond
that assumed in the base \lcdm\ model. Irrespective of the
interpretation, the unusual shape of the low multipole spectrum is at
least partly responsible for some of the $2\,\sigma$ effects seen in the
analysis of extensions to the \lcdm\ model discussed in
Sect. \ref{sec:grid}.

Supplementary information from astrophysical data sets has played an
important role in the analysis of all previous CMB experiments.  For
\planck\ the interpretation of results combined with non-CMB data sets
is not straightforward (as a consequence of the tensions discussed in
Sect. \ref{sec:datasets}).  For the base \lcdm\ model, the statistical
power of the \planck\ data is so high that we find very similar
cosmological parameters if we add the \citet{Riess:2011yx} constraint
on $H_0$, or either of the two SNe samples, to those derived from the
CMB data alone.  In these cases, the solutions simply reflect the
tensions discussed in Sect. \ref{sec:datasets}, for example, including the
$H_0$ measurement with the \planck+\WP\ likelihood
we find $H_0 = (68.6 \pm 1.2) \, {\rm km}\,{\rm s}^{-1}\,{\rm
  Mpc}^{-1}$, discrepant with the direct measurement at the
 $2.2\,\sigma$ level. 

The interpretation becomes more complex for extended models where
astrophysical data is required to constrain parameters that cannot be
determined accurately from CMB measurements alone. As an example, it
is well known that CMB data alone provide weak constraints on the dark
energy equation of state parameter $w$ (see Fig. \ref{fig:wpost}). The
addition of BAO data to the CMB data gives a tight constraint of $w =
-1.13 \pm 0.12$\footnote{The addition of the \planck\ lensing
  measurements tightens this further to $w=-1.08^{+0.11}_{-0.086}$.}.
However, adding the SNLS SNe data gives $w=-1.135 \pm 0.069$ and
adding the $H_0$ measurement gives $w=-1.244 \pm 0.095$. Adding either
of the two data sets which show tension with the CMB measurements for
the base \lcdm\ model,  draws the solutions into the phantom domain ($w
< -1$) at about the $2\,\sigma$ level. In contrast, if we use the BAO
data in addition to the CMB, we find no evidence for dynamical dark
energy;  these data are compatible with a cosmological constant, as
assumed in the base \lcdm\ model.

The impact of additional astrophysical data is particularly complex in
our investigation of neutrino physics (Sect. \ref{sec:neutrino}).  We
will use the effective number of relativistic degrees of freedom,
$N_{\rm eff}$ as an illustration. From the CMB data alone, we find
$N_{\rm eff}=3.36 \pm 0.34$.  Adding BAO data gives $N_{\rm eff} =
3.30 \pm 0.27$. Both of these values are consistent with the standard
value of $3.046$. Adding the $H_0$ measurement to the CMB data gives
$N_{\rm eff} = 3.62 \pm 0.25$ {\it and relieves the tension between
  the CMB data and $H_0$} at the expense of new neutrino-like physics
(at around the $2.3\,\sigma$ level). It is possible to alleviate the
tensions between the CMB, BAO, $H_0$ and SNLS data by invoking new
physics such as an increase in $N_{\rm eff}$. However, {\it none of
  these cases are  favoured significantly over the base \lcdm\ model by the \planck\
  data} (and they are often disfavoured). Any preference for new physics comes almost entirely from the
astrophysical data sets. It is up to the reader to decide how to
interpret such results, but it is simplistic to assume that
all astrophysical data sets have accurately quantified estimates of
systematic errors.  We have therefore tended to place greater weight on the
CMB and BAO measurements in this paper rather than  on  more complex
astrophysical data.

Our overall conclusion is that the \planck\ data are remarkably
consistent with the predictions of the base \lcdm\ cosmology.
However, the mismatch with the temperature spectrum at low multipoles,
evident in Figs. \ref{TTspec} and \ref{lowellfunnies}, and the
existence of other ``anomalies'' at low multipoles, is possibly
indicative that the model is incomplete.  The results presented here
are based on a first, and relatively conservative, analysis of the
\planck\ data. The 2014 data release will use data obtained over the
full mission lifetime of \planck, including polarization data. It
remains to be seen whether these data, together with new astrophysical
data sets and CMB polarization measurements, will offer any convincing
evidence for new physics.


\begin{acknowledgements}
The development of \Planck\ has been supported by: ESA; CNES and
CNRS/INSU-IN2P3-INP (France); ASI, CNR, and INAF (Italy); NASA and DoE (USA);
STFC and UKSA (UK); CSIC, MICINN and JA (Spain); Tekes, AoF and CSC (Finland);
DLR and MPG (Germany); CSA (Canada); DTU Space (Denmark); SER/SSO
(Switzerland); RCN (Norway); SFI (Ireland); FCT/MCTES (Portugal); and PRACE
(EU). A description of the \Planck\ Collaboration and a list of its members,
including the technical or scientific activities in which they have been
involved, can be found at
\url{http://www.sciops.esa.int/index.php?project=planck&page=Planck_Collaboration}. \referee{We thank the referee for a comprehensive and helpful report. We also
thank Jean-Philippe Uzan for his contributions to Sect.~\ref{subsec:alpha}.
We additionally acknowledge useful comments on the first version of this paper
from a large number of scientists who have helped improve the clarity of the
revised version. We mention specifically Jim Braatz, John Carlstrom, Alex
Conley, Raphael Flauger, Liz Humphreys, Adam Riess, Beth Reid, Uros Seljak,
David Spergel, Mark Sullivan, and Reynald Pain. }
\end{acknowledgements}

\bibliography{Planck_bib,introduction,final_params_paper,cosmomc,model_and_params,high_ell,datasets_lensing,datasets_BAO,datasets_H0,datasets_SNe,datasets_additional,grid,grid_DE,grid_early,grid_alpha,grid_PMF,conclusions,appendix_SPT,appendix_tests}{}
\bibliographystyle{aa_arxiv}

\appendix

\section{Comparison of the \planck\  and \textit{WMAP-}9 base \lcdm\ cosmologies}
\label{app:wmap}

\referee{The parameters for the base \lcdm\ cosmology derived from \planck\ differ from those derived from \textit{WMAP}-9. In this appendix,
we address the question of whether the parameter shifts are consistent statistically with the shifts expected from the additional multipole
coverage of \planck.}

\referee{We begin with a direct comparison of the shapes of the
  \planck\ and \textit{WMAP}-9 spectra.  Figure~\ref{wmapvplanck}
  shows our estimate of the combined V+W-band \textit{WMAP}-9 power
  spectrum\footnote{The spectrum is a combination of all of the
    cross-spectra computed from the nine-year coadded maps per
    differencing assembly. Cross-spectra are first combined by band
    into VV, VW and WW spectra and the beam corrected spectra are then
    corrected for unresolved point sources, i.e., a Poisson term is
    removed to minimise residuals with respect to the \WMAP\ best-fit
    \LCDM\ spectrum. The spectra are then coadded with inverse noise
    weighting to form a single V+W spectrum.  } computed on the same
  mask used for the $100\times100\,{\rm GHz}$ \planck\ spectrum in the
  main text.  Here we use a combined \textit{WMAP}+\planck\ mask for
  point sources. The magenta points show the \planck\ $100\times
  100$\,GHz spectrum corrected for extra-Galactic foregrounds with the
  best-fit \planck+\WP+\highL\ parameters from
  Table~\ref{LCDMForegroundparams}. The \textit{WMAP} points have been
  rescaled by a multiplicative factor of $0.974$ and agree to high
  precision point-by-point with the \planck\ spectrum. (Note that the
  errors plotted for the \WMAP\ points show the noise errors and the
  cross-term between signal and noise computed from Monte Carlo
  simulations; they do not include CMB-foreground cross-correlations
  and correlated beam errors.)  The rms scatter between the
  \planck\ and \textit{WMAP} points over the multipole range $50 \le
  \ell \le 400$ is only $16\,\mu {\rm K}^2$, i.e., after a
  multiplicative scaling the two spectra are consistent to within
  about 0.5\% of the primary CMB spectrum.  Similar tests are
  described in greater detail in \citet{planck2013-p01a}, including
  comparisons with the LFI $70$\,GHz spectrum. The reason for the
  multiplicative factor (amounting to a 1.3\% difference in the
  calibrations of the HFI and \textit{WMAP} maps) is not fully
  understood and is the subject of ongoing investigations.  For the
  purposes of this appendix, we treat the rescaling as an empirical
  result, i.e., after accounting for a multiplicative calibration
  factor, the \planck\ and \textit{WMAP}-9 power spectra agree to high
  precision, with little evidence for any significant variation of the
  spectra with multipole.}

\begin{figure}
\centering
\includegraphics[width=88mm,angle=0]{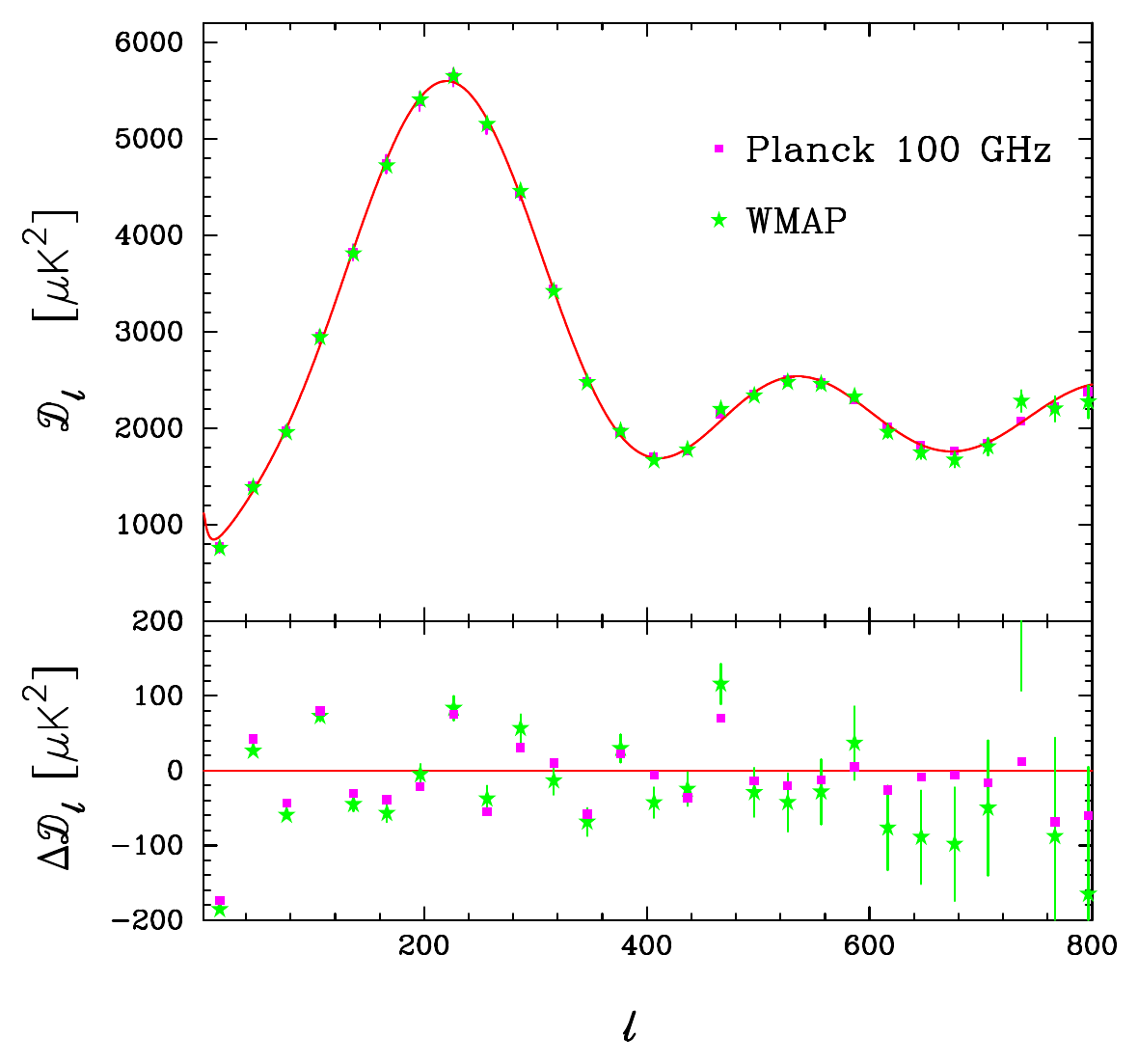}
\caption {\referee{Comparison of the \planck\ and \WMAP-9 power
    spectra.  The green points show the combined \WMAP-9 V+W-band
    spectrum computed on the same mask used for the
    $100\times100$\,GHz \planck\ spectrum (with a combined
    \textit{WMAP}+\planck\ mask for point sources) after rescaling the
    \WMAP\ power spectrum by a multiplicative factor of $0.974$. The
    magenta points show the \planck\ $100\times 100$\,GHz spectrum
    computed on the same mask. The red line shows the best-fit
    \planck+\WP+\highL\ base \lcdm\ model. The lower panel shows the
    residuals with respect to this model.  The error bars on the
    \textit{WMAP} points show the instrumental noise together with the
    noise-signal errors as discussed in the text; errors are not shown
    for \planck.}}
\label{wmapvplanck}
\end{figure}

\begin{table*}[tmb]                 
\begingroup
\newdimen\tblskip \tblskip=5pt
\caption{\referee{Comparison of base \LCDM\ parameters from \WMAP-9 with \planck. The
second column gives parameters derived from the \WMAP-9 likelihood. The third
column gives results for \planck+WP, with the 
\planck\ likelihood restricted to multipoles $\ell \le 1000$. The fourth and
fifth columns show results for the full \planck+WP and \WMAP-9 likelihoods
combined with the BAO data discussed in Sect. \ref{sec:BAO}.  As in the main body of the paper, we have assumed a neutrino mass of $0.06\; {\rm eV}$. }   }                       
\label{WMAPparams}         
\nointerlineskip
\vskip -3mm
\footnotesize
\setbox\tablebox=\vbox{
    \newdimen\digitwidth
    \setbox0=\hbox{\rm 0}
    \digitwidth=\wd0
    \catcode`"=\active
    \def"{\kern\digitwidth}
    \newdimen\signwidth
    \setbox0=\hbox{+}
    \signwidth=\wd0
    \catcode`!=\active
    \def!{\kern\signwidth}
\halign{
\hbox to 0.9in{$#$\leaderfil}\tabskip=1.5em&\hfil$#$\hfil&\hfil$#$\hfil&\hfil$#$\hfil&\hfil$#$\hfil\tabskip=0pt\cr
\noalign{\doubleline}
\multispan1\hfil \hfil&\multispan1\hfil \WMAP-9\hfil&\multispan1\hfil \emph{Planck} $l \leq 1000$+WP\hfil&\multispan1\hfil \emph{Planck}+WP+BAO\hfil&\multispan1\hfil \WMAP-9+BAO\hfil\cr
\omit\hfil Parameter\hfil&\omit\hfil 68\% limits\hfil&\omit\hfil 68\% limits\hfil&\omit\hfil 68\% limits\hfil&\omit\hfil 68\% limits\hfil\cr
\noalign{\vskip 3pt\hrule\vskip 5pt}
\Omega_{\mathrm{b}} h^2&0.02265\pm 0.00051&0.02256\pm 0.00044&0.02212\pm 0.00025&0.02249\pm 0.00044\cr
\noalign{\vskip 2pt}
\Omega_{\mathrm{c}} h^2&0.1137\pm 0.0046&0.1142\pm 0.0035&0.1187\pm 0.0017&0.1160\pm 0.0025\cr
\noalign{\vskip 2pt}
100\theta_{\mathrm{MC}}&1.0402\pm 0.0023&1.0411\pm 0.0011&1.04146\pm 0.00057&1.0396\pm 0.0021\cr
\noalign{\vskip 2pt}
\tau&0.089^{+0.013}_{-0.015}&0.091^{+0.013}_{-0.015}&0.091\pm 0.013&0.086^{+0.012}_{-0.014}\cr
\noalign{\vskip 2pt}
n_\mathrm{s}&0.974\pm 0.013&0.977\pm 0.012&0.9629\pm 0.0057&0.969\pm 0.010\cr
\noalign{\vskip 2pt}
\ln(10^{10} A_\mathrm{s})&3.092\pm 0.031&3.080\pm 0.027&3.090\pm 0.025&3.093\pm 0.030\cr
\noalign{\vskip 2pt}
\Omega_\Lambda&0.717^{+0.028}_{-0.024}&0.717^{+0.023}_{-0.020}&0.692\pm 0.010&0.703\pm 0.012\cr
\noalign{\vskip 2pt}
\Omega_{\mathrm{m}}&0.283^{+0.024}_{-0.028}&0.283^{+0.020}_{-0.023}&0.308\pm 0.010&0.297\pm 0.012\cr
\noalign{\vskip 2pt}
\sigma_8&0.808\pm 0.023&0.807\pm 0.014&0.826^{+0.011}_{-0.012}&0.816\pm 0.018\cr
\noalign{\vskip 2pt}
H_0&69.7\pm 2.2"&69.7\pm 1.8"&67.79\pm 0.78"&68.45\pm 0.96"\cr
\mathrm{Age}/\mathrm{Gyr}&13.76\pm 0.11"&13.744\pm 0.085"&13.800\pm 0.038"&13.807\pm 0.090"\cr
\noalign{\vskip 5pt\hrule\vskip 3pt}
} 
} 
\endPlancktablewide
\endgroup
\end{table*}

\referee{Given the agreement between the \WMAP-9 and \planck\ spectra shown
in Fig. \ref{wmapvplanck}, we should expect the two experiments to give
similar cosmological parameters if the multipole range of \planck\ is
restricted to $\ell \la 1000$. This is illustrated by the results of
Table~\ref{WMAPparams}, which lists base \LCDM\ parameters for \WMAP-9
and for the \planck+\WP\ likelihood limited to a maximum multipole of
$\ell_{\rm max}=1000$.  (For this restricted multipole range, we keep
the foreground and other nuisance parameters fixed to the best-fit
values derived from the full \planck+\WP\ likelihood.) The cosmological
parameters derived from these two likelihoods are in very good agreement.
(See also \citet{planck2013-p08} and Appendix~\ref{app:test}
for further tests of the variations of cosmological parameters from \planck\
as $\ell_{\rm max}$ is varied.)}

\begin{figure}
\centering
\includegraphics[width=88mm,angle=0]{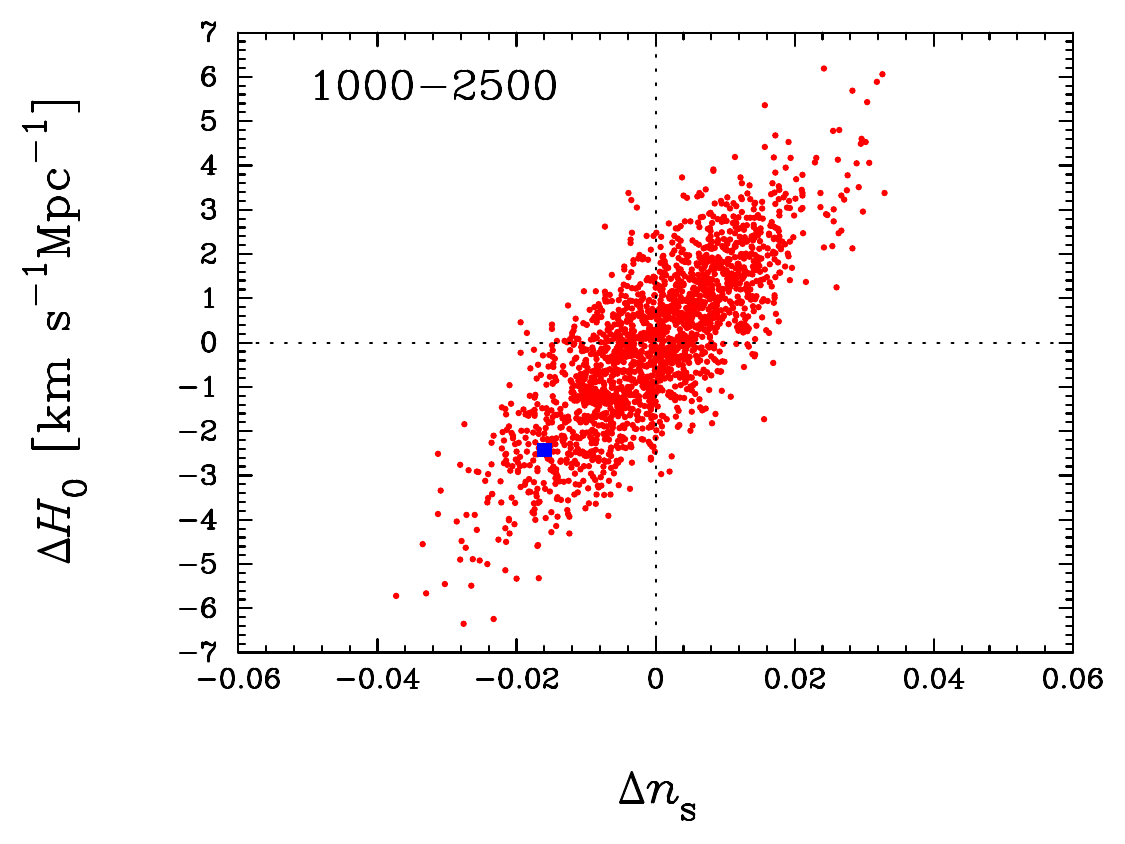}
\caption {\referee{Variations in $H_0$ and $\ns$ as the maximum multipole in the
\planck\ likelihood is increased from $\ell_{\rm max}=1000$ to $2500$. The red
points show the changes in parameters determined from
2000 simulations, as described in the text.
The blue point shows the changes determined from the real data.}}
\label{lmax_sim}
\vspace{-3mm}
\end{figure}

\referee{We should expect the best-fit cosmological parameters to change as the maximum
multipole $\ell_{\rm max}$ is increased, since there is additional cosmological
information at higher
multipoles. As a useful rule-of-thumb, the covariance of the shifts in the
best-fit parameters on adding further independent data should be approximately 
equal to the
difference in the parameter covariances.  To assess more carefully whether
the cosmological parameter shifts seen in the \planck\ analysis of \lcdm\
models are statistically reasonable,
we perform a set of Fisher-matrix-type simulations.
We draw Gaussian realizations of simulated spectra, 
$C^{\rm sim}_\ell$, from the frequency-compressed covariance matrix
$\hat{\mathcal{M}}_{\ell\ell^\prime}$, introduced in Eq.~(\ref{GF2}),
which includes contributions from beam and foreground errors. We
adopt the best-fitting base \LCDM\  model to the \planck+\WP+\highL\ data
as our fiducial model $C^{\rm fid}_\ell$ and form
\be
\chi^2 = \sum_{\ell\ell^\prime} \Delta C_\ell \hat{\mathcal{M}}^{-1}_{\ell \ell^{\prime}} \Delta C_{\ell^\prime} + {(\Delta \tau)^2 \over \sigma^2_\tau},  \label{FM1} 
\ee
where
\be
\Delta C_\ell = C^{\rm sim}_\ell - C^{\rm fid}_{\ell} - \sum_p 
{\partial C^{\rm fid}_\ell \over \partial a_p}
\Delta a_p,  \label{FM2}
\ee
and the $a_p$ are the cosmological parameters of the base model (taken
here to be $\As$, $\omb$, $\omc$, $H_0$, $\ns$ and $\tau$). Since
these simulations are based only on the high multipole \planck\
likelihood, we include a prior on $\tau$ in Eq.~(\ref{FM1}) with
$\sigma_\tau=0.014$.  In addition, since the covariance matrix
$\hat{{\cal M}}_{\ell \ell'}$ includes estimates of foreground and
beam errors, which are highly correlated over a wide multipole range,
we add a ``point source'' amplitude as a catch-all to model
uncertainties from nuisance parameters. With this machinery, we can
quickly calculate the parameter shifts $\Delta a_p$ that minimise the
$\chi^2$ in Eq.~(\ref{FM1}) for different choices of  $\ell_{\rm max}$.
(Note that these simulations reproduce to high precision the parameter errors
and degeneracy directions of the full \planck\ likelihood.)}

\referee{Results for $2000$ simulations are shown in Fig.~\ref{lmax_sim} in the
$H_0$--$\ns$ plane. (The results are similar for the
$\omega_b$--$\omega_c$ plane.) Each red point in Fig. \ref{lmax_sim}
shows the parameter shifts measured from a single simulation as
$\ell_{\rm max}$ is increased from 1000 to 2500.  The blue point shows
the shift in parameters for the real data.  The shifts seen in the
real data follow the degeneracy directions defined by the simulations
(in all parameters) and lie within $1.6\,\sigma$ of the dispersion of
the simulated parameter shifts for any single parameter. We therefore
conclude that the parameter shifts seen between \planck\ and \WMAP-9
are statistically consistent with our expectations based on the
further information contained in the power spectrum at high
multipoles.}

\referee{The last two columns in Table~\ref{WMAPparams} list the base \LCDM\
parameters for \planck+\WP+BAO and for \textit{WMAP}-9+BAO. Adding the
baryon acoustic oscillation data to \WMAP-9 brings the cosmological
parameters closer to the
\planck\ parameters (with or without the addition of the BAO data). This is
what we would expect if the \planck\ base \LCDM\ cosmology is correct and the
\planck, \WMAP-9 and BAO data are largely free of systematic errors.}

\section{Comparison of the \planck\  and SPT S12 base \lcdm\ cosmologies}
\label{app:spt}

\referee{The parameter values derived from \planck\ for the base \lcdm\ cosmology 
differ from those
inferred by combining S12 with \textit{WMAP}-7; e.g., the best-fit
values of $H_0$ and $\Omega_\Lambda$  differ by  $2.7\,\sigma$ and $3.2\,\sigma$
respectively, where $\sigma$ is the uncertainty in the
\textit{WMAP}-7+S12 determination.  Furthermore, in \citet[herefter H12,
a companion paper to S12]{hou12} a trend in the S12
band-powers was identifed relative to the best-fit base \lcdm\ spectrum,
which they tentatively reported as evidence for new physics.  This
again differs from the results of Sect.~\ref{sec:grid}, in which we found
that the \planck\ data provide no evidence for any new physics beyond
that incorporated in the base \LCDM\ model.  The purpose of this
appendix is to investigate (as far as we can)  the origin of these parameter differences
and to comment on the
trend identified in H12 in light of the more precise data we now have
from \planck.}

\referee{Note that the S12 result extends the earlier work of  K11 (a subset of which is used in the \highL\ data combination in the main body of this paper)
from an
analysis of $790\,{\rm deg}^2$ of sky to a total field area of $2540\,{\rm deg}^2$. S12 and H12 present constraints on the base \LCDM\ model
and extensions. Certain extended models are favoured when
\textit{WMAP}-7 and S12 are combined. For example, a running spectral
index is favoured over a constant spectral index at the $2.2\,\sigma$
level.}

\begin{figure}
\centering
\includegraphics[width=80mm,angle=0]{Figures_pdf/appendix_spt/pgBAOspt}
\caption{\referee{The acoustic scale distance ratio $r_{\rm s}/D_V(z)$ divided by
  the distance ratio of the best fit \textit{WMAP}-7+SPT base \LCDM\ cosmology of
  S12.  The points are colour coded as follows:
  green star (6dF); purple squares (SDSS DR7 as analysed by
  \citealt{Percival:10}); black star (SDSS DR7 as analysed by
  \citealt{Padmanabhan:2012hf}); blue cross (BOSS DR9); and blue circles
  (WiggleZ). Error bars show $1\,\sigma$ errors on the data points. The grey
  band shows the $\pm 1\,\sigma$ range allowed by the \textit{WMAP}-7+SPT data.}}
\label{BAOspt}
\end{figure}

\begin{figure*}
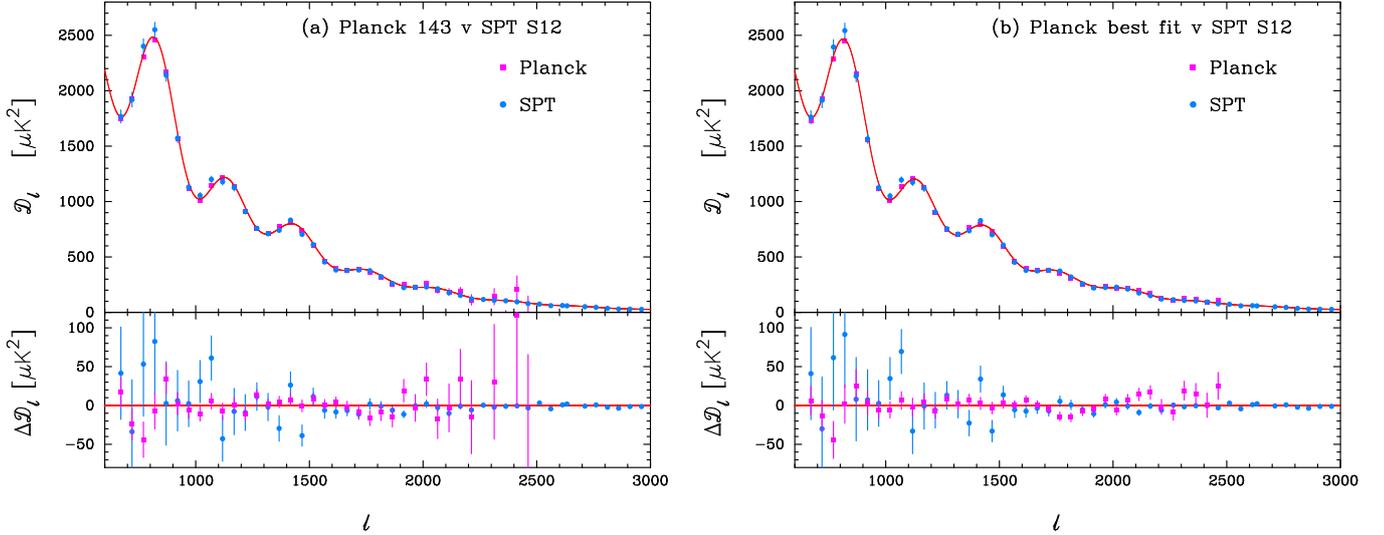

\centering
\includegraphics[width=90mm,angle=0]{Figures_pdf/appendix_spt/pg143S12}
\includegraphics[width=90mm,angle=0]{Figures_pdf/appendix_spt/pgPlanckS12}
\\
\caption {\referee{Fits to the joint likelihoods for \planck\ and SPT
  S12 spectra. (a) Fits using only the
  $143\times143\, {\rm GHz}$ spectrum in the \planck\ likelihood. The blue points
  show the SPT data after recalibration and foreground subtraction,
  using the best-fit solution from the joint likelihood analysis.  The
  magenta points show the foreground-subtracted \planck\ $143\times143\, {\rm GHz}$
  spectrum. The lower panel show the residuals with respect to the
  best-fit \LCDM\ model to the \planck+SPT combined likelihoods (shown
  by  the red line in the top panel) .  
  (b) Foreground-subtracted and recalibrated SPT spectra using the
  best-fit parameters from the likelihood analysis of the full
  \planck\ likelihood combined with the SPT S12 likelihood. The magenta
  points show the best-fit \planck\ \LCDM\ spectrum from
  Fig.~\ref{Planckbestfitcl} and the red line shows the best-fit
  \planck+\WP+\highL\ base \LCDM\ model from the full
  \planck\ likelihood. The residuals with respect to this model are
  plotted in the lower panel.}}
\label{PlanckvSPT}
\end{figure*}

\referee{The differences between the S12 and \planck\ base \LCDM\ cosmologies
lead to different types of tension with non-CMB data. Whereas
\planck\ is consistent to high precision with the BAO data (see Fig.
\ref{BAO}) and shows some tension with the \citet{Riess:2011yx}
measurement of $H_0$, the \textit{WMAP}-7+S12 best-fit cosmology is
consistent with the $H_0$ measurement but in tension with the BAO
measurements. The latter point is illustrated by Fig.~\ref{BAOspt},
which is equivalent to Fig.~\ref{BAO} but uses the \textit{WMAP}-7+SPT
cosmology as a reference. All of the BAO measurements lie
systematically low compared to the best-fit \textit{WMAP}-7+S12
$\Lambda$CDM cosmology.\footnote{H12 quote a 2.3\% probability of
  compatibility between the BOSS measurement and the
  \textit{WMAP}-7+S12 $\Lambda$CDM cosmology.} This discrepancy was
further motivation for the study in H12 of extensions to the standard
cosmological model.}

\referee{Appendix~\ref{app:wmap} shows that the \planck\ and \textit{WMAP}-9 power
spectra are in good agreement with each other after correction for a
multiplicative calibration factor, and lead to closely similar
cosmological parameters when the \planck\ likelihood is restricted
to multipoles less than $1000$. A systematic difference
between \planck\ and \textit{WMAP}-7 band-powers is therefore unlikely to be the
primary  reason for
the discrepancy between the \planck\ and \textit{WMAP}-7+S12 cosmologies.
Alternative explanations might involve a  systematic difference between the
\planck\ and S12 band-powers at high multipoles, or a systematic
problem related to the {\it matching} of the SPT and \WMAP\ spectra, 
i.e., with their relative calibration.}

\begin{table*}[tmb]                 
\begingroup
\newdimen\tblskip \tblskip=5pt
\caption{\referee{Parameter constraints in \lcdm\ models for various likelihood
 combinations as described in the text. The \textit{WMAP} nine-year
 polarization likelihood is used in all of these fits.  For \Planck\ and SPT
 we use the standard foreground model, as described in Sect.~\ref{sec:highell}.
 For \WMAP, we follow the foreground treatment in Appendix~\ref{app:wmap},
 removing only a Poisson-like term from the power spectrum.
 The last row of the table lists the SPT
 $\chi^2$ value for the best-fit parameters ($47$ data points).} }                         
\label{SPTparams2}         
\nointerlineskip
\vskip -3mm
\footnotesize
\setbox\tablebox=\vbox{
   \newdimen\digitwidth
   \setbox0=\hbox{\rm 0}
   \digitwidth=\wd0
   \catcode`*=\active
   \def*{\kern\digitwidth}
   \newdimen\signwidth
   \setbox0=\hbox{+}
   \signwidth=\wd0
   \catcode`!=\active
   \def!{\kern\signwidth}
\halign{\hbox to 1.0in{#\leaderfil}\tabskip 2.2em&
        \hfil#\hfil&
        \hfil#\hfil&
        \hfil#\hfil\tabskip=0pt\cr                
\noalign{\doubleline}
\omit&\textit{WMAP}-9+S12&\textit{Planck} 143+S12&\planck+S12\cr
\omit\hfil Parameter\hfil&68\%\ limit&68\%\ limit&68\%\ limit\cr
\noalign{\vskip 3pt\hrule\vskip 5pt}
$100\Omega_{\rm b}h^2$&$2.239 \pm 0.035$&$2.232\pm0.031$&$2.203 \pm 0.026$\cr
$\Omega_{\rm c}h^2$&$0.1126 \pm 0.0037$&$0.1170\pm0.0027$&  $0.1192 \pm 0.0024$\cr
$10^9A_{\rm s}$&$2.167\pm 0.056$&$2.167\pm 0.054$&$2.177 \pm 0.053$\cr
$n_{\rm s}$&$0.968\pm 0.009$&$0.971 \pm 0.008$&$0.961\pm0.007$\cr
$\tau$&$0.083 \pm 0.013$&$0.085 \pm 0.013$&$0.085\pm0.013$\cr
$100\theta_{\ast}$&$1.0426 \pm 0.0010$&$1.0422 \pm 0.0006$&$1.0417\pm0.0006$\cr
$\Omega_\Lambda$&$0.727 \pm 0.020$&$0.704 \pm 0.016$&$0.689\pm0.015$\cr
$H_0$&$70.7 \pm 1.7*$&$68.8 \pm 1.2*$&$67.6\pm1.1*$\cr
$y_{150}^{\rm SPT}$ &\dots&$0.995\pm 0.004$&$0.994 \pm 0.003$\cr
$y_{150}^{\rm SPT/WMAP}$&$0.999 \pm 0.006$&\dots&\dots\cr
$(\chi^2_{\rm SPT})_{\rm min}$ &$53.0$&$55.7$&$56.3$\cr 
\noalign{\vskip 3pt\hrule\vskip 4pt}}}
\endPlancktablewide                 
\endgroup
\end{table*}                        

\begin{figure}
\centering
\includegraphics[width=88mm]{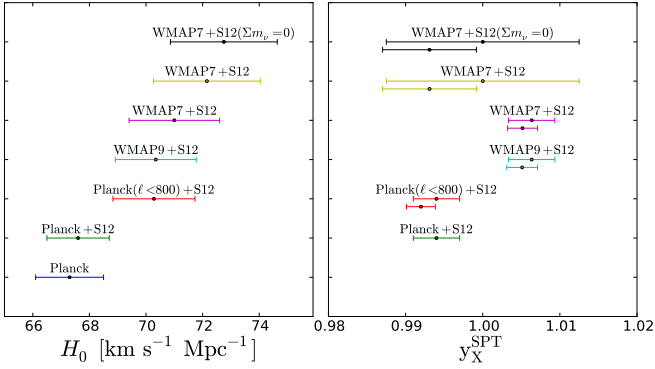}
\caption{\referee{A number of separate effects contribute to the difference in
  $H_0$ inferred from \textit{WMAP}-7+S12 (top of left panel) and $H_0$ inferred
  from \planck+\WP\ (bottom of left panel), all going in the same
  direction. These include assumptions about neutrino masses, calibration
  procedures, differences between \textit{WMAP}-7 and \textit{WMAP}-9, and
  differences in the relative calibrations between SPT and \textit{WMAP}
  (as explained in the text).  The right panel shows
  calibration parameter priors (top lines of each pair) and posteriors
  (bottom lines of each pair). The tighter of the priors shown for \WMAP-7+S12, and that shown for \WMAP-9+S12,  come from using
  \planck\ to provide the relative calibration between \textit{WMAP} and S12.
 We plot only the posterior for the Planck+S12 relative calibration. Note that
 the relative-calibration parameter $y^{\rm SPT}_{\rm X}$ is between S12 and
 the other indicated data set (i.e., \WMAP\ or \Planck).}}
\label{fig:H0shift}
\end{figure}

\referee{We consider first a comparison of the \planck\ and S12 spectra. 
Since these spectra  have a large overlap range at
high multipoles, where both experiments have high signal-to-noise,
there is no need to use \textit{WMAP} as an intermediary to establish a
relative calibration. We can compare the spectra directly via a joint
likelihood analysis using the same foreground model that is used in
the main body of this paper. Since the S12 spectrum is measured at a frequency
of 150\,GHz, we first present results using only the
\planck\ $143\times143$\,GHz spectrum in the \planck\ likelihood. This
reduces sensitivity to the details of the foreground modelling.  Apart
from small colour corrections, the foregrounds are identical, except
for differences in the Poisson point source amplitudes.}

\referee{Absolute calibration of the SPT spectra is determined by comparing
with the \textit{WMAP}-7 spectrum in the multipole range $600 \le \ell
\le 1000$.  Since the spectra from both experiments are noisy in this
multipole range, there is a large (roughly 3\% in power) uncertainty in the absolute
calibration of the S12 data. Here we use a version of the SPT S12
likelihood {\it that does not include} marginalization over
calibration uncertainties. Instead, we self-consistently solve for a map
calibration factor $y^{\rm SPT}_{150}$ between SPT and \planck.  (This
differs from the analyses of S12, H12 and \citealt{Calabrese:13}, which use an SPT
covariance matrix that includes marginalization over calibration
errors,  and combine with other experiments without solving for a
relative calibration factor.)}

\referee{The results are shown in Fig.~\ref{PlanckvSPT}
(a).\footnote{In Fig.~\ref{PlanckvSPT} we use the window functions provided
by S12 to band-average the \planck\ and theory data points at high
  multipoles.} The agreement between the two sets of band-powers is
most easily seen in the lower panel in which the best-fit model has
been subtracted.  The best-fit calibration factor is $y^{\rm
  SPT}_{150} = 0.995$, well within the prior 1.3\% calibration
uncertainty.  The model with minimum $\chi^2$ in this joint analysis
has $\chi^2_{\rm SPT}=55.7$.  To quantify the probability to exceed
(PTE) this value of $\chi^2$ we need to determine the effective number
of degrees of freedom.  The SPT data have 47 band-powers and only two
parameters that were heavily influenced by them: the Poisson point source
amplitude and $y^{\rm SPT}_{150} $.  Taking 45 as the number of
degrees of freedom, we find a PTE of 13\%.}

\referee{We find similar results when we combine the S12 likelihood with the
full \planck+\WP+\highL\ likelihood.  This is illustrated in
Fig.~\ref{PlanckvSPT} (b).  Note that the \planck\ spectrum sits high compared
to the best-fit spectrum at $\ell \ga 2300$, but in this region of
the spectrum foreground and beam errors 
become significant and introduce large correlations between
the data points.  We find a minimum $\chi^2$ value of
$\chi^2_{\rm SPT}=56.3$ for the best-fit cosmological model. Again
assuming 45 degrees of freedom we find a PTE of 12\%.  Based on these
$\chi^2$ values, we see no evidence of any inconsistency between the
S12 band-powers and the best-fit
\planck\ cosmological model. 
The parameter values for the \planck+S12 fits are listed in Table
\ref{SPTparams2}. We also include the parameter values from our own \WMAP-9+S12 analysis. In this latter case, we do not include \Planck-based (re)calibrations of \WMAP\ or SPT, but allow the relative calibration between SPT and \WMAP\ ($y_{150}^{\rm SPT/WMAP}$) to vary.}

\referee{If the \planck\ and SPT power spectra are broadly consistent with each other,
then why do the \textit{WMAP}-7+S12 and \Planck\ \lcdm\ parameter estimates
differ by so much?  The bulk of the difference can
be captured by just one parameter, which we choose here as $H_0$.  The
shifts in other parameters are highly correlated with the shift in
$H_0$.}

\referee{Some factors  contributing to the difference in $H_0$ are
summarized in Fig.~\ref{fig:H0shift}.  We start at the top
with the \textit{WMAP}-7+S12 result, which assumed zero neutrino mass.
Progressing downwards in the plot, we have repeated the \textit{WMAP}-7+S12
analysis assuming a neutrino mass of  $0.06\,{\rm eV}$ as  in the 
\planck\ analysis described here. This lowers $H_0$ slightly.
A further
reduction in $H_0$ comes from using the \planck\ data to reduce the
uncertainty in the \textit{WMAP}-SPT relative calibration. By combining the
\planck-\textit{WMAP} 1.3\% rescaling (see Appendix~\ref{app:wmap}) and the \planck-S12 calibration, we can
place a tight prior on the \textit{WMAP}-7-S12 relative calibration.
Fig.~\ref{fig:H0shift} shows that this prior is
roughly $1.5\,\sigma$ higher than the posterior from the \textit{WMAP}-7+S12 chain
that uses the nominal S12 calibration.
Switching from \textit{WMAP}-7+S12 to \textit{WMAP}-9+S12, in 
the next step in our
progression, we  again see a small shift to lower $H_0$, with $H_0 = (70.4 \pm 1.6)
\,{\rm km}\,{\rm s}^{-1}\,{\rm Mpc}^{-1}$.  This latter value
is very similar to that obtained if we replace the \textit{WMAP}-9
data with the \planck+WP likelihood limited to $\ell_{\rm max}=800$ (as shown in  Fig.~\ref{fig:H0shift}). The Planck+S12 results plotted in Fig.~\ref{fig:H0shift}
are from  the last column of Table~\ref{SPTparams2}.}

\referee{Each of the changes described above brings the base \LCDM\
cosmological parameter values from SPT closer to those derived
from \planck. Our results suggest that part of the discrepancy between
the \textit{WMAP}-7+S12 and \planck\ parameters arises from difficulties in
self-consistently matching the SPT to the \textit{WMAP} power
spectra  over a limited range
of multipoles.  This illustrates the advantages of having a
single experiment, such as \planck, covering both low and high
multipoles.}

\section{Dependence of cosmological parameters in extended models on foreground modelling and likelihood choices}
\label{app:test}

A large number of likelihood comparison tests on parameters in the
base \lcdm\ cosmology are discussed in~\citet{planck2013-p08}.  In the
main body of this paper, we report constraints on a wide variety of
extended models. In many of these models the cosmological
parameters are strongly degenerate with each other and are therefore more
sensitive to the detailed modelling of foregrounds, frequency choices, 
and likelihood methodology. In this Appendix we discuss briefly how one-parameter
extensions of the \lcdm\ model are affected by various choices.

\subsection{Impact of foreground priors}

Throughout this paper we have used a particular parameterization of the
foreground model, and marginalized over the free parameters using
relatively wide priors. As discussed in Sect.~\ref{sec:highell}, the choice
of these priors is subjective and was guided by theoretical expectations
and by other data, particularly results from high-resolution CMB experiments
and the early \planck\ analysis of the CIB power spectrum 
\citep{planck2011-6.6}. As discussed in Sect.~\ref{sec:highell}, for \planck\
the dominant foregrounds are the Poisson contributions from unresolved
point sources and the clustered CIB component at $217\,$GHz.
The other components are of much lower amplitude and poorly constrained
by \planck\ data alone. 

For the thermal and kinetic SZ amplitudes we have
imposed uniform priors of $0\le A^{\mathrm{tSZ}}\le 10$ and
$0\le A^{\mathrm{kSZ}}\le 10$. These priors have little impact
on the parameters derived for the base \lcdm\ model, or
on the parameters of  extended cosmologies if \planck\ is
combined with  ACT and SPT data at high multipoles.
However, for extended cosmologies the priors on these `minor'
components do have a small impact on the cosmological parameters.
Table~\ref{tab:grid_1paramext_tests} gives results obtained from doubling
the width of the  SZ priors.  The constraints on the extended parameters
change by small amounts compared to the \planck+\WP\ entries in Table
~\ref{tab:grid_1paramext}, giving an impression of 
the sensitivity of \planck+\WP\ numbers to minor foregrounds\footnote{%
The constraint on $\Alens$ for \planck+\WP\ is not given in
Table~\ref{tab:grid_1paramext_tests}; the result is $\Alens = 1.22^{+0.25}_{-0.22}$ (95\%\ CL).}.

The use of additional high-$\ell$ CMB data to constrain the
foreground parameters depends on having a foreground model that can
reliably extrapolate between the scales relevant for \planck\ and the
smaller scales where the high-resolution experiments have the tightest
constraints. As a simple test of the model used in the main body of
the paper, we relax here our assumption that the CIB spectral index is
constant with a Gaussian prior $\ncib = 0.7\pm 0.2$. Any change in CIB
index between small and larger scales could lead to a bias in the
foreground model subtracted from the \planck\ spectra, particularly in
the $217\times217\,\mathrm{GHz}$ spectrum where the CIB is the dominant
foreground component.  However, as shown in Fig.~\ref{cibCompare}, and Table~\ref{tab:grid_1paramext_tests}, the
inferred cosmological parameters are actually extremely insensitive
to the details of the model, with very similar results obtained with
no $\ncib$ prior and allowing a free running of the spectral index
through the parameter $d \gamma^{\mathrm{CIB}} / d \ln \ell$.
(Note that we have assumed here that the $\ell$-dependence of the CIB
is the same,  up to an amplitude,  at the different frequencies.) The
interpretation of the foregrounds does change significantly,
and indeed there is mild evidence for running of the CIB spectral
index, but it has almost no impact on the cosmology. This should not
be too surprising, since the CIB signal is frequency-dependent unlike
the cosmological signal, but nonetheless it is reassuring that
degeneracies with, for example, the SZ amplitudes do not indirectly
cause biases in cosmological parameters.

\begin{figure}
\centering
\includegraphics[width=88mm,angle=0]{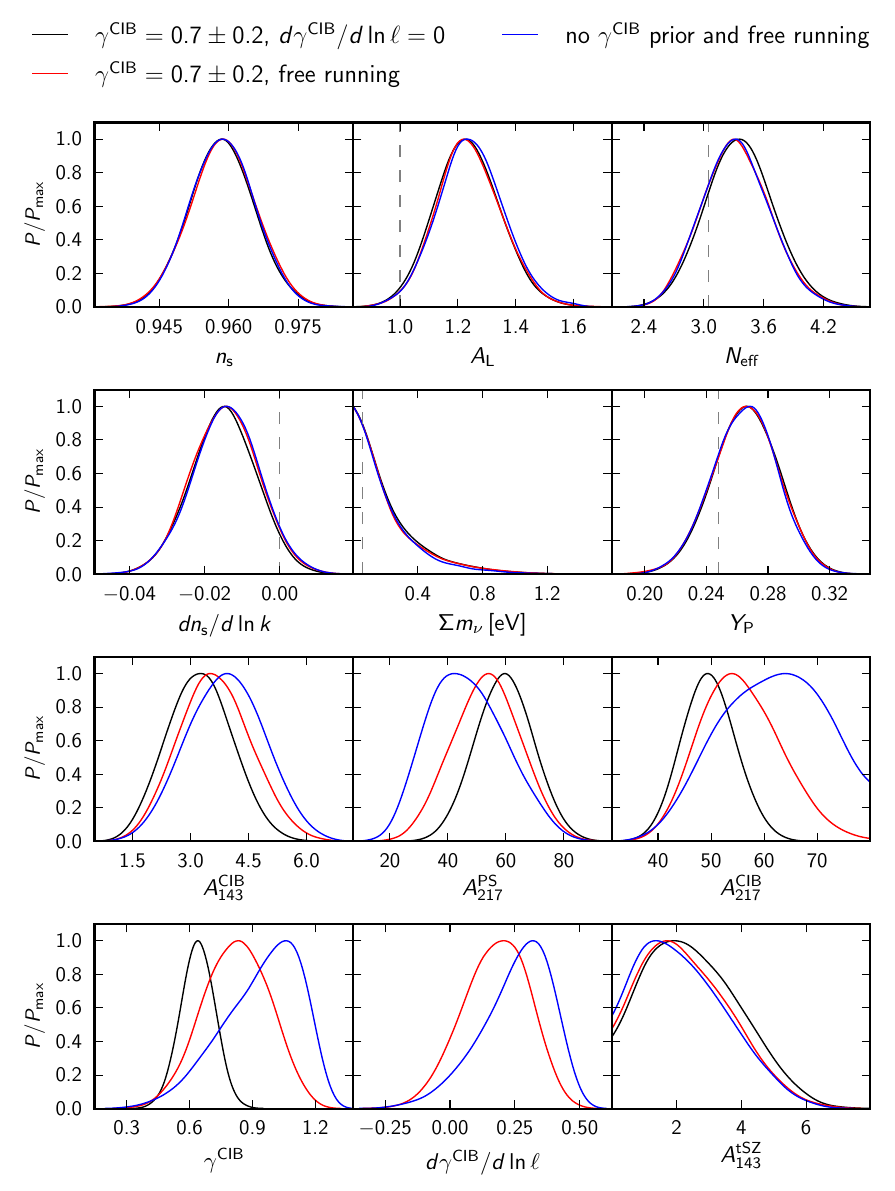}
\caption{Comparison of parameter constraints from
  \planck+\WP+\highL\ for three CIB foreground models with different
  restrictions on the CIB spectral index $\ncib$ (assumed to be the
  same in the $143$ and $217\,\mathrm{GHz}$ channels). The top six panels show
cosmological parameter constraints on $\ns$ (top left) in the base \lcdm\ model and on single-parameter extensions of the \lcdm\ model. These are very stable to the modelling of the CIB. Each sub-plot is obtained from an independent analysis of that model with \COSMOMC. The lower six panels show the  constraints on a subset of the foreground parameters in the base \lcdm\ model, some of which change significantly.}
\label{cibCompare}
\end{figure}

We have not investigated extensively the impact of varying the tSZ and
kSZ templates. A variety of different approaches (analytic,
semi-analytic and numerical) have been used to estimate tSZ templates
\citep[e.g.,][]{Komatsu:02, Shaw:10, Sehgal:10, Trac:11, Battaglia:10,
Battaglia:12}. These have similar shapes at multipoles $\ell
\la 3000$, relevant to \planck\ and to the tSZ template used
here. The shape of our template is also a good match to the power
spectrum of the \planck\ Compton-$y$ map over the multipole range $100
\la \ell \la 1000$ \citep{planck2013-p05b}. The normalization of
the tSZ templates (i.e., their dependence on $\sigma_8$) and their
shapes at multipoles $\ell \ga 3000$, depend on uncertain gas
physics (including energy injection from AGN). For this reason, we
have not attempted to link the amplitude of the tSZ template to the
amplitude of the matter power spectrum in the parameter analyses.  The
tSZ template used here is similar in shape to the~\citet{Battaglia:10}
template that has been used extensively in the analysis of ACT and SPT
data.  The effects of varying tSZ and kSZ templates on high-resolution
CMB experiments have been investigated by
\citet{dunkley11}, \citet{Reichardt:12}, and \citet{Dunkley:13} who find
very little effect on cosmological parameters.

\begin{table*}
\begin{center}
\caption{
Constraints on one-parameter extensions of the \lcdm\ model from 
\planck\ with various likelihood variations.  \Planck+\WP+\highL\ is used in
all cases except for the column listing results for  $\tau=0.07\pm0.013$,
where the  \WMAP\ polarization likelihood is replaced by this $\tau$ prior, and the ninth column, which does not include the high-$\ell$ experiments and
doubles the default width of the 
flat priors on the two SZ amplitudes $A^{\mathrm{tSZ}}$ and $A^{\mathrm{kSZ}}$. 
The running CIB model has no prior on $\gamma^{\mathrm{CIB}}$ and allows for spectral curvature through the parameter $d \gamma^{\mathrm{CIB}}/d \ln \ell$. \referee{The final column in the table shows the  results of modelling a small systematic feature
in the $217\times217\,{\rm GHz}$ spectrum, as described in Sect.~\ref{subsec:systematic}.}
\label{tab:grid_1paramext_tests}
}

\begingroup
\newdimen\tblskip \tblskip=5pt
\nointerlineskip
\vskip -3mm
\scriptsize
\setbox\tablebox=\vbox{
    \newdimen\digitwidth
    \setbox0=\hbox{\rm 0}
    \digitwidth=\wd0
    \catcode`"=\active
    \def"{\kern\digitwidth}
    \newdimen\signwidth
    \setbox0=\hbox{+}
    \signwidth=\wd0
    \catcode`!=\active
    \def!{\kern\signwidth}
\halign{
\hbox to 0.9in{#\leaderfil}\tabskip=0.9em&
\hfil#\hfil&
\hfil#\hfil&
\hfil#\hfil&
\hfil#\hfil&
\hfil#\hfil&
\hfil#\hfil&
\hfil#\hfil&
\hfil#\hfil&
\hfil#\hfil\tabskip=0pt\cr
\noalign{\doubleline}
\omit&\hfil \camspec\hfil&\hfil \plik\hfil&\hfil $\lmax=2000$ \hfil&\hfil  $\lmin=1200$\hfil&\hfil no $217 \times\ 217$ \hfil&\hfil  $\tau=0.07\pm0.013$ \hfil&\hfil  Running CIB\hfil&\hfil $0\leq A^{\mathrm{tSZ}} \leq 20$\hfil& \hfil 217 systematic\hfil\cr
\noalign{\vskip 2pt}
 \omit\hfil Parameter\hfil & \omit\hfil  95\%\ limits\hfil & \omit\hfil 95\%\ limits\hfil
 & \omit\hfil 95\%\ limits\hfil
 & \omit\hfil 95\%\ limits\hfil
 & \omit\hfil 95\%\ limits\hfil
 & \omit\hfil 95\%\ limits\hfil
& \omit\hfil 95\%\ limits\hfil
& \omit\hfil 95\%\ limits\hfil
  & \omit\hfil 95\%\ limits\hfil\cr
\noalign{\vskip 3pt\hrule\vskip 5pt}
$\Sigma m_\nu\,[\mathrm{eV}]$&$< 0.663$&$< 0.691$&$< 0.398$&$< 0.600$&$<0.485$&$< 0.768$&$<0.581$&$<0.999$& $< 0.663$\cr
\noalign{\vskip 3pt}
$N_{\mathrm{eff}}$&$3.36^{+0.68}_{-0.64}$&$3.36^{+0.78}_{-0.68}$&$2.89^{+0.67}_{-0.63}$&$3.30^{+0.72}_{-0.70}$&$2.99^{+0.69}_{-0.64}$&$3.23^{+0.63}_{-0.61}$&$3.32^{+0.66}_{-0.63}$&$3.67^{+0.86}_{-0.83}$ & $3.43^{+0.74}_{-0.71}$\cr
\noalign{\vskip 3pt}
$Y_{\mathrm{P}}$&$0.266^{+0.040}_{-0.042}$&$0.254^{+0.046}_{-0.048}$&$0.233^{+0.047}_{-0.050}$&$0.262^{+0.045}_{-0.047}$&$0.232^{+0.044}_{-0.047}$&$0.259^{+0.040}_{-0.039}$&$0.264^{+0.041}_{-0.043}$&$0.293^{+0.046}_{-0.048}$ & $0.272^{+0.047}_{-0.050}$\cr
\noalign{\vskip 3pt}
$dn_{\mathrm{s}}/d\ln k$&$-0.015^{+0.017}_{-0.017}$&$-0.013^{+0.019}_{-0.019}$&$-0.007^{+0.018}_{-0.018}$&$-0.017^{+0.020}_{-0.020}$&$-0.005^{+0.017}_{-0.017}$&$-0.012^{+0.016}_{-0.016}$&$-0.014^{+0.017}_{-0.017}$&$-0.016^{+0.017}_{-0.017}$& $-0.011^{+0.017}_{-0.018}$\cr
\noalign{\vskip 3pt}
$A_{\mathrm{L}}$&$1.23^{+0.22}_{-0.21}$&$1.26^{+0.26}_{-0.25}$&$1.38^{+0.26}_{-0.25}$&$1.31^{+0.24}_{-0.23}$&$1.30^{+0.24}_{-0.22}$&$1.26^{+0.24}_{-0.24}$&$1.24^{+0.23}_{-0.21}$&$1.20^{+0.25}_{-0.24}$& $1.21^{+0.24}_{-0.24}$\cr
\noalign{\vskip 5pt\hrule\vskip 3pt}
} 
} 
\endPlancktable
\endgroup
\end{center}
\end{table*}

\subsection{WMAP low-$\ell$ polarization likelihood}

The large-scale polarization from nine years of
\WMAP\ observations~\citep{bennett2012} provides our most powerful
constraint on the reionization optical depth $\tau$. As shown in
Sect.~\ref{sec:planck_lcdm} it is not essential to use
\WMAP\ polarization information to obtain tight constraints on
cosmological parameters from \planck.  Using \WMAP\ does, however,
improve the constraints on the amplitude of the power spectrum and, via
the partial parameter degeneracies sensitive to the relative amplitude
of large and small-scale power and the amount of lensing, \WMAP\ polarization
also has an impact on other cosmological parameters. Most directly, since the
small-scale CMB power scales roughly with $e^{-2\tau}\sigma_8^2$, the
inferred value of $\sigma_8$ is approximately proportional to $e^{\tau}$ as
discussed in Sect.~\ref{subsec:tau}.

The polarization measurement at low multipoles is challenging because
of the high level of polarized Galactic foregrounds, so it is
important to assess the impact if the assumed constraint were slightly
wrong. Figure~\ref{tauCompare} shows how cosmological parameters shift
if instead of using \WMAP\ polarization we impose a prior
$\tau=0.07\pm 0.013$ ($1\sigma$), which is about the same width but
about $0.02$ lower than the posterior obtained from \WMAP\ (as
for example might be obtained if there were some residual foreground
contamination or instrument systematic). Table.~\ref{tab:grid_1paramext_tests} shows the
corresponding impact on parameter constraints with high-$\ell$ CMB
data added to \planck.  The shifts are consistent with the known parameter
degeneracies, and the \lcdm\ constraints on $\sigma_8$ would shift
downwards by a factor of approximately $e^{0.02}$, or about $2\%$. For this reason, in
Sect.~\ref{subsec:tau} we have quoted a constraint on
$e^{-\tau}\sigma_8$, which is insensitive to possible errors in
the large-scale polarization likelihood. A change in the
\planck\ calibration would also have a similar direct effect on the
inferred physical amplitudes.

\begin{figure}
\centering
\includegraphics[width=88mm,angle=0]{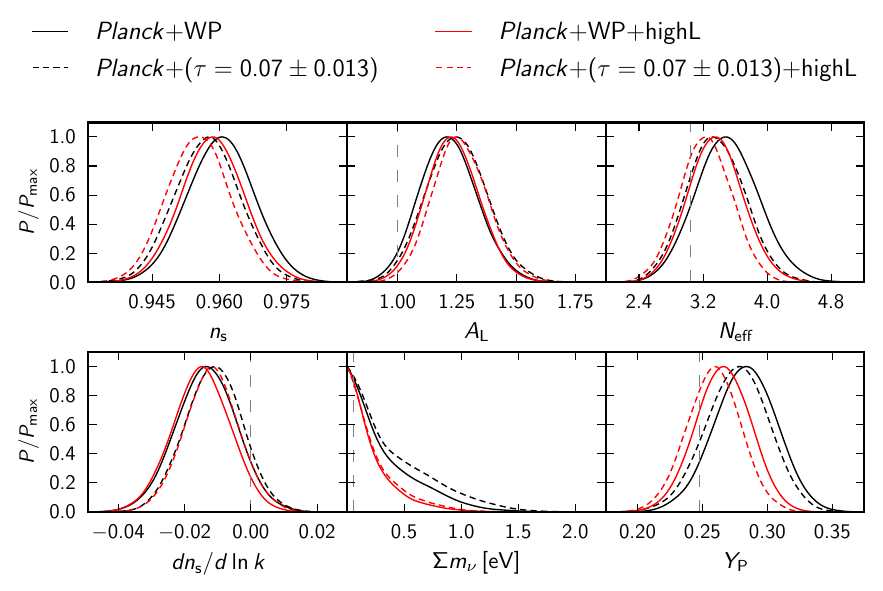}
\caption{Effect on cosmological parameter constraints of replacing the 
\WMAP\ low-$\ell$ polarization likelihood with a prior of
$\tau=0.07\pm 0.013$, which prefers lower values of the optical depth.
The top-left sub-plot is $\ns$ in the base \lcdm\ model, while the others are for one-parameter extensions. Each sub-plot shows results from independent \COSMOMC\ analyses of the corresponding model.}
\label{tauCompare}
\end{figure}

\subsection{\planck\ likelihood}

The results of this paper are based on the \camspec\ likelihood, which
includes information from 100, 143 and $217\,\mathrm{GHz}$ channels, with a range
of multipole cuts as reviewed briefly in Sect.~\ref{subsec:likelihood}
and summarized in Table~\ref{planckchi2}. The combination of channels
used allows a number of foreground parameters to be partially
determined from \planck\ data alone, and as discussed in
Sect.~\ref{sec:highell} the foreground parameters can be determined more precisely
by including additional information from other high-$\ell$ data sets.
\citet{planck2013-p08} discusses and compares the \camspec\ likelihood
with an alternative cross- and auto-spectrum likelihood, \plik. The
\plik\ likelihood uses identical masks over the frequency range
$100$--$217\,\mathrm{GHz}$ (retaining $48\%$ of the sky), ignores correlations between
multipoles and uses different assumptions to estimate instrument noise
and to correct for Galactic dust, as described in
\citet{planck2013-p08}. For the base \lcdm\ model, these two
likelihood codes give almost identical results. In this section, we
investigate briefly how cosmological parameters in extended
\lcdm\ models vary between the two likelihood codes and for some data
cuts in the \camspec\ likelihood.

\begin{figure}
\centering
\includegraphics[width=8.8cm,angle=0]{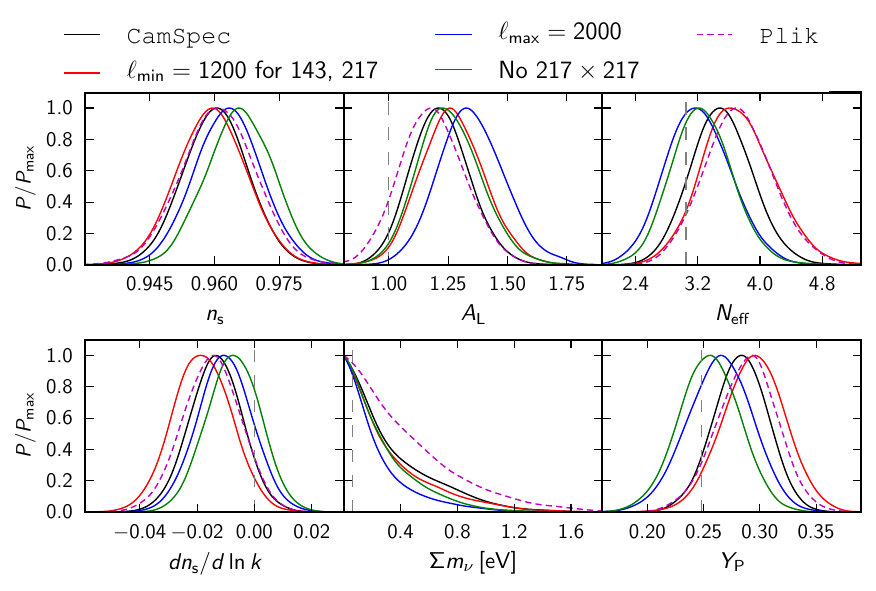}
\\
\includegraphics[width=8.8cm,angle=0]{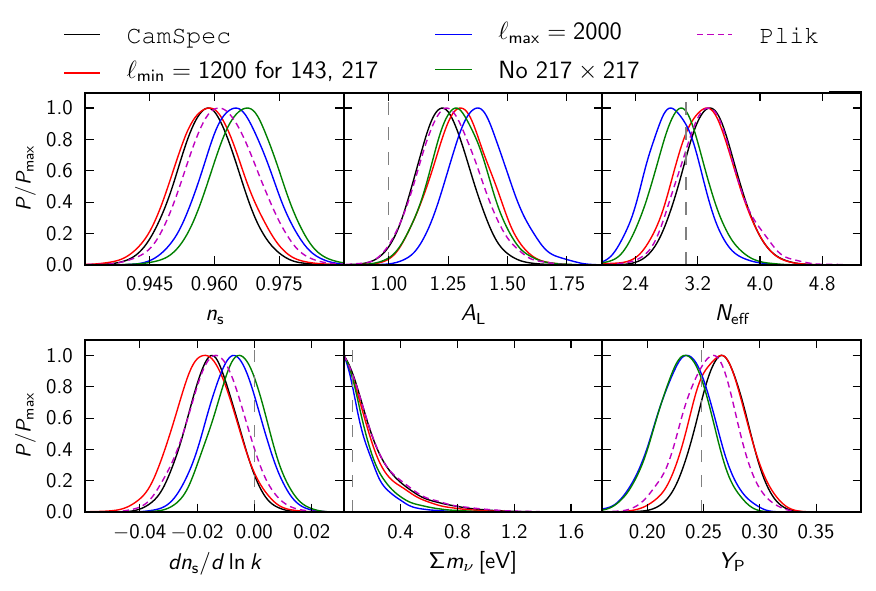}
\caption {\emph{Upper}: comparison of the \lcdm\ constraints on $\ns$ (top-left) and single-parameter extensions of the \lcdm\ model for a
  variety of data cuts for \planck+\WP. Each sub-plot is obtained from a separate
  \COSMOMC\ analysis of the corresponding model.
  The dashed lines show the results from \plik, an
  alternative likelihood discussed in ~\citet{planck2013-p08}, run
  here with the same SZ and CIB foreground priors as for the
  \camspec\ results. For the extended models, the value of the additional parameter in the base \lcdm\ model is shown with the vertical dashed lines.
\emph{Lower}: same as the upper set of panels, but for \planck+\WP+\highL.
Additional data from the high-$\ell$ CMB
  experiments significantly reduce the foreground degeneracies.}
\label{camspecCuts}
\end{figure}

The results are summarized in Fig.~\ref{camspecCuts} and
Table~\ref{tab:grid_1paramext_tests}.  Shifts in parameter values are
expected, since different combinations of data are being used and hence
have differing amounts of noise and cosmic variance. With fewer frequencies 
overlapping at any angular scale, the foreground parameters are less well determined, and
any degeneracies with foreground parameters are expected to open
up. Fig.~\ref{camspecCuts} shows that there are noticeable shifts in
parameter values, but when also including additional high-$\ell$ CMB
data, which constrain the foreground parameters to higher precision,
the relatively small differences between \camspec\ and \plik\
are significantly
reduced.  The two likelihood codes agree well,
even for extended models when the high-$\ell$ CMB data are added to
the likelihoods.

However, the preference from the temperature power spectrum
for $\Alens>1$ actually becomes stronger on adding high-$\ell$ data.
Reducing  $\ell_{\rm max}$ to $2000$ in the \camspec\ likelihood also shifts $\Alens$
to higher values, particularly with the addition of the high-$\ell$ data.
We do not, at this stage, have a full understanding of these shifts in $\Alens$.
We note that the high-$\ell$ ACT data itself favours high
$\Alens$~\citep{Sievers:13}, and this may be part of the reason behind
the shift to high values when high-$\ell$ data are included. (The truncation of
\planck\ spectra at $\lmax=2000$  also limits the accuracy of matching 
high-$\ell$ data to \planck. There is then no overlapping multipole range with the
SPT data and a significantly narrower overlap range for ACT.)

The analysis of extended \lcdm\ models with strong parameter
degeneracies is complex and sensitive to small systematic errors in
the CMB data and to errors in the foreground model. In the
analysis of extended models we have usually quoted results from the
\planck+\WP+\highL\ data combination, using the full
\camspec\ likelihood. With this combination, we utilize the high
signal-to-noise ratio of the \planck\ spectra at $143$ and $217\,\mathrm{GHz}$,
which have a
wide multipole range with which to match high-$\ell$ experiments to
\planck, and gain better control of foregrounds via the inclusion of
high-$\ell$ data. The general ``rule of thumb'' adopted in this paper
has been to use the differences between parameter constraints from
\planck+\WP\ and \planck+\WP+\highL\ as a guide to whether parameters
are sensitive to errors in the foreground model (or other sources of error).  We
do sometimes see shifts of up to around $1\,\sigma$ between these
likelihoods in the parameter values for extended models and this needs
to be borne in mind when interpreting our results. In the
absence of any additional information, we take the cosmological parameters from \planck+\WP+\highL\ as our best estimates for extended models. 

\subsection{$217\, {\rm GHz}$ systematic feature}
\label{subsec:systematic}

\referee{As discussed in Sect.~\ref{sec:introduction}, following the submission of the \planck\ 2013 papers,
we discovered strong evidence that a small dip in the
$217\times 217\,{\rm GHz}$ spectrum at $\ell\approx 1800$ varies between surveys and is
a systematic feature caused by incomplete subtraction of $4\,{\rm K}$ cooler lines
from the time-ordered data. To estimate the impact of such a systematic on cosmology, we test
the sensitivity of our results to adding a dip in the $217\times 217\,{\rm GHz}$
spectrum in the range $1700 \le \ell \le 1860$, which we model as
\begin{equation}
\Delta \mathcal{D}_\ell^{217\times 217} =  -W\sin\left(\frac{(\ell-1700)
\pi}{160}\right) .
\end{equation}
(Note that the tests described in this section were done before the submission of the 2013
papers.})
Here, $W$ is a free amplitude parameter that we marginalize over
using a flat prior. For the base \LCDM\ model we find $W = (26\pm 5)\, \mu {\rm K}^2$
(\planck+\WP), and hence a significant (but highly a posteriori) dip amplitude
is strongly preferred by the data, consistent with a systematic effect. The
impact on the cosmological parameters is small, but not negligible,
typically causing shifts of below $0.5\,\sigma$.
Marginalizing over the dip amplitude $W$ raises
the mean $H_0$ in the base model by approximately $0.3\,\sigma$, and gives
comparable small shifts to one-parameter extensions, as summarized in the final
column (labelled ``217 systematic'') of
Table~\ref{tab:grid_1paramext_tests}. Marginalizing over $W$ does not
significantly change the marginalized value of $\Alens$.}

\referee{In summary, the $\ell=1800$ dip in the $217\times217\,{\rm GHz}$ spectrum
has a non-negligible, but small, impact on  cosmological
parameters, even for extensions to the base \LCDM\ model.\footnote{It is
worth noting that the results presented in this section are consistent with
those derived from a Fisher matrix analysis as described in
Appendix~\ref{app:wmap}, which includes a model for the
$217\times217\,{\rm GHz}$ systematic effect.} The impact on
cosmological parameters is typically less than $0.5\,\sigma$, comparable
to the shifts caused by uncertainties in the modelling of unresolved
foregrounds. However, tests designed to search for localized features in the
\planck\ power spectrum can respond strongly to the systematic effect, as
reported in \citet{planck2013-p17}. Users of the \planck\ likelihood
should bear this in mind.}

\vfill\eject

\raggedright

\end{document}